\documentclass[10pt,a4paper,twoside]{report}

\usepackage[utf8]{inputenc}   

\usepackage[english]{babel} 

\newcommand{\acknowledgments}{@undefined} 
\addto\captionsenglish{\renewcommand{\acknowledgments}{Acknowledgments}}

\addto\captionsportuguese{\renewcommand{\acknowledgments}{Agradecimentos}}
\addto\captionsportuguese{} 

\addto\captionsenglish{}
\addto\captionsenglish{}
\addto\captionsenglish{}
\addto\captionsenglish{}
\addto\captionsenglish{}
\addto\captionsenglish{}
\addto\captionsportuguese{}
\addto\captionsportuguese{}
\addto\captionsportuguese{}
\addto\captionsportuguese{}
\addto\captionsportuguese{}
\addto\captionsportuguese{}

\usepackage{geometry}	
\usepackage{setspace}

\usepackage{graphicx}

\usepackage{amsmath}  
\usepackage{amsthm}   
\usepackage{amsfonts} %
\usepackage{amssymb}
\usepackage{dsfont} 
\usepackage{makecell}

\usepackage{caption}
\usepackage{subcaption}

\usepackage{dcolumn}
\newcolumntype{d}{D{.}{.}{-1}} 
\newcolumntype{e}{D{E}{E}{-1}} 

\usepackage{nomencl}
\makenomenclature
\RequirePackage{ifthen} 
\ifthenelse{\equal{\languagename}{english}}%
    { 
    \renewcommand{\nomgroup}[1]{%
      \ifthenelse{\equal{#1}{R}}{%
        \item[\textbf{Roman symbols}]}{%
        \ifthenelse{\equal{#1}{G}}{%
          \item[\textbf{Greek symbols}]}{%
          \ifthenelse{\equal{#1}{S}}{%
            \item[\textbf{Subscripts}]}{%
            \ifthenelse{\equal{#1}{T}}{%
              \item[\textbf{Superscripts}]}{}}}}}%
    }{
    \renewcommand{\nomgroup}[1]{%
      \ifthenelse{\equal{#1}{R}}{%
        \item[\textbf{Simbolos romanos}]}{%
        \ifthenelse{\equal{#1}{G}}{%
          \item[\textbf{Simbolos gregos}]}{%
          \ifthenelse{\equal{#1}{S}}{%
            \item[\textbf{Subscritos}]}{%
            \ifthenelse{\equal{#1}{T}}{%
              \item[\textbf{Sobrescritos}]}{}}}}}%
    }%

\usepackage[number=none]{glossary}
\setglossary{gloskip={}}
\makeglossary

\usepackage{rotating}

\usepackage[pdftex]{hyperref} 
\hypersetup{colorlinks,       
            linkcolor=black,  
            anchorcolor=black,
            citecolor=black,  
            filecolor=black,  
            menucolor=black,  
            pagecolor=black,  
            urlcolor=black,   
	          bookmarks=true,         
	          bookmarksopen=false,    
	          bookmarksnumbered=true, 
	          pdftitle={Thesis},
            pdfauthor={Andre C. Marta},
            pdfsubject={Thesis Title},
            pdfkeywords={Thesis Keywords},
            pdfstartview=FitV,
            pdfdisplaydoctitle=true}

\usepackage[figure,table]{hypcap}

\usepackage[numbers,sort&compress]{natbib} 

\usepackage{notoccite}

\usepackage{multirow}

\usepackage{booktabs}

\usepackage{pdfpages}

\begin{document}

\pagestyle{plain}



\pagenumbering{gobble}
\includepdf[pages=1-last]{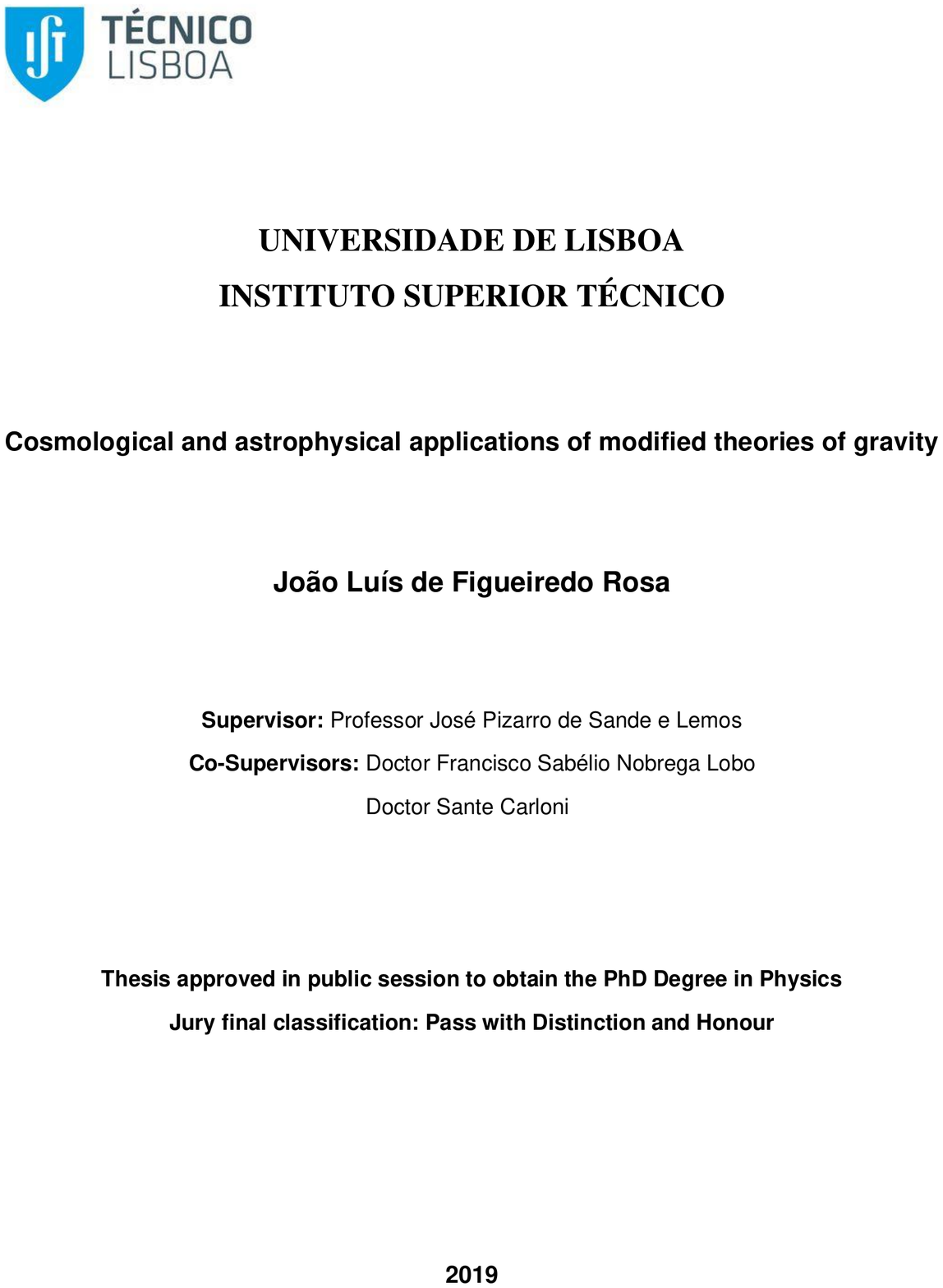} 
\cleardoublepage

\pagenumbering{roman}


\null\vskip5cm%
\begin{flushright}
     To the memory of Líria Guerra
\end{flushright}
\vfill\newpage

\cleardoublepage


\section*{\acknowledgments}

\addcontentsline{toc}{section}{\acknowledgments}

In this section I would like to acknowledge a few people, not only for making this thesis possible but also for standing by my side everytime I needed. I want to start by refering the two men that accepted to supervise me, gave me the opportunity to adventure further into the knowledge of gravitation and astrophysics, and successfully guided me for the past 4 years: Prof. José Sande Lemos and Prof. Francisco Lobo. Also, I want to give a special thanks to Dr. Sante Carloni who was not oficially my supervisor from the beginning but surely was an excellent teacher and coworker with whom I learned a huge variety of topics in physics and mathematics. This thesis would not be possible without him.

The research path of a theoretical physicist is regularly stunned by fundamental mathematical problems and my knowledge of geometry and analysis is often unsufficient to proceed further on my own. Fortunately, the Department of Mathematics of IST has always been kindly available to lend me a helping hand. I thus want to thank Prof. Jorge D. Silva, Prof. José Natário, and Prof. Pedro M. Girão, for helping me to overcome all mathematical barriers I have faced during the development of this work.

The final objective of my academic career is to become a full Professor in a well-known university, and during my research work I was given the opportunity to start teaching classes at IST. For this, I want to thank Prof. João Seixas who was the senior lecturer of the first course I taught and who always motivated me to keep teaching and recommended me to the individuals responsible for selecting lecturers for the different courses. I also want to thank all my students for their extremely important feedback on the quality of my classes and for helping me to break the teacher-student barrier by becoming my friends by the end of the semester.

Surprisingly and unexpectedly, the paramount moment of my adventure came in the final year of my PhD, when I was accepted for a Fulbright Research Scholarship in the Johns Hopkins University in Baltimore. I want to thank not only my Fulbright host Prof. Emanuele Berti for accepting me in his research group and for the interesting discussions we had during these four and a half months, but also all the friends I made in the USA, specially the ones from the Graduate Representative Organization, the Science Fiction and Fantasy Club, the Men's Soccer Club, and of course all the members of the Bloomberg Center for Physics and Astronomy. 

I believe that one can not live a happy life solely based on the success of his career, and thus I want to thank all my friends for every minute they made me happy during the stressful periods of my work. A special thanks to the members of the "Proletariado Internacional" group, my office mates also known as the Gamiades, my besties Miguel Pinto (Sniper) and Mariana Paeta (Mia), my friends from "the other offices", specially Gonçalo Quinta, Rui André, and Miguel Orcinha, my room mate and amazing friend José Cardoso, my friends from the volunteering program "Viver Astronomia" specially João Retrê, and also Beatriz Bento with whom I shared a lot of adventures. Thank you for every day and night out, for our trips to the beach and to the mountains, for our adventures abroad (specially Morocco and Iceland), for the hikes, for the bike rides, for the studio rehearsals with Inconscientia, and for our group dinners.

A huge and special thanks goes to Catarina Bravo, my Taree, for the moments we spent together providing endless fountains of trust and motivation to each other, for always being there to listen everytime I needed to let off some steam from inside, and for always being ready to help me with my problems even when the solutions were out of her knowledge. We understand each other more than anyone else, and that is a huge reason for me to feel happy and confortable. I only know I made it home when I drown in your ghost light.

An overwhelmingly important acknowledge belongs to my family, my parents and my sister, who have always
been supportive and whose sacrifices are the main reason for me to be where I am today. Thank you
for teaching me that with hard work and the pleasure to live for the reason we love most are the main
ingredients for a life of success and happiness. A special thanks to my grandmother who passed away a few months before I started this work and whose dream was to watch me becoming a successful scientist.

To finalize, I would like to acknowledge the financial support of FCT-IDPASC through the grant number PD/BD/114072/2015, to LIP for the finantial support to attend the IDPASC schools in 2016 and 2019 and the IDPASC students workshops from 2016 to 2019, and also Fulbright Comission Portugal for my Fulbright Research Scholarship in the Johns Hopkins University in Baltimore, MD.
\cleardoublepage


\section*{Resumo}

\addcontentsline{toc}{section}{Resumo}

Neste projecto, estudamos aplicações cosmológicas e astrofísicas da teoria de gravidade híbrida métrica-Palatini generalizada recentemente proposta, que combina aspectos de ambos os formalismos métrico e de Palatini do método variacional em gravidade $f\left(R\right)$. Esta teoria nasce como uma generalização natural da teoria de gravidade híbrida métrica-Palatini que provou ser a primeira teoria a unificar a expansão cosmológica acelerada com os constrangimentos do sistema solar, sem recorrer ao efeito camaleão.

Do ponto de vista cosmológico, mostramos usando métodos de reconstrução que os factores de escala com comportamentos da forma de potências e exponenciais em universos FLRW existem para várias distribuições de matéria diferentes, assim como soluções para universos colapsantes. Usando o formalismo de sistemas dinâmicos, mostramos também que não existem atractores globais no espaço de fases cosmológico e que universos estáveis podem ser descritos por factores de escala que divergem em tempo finito ou tendem assimptoticamente para valores constantes. Estudamos ainda o espaço de fases cosmológico de teorias de gravidade com termos de sexta e oitava ordem nas derivadas da métrica e concluimos que os termos de ordem superior não são negligenciáveis.

Na área de astrofísica, mostramos que usando as condições de junção da teoria é possível obter soluções para objectos compactos suportados por cascas finas, tais como cascas auto-gravitantes com e sem fluidos perfeitos no exterior e ainda soluções para wormholes transitáveis que satisfazem a condição de energia nula em todo o espaço-tempo, nao necessitando portanto do suporte de matéria exótica. Mostramos também  que existem formas específicas da acção para as quais o grau de liberdade escalar massivo da teoria é estável no âmbito de buracos negros em rotação descritos pela métrica de Kerr.

\vfill

\textbf{\Large Palavras-chave:} gravidade modificada, cosmologia, sistema dinâmico, cascas finas, wormhole, buraco negro

\cleardoublepage


\section*{Abstract}

\addcontentsline{toc}{section}{Abstract}

In this work, we study cosmological and astrophysical applications of the recently proposed generalized hybrid metric-Palatini gravity theory, which combines features of both the metric and the Palatini approaches to the variational method in $f\left(R\right)$ gravity. This theory arises as a natural generalization of the hybrid metric-Palatini gravity which has been proven to be the first theory to unify the cosmic acceleration with the solar system constraints, without resource to the chameleon mechanism. 

In the cosmological point of view, we show using reconstruction methods that the usual power-law and exponential scale factor behaviors in FLRW universes exist for various different distributions of matter, along with solutions for collapsing universes. Using the dynamical system approach, we also show that no global attractors can exist in the cosmological phase space and that stable universes can either be described by scale factors that diverge in finite time or asymptotically approach constant values. Furthermore, we also study the cosmological phase space of theories of gravity with terms of order six and eight in the derivatives of the metric and we conclude that the higher-order terms are not neglectable.

In the area of astrophysics, we show that using the junction conditions of the theory it is possible to obtain solutions for compact objects supported by thin-shells, such as self-gravitating shells with and without perfect fluids on their exteriors, and also traversable wormhole solutions which satisfy the null energy condition for the whole spacetime, thus not needing the support of exotic matter. Furthermore, we show that there exist specific forms of the action for which the massive scalar degree of freedom of the theory is stable in the scope of rotating black-holes described by the Kerr metric.

\vfill

\textbf{\Large Keywords:} modified gravity, cosmology, dynamical system, thin shell, wormhole, black-hole

\cleardoublepage


%
\tableofcontents
\cleardoublepage 

%
\phantomsection
\addcontentsline{toc}{section}{\listtablename}
\listoftables
\cleardoublepage 

%
\phantomsection
\addcontentsline{toc}{section}{\listfigurename}
\listoffigures
\cleardoublepage 

%
%



%

\chapter*{Preface}

\addcontentsline{toc}{section}{Preface}

Official research presented in this thesis has been carried out at Centro de Astrofísica e Gravitação
(CENTRA) in the Physics Department of Instituto Superior T écnico, and was supported by FCT-IDPASC, through Grant No. PD/BD/114072/2015.

I declare that this thesis is not substantially the same as any that I have submitted for a degree,
diploma or other qualification at any other university and that no part of it has already been or is
concurrently submitted for any such degree, diploma or other qualification.

This work was done in collaboration with my supervisor Professor José Sande Lemos (chapters 3 to 8), my co-supervisor Doctor Francisco Lobo (chapters 3, 6 and 7), my co-supervisor Doctor Sante Carloni (chapters 3, 4, 8 and 9), and also my Fulbright Research Scholarship supervisor Prof. Emanuele Berti (chapter 7). Chapters 3, 6, 8, and 9 have been published, chapter 4 has been submitted and chapters 5 and 7 are being prepared for submission. The works published or to be published presented in this thesis are 7 in total, namely:
\ \\
\ \\
- J. L. Rosa, S. Carloni,  J. P. S. Lemos, F. S. N. Lobo, “Cosmological solutions in generalized hybrid metric-Palatini gravity”, Physical Review D \textbf{95}, 124035 (2017); arxiv:1703.03335 [gr-qc]. (Chapter 3)
\ \\
\ \\
- J. L. Rosa, S. Carloni, J. P. S. Lemos, “The cosmological phase space of generalized hybrid metric-Palatini theories of gravity”, submitted; arxiv:1908.XXXXX [gr-qc]. (Chapter 4)
\ \\
\ \\
- J. L. Rosa, J. P. S. Lemos, “Junction conditions of the generalized hybrid metric-Palatini gravity and their consequences”, to be submitted. (Chapter 5)
\ \\
\ \\
- J. L. Rosa, J. P. S. Lemos, F. S. N. Lobo, “Wormholes in generalized hybrid metric-Palatini gravity obeying the matter null energy condition everywhere”, Physical Review D \textbf{98}, 064054 (2018); arxiv:1808.08975 [gr-qc]. (Chapter 6)
\ \\
\ \\
- J. L. Rosa, J. P. S. Lemos, F. S. N. Lobo, “Stability of Kerr black-holes in generalized hybrid metric-Palatini gravity”, to be submitted. (Chapter 7)
\ \\
\ \\
- S. Carloni, J. L. Rosa, J. P. S. Lemos, "Cosmology of $f\left(R,\Box R\right)$ gravity", Physical Review D \textbf{99}, 104001 (2019); arxiv:1808.07316 [gr-qc]. (Chapter 8)
\ \\
\ \\
- S. Carloni, J. L. Rosa, “Derrick’s theorem in curved spacetime”, Physical Review D \textbf{100}, 025014 (2019); arxiv:1906.00702 [gr-qc]. (Chapter 9)
\ \\
\ \\
\ \\
A parallel work published by the author during the duration of the thesis but not presented here is:
\ \\
\ \\
- E. Berti, R. Brito, C. F. B. Macedo, G. Raposo, J. L. Rosa, “Ultralight boson cloud depletion in binary systems”, Physical Review D \textbf{99}, 104039 (2019); arxiv:1904.03131 [gr-qc].

\cleardoublepage

\setcounter{page}{1}
\pagenumbering{arabic}



\chapter{Introduction}
\label{chapter:introduction}

This chapter is divided into three sections. In the first section we motivate the study of modified theories of gravity as a needed tool to approach the unsolved problems of the theory of General Relativity (GR). We then provide a brief summary of the most important milestones in the subject of successes and failures of GR experimental tests and the analysis of modified theories of gravity that eventually lead to the development of this thesis. Finally, we discuss the structure of the thesis itself and state the contents of each chapter.

\section{Motivation}

Since its publication in 1915, the Theory of GR (Firstly published in \cite{einstein1,einstein2}, later consolidated in \cite{einstein4,einstein5}) has proven to be the most succesful gravitation theory ever developed, being able not only to accurately describe phenomena poorly understood at the time such as the precession of the perihelium of Mercury \cite{einstein3,verrier1}, but also to predict new phenomena related to the effects of gravity on light \cite{einstein3}, e.g., the bending of light in gravitational fields or the gravitational redshift who were tested experimentally afterwards \cite{clemence1,dyson1,pound1,holberg1,popper1}. More recently, the detection of gravitational waves emitted by black-hole mergers provides further evidence of the success of GR \cite{abbott1}. 

However, despite its success, there are still a few flaws and unsolved problems associated with GR. Due to its purely geometrical interpretation of gravity as curvature of space-time, the theory does not assign a definite stress-energy tensor to the gravitational field \cite{carmeli1}. Another common problem is that the theory admits the existence of solutions with mathematical anomalies like singularities and horizons, e.g., the Schwarzschild solution \cite{schwarzschild1,misner1}. Also, GR is not quantizable \cite{zee1,utiyama1}, and thus fails to provide a full quantum field theory description of gravity, which is essential to the unification of gravity with the other fundamental interactions \cite{ross1}.

To solve one or more of these problems, various modified theories of gravity have been proposed. Some of these theories consist of straightforward generalizations of GR, by either a modification of the geometrical sector, like the $f\left(R\right)$ \cite{sotiriou1} and Gauss-Bonnet \cite{lovelock1} theories, or by the addition of extra fields, either scalar, e.g., Brans-Dicke theory \cite{brans1}, vector, e.g., scalar-vector-tensor theories \cite{moffat1}, or tensorial, e.g., bi-metric theories \cite{clifton1}. There are also attempts to construct theories of quantum gravity, e.g., loop quantum gravity \cite{rovelli1} and string theory \cite{mukhi1}, theories that aim to unify gravity with other fundamental forces, e.g., the Kaluza-Klein theory \cite{kaluza1}, and also theories that attempt to solve both of these problems at once such as the M-Theory \cite{berman1}. For reviews on this matter, refer to \cite{capozziello7} and \cite{capozziello8}.

In the recent years, cosmology has entered a "golden age", in which the rapid development of increasingly high precision data has turned it from a speculative to an observationally based science. Recent experiments call upon state of the art technology to provide detailed information about the contents and history of the universe. These experiments include the Hubble Space Telescope \cite{freedman1}, the NASA WMAP probe instrument \cite{spergel1,bennet1} and the Planck satellite \cite{salvatelli1,planck1}, that measure the temperature and polarization of the microwave cosmic background radiation (CMB), and the Sloan Digital Sky Survey (SDSS) \cite{blanton1}, that is automatically mapping the properties and distribution of 1 million galaxies. High precision cosmology has allowed us to tie down the parameters that describe our universe with growing accuracy.

The standard model of cosmology is remarkably successful in accounting for the observed features of the universe. However, there remain a number of fundamental open questions at the foundations of the standard model. In particular, we lack a fundamental understanding of the acceleration of the late universe. Observations of supernovae, together with the WMAP and SDSS data, lead to the remarkable conclusion that our universe is not just expanding, but has begun to accelerate \cite{perlmutter1,riess1}. For GR to accurately predict this behavior, one has to take into account the existence of unknown sources of mass and energy, commonly named dark matter and dark energy, respectively. These unknown sources contribute with aproximately $\sim 95\%$ of the universe's energy density ($\sim 25\%$ dark matter, $\sim 70\%$ dark energy, and $\sim 5\%$ baryonic matter) \cite{spergel1,bennet1,salvatelli1,planck1}. However, one can also argue that this behavior of the scale factor of the universe is due to gravitational effects that are not taken into account by GR, and that the correct expansion rate of the universe can be naturally obtained from a more complex theory of gravity \cite{clifton1}.

The f(R) theories of gravity were shown to accurately describe the expansion rate of the universe in both the metric \cite{carroll1} and the Palatini formalism \cite{olmo1}. However, in the weak-field slow-motion limit, these theories modify the solar system dynamics and also lead to matter instabilities and feature the wrong evolution of cosmological perturbations \cite{koivisto1,koivisto2}. These theories must therefore be excluded. To solve this problem, a new modified theory of gravity known as the hybrid metric-Palatini gravity which combines aspects of both formalisms (hence the name) was proposed \cite{capozziello1} and shown to reproduce correctly the behavior of the scale-factor of the universe without affecting the solar system dynamics \cite{harko1}. This theory was also succesful in describing a wide range of other observable phenomena such as the rotation of galaxies \cite{capozziello2}. 

The hybrid metric-Palatini gravity is described by a Lagrangian $\mathcal L=R+f\left(\mathcal R\right)$, where $R$ is the metric Ricci scalar and $\mathcal R$ is the Palatini scalar \cite{capozziello1}. Recently, the generalized hybrid metric-Palatini gravity, which arises as a natural generalization of the hybrid theory, was proposed \cite{tamanini1}. This theory is described by a general function $\mathcal L=f\left(R,\mathcal R\right)$. In 2015 this was the only available reference on the subject, and exploring other topics within the same theory was the main motivation to the development of this thesis.

\section{Historical introduction and state of the art}

Let us now provide a historical background of the most important milestones attained in our topic. We start by exploring the three experimental tests originally proposed by Albert Einstein upon the development of GR to emphasize the sucess of the theory and how it was responsible for a change of paradigm in the area of gravitation. Then, we explore the history of relativistic cosmology, pointing out the most important observations made and how the cosmological models evolved to be in agreement with these results. Finally, we state how modified theories of gravity approached the unsolved problems of cosmology, focusing specially on $f\left(R\right)$ gravity theories, which are the basic foundations of the generalized hybrid metric-Palatini gravity that we aim to study in this thesis.

\subsection{Experimental tests of GR}

Developed by Albert Einstein in 1915 \cite{einstein1,einstein2,einstein3,einstein4}, the Theory of General Relativity provides a description of gravity as a geometrical property of spacetime. This description states that spacetime is not flat, like it was assumed in special relativity and Newton’s universal law of gravitation, but rather curved by the presence of matter and energy \cite{einstein5}. This interaction is described by a set of partial differential equations for the metric tensor called Einstein’s field equations (EFE), namely
\begin{equation}\label{efe}
G_{ab}=R_{ab}-\frac{1}{2}R g_{ab}=8\pi T_{ab},
\end{equation}
where the Einstein tensor $G_{ab}$, the Ricci tensor $R_{ab}$ and the Ricci scalar $R=g^{ab}R_{ab}$ are purely geometrical entities which can be obtained from the metric tensor $g_{ab}$, and $T_{ab}$ is the stress-energy tensor associated with the presence of matter. In his original papers \cite{einstein5}, Einstein proposed three classical tests of GR, namely the anomalous precession of the perihelion of Mercury, the deflection of light by gravitational fields, and the gravitational redshift. 

The anomalous precession of the perihelion of Mercury was known at the time. In 1859, Urbain Le Verrier analyzed the available data of transits of Mercury over the Sun from 1697 to 1848 and computed a 38 arcseconds per tropical century disagreement of the actual precession rate in comparison to the one predicted by Newton’s theory \cite{verrier1}. In his third paper on GR \cite{einstein3}, Einstein shown that it correctly predicts the observed rate of precession of the perihelion of Mercury. A few decades later, in 1947, more accurate measurements made with radar set the observed rate of precession on $574.10\pm 0.65$ arcseconds per century \cite{clemence1}.

In 1784, in an unpublished manuscript, Henry Cavendish used Newton's gravity to predict a deflection on the direction of lightrays emitted by stars around massive objects. This feature was published a few years later by Johann Georg von Soldner, in 1804 \cite{soldner1}. Einstein was able to reproduce the same result based solely on the equivalence principle in 1911 \cite{einstein7}. However, he realized that in the context of GR, the previous result was not completely right, and the first correct calculation of the bending of light was published in his third paper on GR \cite{einstein3}. To experimentally observe this bending, two expeditions were conducted by Sir Arthur Eddington during the solar eclipse of 29th of May of 1919, one in the city of Sobral in Brazil, and another in São Tomé e Príncipe \cite{dyson1}. These observations confirmed the predictions of GR and were all over the news worldwide.

Einstein was the first to predict a gravitational redshift in 1907, again based solely on the equivalence principle \cite{einstein8}. To observe this effect, the first idea was to use the spectra of white dwarf stars, due to their high gravitational field. Several attempts to observe the spectrum of Sirius B were performed in 1925 by Walter Sydney Adams, but these measurements were considered unusable due to the scattering of light on its neighbour star Sirius A \cite{holberg1}. The first accurate measurement was only obtained in 1954 by Daniel Popper, using the spectrum of 40 Eridani B \cite{popper1}. The same effect was also verified by a terrestrial experiment known as the Pound-Rebka experiment in 1959 using the spectral lines of a $\ ^{57}$Fe gamma source redshifted over a vertical height of $22.5$ meters \cite{pound1}.  

In a subsequent paper published in 1916, Einstein made another prediction of major importance: the emission of gravitational waves by perturbed massive bodies \cite{einstein9}. For decades, these gravitational waves were just a mathematical result with no observations to support their existence whatsoever. In 1974, Russel Hulse and Joseph Taylor discovered a binary pulsar now commonly known as the Hulse-Taylor binary (although its original name was PSR B1913+16). Their observations showed that the orbital period of this binary decreased with time in perfect agreement with the predictions of GR \cite{weisberg1}. This indirect detection of gravitational waves granted Hulse and Taylor the Nobel prize in physics in 1993. It took physicists another two decades until, in 2015, the LIGO collaboration finally made the first direct detection of gravitational waves using a laser interferometer \cite{abbott1}, for which Rainer Weiss, Barry Barish and Kip Thorne were also awarded the Nobel prize in physics in 2017.

\subsection{Cosmological models}

All of the experimental tests discussed above supported strongly the validity of GR as a theory of gravity. However, the first serious problems arised when Einstein tried to develop a cosmological model based on his theory. Back in the beggining of the 20th century, it was believed that the universe was static and not much bigger than the size of the Milky Way itself. In 1917, Einstein presented his cosmological model to support this belief \cite{einstein10}. Einstein's universe was constructed as an attempt to include Mach's principle in GR and also to overcome the boundary conditions of the theory. This model had a spherical (closed) geometry and was static. To counteract the attractive effects of gravity, Einstein added a positive cosmological constant in order to obtain an equillibrium in every spacial direction. The model was thus unstable, because a slight change in the position of a celestial body would cause the universe to collapse or expand forever. 

Einstein's model was soon discarded when Edwin Hubble showed, based on his observations of Cepheids from nebulae (that were accepted to be other galaxies) outside our galaxy and the Doppler effect, that these far galaxies were moving away from us \cite{hubble1}. But the result was even deeper: there was an approximate linear relation between the radial velocity at which the galaxy was moving and the distance it was from us, implying that the universe was expanding. A non-static model of the universe should therefore be considered. In fact, by the time Hubble made this discovery, there were already some non-static models developed. Willem de-Sitter was the first one to present such a model \cite{desitter1,desitter2}, in 1917. The so called de-Sitter universe was unpopulated of matter but had a positive cosmological constant. Also, in 1922, Alexander Friedmann developed a set of mathematical equations that could be used to obtain non-static universe models \cite{friedmann1}. Later on, in 1927, Georges Lemaître also published a solution for an expanding universe, the so called Eddington-Lemaître universe \cite{lemaitre1}. 

The de-Sitter and Lemaître solutions had a very important difference: both the solutions were expanding, but the de-Sitter universe has existed for an infinite amount of time, whereas the Eddington-Lemaître universe must have had a beginning \cite{dinverno1}. Lemaître then suggested that the universe might have been created by the explosion of some primeval ”cosmic egg”, creating an expanding universe, which was later denoted by the "big bang" by Fred Hoyle. This theory was supported by the calculations of George Gamow that, in 1946, computed which should be the relative ammounts of hidrogen and helium to be formed under the conditions of an extremely large density and temperature initial state \cite{gamow1}. His results agreed successfully with the observations from the composition of stars. 

One of the most important discoveries in cosmology came 17 years later, in 1963, when Arno Penzias and Robert Wilson were mapping the sky's microwave noise and accidently detected a microwave background coming from every direction in the sky \cite{penzias1}, now well-known as the Cosmic Microwave Background, or CMB. This discovery granted Penzias and Wilson the Nobel prize in physics in 1978. Two years later, Robert Dicke studied that radiation and interpreted it as a $\sim 3$K radiation left over from the big bang. The existance of this radiation had already been predicted by Gamow, Alpher and Herman in 1948 \cite{alpher1}. From this point on, it was generally accepted that the universe had started from a hot and dense state, had a big bang, and had been expanding ever since. 

During the 1970's, the cosmologists tried to find a big bang model that was coherent with the observations. The first models were pure baryonic models, that is, they considered that all the density of the universe belonged to baryonic matter. These models failed to explain the large scale structure of the universe, namely the formation and distribution of galaxies. Also, baryonic matter could not be the only source of mass in the universe. The analysis of rotational curves of globular clusters by Vera Rubin and Kent Ford in the 1960's revealed that stars placed further from the center of gravity were moving faster than the closer ones \cite{rubin1,rubin2}, which would cause them to be ejected from their orbits if it were only for the gravitational force of visible matter. Other observations based in the phenomenon of gravitational lensing supported this fact. Dark matter should be considered in the models. 

The new so called Cold Dark Matter (CDM) models \cite{davis1} were continuously modified to fit new observations. In the early 1980's, most models considered a critical density of matter, with $\sim 5\%$ baryonic matter and $\sim 95\%$ CDM. These models successfully predicted the formation of galaxies and galaxy clusters. However, these models required a slower rate of expansion than the observational one. Also, between 1988 and 1990, new observations showed more galaxy clustering than predicted at larger scales. In 1992, the COBE satellite provided better observations of the CMB anisotropy and confirmed the same problems for the CDM models \cite{smoot1}.

Another important step towards a consistent cosmological model was made by Alan Guth in 1981, when he suggested that the universe might have had an era at the early stages of its expansion for which the expansion rate was exponentially fast \cite{guth1}. This so-called inflationary period was proposed to solve two problems raised by the current observations of the CMB: the horizon problem, related to the fact that regions in the sky outside each other's cosmological horizon display the same physical properties despite not having any causal contact, and the flatness problem, which states that the universe must have been fine-tuned to have an energy density so close to the critical density needed for its geometry to be flat.  

Probably the most astonishing discovery in the history of cosmology came in 1998, when the observations of Ia supernovae revealed that the universe is expanding at an accerelated rate \cite{riess1}. This discovery led cosmologists to consider cosmological models where density was not only provenient from matter (baryonic and dark), but from vacuum energy as well, also known as dark energy \cite{perlmutter1}. The new so called $\Lambda$CDM models, where $\Lambda$ is the cosmological constant related to dark energy, agreed successfully with the observations.

In 2000, the experiment Balloon Observations Of Millimetric Extragalactic Radiation And Geophysics (BOOMERanG) measured the CMB of a part of the sky during three sub-orbital balloon flights \cite{melchiorri1}, and led to the discovery that the geometry of the universe is approximately flat, which means the density of the universe should be very close to the critical density. One year later, the Two-Degree-Field Redshift Survey (2dFRS) \cite{percival1} conducted by the Anglo-Australian Observatory measured the matter density to be $\sim 25\%$ which implies the dark energy density should be around $\sim 75\%$. Finally, further observations from the WMAP \cite{spergel1} probe and Planck \cite{salvatelli1} satellite in 2003 and 2013 respectively have supported the model so far and helped to tune the parameters of the $\Lambda$CDM model, which are now constrained to uncertainties below $1\%$.

\subsection{Modified theories of gravity}

There is an inherent problem with the $\Lambda$CDM model discussed in the previous section, which is the lack of understanding about the nature of the extra sources of the energy density needed for the model to be consistent with observational data, namely dark matter and dark energy. There are two commonly accepted lines of thought to approach this problem. One can argue that dark matter and dark energy exist as physical fields or particles and can, if they somehow interact with baryonic matter through an interaction that is not purely gravitational, be detected in particle detectors \cite{liu1}. On the other hand, one can also argue that the observed behaviors may be due to a lack of fundamental knowledge of the gravity sector itself, i.e., this analysis might be out of the validity range of GR and a new modified theory of gravity should be taken into account to obtain the correct behavior of the expansion rate of the universe and the galactic rotational curves without the need to assume the existence of extra sources of energy density \cite{clifton1}. 

Many different kinds of modified theories of gravity have been proposed to approach this problem. The most important one for the development of this thesis was first proposed in 1970, by Hans Buchdahl. Buchdahl's idea was to modify the fundamental action $S$ from which one can deduce Einstein's field equations of GR, which depends solely on the Ricci scalar $R$, to a more general arbitrary function $f\left(R\right)$ of the Ricci scalar \cite{buchdahl1}:
\begin{equation}
S_{GR}=\int\sqrt{-g}\ R\ d^4x\ \ \ \ \ \Rightarrow\ \ \ \ \ S=\int\sqrt{-g}\ f\left(R\right)\ d^4x,
\end{equation}
where $g$ is the determinant of the metric $g_{ab}$ and $x$ collectively represents the spacetime coordinates. There are two different main approaches to this action: one can either assume that the Christoffel symbols $\Gamma$ are the Levi-Civita connection of the metric $g_{ab}$, the so-called metric formalism; or one can consider the metric and the connection to be independent variables, that is the Palatini formalism. In GR, both formalisms give rise to the same equations of motion and thus the same results. However, the same is not true in more complex theories of gravity. Using both formalisms, the $f\left(R\right)$ theories have been proven to be of extreme usefulness in accounting for cosmological models not only to be compatible with the current accelerated expansion rate \cite{carroll1,olmo1} but also to account for other phenomena like inflation, as shown by Alexei Starobinski in 1980 \cite{starobinski1}. 

However, $f\left(R\right)$ theories were riddled with difficulties. Despite being successful in reproducing cosmological behaviors, these theories were recently shown to lead to several problems of different natures. In 2004, Eanna Flanagan studied the conformal frame freedom in theories of gravity and shown that $f\left(R\right)$ theories appear to be in conflict with the Standard Model of particle physics\cite{flanagan1}. The following year, Gonzalo Olmo studied the weak-field regime of the theory and concluded that the theory may contradict the observations from solar system experiments \cite{olmo2}. Other problems arose in 2008, when Enrico Barausse, Thomas Sotiriou and John Miller analyzed common polytropic equations of state for nonvacuum spheres in the Palatini formalism of the theory and verified that unwanted singularities appear often, making these solutions unphysical \cite{barausse1}. 

The difficulties of $f\left(R\right)$ were overcome by proposing novel approaches. In 2013, a new idea was proposed Tiberiu Harko, Tomi Koivisto, Francisco Lobo and Gonzalo Olmo: combine the metric and Palatini approaches on the $f\left(R\right)$ theories in a hybrid theory by adding an extra term $f\left(\mathcal R\right)$ to the Einstein-Hilbert action in GR \cite{capozziello1}, where $\mathcal R$ is a geometrical entity analogous to the Ricci scalar $R$ but constructed in terms of an independent connection $\hat\Gamma$. This theory, now commonly called the hybrid metric-Palatini gravity, sucessfully reproduces the accelerated rate of expansion of the universe while keeping the solar system dynamics under control in the weak-field regime \cite{harko1}. 

The success of the hybrid metric-Palatini gravity inspired Nicola Tamanini and Christian Boehmer to generalize the action even further by considering a general function $f\left(R,\mathcal R\right)$ of both the Ricci scalar $R$ and the Palatini scalar $\mathcal R$ \cite{tamanini1}:
\begin{equation}
S=\int\sqrt{-g}\ f\left(R,\mathcal R\right)\ d^4x.
\end{equation} 
The cosmological phase space of this theory, that the authors named generalized hybrid metric-Palatini gravity, was studied under the formalism of dynamical systems and it was shown that cosmological accelerated solutions not only exist but can also be, under specific choices of the parameters, global attractors of the phase space. These solutions undergo an extended period of matter domination before starting to accelerate, allowing for the formation of large scale structures in the universe which might be in agreement with observations. 

As we can see, modified theories of gravity play an important role in the context of cosmology and are still an open and timely topic of research, featuring an extremely rich phenomenology and a wide range of applications. When this thesis started being worked in 2015, Tamanini's and Boehmer's paper was the only reference available in the generalized hybrid metric-Palatini gravity, and many other applications and projects were sitting there waiting to be explored. The objective of this thesis is then to provide a broad study of the generalized hybrid metric-Palatini gravity not only in the context of cosmology but also to explore the existence of stable compact object solutions and their applications (which will be the aim of chapters 3 to 7), to study possible extensions to this theory to higher order versions (chapter 8), and also to provide a basic framework on which one could costraint modified theories of gravity with extra scalar degrees of freedom via the stability analysis of scalar-field solutions (chapter 9).

\section{Thesis outline}

This thesis is organized as follows: In chapter 2, we introduce the metric and Palatini approaches to $f\left(R\right)$ theories of gravity and the equivalent scalar-tensor representation and generalize the analysis to the generalized hybrid metric-Palatini gravity. In chapter 3, we use the scalar-tensor representation of the generalized hybrid metric-Palatini gravity to obtain various cosmological solutions with different scale-factor behaviors. In chapter 4, we use the formalism of dynamical systems to study the structure of the cosmological phase space of the generalized hybrid metric-Palatini gravity. In chapter 5, we deduce the junction conditions for the smooth matching and for the matching with thin shells of two spacetime solutions of the generalized hybrid metric-Palatini gravity and provide two examples of applications. In chapter 6, we use the junction conditions previously deduced to obtain a full analytical traversable wormhole solution in the generalized hybrid metric-Palatini gravity that satisfies the null energy condition for the whole spacetime. In chapter 7, we use first-order perturbation theory to analyse the stability of static and rotating black-hole solutions in the generalized hybrid metric-Palatini gravity and provide specific forms of the action for which the solutions are stable. In chapter 8, we apply again the dyamical system formalism to study the influence of sixth and eight-order terms in the action to the cosmological phase space as a first step to study higher-order hybrid metric-Palatini theories. In chapter 9, we use the 1+1+2 formalism to generalize the Derrick's theorem to curved spacetimes and study applications to modified gravity theories. Finally, in Chapter 10 we present our conclusions. 
\cleardoublepage

\chapter{The generalized hybrid metric-Palatini gravity theory}
\label{chapter:chapter2}

The generalized hybrid metric-Palatini gravity arises as a natural generalization of the hybrid metric-Palatini gravity which is, in turn, a generalization of the $f\left(R\right)$ theories of gravity. In this chapter, we start by introducing $f\left(R\right)$ theories of gravity and the two formalisms one can use to obtain the equations of motion via the variational method. The equivalent scalar-tensor representation of the $f\left(R\right)$ theory is obtained via the use of auxiliary fields. The same procedures are then applied to the generalized hybrid metric-Palatini gravity and we compare the results with the ones obatined in both formalisms of $f\left(R\right)$.

\section{Variational methods to $f\left(R\right)$ gravity}

When one wants to derive the equations of motion of a given physical theory from the action function, the variational method must be applied. In this method, one computes the variation of the action with respect to its independent variables and extremizes it by equaling the resultant condition to zero. There are two different ways of doing the variation: one either assumes that the connection $\Gamma_{ab}^c$ is the Levi-Civita connection of the metric $g_{ab}$, for which we have the relation
\begin{equation}\label{deflevicivita}
\Gamma_{ab}^c=\frac{1}{2}g^{cd}\left(\partial_ag_{cb}+\partial_bg_{ac}-\partial_cg_{ab}\right),
\end{equation}
where $\partial_a\equiv \partial/\partial x^a$ denotes a partial derivative with respect to the coordinate $x^a$ and roman indi	ces run from $0$ to $3$; or one assumes the metric $g_{ab}$ and the connection $\hat\Gamma_{ab}^c$ to be independent variables and thus the variational principle yields two different equations of motion. In GR, both approaches give rise to the same field equations since the equation of motion for the independent connection in the Palatini approach forces the connection to be Levi-Civita. However, in more general theories with more complicated actions one can have in general different field equations in the metric and the Palatini approaches, as we shall see in the particular example of $f\left(R\right)$ theories.

\subsection{Metric approach}

Let us start by considering the metric approach to the $f\left(R\right)$ theories. The action for these theories is given by
\begin{equation}\label{frmaction}
S=\frac{1}{2\kappa^2}\int\sqrt{-g} f\left(R\right)d^4x+S_m,
\end{equation}
where $\kappa^2=8\pi$ in units for which $G=c=1$, and $S_m$ denotes the matter action. In the metric approach, the metric $g_{ab}$ is the only independent variable in the action and thus we only have one equation of motion. Varying the action given in Eq.\eqref{frmaction} yields
\begin{equation}\label{frmfield}
f'\left(R\right)R_{ab}-\frac{1}{2}f\left(R\right)g_{ab}-\left[\nabla_a\nabla_b-\Box g_{ab}\right]f'\left(R\right)=\kappa^2T_{ab},
\end{equation}
where a prime $'$ denotes a derivative with respect to the argument of the function, in this case $R$, $\nabla_a$ denotes a covariant derivative, $\Box\equiv \nabla_a\nabla^a$ is the D'Alembert operator, and we define the stress-energy tensor $T_{ab}$ as
\begin{equation}\label{deftab}
T_{ab}=-\frac{2}{\sqrt{-g}}\frac{\delta S_m}{\delta g_{ab}}.
\end{equation}
Note that in the limit $f\left(R\right)=R$ the fourth order terms vanish and one recovers the EFE given in Eq.\eqref{efe}. Taking the trace of Eq.\eqref{frmfield} and defining the trace of the stress-energy tensor as $T\equiv g^{ab}T_{ab}$ we obtain
\begin{equation}\label{frmtrace}
f'\left(R\right)R-2f\left(R\right)+3\Box f'\left(R\right)=\kappa^2 T.
\end{equation}
This equation provides a differential relation between $R$ and $T$ unlike the GR case where they are related by $R=\kappa^2T$. Being a partial differential equation (PDE), this implies that the relation between $R$ and $T$ is no longer unique and these theories should admit a wider variety of solutions. 

Finally, let us write the field equations in a form that resembles the EFE with an effective stress-energy tensor. To do so, we force the appearence of the Einstein tensor, $G_{ab}\equiv R_{ab}-\frac{1}{2}Rg_{ab}$, and transfer the extra terms to the right hand side of the field equation. The results are thus
\begin{equation}\label{deftabeff}
G_{ab}=\kappa^2T_{ab}^{(eff)},
\end{equation}
where the effective stress-energy tensor $T_{ab}^{(eff)}$ in this case is given by
\begin{equation}\label{frmteff}
T_{ab}^{(eff)}=\frac{1}{f'\left(R\right)}\left\{T_{ab}+\frac{1}{\kappa^2}\left[\frac{1}{2}g_{ab}\left(f\left(R\right)-R f'\left(R\right)\right)+\left(\nabla_a\nabla_b-g_{ab}\Box\right)f'\left(R\right)\right]\right\}.
\end{equation}
This representation of the field equations is particularly useful when one considers the scalar-tensor representations of the theory, as we shall see further on. Again, note that in the limit $f\left(R\right)=R$ one recovers $T_{ab}^{(eff)}=T_{ab}$ and Eq.\eqref{deftabeff} reduces to the EFE in Eq.\eqref{efe}.

\subsection{Palatini approach}

Let us now turn to the Palatini approach. In this case, the action becomes 
\begin{equation}\label{frpaction}
S=\frac{1}{2\kappa^2}\int\sqrt{-g} f\left(\mathcal R\right)d^4x+S_m,
\end{equation}
where $\mathcal R=g^{ab}\mathcal R_{ab}$ is the Palatini scalar defined in terms of the Palatini Ricci tensor $\mathcal R_{ab}$ which is given in the same form as the Ricci scalar but as a function of the independent connection $\hat\Gamma^c_{ab}$ instead of the Levi-Civita connection $\Gamma^c_{ab}$:
\begin{equation}\label{defpalatini}
\mathcal R_{ab}=\partial_c\hat\Gamma^c_{ab}-\partial_b\hat\Gamma^c_{ac}+\hat\Gamma^c_{cd}\hat\Gamma^d_{ab}-\hat\Gamma^c_{ad}\hat\Gamma^d_{cb}.
\end{equation}
The action in Eq.\eqref{frpaction} is a function of two independent variables, namely the metric $g_{ab}$ and the independent connection $\hat\Gamma$. We thus have to apply the variational method twice, by varying the action with respect to $g_{ab}$ and $\hat\Gamma$ respectively, and obtain two different equations of motion, namely:
\begin{equation}\label{frpfield}
f'\left(\mathcal R\right)\mathcal R_{(ab)}-\frac{1}{2}f\left(\mathcal R\right)g_{ab}=\kappa^2T_{ab},
\end{equation}
\begin{equation}\label{frpeomcon}
\hat\nabla_c\left[\sqrt{-g}f'\left(\mathcal R\right)g^{ab}\right]=0,
\end{equation}
where we define the symmetric tensor $2X_{(ab)}=X_{ab}+X_{ba}$, and $\hat\nabla$ is the covariant derivative written in terms of the independent connection $\hat\Gamma^c_{ab}$. At this point we can see that in the limit $f\left(\mathcal R\right)=\mathcal R$, we not only recover Eq.\eqref{efe} from Eq.\eqref{frpfield}, but also Eq.\eqref{frpeomcon} becomes the definition of the Levi-Civita connection for the metric $g_{ab}$, and hence the antecipated result that in GR both the metric and the Palatini approach give rise to the same equations of motion. Note that the fact that in GR the connection is Levi-Civita in the Palatini approach is not an assumption like in the metric approach but rather a dynamical result.

Let us now analyse Eq.\eqref{frpeomcon} and its implications. To do so, we define a new metric tensor $h_{ab}=f'\left(\mathcal R\right)g_{ab}$ conformally related to the metric $g_{ab}$. For these two metrics, we have $\sqrt{-h}h^{ab}=\sqrt{-g}f'\left(\mathcal R\right)g^{ab}$, where $h$ is the determinant of the metric $h_{ab}$. This implies that Eq.\eqref{frpeomcon} is the definition of the Levi-Civita connection of the metric $h_{ab}$ and thus the connection $\hat\Gamma$ can be written in the same form as Eq.\eqref{deflevicivita} as:
\begin{equation}\label{deflevicivitap}
\hat\Gamma_{ab}^c=\frac{1}{2}h^{cd}\left(\partial_ah_{cb}+\partial_bh_{ac}-\partial_ch_{ab}\right).
\end{equation}
As the two metrics $h_{ab}$ and $g_{ab}$ are related to each other, then $R_{ab}$ and $\mathcal R_{ab}$, and also $R$ and $\mathcal R$ must also be related to each other, and these relations are (see Sec. \ref{Sec:Brelproof} for details)
\begin{equation}\label{relricten}
\mathcal R_{ab}=R_{ab}+\frac{3}{2f'\left(\mathcal R\right)^2}\partial_af'\left(\mathcal R\right)\partial_bf'\left(\mathcal R\right)-\frac{1}{f'\left(\mathcal R\right)}\left(\nabla_a\nabla_b+\frac{1}{2}g_{ab}\Box\right)f'\left(\mathcal R\right),
\end{equation}
\begin{equation}\label{relricsca}
\mathcal R=R+\frac{3}{2f'\left(\mathcal R\right)^2}\partial_af'\left(\mathcal R\right)\partial^af'\left(\mathcal R\right)-\frac{3}{f'\left(\mathcal R\right)}\Box f'\left(\mathcal R\right)
\end{equation}

Note that $\mathcal R$ is not the Ricci scalar for the metric $h_{ab}$ because it is contracted using the inverse metric $g^{ab}$ instead of $h^{ab}$. 

Taking the trace of Eq.\eqref{frpfield} leads to the relation
\begin{equation}\label{frptrace}
f'\left(\mathcal R\right)\mathcal R-2f\left(\mathcal R\right)=\kappa^2T.
\end{equation}
This relation seems to indicate that, unlike the metric approach, there is a direct algebraic relation between curvature and the trace of the stress-energy tensor $T$. However, if we use Eq.\eqref{relricsca} to cancel the term $\mathcal R$ in Eq.\eqref{frptrace}, we verify that the relation between $R$ and $T$ is still a differential relation and hence the solution is not unique.

To finalize, let us again write Eq.\eqref{frpfield} in the form of Eq.\eqref{deftabeff}. Using Eq.	\eqref{relricten} to write the term $\mathcal R_{ab}$ as a function of $R_{ab}$ and then using Eq.\eqref{relricsca} to write $R$ in terms of $\mathcal R$ one obtains
\begin{eqnarray}
T_{ab}^{(eff)}&=&\frac{1}{f'\left(R\right)}\left\{T_{ab}+\frac{1}{\kappa^2}\left[\frac{1}{2}g_{ab}\left(f\left(\mathcal R\right)-\mathcal R f'\left(\mathcal R\right)\right)+\left(\nabla_a\nabla_b-g_{ab}\Box\right)f'\left(\mathcal R\right)-\right.\right.\nonumber \\
&-&\left.\left.\frac{3}{2f'\left(\mathcal R\right)}\left(\nabla_af'\left(\mathcal R\right)\nabla_bf'\left(\mathcal R\right)-\frac{1}{2}g_{ab}\nabla_cf'\left(\mathcal R\right)\nabla^cf'\left(\mathcal R\right)\right)\right]\right\}\label{frpteff}
\end{eqnarray}
Again, note that in the limit $f\left(\mathcal R\right)=\mathcal R$ we recover $T_{ab}^{(eff)}=T_{ab}$ and Eq.\eqref{frpfield} reduces to Eq.\eqref{efe}, as expected.

\section{Equivalent scalar-tensor representation}

It is sometimes useful to rewrite the action of a given theory in terms of new coordinates and variables in order for the equations of motion to be in more suitable forms for the system in study. In the case of $f\left(R\right)$ theories, using Legendre transformations one can rewrite the actions in Eqs.\eqref{frmaction} and \eqref{frpaction} in terms of a scalar-tensor theory of the form of Brans-Dicke theory with a scalar-field potential $V$:
\begin{equation}\label{bdaction}
S=\frac{1}{2\kappa^2}\int\sqrt{-g}\left[\phi R+\frac{\omega}{\phi}\partial_a\phi\partial^a\phi-V\left(\phi\right)\right]d^4x,
\end{equation}
where $\phi$ is a scalar-field and $\omega$ is a free parameter of the theory. In the following, we show that the difference between the scalar-tensor representations of both the metric and the Palatini approaches to $f\left(R\right)$ deppends on the value of $\omega$, which is $\omega=0$ for the metric approach and $\omega=-3/2$ for the Palatini approach.

\subsection{Metric scalar-tensor}

The first step needed to obtain the scalar-tensor representation of the theory is to define an auxiliary field $\alpha$ and rewrite Eq.\eqref{frmaction} in the form
\begin{equation}\label{frmstauxaction}
S=\frac{1}{2\kappa^2}\int \sqrt{-g}\left[f\left(\alpha\right)+f'\left(\alpha\right)\left(R-\alpha\right)\right]d^4x+S_m.
\end{equation}
This action is now a function of two variables, the metric $g_{ab}$ and the auxiliary field $\alpha$. Varying this action with respect to $\alpha$ yields the condition
\begin{equation}\label{frmstalpha}
f''\left(\alpha\right)\left(R-\alpha\right)=0,
\end{equation}
which implies that if $f''\left(\alpha\right)\neq 0$, then $\alpha=R$ and we recover Eq.\eqref{frmaction}. At this point, we can define the scalar field $\phi=f'\left(R\right)$ and rewrite the action in Eq.\eqref{frmstauxaction} in the form
\begin{equation}\label{frmstaction}
S=\frac{1}{2\kappa^2}\int\sqrt{-g}\left[\phi R-V\left(\phi\right)\right]d^4x+S_m,
\end{equation}
\begin{equation}\label{frmstpotential}
V\left(\phi\right)=\alpha \phi-f\left(\alpha\right),
\end{equation}
where the function $V\left(\phi\right)$ plays the role of a scalar potential. Comparing Eq.\eqref{frmstaction} with Eq.\eqref{bdaction}, we see that the action for the scalar-tensor representation of the metric formalism of $f\left(R\right)$ corresponds to a Brans-Dicke theory with a parameter $\omega=0$ and a potential $V\left(\phi\right)$. 

Note that if $f''\left(\alpha\right)=0$, then the equivalence between the scalar-tensor and the geometrical representations of the theory is no longer guaranteed. This condition can be thought of as the condition for the relation $\phi=\phi\left(R\right)$ to be invertible, that is $d\phi/dR\neq 0$ and we should be able to write $R\left(\phi\right)$. Also, after the potential has been defined, we can rewrite again $\phi=f'\left(R\right)$ and Eq.\eqref{frmstpotential} can be regarded as a PDE for the function $f\left(R\right)$. Since this is a partial differential equation, the solutions are not unique and one should expect that for the same form of the potential $V\left(\phi\right)$ there are many solutions for the function $f\left(R\right)$.

The action given in Eq.\eqref{frmstaction} depends on two independent variables, the metric $g_{ab}$ and the scalar field $\phi$. Thus, we can obtain two different equations of motion, namely
\begin{equation}\label{frmstfield}
\phi G_{ab}=\kappa^2 T_{ab}-\frac{1}{2\phi}g_{ab}V\left(\phi\right)+\left(\nabla_a\nabla_b-g_{ab}\Box\right)\phi,
\end{equation}
\begin{equation}\label{frmsteomphi}
R=V'\left(\phi\right).
\end{equation}
Eq.\eqref{frmstfield} could be obtained directly from Eq.\eqref{frmfield} by performing the same field and potential refinitions that we did in the action. A final equation that specifies the dynamics of the scalar field $\phi$ given the matter sources can be obtained by taking the trace of Eq.\eqref{frmstfield} and using Eq.\eqref{frmsteomphi} to cancel the terms depending on $R$, and the result is
\begin{equation}\label{frmstkg}
3\Box\phi+2V\left(\phi\right)-\phi V'\left(\phi\right)=\kappa^2 T.
\end{equation}
This equation plays the role of a modified Klein-Gordon equation for the scalar-field $\phi$. Eqs.\eqref{frmstfield} and \eqref{frmstkg} are then a system of two coupled PDEs for the two unknowns $g_{ab}$ and $\phi$.

\subsection{Palatini scalar-tensor}

The method for obtaining a scalar-tensor representation for the Palatini approach to $f\left(\mathcal R\right)$ theories of gravity is identical to the previous case: we first define an auxiliary field $\beta$ and rewrite Eq.\eqref{frpfield} as
\begin{equation}\label{frpstauxaction}
S=\frac{1}{2\kappa^2}\int \sqrt{-g}\left[f\left(\beta\right)+f'\left(\beta\right)\left(\mathcal R-\beta\right)\right]d^4x+S_m.
\end{equation}
Again, this action is a function of both the metric $g_{ab}$ and the auxiliary field $\beta$. A variation with respect to $\beta$ implies again that if $f''\left(\mathcal R\right)\neq 0$, then we have $\beta=\mathcal R$ and we recover Eq.\eqref{frpaction}. If $f''\left(\mathcal R\right)=0$, the equivalence of the two representations is not guaranteed. We now define the scalar field $\phi=f'\left(\mathcal R\right)$ for which Eq.\eqref{frpstauxaction} becomes
\begin{equation}\label{frpstauxaction2}
S=\frac{1}{2\kappa^2}\int\sqrt{-g}\left[\phi \mathcal R-V\left(\phi\right)\right]d^4x+S_m,
\end{equation}
\begin{equation}\label{frpstpotential}
V\left(\phi\right)=\beta \phi-f\left(\beta\right),
\end{equation}
like before. Finally, the last step consists of using the relation between $\mathcal R$ and $R$ given in Eq.\eqref{relricsca} to cancel the factor $\mathcal R$ in Eq.\eqref{frpstauxaction2} for which the final action becomes
\begin{equation}\label{frpstaction}
S=\frac{1}{2\kappa^2}\int\sqrt{-g}\left[\phi R+\frac{3}{2\phi}\partial_a\phi\partial^a\phi-V\left(\phi\right)\right]d^4x+S_m,
\end{equation}
and a comparison with Eq.\eqref{bdaction} immediatly tells us that the scalar-tensor representation of the Palatini $f\left(\mathcal R\right)$ gravity is dynamically equivalent to a Brans-Dicke theory with parameter $\omega=-3/2$ and a potential $V\left(\phi\right)$. Again, Eq.\eqref{frpstaction} depends on the two variables $g_{ab}$ and $\phi$, for which we obtain the two equations of motion
\begin{equation}\label{frpstfield}
\phi G_{ab}=\kappa^2 T_{ab}-\frac{3}{2\phi}\left(\nabla_a\phi\nabla_b\phi-\frac{1}{2}g_{ab}\nabla_c\phi\nabla^c\phi\right)+\left(\nabla_a\nabla_b-g_{ab}\Box\right)\phi-\frac{V}{2}g_{ab},
\end{equation}
\begin{equation}\label{frpsteomphi}
\Box\phi=\frac{\phi}{3}\left[R-V'\left(\phi\right)\right]+\frac{1}{2\phi}\nabla_c\phi\nabla^c\phi.
\end{equation}
Eq.\eqref{frpstfield} could be directly obtained by applying the definitions of $\phi$ and $V\left(\phi\right)$ in Eq.\eqref{frpfield}, as expected. Taking the trace of Eq.\eqref{frpstfield} and using the result to cancel the term depending on $R$ in Eq.\eqref{frpsteomphi} yields an equation that relates directly the dynamics of the scalar field $\phi$ with the matter distribution as
\begin{equation}\label{frpstkg}
\phi V'\left(\phi\right)-2V\left(\phi\right)=-\kappa^2 T.
\end{equation}
Eqs.\eqref{frpstfield} and \eqref{frpstkg} are thus a system of two equations for the two variables $\phi$ and $g_{ab}$, the difference being that, in this case, once one specifies the potential and the matter sources, the scalar field $\phi$ is directly defined whereas in the metric scalar-tensor case this is still a PDE coupled to the metric $g_{ab}$.

\section{Generalization to the hybrid theory}

In this section, we want to generalize the metric and the Palatini formalisms of $f\left(R\right)$ gravity to a more general theory which contains features of the two approaches simultaneously, hence the designation "hybrid". To do so, one writes an action function that is both a function of the Ricci scalar $R$ and the Palatini scalar $\mathcal R$ as
\begin{equation}\label{ghmpgaction}
S=\frac{1}{2\kappa^2}\int\sqrt{-g} f\left(R,\mathcal R\right)d^4x+S_m.
\end{equation}
The Ricci scalar $R$ is written in terms of the connection $\Gamma^c_{ab}$ which is the Levi-Civita connection of the metric $g_{ab}$ given in Eq.\eqref{deflevicivita}, and the Palatini scalar $\mathcal R$ is written in terms of an independent connection $\hat\Gamma^c_{ab}$, see Eq.\eqref{defpalatini}. In the upcoming sections, we will study both the geometrical and the scalar-tensor representations of this theory and compare the results with the particular metric and Palatini approaches to $f\left(R\right)$ gravity.

\subsection{Geometrical representation}

Let us start by analysing Eq.\eqref{ghmpgaction}. This action is a function of two independent variables, the metric $g_{ab}$ and the connection $\hat\Gamma^c_{ab}$, and thus two equations of motion can be obtained from the variational method. These equations are
\begin{equation}\label{ghmpgfield}
\frac{\partial f}{\partial R}R_{ab}+\frac{\partial f}{\partial \mathcal{R}}\mathcal{R}_{ab}-\frac{1}{2}g_{ab}f\left(R,\cal{R}\right)-\left(\nabla_a\nabla_b-g_{ab}\Box\right)\frac{\partial f}{\partial R}=\kappa^2 T_{ab},
\end{equation}
\begin{equation}\label{ghmpgeomcon}
\hat\nabla_c\left(\sqrt{-g}\frac{\partial f}{\partial \cal{R}}g^{ab}\right)=0,
\end{equation}
for the metric $g_{ab}$ and the connection $\hat\Gamma^c_{ab}$, respectively. From this point on, to simplify the notation, we shall denote the derivatives of $f\left(R,\mathcal R\right)$ with respect to $R$ and $\mathcal R$ with the subscripts $f_R$ and $f_\mathcal R$, respectively.

At this point, it is useful to note that in the limit $f\left(R,\mathcal R\right)=f\left(R\right)$, Eq.\eqref{ghmpgfield} reduces to Eq.\eqref{frmfield} and Eq.\eqref{ghmpgeomcon} becomes an identity. On the other hand, in the limit $f\left(R,\mathcal R\right)=f\left(\mathcal R\right)$, Eq.\eqref{ghmpgfield} reduces instead to Eq.\eqref{frpfield} and Eq.\eqref{ghmpgeomcon} remains the same, which is essentially Eq.\eqref{frpeomcon}. Finally, in the limit $f\left(R,\mathcal R\right)=\left\{R,\mathcal R\right\}$, Eq.\eqref{ghmpgfield} yields the EFE from Eq.\eqref{efe} and Eq.\eqref{ghmpgeomcon} becomes the definition of the Levi-Civita connection for the metric $g_{ab}$, as expected.

Via the same procedure as for the Palatini $f\left(\mathcal R\right)$ theories, one can define a new metric $h_{ab}=f_\mathcal R g_{ab}$ conformally related to the metric $g_{ab}$ and show that Eq.\eqref{ghmpgeomcon} corresponds to the definition of the Levi-Civita connection for the metric $h_{ab}$, This implies that despite being considered independent \textit{a priori}, the connection $\hat\Gamma^c_{ab}$ can be written in terms of $h_{ab}$ (and hence $g_{ab}$) as given in Eq.\eqref{deflevicivitap}, and thus the tensors $R_{ab}$ and $\mathcal R_{ab}$, as well as the scalars $R$ and $\mathcal R$, are related via the expressions
\begin{equation}\label{ghmpgrelricten}
\mathcal R_{ab}=R_{ab}+\frac{3}{2f_\mathcal R^2}\partial_af_\mathcal R\partial_bf_\mathcal R-\frac{1}{f_\mathcal R}\left(\nabla_a\nabla_b+\frac{1}{2}g_{ab}\Box\right)f_\mathcal R,
\end{equation}
\begin{equation}\label{ghmpgrelricsca}
\mathcal R=R+\frac{3}{2f_\mathcal R^2}\partial_af_\mathcal R\partial^af_\mathcal R-\frac{3}{f_\mathcal R}\Box f_\mathcal R.
\end{equation}

For completeness, let us also take the trace of Eq.\eqref{ghmpgfield}:
\begin{equation}\label{ghmpgtrace}
f_R R+f_\mathcal R \mathcal R -2f\left(R,\mathcal R\right)+3\Box f_R=\kappa^2T,
\end{equation}
from which one can again verify that, in the limits $f\left(R,\mathcal R\right)=f\left(R\right)$ and $f\left(R,\mathcal R\right)=f\left(\mathcal R\right)$, Eq.\eqref{ghmpgtrace} reduces to Eqs.\eqref{frmtrace} and \eqref{frptrace}, respectively. As before, we still have a differential relation between $R$ and $T$.

Finally, the field equations Eq.\eqref{ghmpgfield} can also be written in the form of the EFE with an effective stress-energy tensor as in Eq.\eqref{deftabeff}. After some manipulations, one arrives at the rather useful form
\begin{eqnarray}
T_{ab}^{(eff)}&=&\frac{1}{f_R+f_\mathcal R}\left\{T_{ab}+\frac{1}{\kappa^2}\left[\frac{1}{2}g_{ab}\left(f\left(R,\mathcal R\right)-Rf_R-\mathcal Rf_\mathcal R\right)+\right.\right.\nonumber\\
&+&\left.\left.\left(\nabla_a\nabla_b-g_{ab}\Box\right)\left(f_R+f_\mathcal R\right)-\frac{3}{2f_\mathcal R}\left(\nabla_af_\mathcal R\nabla_bf_\mathcal R-\frac{1}{2}g_{ab}\nabla_cf_\mathcal R\nabla^c\mathcal R\right)\right]\right\}.
\end{eqnarray}
As expected, in the limits $f\left(R,\mathcal R\right)=f\left(R\right)$ and $f\left(R,\mathcal R\right)=f\left(\mathcal R\right)$ one recovers Eqs.\eqref{frmteff} and \eqref{frpteff} respectively, and also in the limit $f\left(R,\mathcal R\right)=R$ we have $T_{ab}^{(eff)}=T_{ab}$ and we recover Eq.\eqref{efe}. Despite being useful to verity that the previous limits hold, when one deals with the scalar-tensor representation of the theory it is going to be more useful to use Eq.\eqref{ghmpgrelricsca} to cancel the term $\mathcal R$ and write the results in terms of $R$ and the scalar fields.

\subsection{Scalar-tensor representation}

Let us now derive the scalar-tensor representation of the generalized hybrid metric-Palatini gravity. In this case, the action is a function of both $R$ and $\mathcal R$, and hence we need to define two auxiliary fields $\alpha$ and $\beta$ to rewrite Eq.\eqref{ghmpgaction} into the form
\begin{equation}\label{ghmpgstauxaction}
S=\frac{1}{2\kappa^2}\int\sqrt{-g}\left[f\left(\alpha,\beta\right)+\frac{\partial f}{\partial \alpha}\left(R-\alpha\right)+\frac{\partial f}{\partial \beta}\left(\mathcal R-\beta\right)\right]d^4x.
\end{equation}
This action is a function of three variables, namely the metric $g_{ab}$ and the two auxiliary fields $\alpha$ and $\beta$. The variations of Eq.\eqref{ghmpgstauxaction} with respect to the auxiliary fields leads to the equations
\begin{equation}
\frac{\partial^2 f}{\partial\alpha^2}\left(R-\alpha\right)+\frac{\partial^2 f}{\partial\alpha\partial\beta}\left(\cal{R}-\beta\right)=0,
\end{equation}
\begin{equation}
\frac{\partial^2 f}{\partial\beta\partial\alpha}\left(R-\alpha\right)+\frac{\partial^2 f}{\partial\beta^2}\left(\cal{R}-\beta\right)=0.
\end{equation}
These two coupled equations can be rewritten in a matrix form $\mathcal M \textbf{x}=0$ as
\begin{equation}\label{ghmpgstmatrix}
\mathcal{M}\textbf{x}=
\begin{bmatrix}
\frac{\partial^2 f}{\partial\alpha^2} & \frac{\partial^2 f}{\partial\alpha\partial\beta} \\[0.8em]
\frac{\partial^2 f}{\partial\beta\partial\alpha} & \frac{\partial^2 f}{\partial\beta^2} 
\end{bmatrix}
\begin{bmatrix}
R-\alpha \\[0.8em]
\cal{R}-\beta
\end{bmatrix}
=0 .
\end{equation}
The solution for Eq.\eqref{ghmpgstmatrix} is unique if and only if the determinant of $\mathcal M$ does not vanish, i.e. $\det \mathcal M\neq 0$. This condition yields the relation
\begin{equation}\label{ghmpgstequiv}
\frac{\partial^2 f}{\partial\alpha^2}\frac{\partial^2 f}{\partial\beta^2}\neq \left(\frac{\partial^2 f}{\partial\alpha\partial\beta}\right)^2.
\end{equation}
If Eq.\eqref{ghmpgstequiv} is satisfied, then the solution for Eq.\eqref{ghmpgstmatrix} is unique and is given by $\alpha=R$ and $\beta=\mathcal R$. Inserting these results into Eq.\eqref{ghmpgstauxaction}, one recovers Eq.\eqref{ghmpgaction} and the two representations are equivalent. If Eq.\eqref{ghmpgstequiv} is not satisfied, then the equivalence between the two representations of the theory is not guaranteed. Defining two scalar fields as $\varphi=f_R$ and $\psi=-f_\mathcal R$, where the negative sign in $\psi$ is chosen for convenience, we can rewrite Eq.\eqref{ghmpgstauxaction} in the form
\begin{equation}\label{ghmpgstauxaction2}
S=\frac{1}{2\kappa^2}\int \sqrt{-g}\left[\varphi R-\psi\mathcal{R}-V\left(\varphi,\psi\right)\right]d^4x,
\end{equation}
\begin{equation}\label{ghmpgstpotential}
V\left(\varphi,\psi\right)=-f\left(\alpha,\beta\right)+\varphi\alpha-\psi\beta.
\end{equation}
Note that Eq.\eqref{ghmpgstequiv} will be satisfied if and only if the definitions of the scalar fields $\varphi$ and $\psi$ as functions of $R$ and $\mathcal R$ are invertible to obtain $R$ and $\mathcal R$ as functions of the scalar fields. Finally, one can use the relation between $R$ and $\mathcal R$ given by Eq.\eqref{ghmpgrelricsca} along with the definition for the scalar field $\psi=-f_\mathcal R$ to write the scalar-tensor action as
\begin{equation}\label{ghmpgstaction}
S=\frac{1}{2\kappa^2}\int \sqrt{-g}\left[\left(\varphi-\psi\right) R-\frac{3}{2\psi}\partial^a\psi\partial_a\psi-V\left(\varphi,\psi\right)\right]d^4x.
\end{equation}
Comparing Eq.\eqref{ghmpgstaction} with \eqref{bdaction} one verifies that the scalar-tensor representation of the generalized hybrid metric-Palatini gravity corresponds to a hybrid Brans-Dicke theory with two scalar fields, for which the parameters are $\omega_\varphi=0$ and $\omega_\psi=3/2$, and the presence of an interaction potential $V\left(\varphi,\psi\right)$. Taking the limit $\psi=0$, which corresponds to $f\left(R,\mathcal R\right)=f\left(R\right)$, Eq.\eqref{ghmpgstaction} reduces to Eq.\eqref{frmstaction}, whereas taking the limit $\varphi=0$, which corresponds to $f\left(R,\mathcal R\right)=f\left(\mathcal R\right)$, Eq.\eqref{ghmpgstaction} reduces to Eq.\eqref{frpstaction} with $\phi=-\psi$, as expected.

The action given in Eq.\eqref{ghmpgstaction} is now a function of three variables, the metric $g_{ab}$ and the two scalar fields $\varphi$ and $\psi$. The variational method yields thus three equations of motion, 
\begin{equation}\label{ghmpgstfield}
\left(\varphi-\psi\right) G_{ab}=\kappa^2T_{ab}-\left[\Box\left(\varphi-\psi\right)+\frac{1}{2}V+\frac{3}{4\psi}\partial^c\psi\partial_c\psi\right]g_{ab}+\frac{3}{2\psi}\partial_a\psi\partial_b\psi+\nabla_a\nabla_b\left(\varphi-\psi\right),
\end{equation}
\begin{equation}\label{ghmpgsteomphi}
R=V_\varphi,
\end{equation}
\begin{equation}\label{ghmpgsteompsi}
-R-\frac{3}{2\psi^2}\partial_a\psi\partial^a\psi+\frac{3}{\psi}\Box\psi-V_\psi=0,
\end{equation}
respectively, where the subscripts $V_\varphi$ and $V_\psi$ denote derivatives with respect to $\varphi$ and $\psi$ respectively. These equations could be obtained after some manipulations from Eq.\eqref{ghmpgfield} by applying the definitions of the scalar fields and the potential. To find dynamical equations for $\varphi$ and $\psi$ with no dependence on $R$ one can take the trace of Eq.\eqref{ghmpgstfield}, to cancel the terms $R$ in Eqs.\eqref{ghmpgsteomphi} and \eqref{ghmpgsteompsi} and after some algebra one arrives to
\begin{equation}\label{ghmpgstkgphi}
\Box\varphi+\frac{1}{3}\left(2V-\psi V_\psi-\varphi V_\varphi\right)=\frac{\kappa^2T}{3},
\end{equation}
\begin{equation}\label{ghmpgstkgpsi}
\Box\psi-\frac{1}{2\psi}\partial^a\psi\partial_a\psi-\frac{\psi}{3}\left(V_\varphi+V_\psi\right)=0.
\end{equation}
Eqs.\eqref{ghmpgstkgphi} and \eqref{ghmpgstkgpsi} take the role of the modified Klein-Gordon equations for the scalar fields $\varphi$ and $\psi$, respectively. The main difference between the generalized hybrid theory and the metric and Palatini approaches to $f\left(R\right)$ stands on the coupling of the fields to matter. In this case, only the scalar field $\varphi$ couples to matter, whereas the field $\psi$ does not. This difference comes from the couplings between the two scalar fields via the potential terms, and thus the limits for $\varphi=0$ and $\psi=0$ are not evident in these equations. However, these limits can be found in the following manipulated versions of Eqs.\eqref{ghmpgstkgphi} and \eqref{ghmpgstkgpsi}:
\begin{equation}\label{ghmpgstkg1}
\Box\varphi-\Box\psi+\frac{1}{2\psi}\partial^a\psi\partial_a\psi+\frac{1}{3}\left[2V-\left(\varphi-\psi\right)V_\varphi\right]=\frac{\kappa^2}{3}T,
\end{equation}
\begin{equation}\label{ghmpgstkg2}
\varphi\Box\psi-\psi\Box\varphi-\frac{\varphi}{2\psi}\partial^a\psi\partial_a\psi-\frac{\psi}{3}\left[2V+\left(\varphi-\psi\right)V_\psi\right]=-\frac{\kappa^2\psi}{3}T.
\end{equation}
From these equations, one can see that in the limit $\psi=0$, for which $f\left(R,\mathcal R\right)=f\left(R\right)$, Eq.\eqref{ghmpgstkg1} reduces to Eq.\eqref{frmstkg} and Eq.\eqref{ghmpgstkg2} becomes an identity, and also that in the limit $\varphi=0$, which corresponds to $f\left(R,\mathcal R\right)=f\left(\mathcal R\right)$, Eq.\eqref{ghmpgstkg2} reduces to Eq.\eqref{frpstkg} and Eq.\eqref{ghmpgstkg1} becomes an identity. The behaviors of the non generalized limits are thus obtained as expected.
\cleardoublepage

\chapter{Cosmological solutions in the generalized hybrid metric-Palatini gravity}
\label{chapter:chapter3}

We construct exact solutions representing a Friedmann-Lemaître-Robsertson-Walker (FLRW) universe in a generalized hybrid metric-Palatini theory. By writing the gravitational action in a scalar-tensor representation, the new solutions are obtained by either making an ansatz on the scale factor or on the effective potential. Among other relevant results, we show that it is possible to obtain exponentially expanding solutions for flat universes even when the cosmology is not purely vacuum. We then derive the classes of actions for the original theory which generate these solutions.

\section{Introduction}

The late-time cosmic accelerated expansion \cite{perlmutter1,riess1} has posed important and
challenging problems to theoretical cosmology. Although the standard model of cosmology has favored dark energy models \cite{copeland1} as fundamental candidates responsible for the accelerated cosmic expansion, it is also viable that this expansion is due to modifications of general relativity \cite{clifton1,nojiri1}, which introduce new degrees of freedom to the gravitational sector itself.  

The phenomenology of $f(R)$ gravity, where $R$ is the metric Ricci curvature scale and $f$ a general function, has been scrutinized motivated by the possibility to account for the self-accelerated cosmic expansion without invoking dark energy sources \cite{sotiriou1,defelice1}. In this approach of $f(R)$ gravity, and using its equivalent scalar-tensor representation, one can show that in order to satisfy local, i.e., solar system, observational constraints a large mass of the scalar field is required, 
which scales with the curvature through the chameleon mechanism \cite{khoury1,khoury2}. In turn, this has undesirable effects at cosmological scales. A Palatini version of the  $f(R)$ gravity theory, where the connection rather than the metric represents the fundamental gravitational field, has been proposed \cite{olmo1}, and it has been established that it has interesting features but deficiencies and downsides of the metric $f(R)$ gravity also appear.

A hybrid combination of the two versions of the  $f(R)$ gravity theories, containing elements from both of
the two formalisms, i.e., a hybrid metric-Palatini theory, consists of adding a $f({\cal R})$ term
constructed à la Palatini to the Einstein-Hilbert Lagrangian \cite{harko1}. It turns out to be very successful in accounting for the observed phenomenology and is able to avoid some of the shortcommings of the original approaches \cite{capozziello3}. In the scalar-tensor representation of the hybrid metric-Palatini theory there is a long-range light mass scalar field, which is able to
modify, in a way consistent with the observations, the cosmological and galactic dynamics, but leaves the solar system unaffected \cite{capozziello4}. This light scalar field thus allows to evade the screening chameleon mechanism.

The understanding and the building of solutions in complex theories like the generalized hybrid metric-Palatini gravity, and its scalar-tensor representation,  is a difficult task. Among other methods to find solutions, the reconstruction technique method has been often valuable in this search. This method consists in imposing an ansatz for the solution \textit{a priori} and verifying which are the conditions for the other variables e.g. the scalar fields that support this solution, instead of the other way around. The method was first employed in \cite{lucchin1} in order to select the form for the inflation potential able to resolve the open problems of the inflationary paradigm, e.g., the graceful exit. Also other elegant attempts were made to generalize this approach to nonminimally coupled scalar-tensor theories with a single scalar field \cite{ellis1}.

In this chapter, we aim to find cosmological solutions in the generalized hybrid metric-Palatini gravity proposed in \cite{tamanini1} through the use of its scalar-tensor representation.  We will then
devise a reconstruction technique algorithm for scalar-tensor theories in a Friedmann-Lema\^itre-Robsertson-Walker (FLRW) universe and we will show that the generalized theory through its scalar-tensor
representation provides a very rich structure in astrophysical and cosmological applications.  In particular, we will find that the cosmology of the generalized hybrid metric-Palatini gravity can differ
in subtle ways from both general relativity and $f(R)$ gravity. In principle these differences can be used in combination with observational data to verify the viability of this class of theories.

\section{Cosmological equations}

In this section, we look for homogeneous and isotropic cosmological solutions described by the Friedmann-Lemaître-Robertson-Walker (FLRW) spacetime. In spherical coordinates $x^a=\left(t,r,\theta,\phi\right)$, the line element can be written as
\begin{equation}\label{metricflrw}
ds^2=-dt^2+a^2\left(t\right)\left[\frac{dr^2}{1-kr^2}+r^2d\theta^2+r^2\sin^2\theta d\phi^2\right],
\end{equation} 
where $a\left(t\right)$ is the scale factor of the universe. We also assume that matter can be described by an isotropic perfect fluid, i.e. $T^a_{b}=\text{diag}\left(-\rho,p,p,p\right)$, so that the trace is given by $T=-\rho+3p$, where $\rho\left(t\right)$ is the energy density and $p\left(t\right)$ is the pressure of the fluid, assumed to be only functions of the time $t$. We shall be working with the scalar-tensor representation of the theory, i.e. the field equations are given by Eq.\eqref{ghmpgstfield} and the scalar field equations are given by Eqs.\eqref{ghmpgstkgphi} and \eqref{ghmpgstkgpsi}, where $\varphi\left(t\right)$ and $\psi\left(t\right)$ are also assumed to be only functions of the time $t$.

The modified cosmological equations can be obtained by computing the two independent components of the field equations Eq.\eqref{ghmpgstfield} for the metric in Eq.\eqref{metricflrw}. We thus obtain:
\begin{equation}\label{cosfield1}
3\left(\frac{\dot a}{a}\right)^2+\frac{3k}{a^2}=\frac{1}{\varphi-\psi}\left\{\kappa^2\rho+\frac{V}{2}
+3\left[\frac{\dot\psi^2}{4\psi}-\left(\frac{\dot a}{a}\right)\left(\dot\varphi-\dot\psi\right)\right]\right\},
\end{equation}
\begin{equation}\label{cosfield2}
2\frac{d}{dt} {\left(\frac{\dot a}{a}\right)}-\frac{2k}{a^2}=\frac{1}{\varphi-\psi}\left[-\kappa^2\left(\rho+p\right)-\frac{3\dot\psi^2}{2\psi}+\left(\frac{\dot a}{a}\right)\left(\dot\varphi-\dot\psi\right)-\left(\ddot\varphi-\ddot\psi\right)\right],
\end{equation}
where a dot ($\ \dot{}\ $) denotes a derivative with respect to time. Eq.\eqref{cosfield1} is called the modified Friedmann equation, and Eq.\eqref{cosfield2} is called the modified Raychaudhuri equation. The evolution equations for the scalar fields given by Eqs.\eqref{ghmpgstkgphi} and \eqref{ghmpgstkgpsi} become then
\begin{equation}\label{coskgphi}
\ddot\varphi+3\left(\frac{\dot a}{a}\right)\dot\varphi-\frac{1}{3}\left[2V-\psi V_\psi-\varphi V_\varphi\right]=-\frac{\kappa^2T}{3},
\end{equation}
\begin{equation}\label{coskgpsi}
\ddot\psi+3\left(\frac{\dot a}{a}\right)\dot\psi-\frac{\dot\psi^2}{2\psi}+\frac{\psi}{3}\left(V_\varphi+V_\psi\right)=0,
\end{equation}
respectively.

Now, an equation specifically for the potential can be obtained as follows: multiplying Eq.\eqref{cosfield2} by $3/2$, summing it to Eq.\eqref{cosfield1}, and using Eqs.\eqref{coskgphi} and \eqref{coskgpsi} to cancel the terms depending on $\ddot \varphi$ and $\ddot\psi$. The resultant equation is
\begin{equation}\label{cospotential}
V_\varphi=6\left[{\frac{d}{dt} \left(\frac{\dot a}{a}\right)} +2\left(\frac{\dot a}{a}\right)^2+\frac{k}{a^2}\right].
\end{equation}
Note that the right-hand side of Eq.\eqref{cospotential} corresponds to the Ricci scalar $R$ for the metric in Eq.\eqref{metricflrw}, and thus this result is in agreement with Eq.\eqref{ghmpgsteomphi}. This equation is useful to replace Eq.\eqref{cosfield1} in the final system of equations due to its simplicity in comparison.

Finally, we need the equations that describe the evolution of matter. These equations are the equation of state and the conservation law for the stress-energy tensor $\nabla_aT^{ab}=0$, which are given by
\begin{equation}\label{coseos}
p=w\rho,
\end{equation}
\begin{equation}\label{costabcons}
\dot\rho=-3\,\frac{\dot a}{a}\,\rho\left(1+w\right),
\end{equation}
respectively, where $w$ is a dimensionless parameter that describes the type of matter and Eq.\eqref{coseos} was used to cancel the terms depending on $p$ in Eq.\eqref{costabcons}. Eqs.\eqref{costabcons} and \eqref{cosfield2} are not independent and thus we shall work with the first one due to its simpler structure. We now have a system of five independent equations, namely Eqs.\eqref{coskgphi} to \eqref{costabcons} for eight independent variables $k, a,\varphi, \psi, V, p, \rho$ and $w$. In what follows, we choose a specific geometry i.e. we set a value for $k$, and leave $w$ free. This implies that we can still impose an extra constraint to close the system and have a unique solution.

It is also useful to define the Hubble function or parameter $H$ as usual
\begin{equation}\label{defhubble}
H=\frac{\dot a}{a},
\end{equation}
which is a function of the time $t$ only. With Eq.\eqref{defhubble} we can exchange $a\left(t\right)$ for $H\left(t\right)$ and vice versa whenever it is appropriate. 

\section{Cosmological solutions}

In this section we shall use the reconstruction methods to compute cosmological solutions. We start by considering the simplest case of flat ($k=0$) and vacuum ($\rho=p=0$) solutions and impose constraints on both the scale factor $a\left(t\right)$ or the potential $V$ that simplify the equations. We then generalize our approach to an arbitrary geometry $k$ and still impose constraints on the scale factor. To finalize, we use the fact that only the scalar field $\varphi$ is coupled to matter (see Eq.\eqref{coskgphi} to provide a method of generalizing the vacuum solutions obtained to solutions with perfect-fluid matter).

\subsection{Flat and vacuum cosmological solutions}
\label{sec:cosflat}

Let us start by considering the simplest possible solutions, for which the geometry is flat ($k=0$) and the matter distribution is vacuum ($\rho=p=0$). In this case, Eqs.\eqref{coseos} and \eqref{costabcons} become identities and we are left with a system of three independent equations for the four independent variables $a, \varphi, \psi$ and $V$. Eqs.\eqref{coskgphi} to \eqref{cospotential} become then
\begin{equation}\label{cosfvkgphi}
\ddot\varphi+3\left(\frac{\dot a}{a}\right)\dot\varphi-\frac{1}{3}\left[2V-\psi V_\psi-\varphi V_\varphi\right]=0,
\end{equation}
\begin{equation}\label{cosfvkgpsi}
\ddot\psi+3\left(\frac{\dot a}{a}\right)\dot\psi-\frac{\dot\psi^2}{2\psi}+\frac{\psi}{3}\left(V_\varphi+V_\psi\right)=0
\end{equation}
\begin{equation}\label{cosfvpotential}
V_\varphi=6\left[{\frac{d}{dt} \left(\frac{\dot a}{a}\right)} +2\left(\frac{\dot a}{a}\right)^2\right]
\end{equation}
We provide four cosmological solutions for this system of equations. We start by imposing a de-Sitter solution as a constraint, then we choose particular combinations of the scale factor terms in the potential equation for which $V_\varphi$ is simplified, and we finalize by considering a specific form of the potential $V$ for which the scalar field equations are simplified.

\subsubsection{The de-Sitter solution}

\begin{figure}
\centering 
\begin{subfigure}{0.48\textwidth}
\includegraphics[scale=0.65]{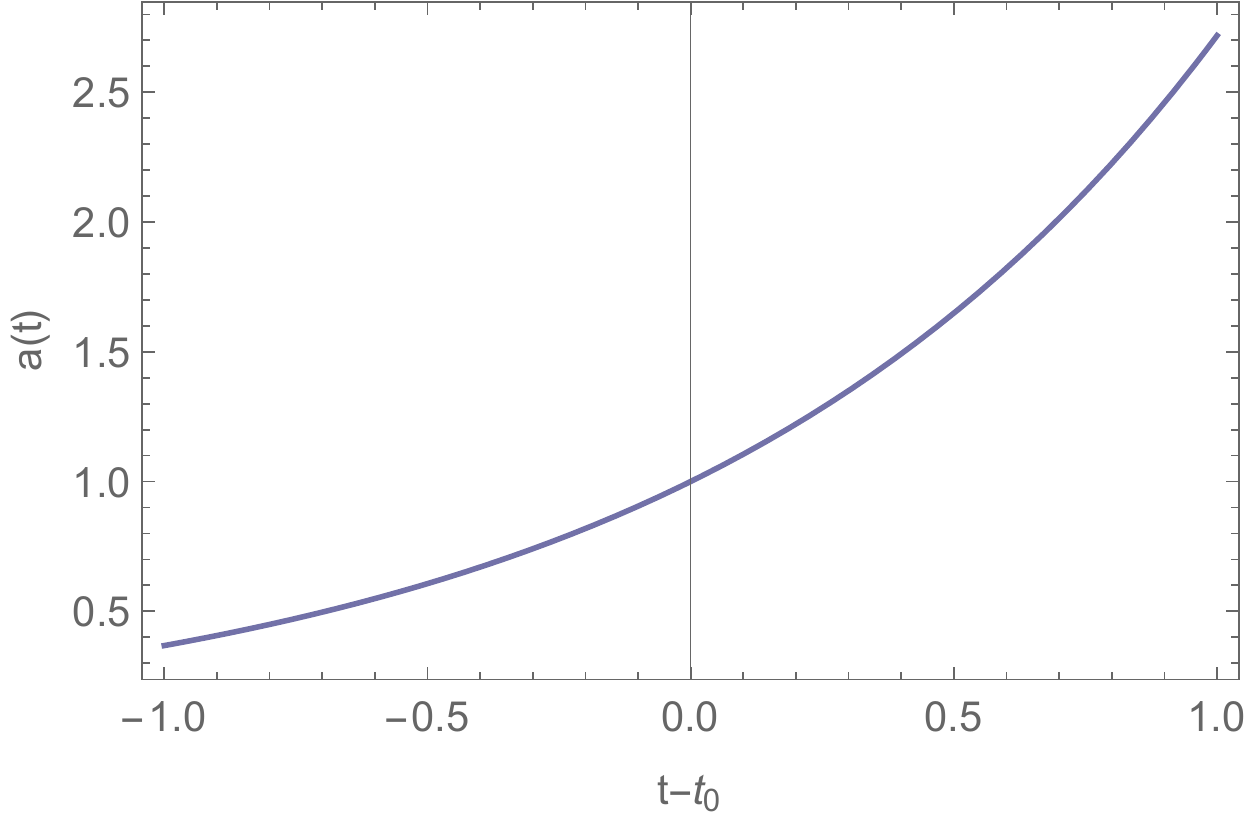}
\caption{Scale factor $a(t)$}
\label{fig:cosscale1}
\end{subfigure}
\ \\
\ \\
\ \\
\begin{subfigure}{0.48\textwidth}
\includegraphics[scale=0.65]{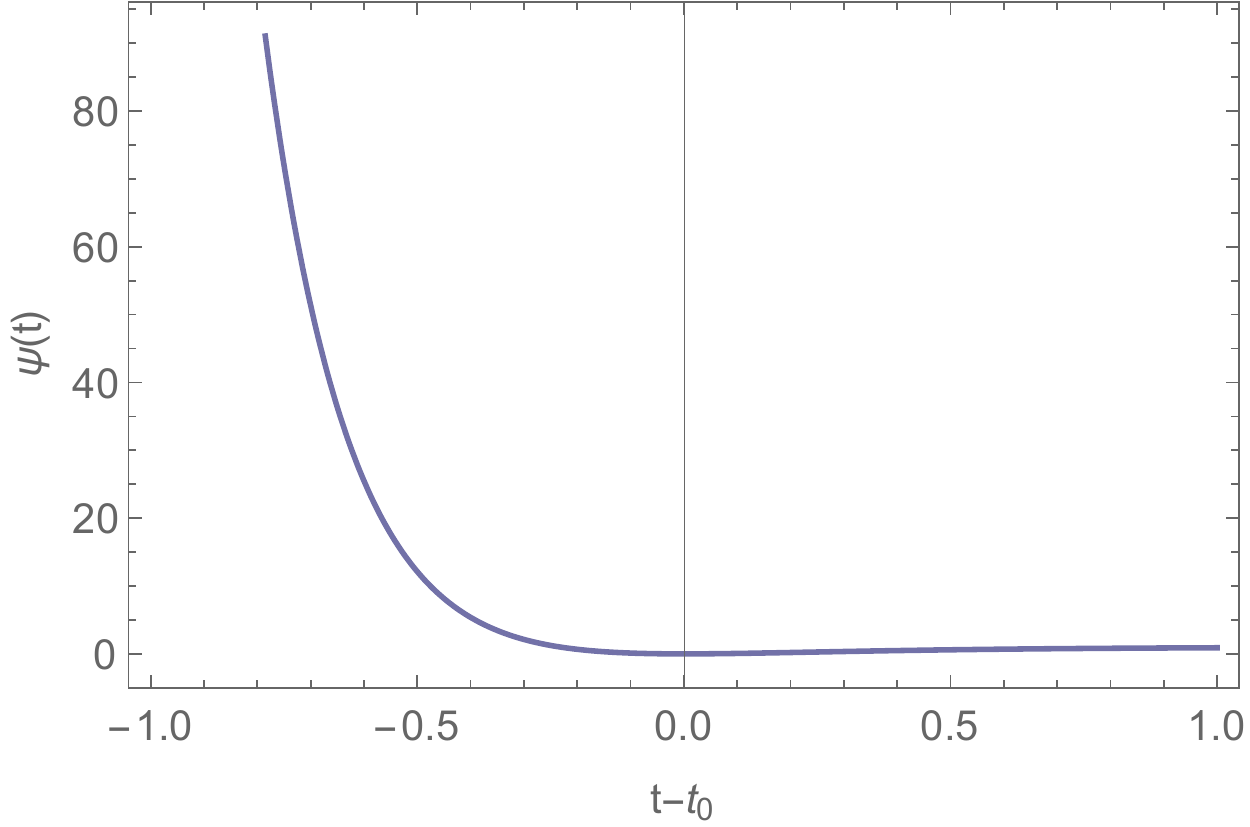}
\caption{Scalar field $\psi(t)$}
\label{fig:cospsi1}
\end{subfigure}
\ \ \ \ \ 
\begin{subfigure}{0.48\textwidth}
\includegraphics[scale=0.65]{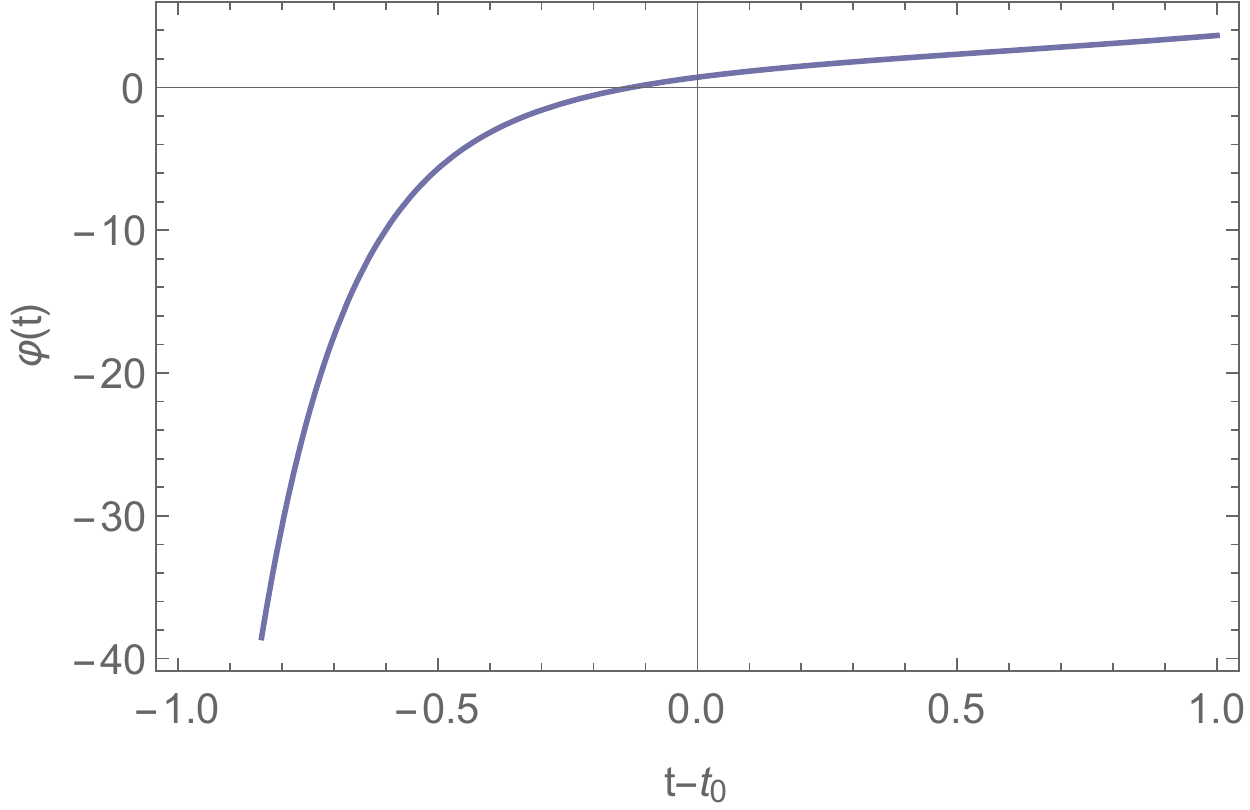}
\caption{Scalar field $\varphi(t)$}
\label{fig:cosphi1}
\end{subfigure}
\caption{Solutions for the scale factor $a\left(t\right)$ in Eq.~\eqref{cosscale1}, the scalar field $\psi\left(t\right)$ in Eq.\eqref{cospsi1}, and the scalar field $\varphi\left(t\right)$ in Eq.\eqref{cosphi1}, with $a_0=\Lambda=\psi_0=\varphi_0=\varphi_1=1$, $t_0=0$ }
\end{figure}

In this section, we try to find a solution of the form of a de-Sitter expansion. The scale factor for this solution behaves as an exponential function of the form
\begin{equation}\label{cosscale1}
a\left(t\right)=a_0e^{\sqrt\Lambda\, \left(t-t_0\right)}\,,
\end{equation}
where $a_0$ and $\Lambda$ are free parameters. The parameter $a_0$ represents the scale factor at the initial time $t=t_0$, and $\Lambda$ can be seen as playing the role of a cosmological constant. This scale factor is plotted in Fig. \ref{fig:cosscale1}.

With these assumptions, Eq.\eqref{cosfvpotential} becomes simply $V_\varphi=12\Lambda$. This equation can be directly integrated with respect to $\varphi$ to obtain a potential of the form
\begin{equation}
V\left(\varphi,\psi\right)=12\Lambda\varphi+b\left(\psi\right),
\end{equation}
where $b\left(\psi\right)$ is an arbitrary function of $\psi$ which arises from the fact that the potential is a function of both $\varphi$ and $\psi$. Inserting this potential back into Eqs.\eqref{cosfvkgphi} and \eqref{cosfvkgpsi} leads to the equations
\begin{equation}\label{coskgphi1}
\ddot\varphi+3\sqrt\Lambda\,\dot\varphi-\frac{1}{3}\left[12\Lambda\,\varphi+2b\left(\psi\right)-\psi b'\left(\psi\right)\right]=0,
\end{equation}
\begin{equation}\label{coskgpsi1}
\ddot\psi+3\sqrt\Lambda\,\dot\psi-\frac12\frac{\dot\psi^2}{\psi}+\frac13\,\psi\left[12\Lambda+b'\left(\psi\right)\right]=0,
\end{equation}
respectively. Since the coupling between the scalar fields $\varphi$ and $\psi$ with curvature is given by $\left(\varphi-\psi\right)R$ (see Eq.\eqref{ghmpgstaction}), then one could assume that the potential, being an alternative representation of the $f\left(R,\mathcal R\right)$ function, depends on both fields in the same way as the coupling to gravity. This is not a mandatory step but it is instead a choice made to simplify the resultant equations and impose a symmetry on the potential $V$. In this particular case, we choose $b\left(\psi\right)= -12\Lambda\psi$, the potential takes the final form
\begin{equation}\label{cospot1}
V\left(\varphi,\psi\right)=12\Lambda\left(\varphi-\psi\right),
\end{equation}
and Eq. \eqref{coskgpsi1} can be directly integrated over time $t$ to obtain the solution for $\psi$ as
\begin{equation}\label{cospsi1}
\psi\left(t\right)=\psi_0 e^{-6 \sqrt\Lambda \, t} \left[e^{3\sqrt \Lambda \, \left(t-t_0\right)}-1\right]^2,
\end{equation}
where $t_0$ and $\psi_0$ are constants of integration. Inserting Eq.\eqref{cospsi1} into Eq.\eqref{coskgphi1} and considering the same form of the potential $V$ one can integrate directly the result over time $t$ to obtain the solution
\begin{equation}\label{cosphi1}
\varphi\left(t\right)=\psi_0 e^{-6  \sqrt\Lambda \,t_0}-\frac{2}{7} \psi_0 e^{-6 \sqrt\Lambda  t}-2 \psi_0 e^{-3 \sqrt\Lambda  \left(t+t_0\right)}+\varphi_0 e^{-4 \sqrt\Lambda  t}+\varphi_1 e^{\sqrt\Lambda  t},
\end{equation}
where $\varphi_0$ and $\varphi_1$ are constants of integration. The solutions for the scalar fields $\psi\left(t\right)$ and $\varphi\left(t\right)$ are plotted in Figs. \ref{fig:cospsi1} and \ref{fig:cosphi1}, respectively. The solution is complete since all the variables are known.

\subsubsection{Solution with a simplified potential equation I}

\begin{figure}
\centering 
\begin{subfigure}{0.48\textwidth}
\includegraphics[scale=0.65]{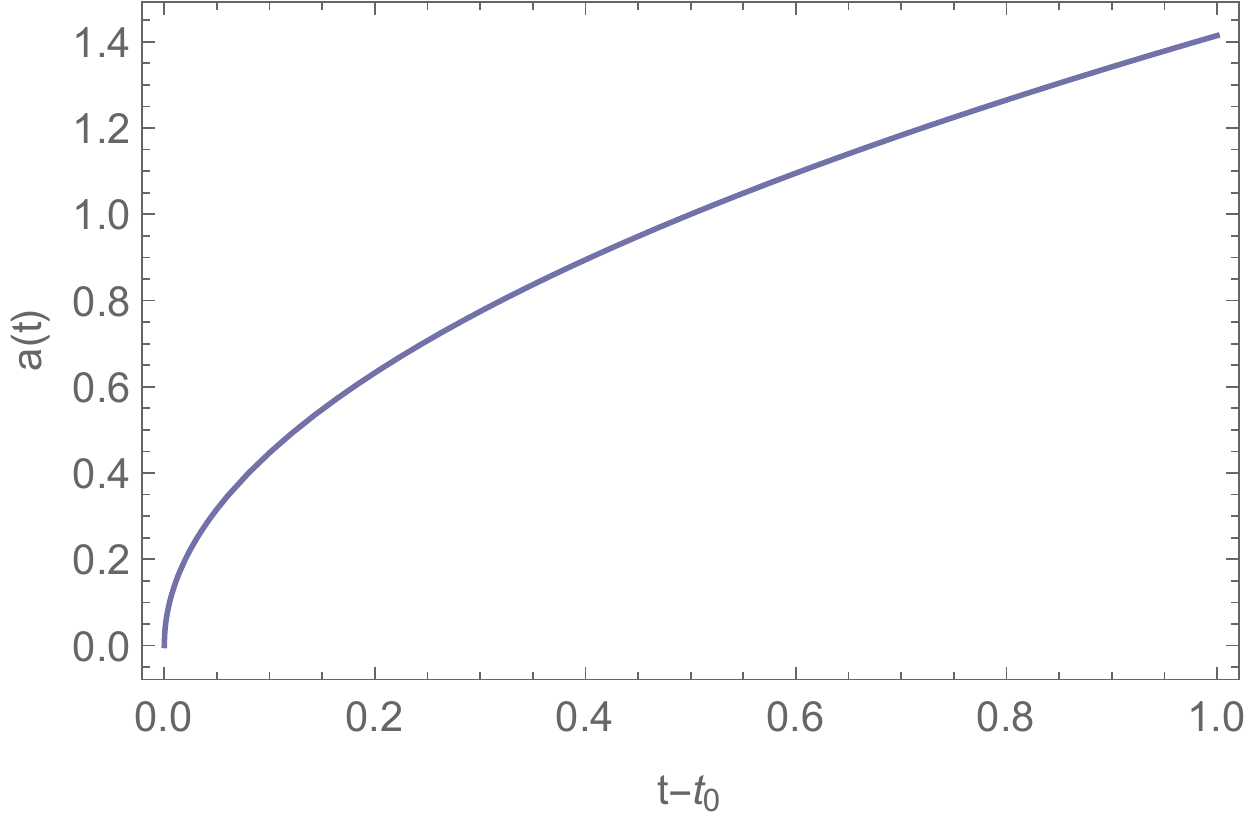}
\caption{Scale factor $a(t)$}
\label{fig:cosscale2}
\end{subfigure}
\ \\
\ \\
\ \\
\begin{subfigure}{0.48\textwidth}
\includegraphics[scale=0.65]{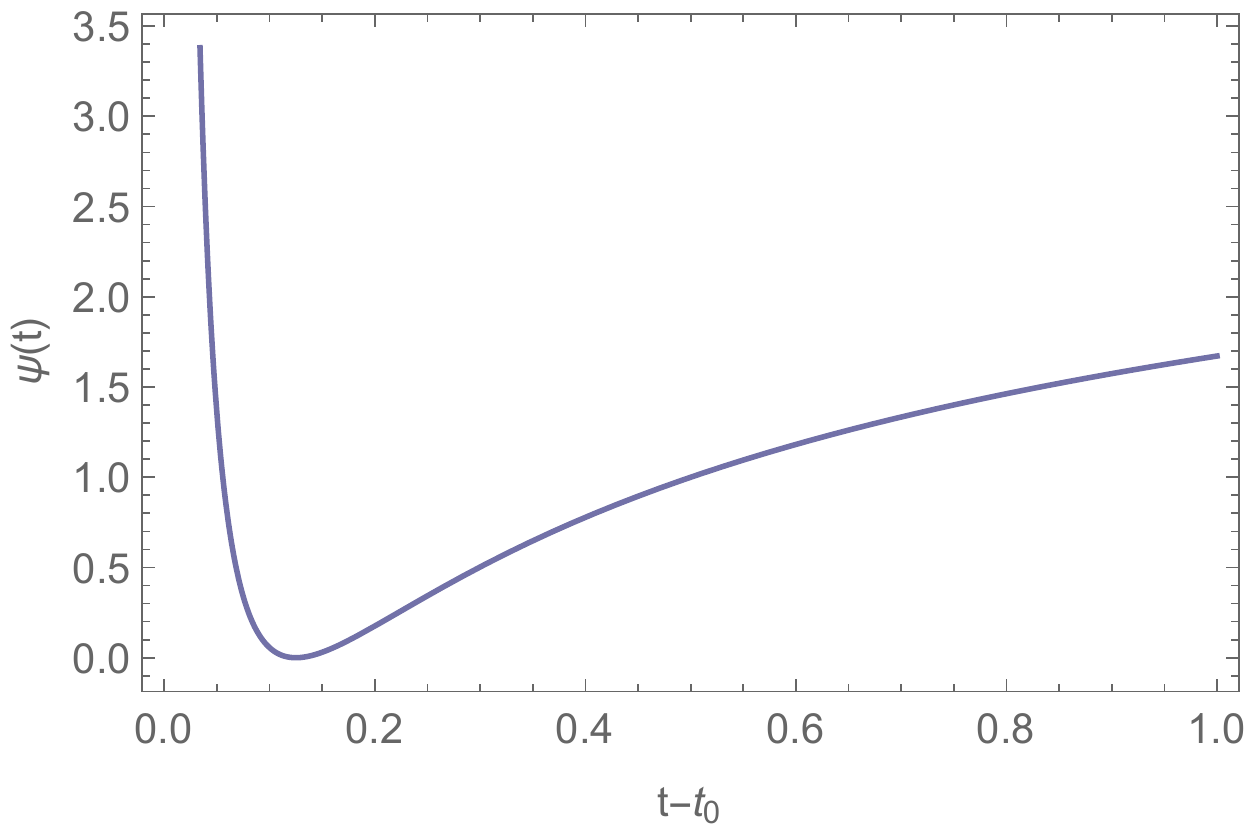}
\caption{Scalar field $\psi(t)$}
\label{fig:cospsi2}
\end{subfigure}
\ \ \ \ \ 
\begin{subfigure}{0.48\textwidth}
\includegraphics[scale=0.65]{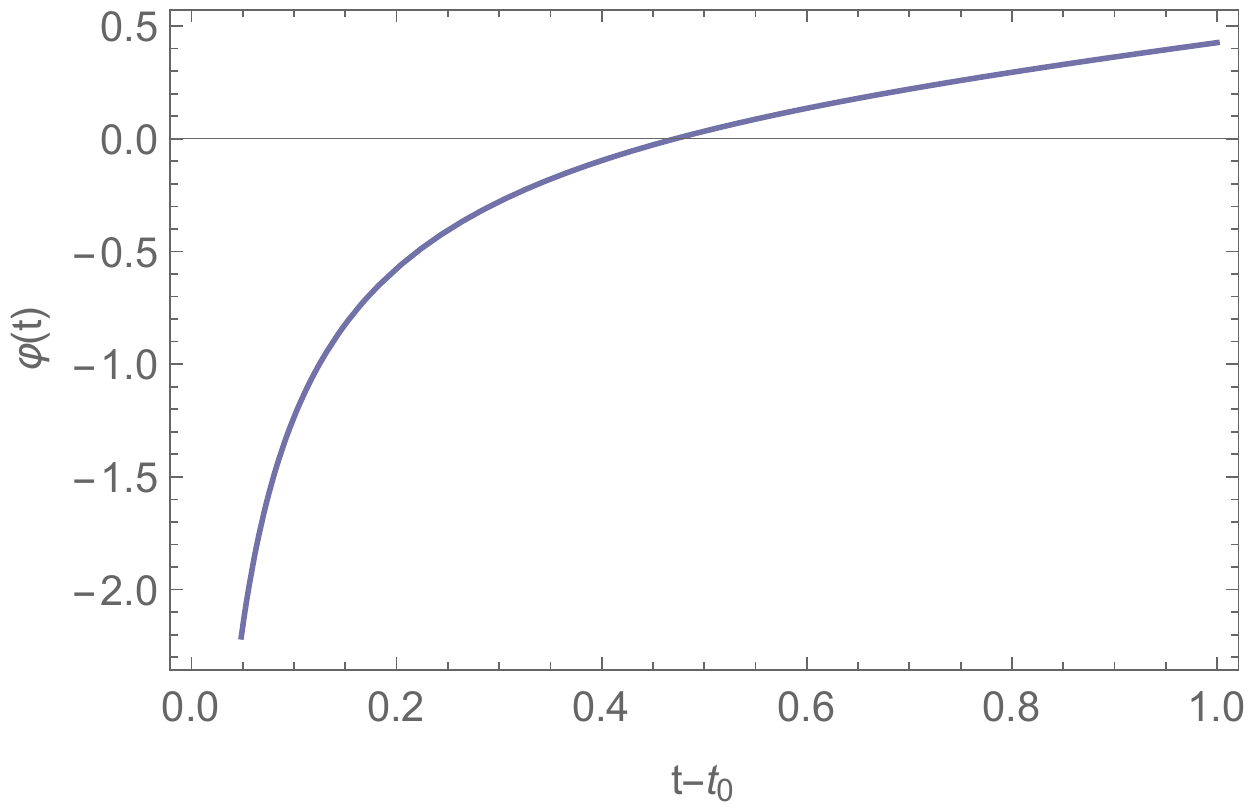}
\caption{Scalar field $\varphi(t)$}
\label{fig:cosphi2}
\end{subfigure}
\caption{Solutions for the scale factor $a\left(t\right)$ in Eq.~\eqref{cosscale2}, the scalar field $\psi\left(t\right)$ in Eq.\eqref{cospsi2}, and the scalar field $\varphi\left(t\right)$ in Eq.\eqref{cosphi2}, with $a_0=\psi_0=\sqrt{2}$, $\psi_1=1/2$, $t_0=0$}
\end{figure}

Let us now try to find a solution by constraining the scale factor indirectly. Inspecting Eq.\eqref{cosfvpotential}, we can see that it simplifies if the first two terms within the brackets cancel each other, i.e., 
\begin{equation}\label{cosconst2}
\dot H+2H^2=0,
\end{equation}
where we have used the definition for the Hubble function $H$, Eq.~\eqref{defhubble}. Note that Eq.\eqref{cosconst2} is a second order differential equation for the scale factor $a\left(t\right)$ and can be directly integrated twice with respect to the time $t$ in order to obtain
\begin{equation}\label{cosscale2}
a\left(t\right)=a_0\sqrt{t-t_0},
\end{equation}
where $a_0$ and $t_0$ are integration constants. This solution is plotted in Fig.\ref{fig:cosscale2}.
The interest in this solution resides in the fact that, for a universe populated by radiation, we expect that the behavior of the scale factor is proportional to $\sqrt{t}$, but in this case the same behavior can be obtained in vacuum, i.e., with $\rho=p=0$.

With this set of assumptions, the equation for the potential, Eq.~\eqref{cosfvpotential}, becomes simply $V_\varphi=0$, which can be directly integrated with respect to $\varphi$ to obtain
\begin{equation}
V\left(\varphi,\psi\right)=b\left(\psi\right),
\end{equation}
where $b\left(\psi\right)$ is again an arbitrary function on $\psi$ that arises from the fact that the potential is a function of both $\varphi$ and $\psi$ in general. Inserting this potential along with the solution for the scale factor given in Eq.\eqref{cosscale2} into Eq.\eqref{cosfvkgphi} and \eqref{cosfvkgpsi} yields
\begin{equation}\label{coskgphi2}
\ddot\varphi+\frac{3\dot\varphi}{2\left(t-t_0\right)}-\frac{1}{3}\left[2b\left(\psi\right)-\psi b'\left(\psi\right)\right]=0,
\end{equation}
\begin{equation}\label{coskgpsi2}
\ddot\psi+\frac{3\dot\psi}{2\left(t-t_0\right)}-\frac{\dot\psi^2}{2\psi}+\frac{\psi}{3}b'\left(\psi\right)=0,
\end{equation}
respectively. Again, we shall assume that the dependence of the potential $V$ in the scalar fields is of the form $\left(\varphi-\psi\right)$, and thus select a specific form for the function $b\left(\psi\right)$ and $b\left(\psi\right)=V_0$, for some constant $V_0$, and the potential takes the final form $V\left(\varphi,\psi\right)=V_0$. Under these conditions, one can directly integrate Eq.\eqref{coskgpsi1} and obtain the solution
\begin{equation}\label{cospsi2}
\psi\left(t\right)=\frac{\psi_1 \left(\psi_0 \sqrt{t-t_0}-1\right)^2}{t-t_0}
\end{equation}
where $\psi_0$ and $\psi_1$ are constants of integration. Under the same assumptions, Eq.\eqref{coskgphi2} decouples completely from $\psi$ and one can directly integrate the result over the time $t$ and get
\begin{equation}\label{cosphi2}
\varphi\left(t\right)=\frac{2}{15}V_0t\left(t-2t_0\right)-\frac{\varphi_0}{\sqrt{t-t_0}}+\varphi_1,
\end{equation}
where $\varphi_0$ and $\varphi_1$ are constants of integration. The solutions for the scalar field $\psi\left(t\right)$ and $\varphi\left(t\right)$ are plotted in Figs. \ref{fig:cospsi2} and \ref{fig:cosphi2}, respectively. The solution is complete since all the variables have been obtained.

\subsubsection{Solution with a simplified potential equation II}

\begin{figure}
\centering 
\begin{subfigure}{0.48\textwidth}
\includegraphics[scale=0.65]{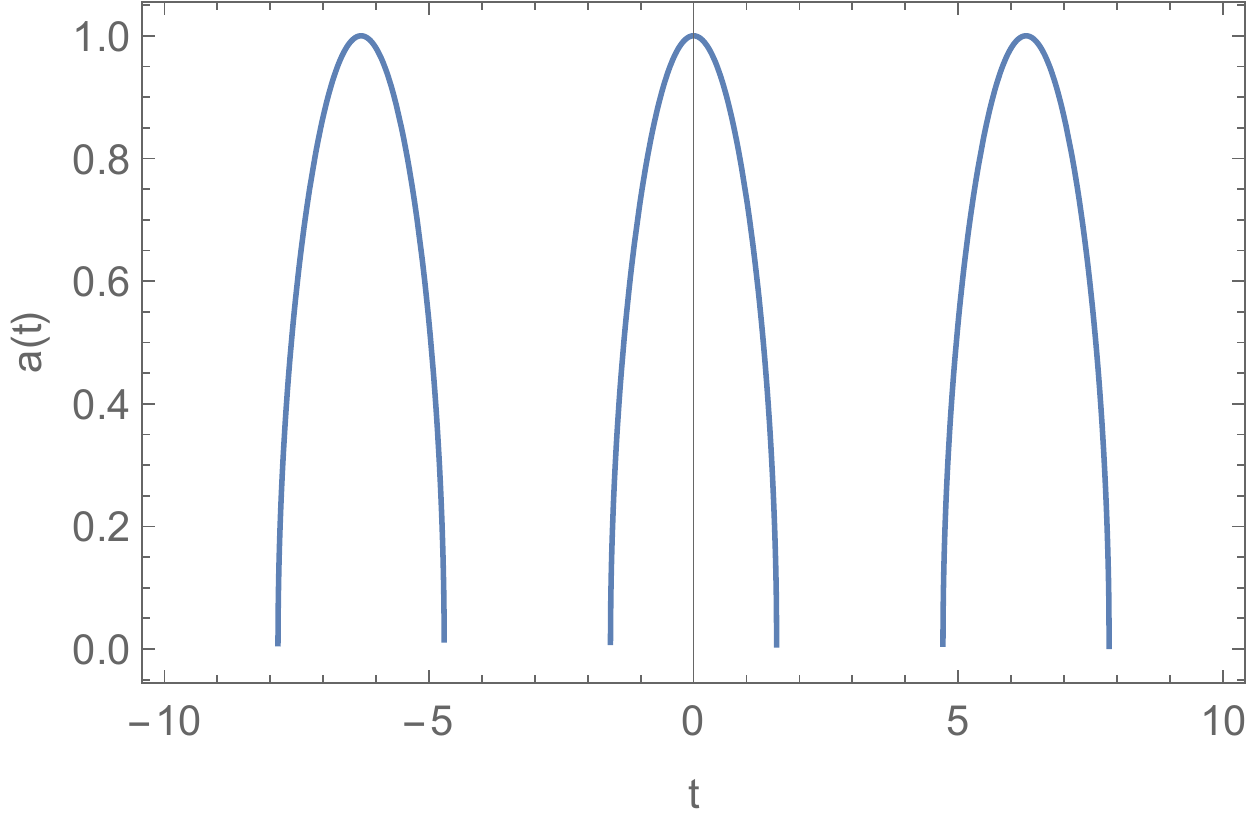}
\caption{Scale factor $a(t)$}
\label{fig:cosscale3}
\end{subfigure}
\ \\
\ \\
\ \\
\begin{subfigure}{0.48\textwidth}
\includegraphics[scale=0.65]{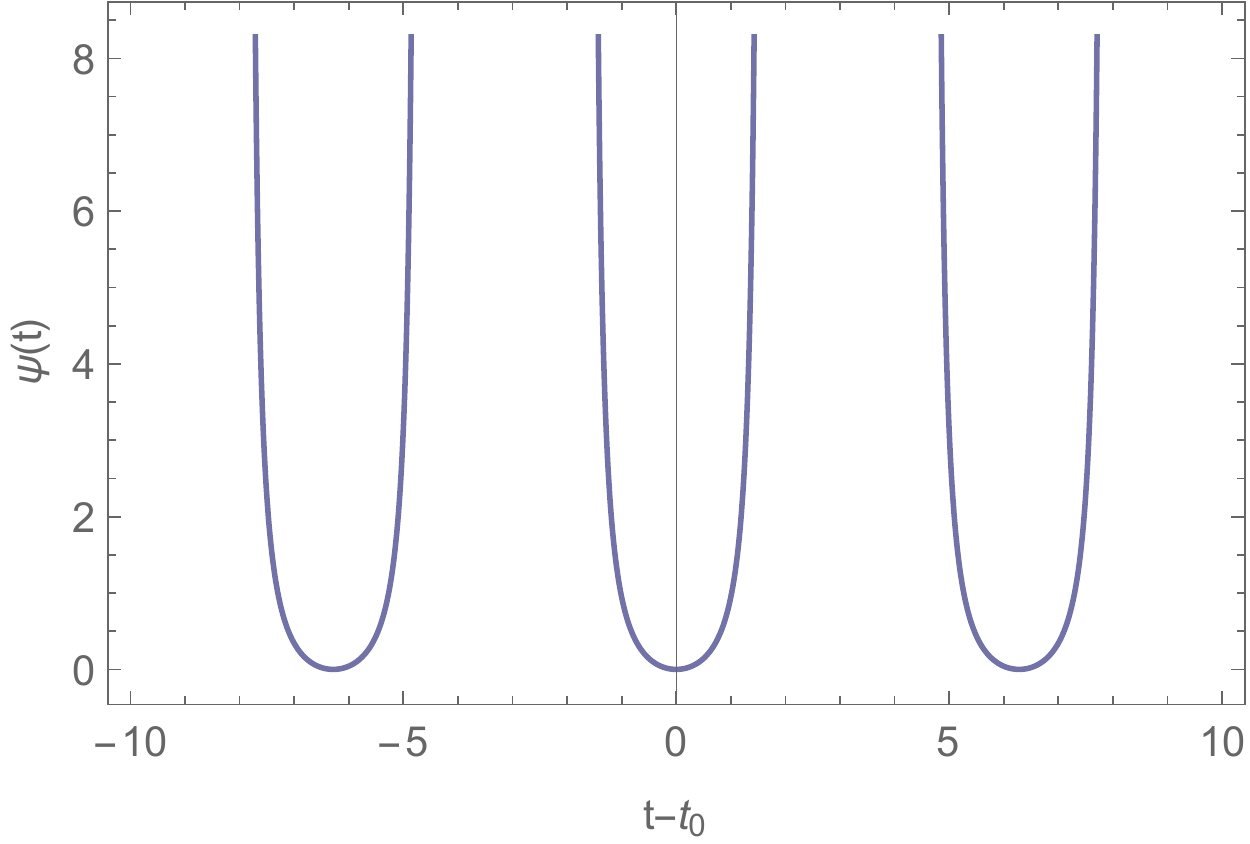}
\caption{Scalar field $\psi(t)$}
\label{fig:cospsi3}
\end{subfigure}
\ \ \ \ \ 
\begin{subfigure}{0.48\textwidth}
\includegraphics[scale=0.65]{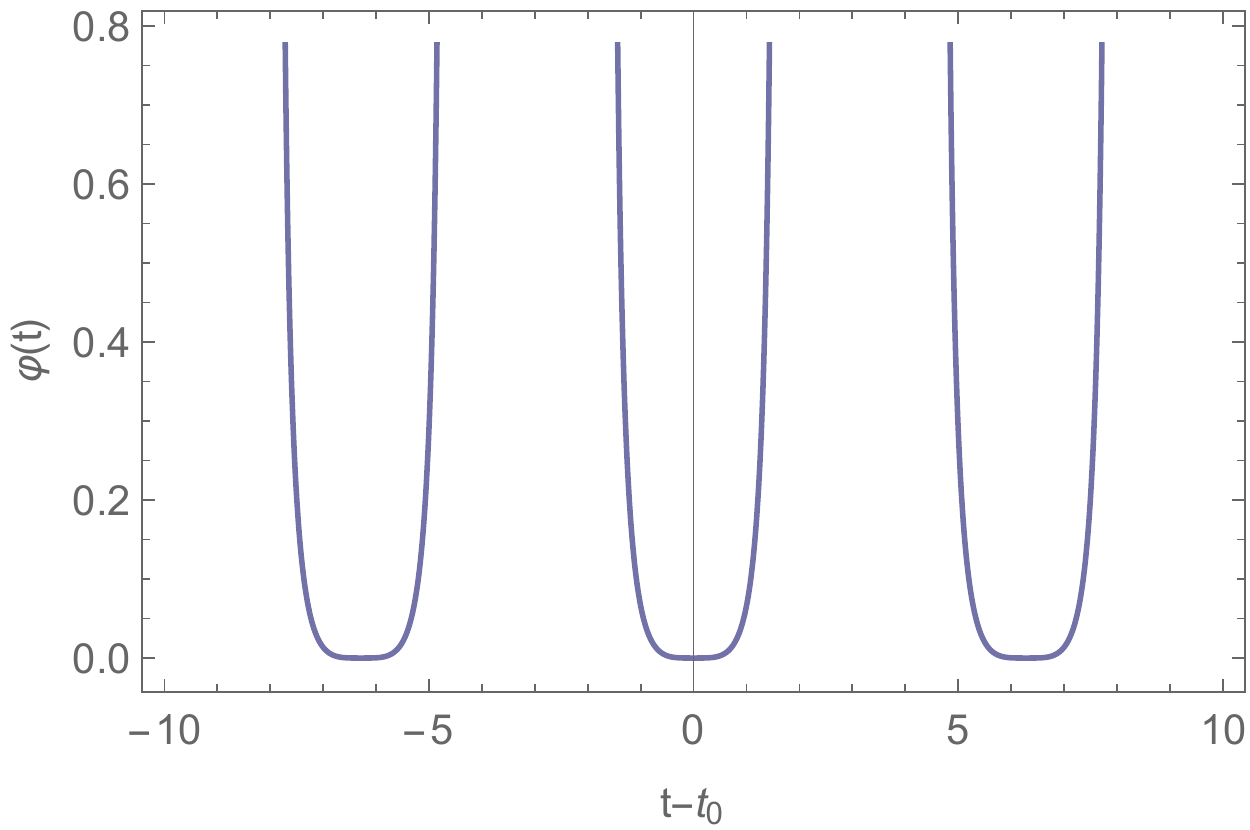}
\caption{Scalar field $\varphi(t)$}
\label{fig:cosphi3}
\end{subfigure}
\caption{Solutions for the scale factor $a\left(t\right)$ in Eq.~\eqref{cosscale3}, the scalar field $\psi\left(t\right)$ in Eq.\eqref{cospsi3}, and the scalar field $\varphi\left(t\right)$ in Eq.\eqref{coskgphi3}, with $a_0=\Omega=\psi_1=1$, $\psi_0=\varphi_0=\dot{\varphi_0}=t_0=0$}
\end{figure}

In this section we repeat the process of indirectly constraining the scale factor using Eq.\eqref{cosfvpotential} but instead of considering that the two terms inside the brackets cancel each other, let us assume that they are constant, as follows:
\begin{equation}\label{cosconst3}
\dot H+2H^2+\frac{\Omega^2}{2}=0,
\end{equation}
for some constant $\Omega$, where the exponent and the factor $2$ appear for later convenience, and where we have used the definition for the Hubble function given in Eq.\eqref{defhubble}. Eq.\eqref{cosconst3} is again a second order differential equation for the scale factor $a\left(t\right)$ and can be directly integrated twice over the time $t$ to yield a solution of the form
\begin{equation}\label{cosscale3}
a\left(t\right)=a_0\sqrt{\cos\left[\Omega\left(t-t_0\right)\right]},
\end{equation}
where $a_0$ and $t_0$ are constants of integration related by $a\left(t=t_0\right)=a_0$, and $a_0>0$ is defined positive. The behavior of $a(t)$ in Eq.~\eqref{cosscale3} is plotted in Fig.~\ref{fig:cosscale3}. This solution is valid in between times 
$t=-\frac{\pi}{2\Omega}+t_0$ and $t=\frac{\pi}{2\Omega}+t_0$, or times that are translated from these ones
by $2\pi n$ with integer $n$. In between  times $t=\frac{\pi}{2\Omega}+t_0$ and $t=\frac{3\pi}{2\Omega}+t_0$, or times that are translated from these ones by $2\pi n$, there are no physical solutions, since the scale factor in Eq.~\eqref{cosscale3} becomes a complex number. The solution thus represents a universe which starts expanding at $t=-\frac{\pi}{2\Omega}+t_0$, attains a maximum value at $t=t_0$, and then collapses again at $t=\frac{\pi}{2\Omega}+t_0$. The importance of this solution for $a\left(t\right)$ stands on the fact that in the later times of the universe expansion, where we might approximate the distribution of matter to be vacuum and the geometry to be flat, it is possible to obtain solutions for the scale factor with a dependence on time different than the usual $t^{1/2}$ power-law expected in standard general relativity.

The assumption given in Eq. \eqref{cosconst3} leaves the equation for the potential, Eq.~\eqref{cosfvpotential}, in the form $V_\varphi=-3\Omega^2$, which can be directly integrated over $\varphi$ to obtain
\begin{equation}
V\left(\varphi,\psi\right)=-3\Omega^2\varphi+b\left(\psi\right),
\end{equation}
where $b\left(\psi\right)$ is an arbitrary function of $\psi$ which arises from the fact that the potential is a function of both $\varphi$ and $\psi$ in general. Inserting this potential into the scalar field equations Eqs.\eqref{cosfvkgphi} and \eqref{cosfvkgpsi}, one obtains respectively
\begin{equation}\label{coskgphi3}
\ddot\varphi-\frac32\Omega \tan\left[\Omega\left(t-t_0\right)\right]\dot\varphi+\frac{\psi}{3}b'\left(\psi\right)-\frac{2}{3}b\left(\psi\right)+\Omega^2\varphi=0,
\end{equation}
\begin{equation}\label{coskgpsi3}
\ddot\psi-\frac32\Omega \tan\left[\Omega\left(t-t_0\right)\right]\dot\psi-\frac{\dot\psi^2}{2\psi}+\frac{\psi}{3}\left[b'\left(\psi\right)-3\Omega^2\right]=0.
\end{equation}
Assuming again that the dependence of the potential $V$ in the scalar fields $\varphi$ and $\psi$ is of the form $\left(\varphi-\psi\right)$ to be consistent with the coupling between the fields and $R$, then one choses the particular form $b\left(\psi\right)=3\Omega^2\psi$ such that the potential becomes
\begin{equation}\label{cospot3}
V\left(\varphi,\psi\right)=-3\Omega^2\left(\varphi-\psi\right).
\end{equation}
Under these conditions, Eq.\eqref{coskgpsi3} can be directly integrated over time $t$ and an analytical solution for $\psi\left(t\right)$ arises as
\begin{equation}\label{cospsi3}
\psi\left(t\right)=\psi_1\sec\left[\Omega\left(t-t_0\right)\right]\left\{\sqrt{\cos\left[\Omega\left(t-t_0\right)\right]}\left[\psi_0\Omega-\sqrt{2}E\left({\frac{\Omega}{2}}\left(t-t_0\right)|2\right)\right]+\sqrt{2}\sin\left[\Omega\left(t-t_0\right)\right]\right\}^2
\end{equation}
where $\psi_0$ and $\psi_1$ are arbitrary constants of integration and $E\left(a|b\right)$ is the incomplete elliptic integral of the second kind defined as $E\left(a|b\right)=\int_0^a\sqrt{1-b\sin^2\theta}d\theta$. The solution in Eq.\eqref{cospsi3} is plotted in Fig. \ref{fig:cospsi3}. As for the scale factor $a(t)$ , the solution for $\psi(t)$ is valid in between times $t=-\frac{\pi}{2\Omega}+t_0$ and $t=\frac{\pi}{2\Omega}+t_0$, or times that are translated from these ones by $2\pi n$ with integer $n$. 

The equation for the scalar field $\varphi$ given in Eq.\eqref{coskgphi3}, using the potential given in Eq.\eqref{cospot3} and the solution for $\psi$ given in Eq.\eqref{cospsi3}, does not have an analytical solution. However, it can be integrated numerically. In Fig.~\ref{fig:cosphi3} we plot the  solution  for $\varphi(t)$ for a specific choice of the parameters and with $\varphi(0)\equiv\varphi_0=0$ and  $\dot{\varphi}(0)\equiv\dot{\varphi_0}=0$. \footnote{In doing so we note that, being Eq.~\eqref{coskgphi3} a second order ordinary linear differential equation, it might present complex solutions. By the superposition principle, however, we can always restrict ourselves to the particular solution associated to the real part of $\varphi$ without loss of generality ($\psi$ is defined real in the domain of integration). These solutions are the ones plotted in Fig.~\ref{fig:cosphi3}.} The solution is now complete since all the variables have been obtained.

\subsubsection{Solution with simplified scalar-field equations}

Let us now try to choose a constraint that simplifies the scalar field equations instead of the potential equation. Consider the potential depending terms in Eqs.\eqref{cosfvkgphi} and \eqref{cosfvkgpsi}. These terms vanish if we choose an ansatz for the potential of the form
\begin{equation}\label{cospot4}
V\left(\varphi,\psi\right)=V_0\left(\varphi-\psi\right)^2,
\end{equation}
for some constant $V_0$. Inserting Eq.\eqref{cospot4} into Eqs.\eqref{cosfvkgphi} and \eqref{cosfvkgpsi} and using Eq.\eqref{defhubble} we obtain
\begin{equation}\label{coskgphi4}
\ddot\varphi+3H\dot\varphi=0,
\end{equation}
\begin{equation}\label{coskgpsi4}
\ddot\psi+3H\dot\psi-\frac{\dot\psi^2}{2\psi}=0,
\end{equation}
respectively. Both Eqs.\eqref{coskgphi4} and \eqref{coskgpsi4} can be integrated directly over time to obtain both $\dot\varphi$ and $\dot\psi$ as functions of the scale factor $a\left(t\right)$ as
\begin{equation}\label{cosdphi4}
\dot\varphi=\frac{\varphi_0}{a^3},
\end{equation}
\begin{equation}\label{cosdpsi4}
\frac{\dot\psi}{\sqrt{\psi}}=\frac{\psi_0}{a^3},
\end{equation}
where $\varphi_0$ and $\psi_0$ are constants of integration. Solving Eqs.\eqref{cosdphi4} and \eqref{cosdpsi4} with respect to $a$ and equating the two equations one obtains a relation between $\dot\varphi$ and $\dot\psi$ as
\begin{equation}
\frac{\dot\psi}{\psi_0\sqrt{\psi}}=\frac{\dot\varphi}{\varphi_0},
\end{equation}
The previous relation can be integrated directly over time to provide a relationship between the scalar fields $\varphi$ and $\psi$
\begin{equation}\label{cospsiphi4}
\psi=\left(\frac{\psi_0}{2\varphi_0}\right)^2\left(\varphi-\varphi_1\right)^2,
\end{equation}
where $\varphi_1$ is a constant of integration. 

\begin{figure}
\centering 
\begin{subfigure}{0.48\textwidth}
\includegraphics[scale=0.65]{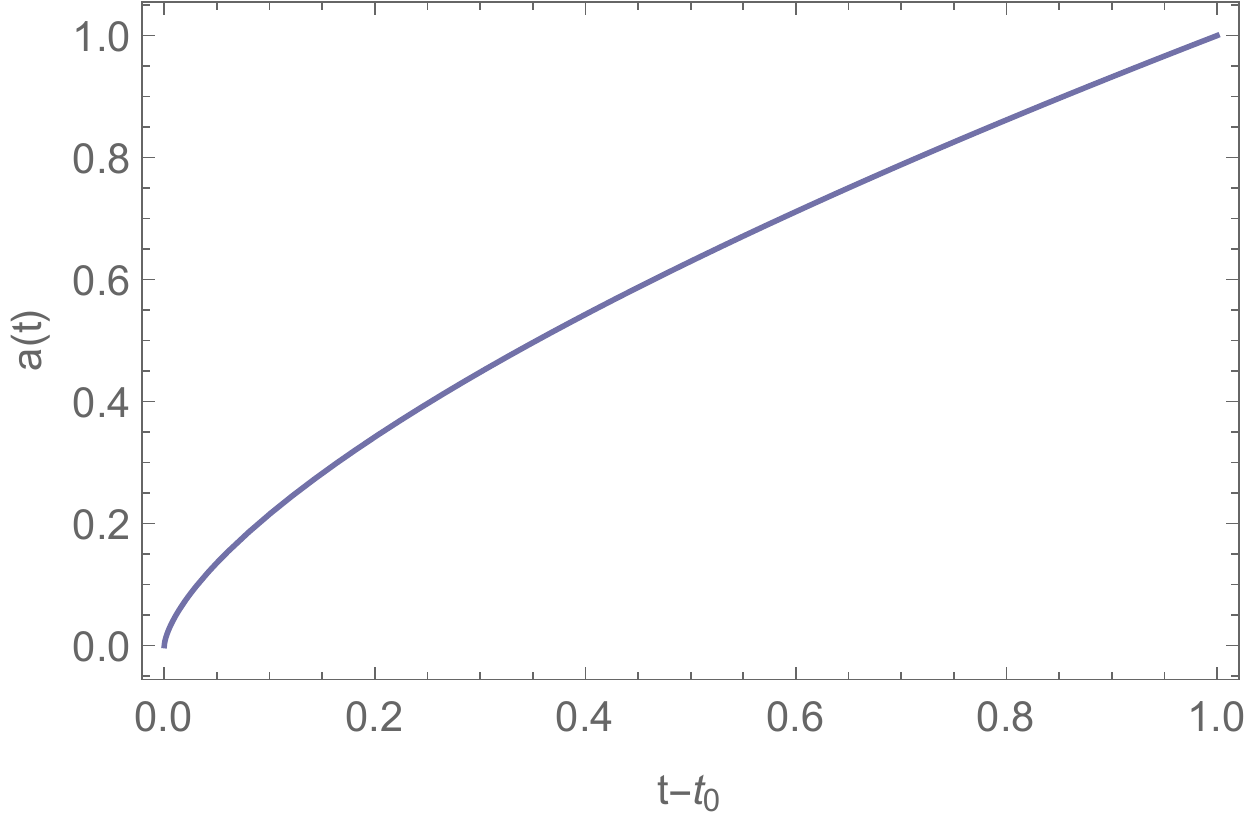}
\caption{Scale factor $a(t)$}
\label{fig:cosscale4}
\end{subfigure}
\ \\
\ \\
\ \\
\begin{subfigure}{0.48\textwidth}
\includegraphics[scale=0.65]{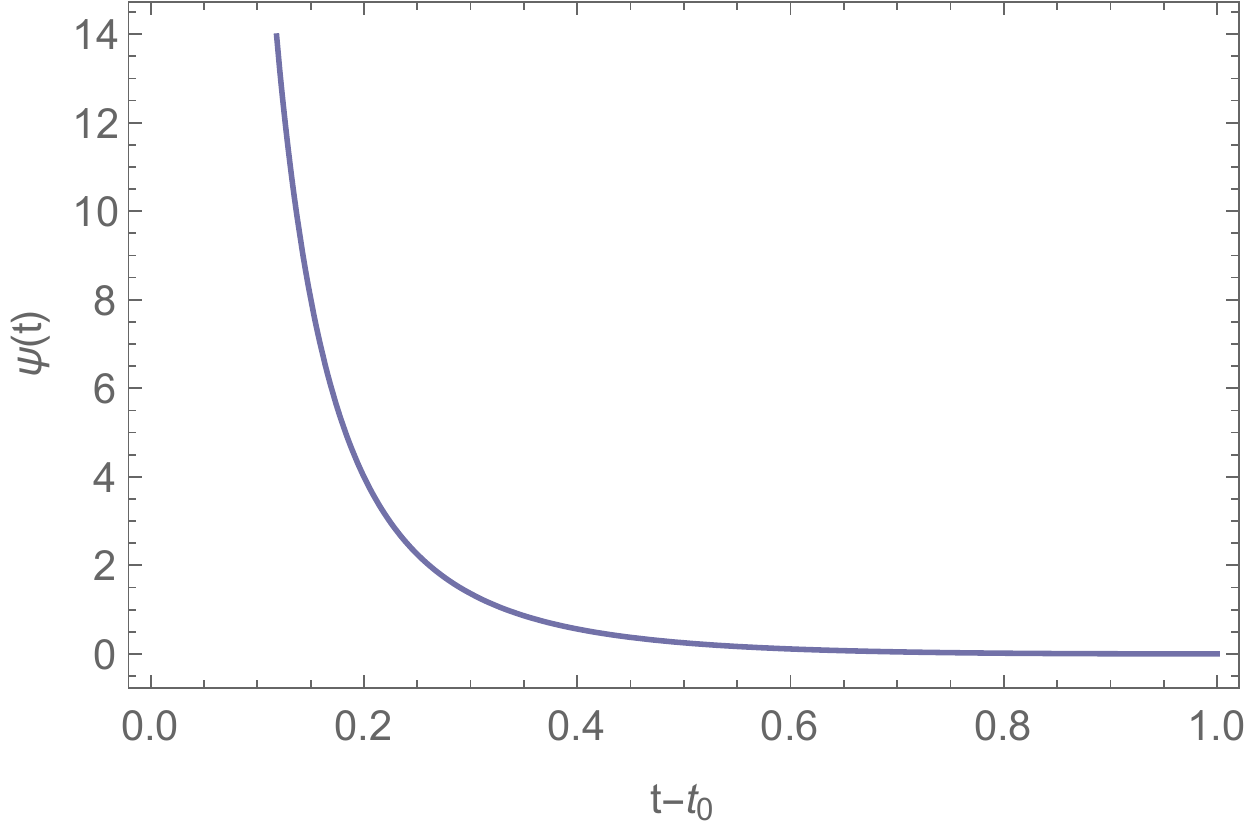}
\caption{Scalar field $\psi(t)$}
\label{fig:cospsi4}
\end{subfigure}
\ \ \ \ \ 
\begin{subfigure}{0.48\textwidth}
\includegraphics[scale=0.65]{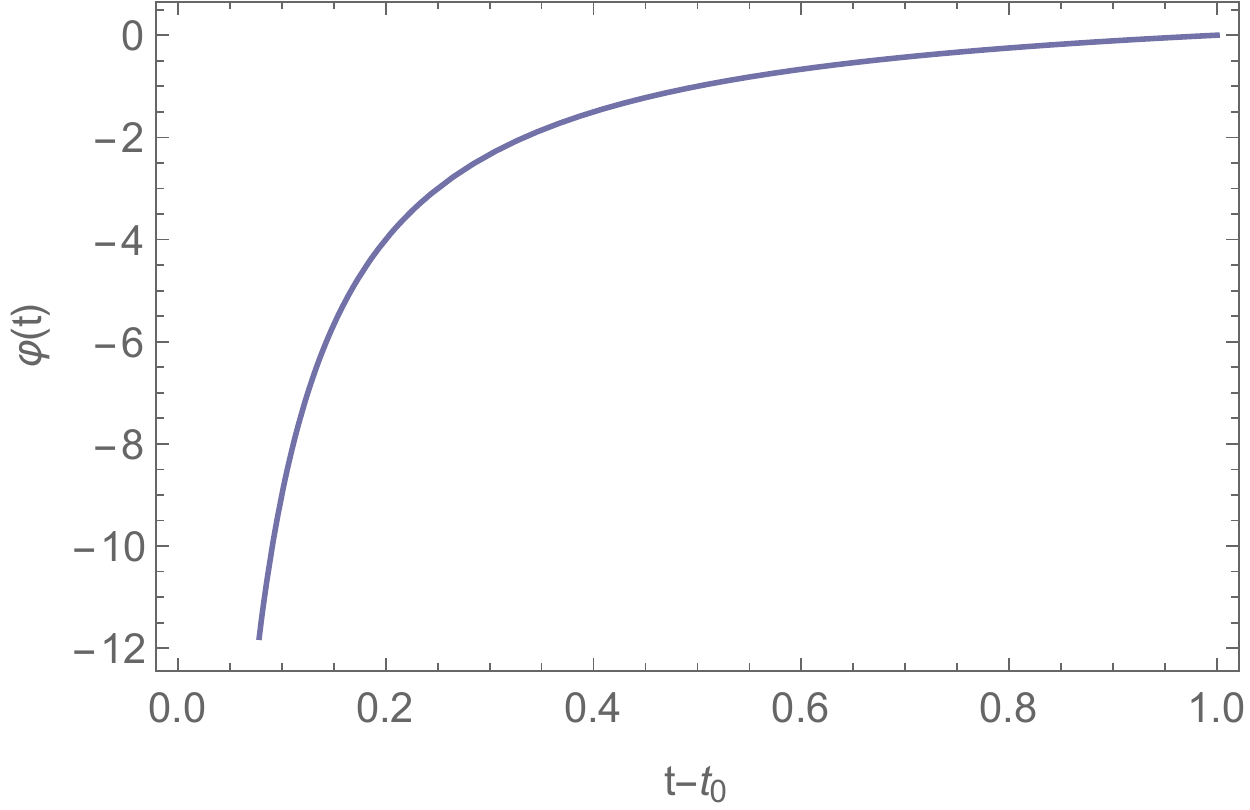}
\caption{Scalar field $\varphi(t)$}
\label{fig:cosphi4}
\end{subfigure}
\caption{Solutions for the scale factor $a\left(t\right)$ in Eq.~\eqref{cosscale4}, the scalar field $\psi\left(t\right)$ in Eq.\eqref{cospsi4}, and the scalar field $\varphi\left(t\right)$ in Eq.\eqref{cosphi4}, with $a_0=\psi_0=\varphi_0=\varphi_1=1$, $t_0=0$}
\end{figure}

Let us now turn to the potential equation. Inserting the potential from Eq.\eqref{cospot4} and the relation between $\varphi$ and $\psi$ from Eq.\eqref{cospsiphi4} into Eq.\eqref{cosfvpotential} one obtains a relationship between the scalar fields and the scale factor as
\begin{equation}\label{cosaphi4}
\dot H+2H^2=\frac{V_0}{3}\left[\varphi-\left(\frac{\psi_0}{2\varphi_0}\right)^2\left(\varphi-\varphi_1\right)^2\right].
\end{equation}
An equation for the scale factor $a\left(t\right)$ only can be obtained from Eq.\eqref{cosaphi4} by differentiating with respect to time once, using Eq.\eqref{cosdphi4} to cancel the terms depending on $\dot\varphi$, and then differentiating again with respect to time. The result is a 4th order ODE for $a\left(t\right)$ of the form
\begin{equation}\label{cosdifa4}
\dddot H+7\ddot H H+4\dot H^2+12 \dot H H^2=-\frac{V_0\psi_0^2}{6a^6}.
\end{equation}
Due to the complicated non-linearity of Eq.\eqref{cosdifa4}, one may expect that no general analytical solutions exist. However, one can look for particular solutions through an ansatz of the form of a power-law as $a\left(t\right)=a_0 t^\alpha$, for some constants $a_0$ and $\alpha$. Inserting this ansatz into Eq.\eqref{cosdifa4} yields the relation
\begin{equation}\label{cossoldifa4}
\left(-12\alpha^3+18\alpha^2-6\alpha\right)\frac{1}{t^4}=-\frac{V_0\psi_0^2}{6a_0^6}\frac{1}{t^{6\alpha}}.
\end{equation} 
Equating the exponents on both sides of Eq.\eqref{cossoldifa4} one obtains $\alpha=2/3$, and thus the scale factor behaves as
\begin{equation}\label{cosscale4}
a\left(t\right)=a_0t^{\frac{2}{3}}.
\end{equation}
The scale factor for this solution is plotted in Fig.\ref{fig:cosscale4}. This solution is interesting because, for a universe dominated by dark matter, i.e., $\rho \neq 0$ and $p=0$, one expects the scale factor to behave as $t^\frac{2}{3}$, but in this case the same behavior can be obtained by considering
$\rho=0$. Moreover, equaling the numerical factors one obtains a constraint for $V_0$ as a function of $a_0$ and $\psi_0$ as
\begin{equation}
V_0=-\frac{8}{3}\frac{a_0^6}{\psi_0^2},
\end{equation}
from which one verifies that since $\psi_0^2>0$ and $a_0^6>0$ then $V_0<0$. The solution for the scale factor given in Eq.\eqref{cosscale4} allows us to directly integrate Eqs.\eqref{cosdphi4} and \eqref{cosdpsi4} and to obtain the solutions for the scalar fields $\varphi$ and $\psi$ in the form
\begin{equation}\label{cosphi4}
\varphi\left(t\right)=-\frac{\varphi_0}{a_0^3t}+\varphi_1,
\end{equation}
\begin{equation}\label{cospsi4}
\psi\left(t\right)=\left(-\frac{\psi_0}{2a_0^3t}+\psi_1\right)^2,
\end{equation}
respectively, where $\psi_1$ is a constant of integration. The solutions for $\varphi$ and $\psi$ are plotted in Figs.\ref{fig:cosphi4} and \ref{fig:cospsi4}, respectively. The solution is now complete since all the variables have been obtained.

\subsection{Solutions with non-flat geometry}
\label{sec:coscurv}

\begin{figure}
\centering 
\begin{subfigure}{0.48\textwidth}
\includegraphics[scale=0.65]{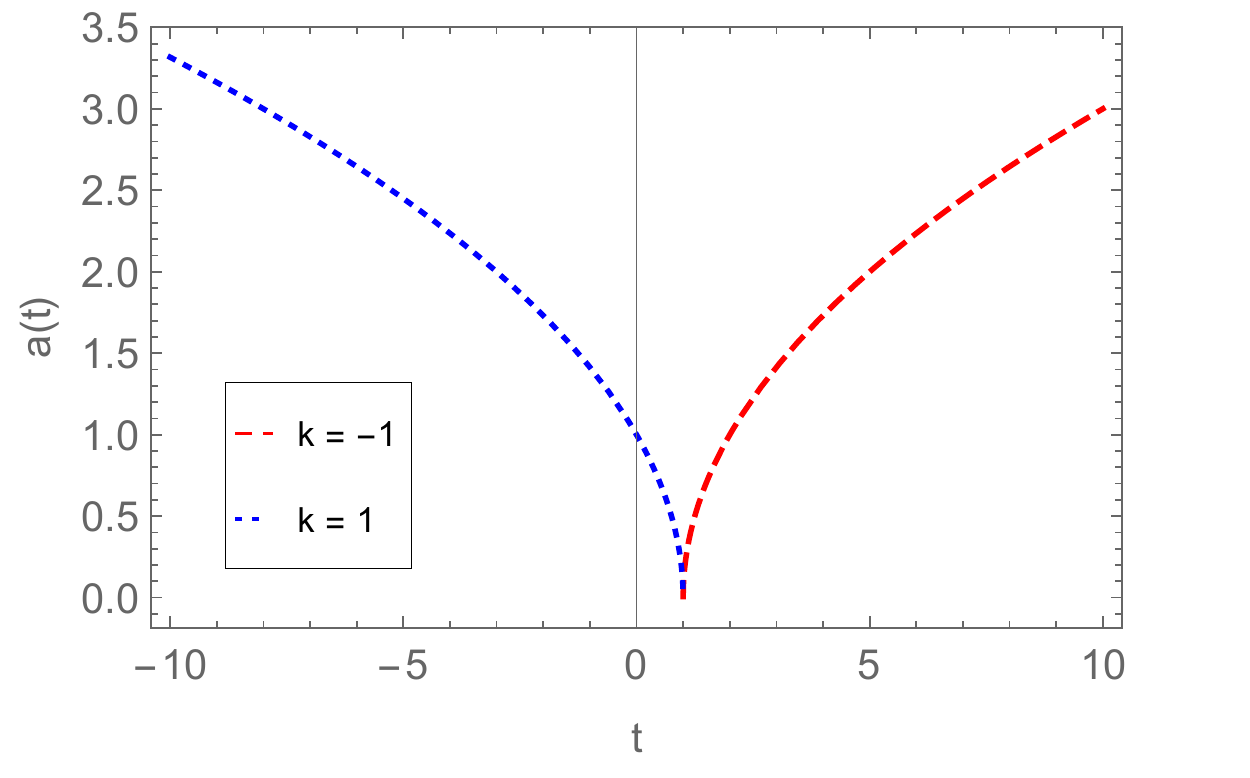}
\caption{Scale factor $a(t)$}
\label{fig:cosscale5}
\end{subfigure}
\ \\
\ \\
\ \\
\begin{subfigure}{0.48\textwidth}
\includegraphics[scale=0.65]{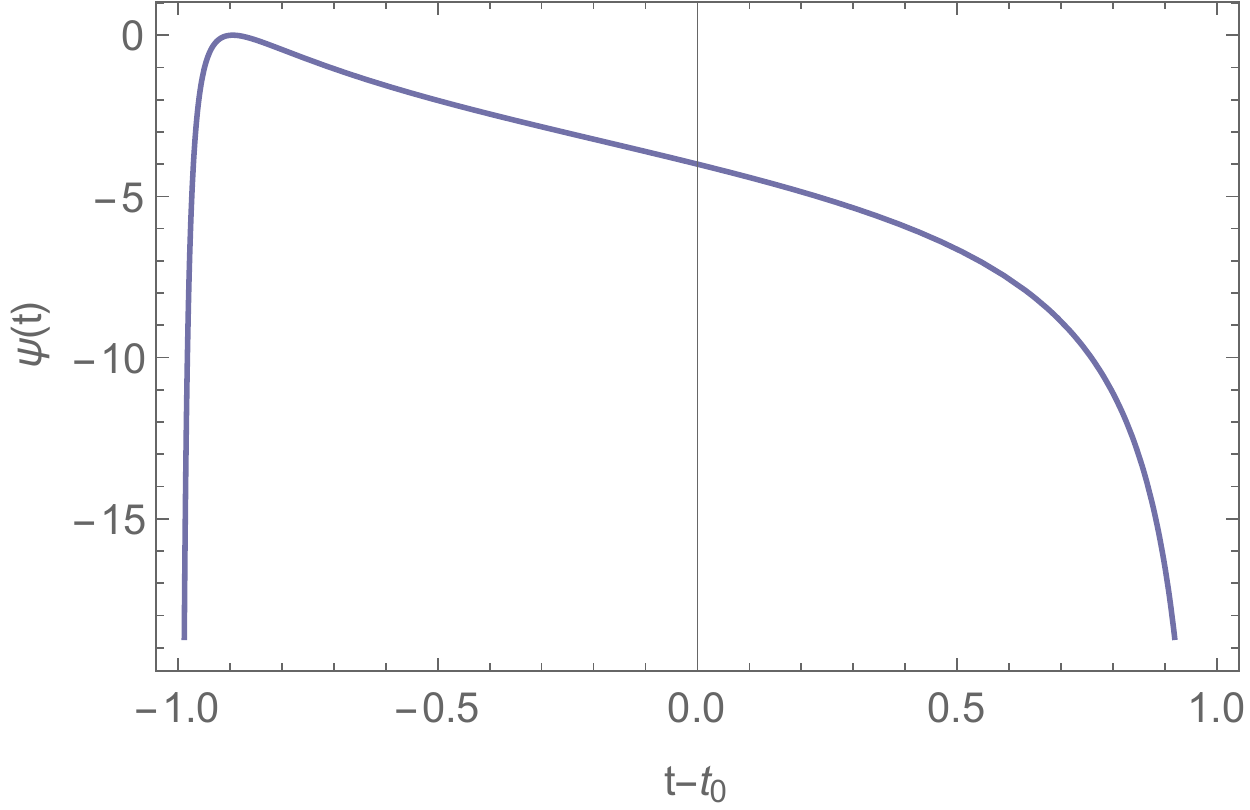}
\caption{Scalar field $\psi(t)$}
\label{fig:cospsi5}
\end{subfigure}
\ \ \ \ \ 
\begin{subfigure}{0.48\textwidth}
\includegraphics[scale=0.65]{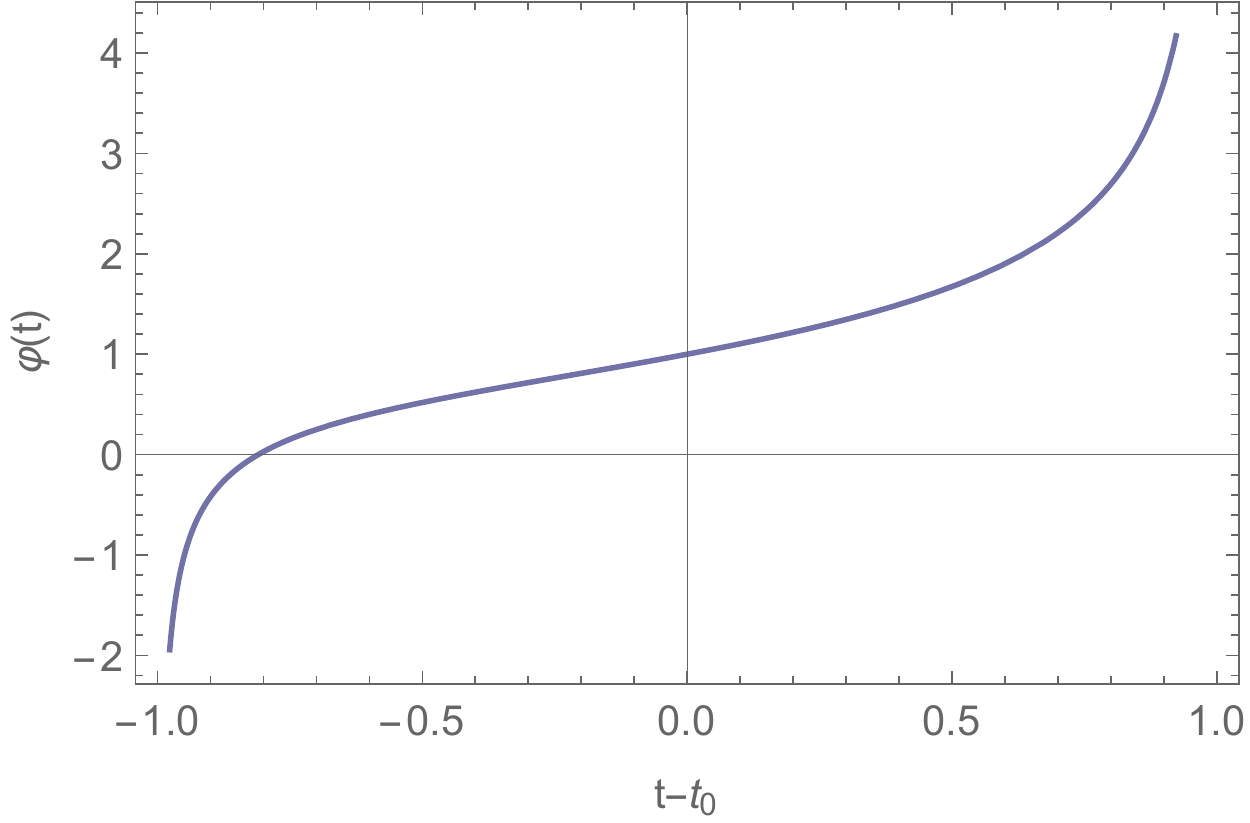}
\caption{Scalar field $\varphi(t)$}
\label{fig:cosphi5}
\end{subfigure}
\ \\
\ \\
\ \\
\begin{subfigure}{0.48\textwidth}
\includegraphics[scale=0.65]{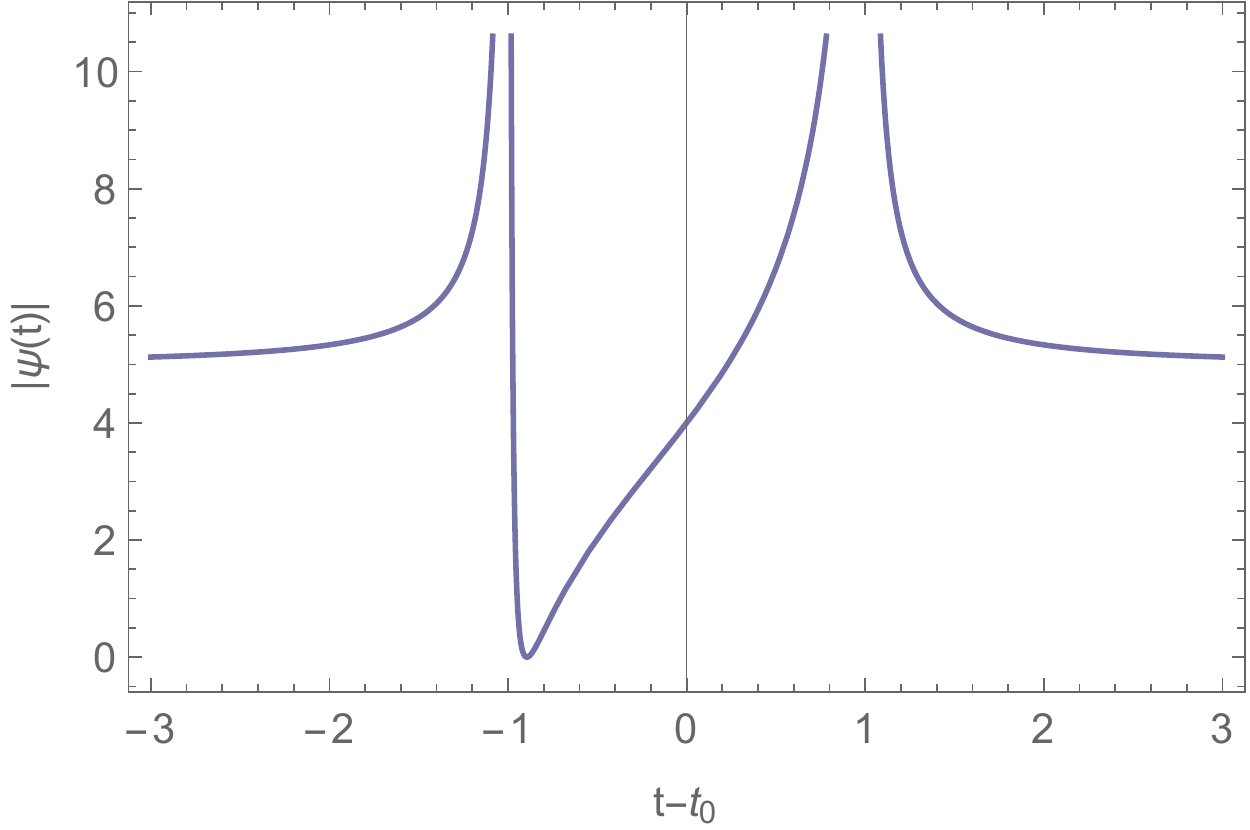}
\caption{Absolute value of the scalar field $\psi(t)$}
\label{fig:cospsi5abs}
\end{subfigure}
\ \ \ \ \ 
\begin{subfigure}{0.48\textwidth}
\includegraphics[scale=0.65]{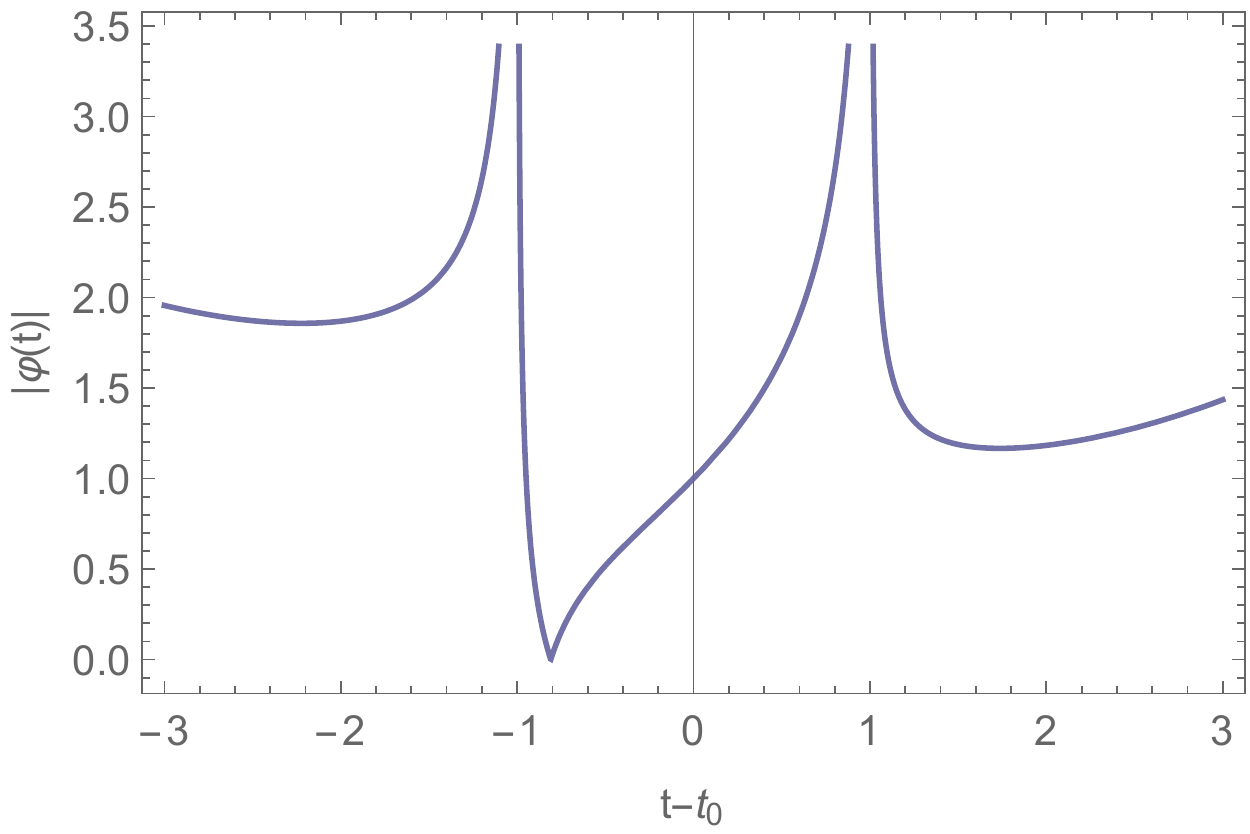}
\caption{Absolute value of the scalar field $\varphi(t)$}
\label{fig:cosphi5abs}
\end{subfigure}
\caption{Solutions for the scale factor $a\left(t\right)$ in Eq.~\eqref{cosscale5}, the scalar field $\psi\left(t\right)$ in Eq.\eqref{cospsi5}, and the scalar field $\varphi\left(t\right)$ in Eq.\eqref{cosphi5}, with $a_0=\psi_0=\psi_1=\varphi_0=\varphi_1=V_0=1$, $t_0=0$}
\end{figure}

In this section we still consider the distribution of matter to be vacuum, that is $\rho=p=0$, for which we still have Eqs.\eqref{coseos} and \eqref{costabcons} automatically satisfied, but we now admit an arbitrary geometry $k$. For these assumptions, Eqs.\eqref{coskgphi} to \eqref{cospotential} become respectively
\begin{equation}\label{cosnfkgphi}
\ddot\varphi+3\left(\frac{\dot a}{a}\right)\dot\varphi-\frac{1}{3}\left[2V-\psi V_\psi-\varphi V_\varphi\right]=0,
\end{equation}
\begin{equation}\label{cosnfkgpsi}
\ddot\psi+3\left(\frac{\dot a}{a}\right)\dot\psi-\frac{\dot\psi^2}{2\psi}+\frac{\psi}{3}\left(V_\varphi+V_\psi\right)=0,
\end{equation}
\begin{equation}\label{cosnfpotential}
V_\varphi=6\left[{\frac{d}{dt} \left(\frac{\dot a}{a}\right)} +2\left(\frac{\dot a}{a}\right)^2+\frac{k}{a^2}\right].
\end{equation}
We thus have a system of three independent equations for the five independent variables $k, a, \varphi,\psi$ and $V$. However, as $k$ must have specific values $k=\pm 1$, we still can only impose one constraint to close the system. Let us use the previous reconstruction methods to provide a non-flat solution for these equations: we assume that the terms inside the bracets of Eq.\eqref{cosnfpotential} cancel each other, i.e.
\begin{equation}\label{cosconst5}
\dot H+2H^2+\frac{k}{a^2}=0,
\end{equation}
where we used the definition of the Hubble parameter in Eq.\eqref{defhubble}. Note that Eq.\eqref{cosconst5} is a non-linear second order differential equation for the scale factor $a\left(t\right)$ and can be integrated twice over time to yield a solution of the form
\begin{equation}\label{cosscale5}
a\left(t\right)=\sqrt{\frac{a_0^2-\left(t-t_0\right)}{k}},
\end{equation} 
where $k=\pm 1$, $a_0$ and $t_0$ are constants of integration, and we have not considered the solution with negative sign for the scale factor $a(t)$. Notice that for $k=1$, there is an end value of $t$, given by $t_e=a_0^2+t_0$, for which the scale factor drops to zero, ending the solution, as for $t>t_e$ it becomes a pure imaginary number. The same happens for $k=-1$ when $t<t_e$. Therefore, the solution for $k=1$ is only defined for $-\infty<t<t_e$ and the solution for $k=-1$ is only defined for $t_e<t<+\infty$, see Fig.\ref{fig:cosscale5}. We know from observations that the universe is approximately flat, and thus $k=0$ is a very good approximation. However, the field equations depend on $k$ in such a way that if the geometry varies even slightly from flat, i.e., if $k\neq 0$, the structure of the equations changes dramatically. The relevance of the solution we have found, therefore, is to provide a glimpse of the role of spatial curvature in these theories.

We now proceed as before. Under these assumptions, Eq.\eqref{cosnfpotential} becomes simply $V_\varphi=0$ which can be directly integrated over $\varphi$ to yield $V\left(\varphi,\psi\right)=b\left(\psi\right)$, where $b\left(\psi\right)$ is an arbitrary function of $\psi$ that arises from the fact that the potential is, in general, a function of $\varphi$ and $\psi$. Assuming that the potential should depend on the scalar fields in the combination $\left(\varphi-\psi\right)$, then we must choose $b\left(\psi\right)=V_0$, for some constant $V_0$, and the potential becomes simply $V\left(\varphi,\psi\right)=V_0$. Inserting these results into Eqs.\eqref{cosnfkgphi} and \eqref{cosnfkgpsi}, one obtains
\begin{equation}\label{coskgphi5}
\ddot \varphi-\frac{3k^2\left(t-t_0\right)}{a_0^2-k^2\left(t-t_0\right)^2}\dot\varphi
-\frac{2}{3}V_0=0,
\end{equation}
\begin{equation}\label{coskgpsi5}
\ddot\psi+\frac{2k^2\left(t-t_0\right)}{k^2\left(t-t_0\right)^2-a_0^2}\dot\psi-\frac{\dot\psi^2}{2\psi}=0,
\end{equation}
respectively. Again, Eqs.\eqref{coskgphi5} and \eqref{coskgpsi5} are decoupled and can be integrated independently over time to obtain the general solutions
\begin{eqnarray}
\varphi\left(t\right)&=&\varphi_1+\frac{ t-t_0}{12a_0^2k \sqrt{a_0^2-k^2 \left(t-t_0\right)^2}}
\left\{3 a_0^4 V_0 \tan ^{-1}\left[\frac{k\left(t-t_0\right)}{\sqrt{a_0^2-k^2 \left(t-t_0\right)^2}}\right]+\right.\nonumber\\
&&\left. +k \left[a_0^2 V_0 \left(t-t_0\right) \sqrt{a_0^2-k^2 \left(t-t_0\right)^2}+\varphi_0\right]\right\},\label{cosphi5}
\end{eqnarray}
\begin{equation}\label{cospsi5}
\psi\left(t\right)=\frac{\psi_1}{a_0^2-k^2\left(t-t_0\right)^2} \left[\left(a_0^4 \psi_0 k^2+1\right) \left(t-t_0\right)^2-a_0^6 \psi_0\right] e^{ 2 \tanh^{-1}\left[\frac{t-t_0}{a_0^2 \sqrt{\psi_0} \sqrt{a_0^2-k^2 \left(t-t_0\right)^2}}\right]},
\end{equation}
respectively, where $\varphi_0, \varphi_1, \psi_0$ and $\psi_1$ are constants of integration. Note that there are values of $t$, namely $t=\pm\frac{a_0}{k}+t_0$, for which the denominators of both scalar fields vanish and the scalar fields diverge. This implies that if one wants to consider the particular case for which the scalar fields are real, then the solution is only valid in the region $-\frac{a_0}{|k|}<t-t_0<\frac{a_0}{|k|}$, which corresponds to a region where the solution with $k=-1$ exists. For one to obtain a solution valid for the entire region $-\infty<t<\frac{a_0}{|k|}$, one would have to compute another solution valid in the region $-\infty<t_s$, with $-\frac{a_0}{|k|}<t_s<\frac{a_0}{|k|}$, and perform a matching with this solution using a timelike hypersurface at $t_s$. On the other hand, the solution with $k=1$ only exists in the region $t>\frac{a_0}{k}+t_0$, region where both the scalar fields are complex and with finite absolute value. 

The solutions for the scalar fields $\varphi$ and $\psi$ are plotted in Figs.\ref{fig:cosphi5} and \ref{fig:cospsi5} for the region where they are real. Figs.\ref{fig:cosphi5abs} and \ref{fig:cospsi5abs} show the absolute value of the scalar fields $\varphi$ and $\psi$, respectively, including the regions where they are complex. The solution is now complete since all variables are known.

\subsection{Solutions with perfect-fluid matter}
\label{sec:cosmat}
To finalize our analysis of cosmological solutions, let us now consider a distribution of matter given by a perfect fluid with an equation of state given by Eq.\eqref{coseos} and a conservation equation given by Eq.\eqref{costabcons}. These equations are decoupled from the scalar fields and the potential, in such a way that the solutions for $\rho\left(r\right)$ and $p\left(t\right)$ are independent of any other assumption and are given by
\begin{equation}\label{cosmatter6}
\rho=\rho_0a^{-3\left(1+w\right)},\ \ \ \ \ p=p_0a^{-3\left(1+w\right)},
\end{equation}
where $\rho_0$ and $p_0$ are constants of integration that are related by $p_0=w \rho_0$, as can be seen from Eq.\eqref{coseos}. The solutions in Eq.\eqref{cosmatter6} allow us to write the trace of the stress-energy tensor as
\begin{equation}\label{costrace6}
T=-\rho_0\left(1-3w\right)a^{-3\left(w+1\right)}.
\end{equation}

The solutions in Eq.\eqref{cosmatter6} define the matter sources $\rho$ and $p$, thus reducing the number of unknowns back to five, that are $k, a, \varphi, \psi$, and $V$, for the system of three independent equations, Eqs.\eqref{coskgphi} to \eqref{cospotential}. Again, since $k=\pm 1$, it does not represent a freedom of choice, and thus only one constraint can be imposed to close the system.

Notice now that in the system formed by Eqs.\eqref{coskgphi} to \eqref{cospotential}, the only equation that depends explicitly on matter is Eq.\eqref{coskgphi}, via the term depending on $T$ on the right hand side. In the previous sections, except for the solution obtained via the choice in Eq.\eqref{cospot4}, all solutions for the scalar field $\varphi$ were obtained after having already computed $a\left(t\right)$, $V\left(\varphi,\psi\right)$ and $\psi\left(t\right)$, because the equation for $\varphi$ was either decoupled or dependent on the solution for $\psi$. This means that the process to generalize these vacuum solutions to solutions with perfect-fluid matter consist of taking the same $a$, $\psi$ and $V$ and simply re-compute $\varphi$ using Eq.\eqref{coskgphi}, which in this case, using Eq.\eqref{costrace6} becomes
\begin{equation}\label{coskgphi6}
\ddot\varphi+3H\dot\varphi-\frac{1}{3}\left[2V-\psi V_\psi-\varphi V_\varphi\right]=\frac{\kappa^2\rho_0}{3}\left(1-3w\right)a^{-3\left(w+1\right)}.
\end{equation}

\begin{figure}
\centering 
\begin{subfigure}{0.48\textwidth}
\includegraphics[scale=0.65]{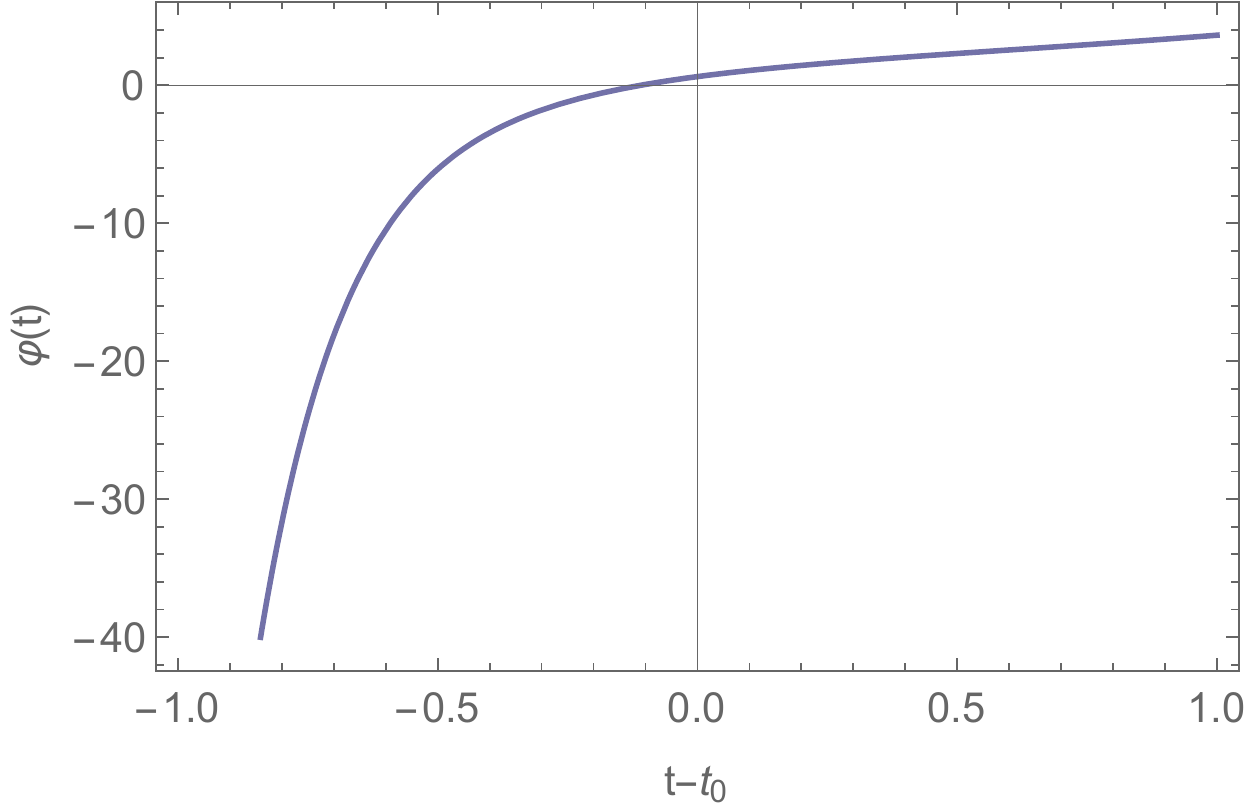}
\caption{Scalar field $\varphi(t)$}
\label{fig:cosphi6a}
\end{subfigure}
\ \ \ \ \ 
\begin{subfigure}{0.48\textwidth}
\includegraphics[scale=0.65]{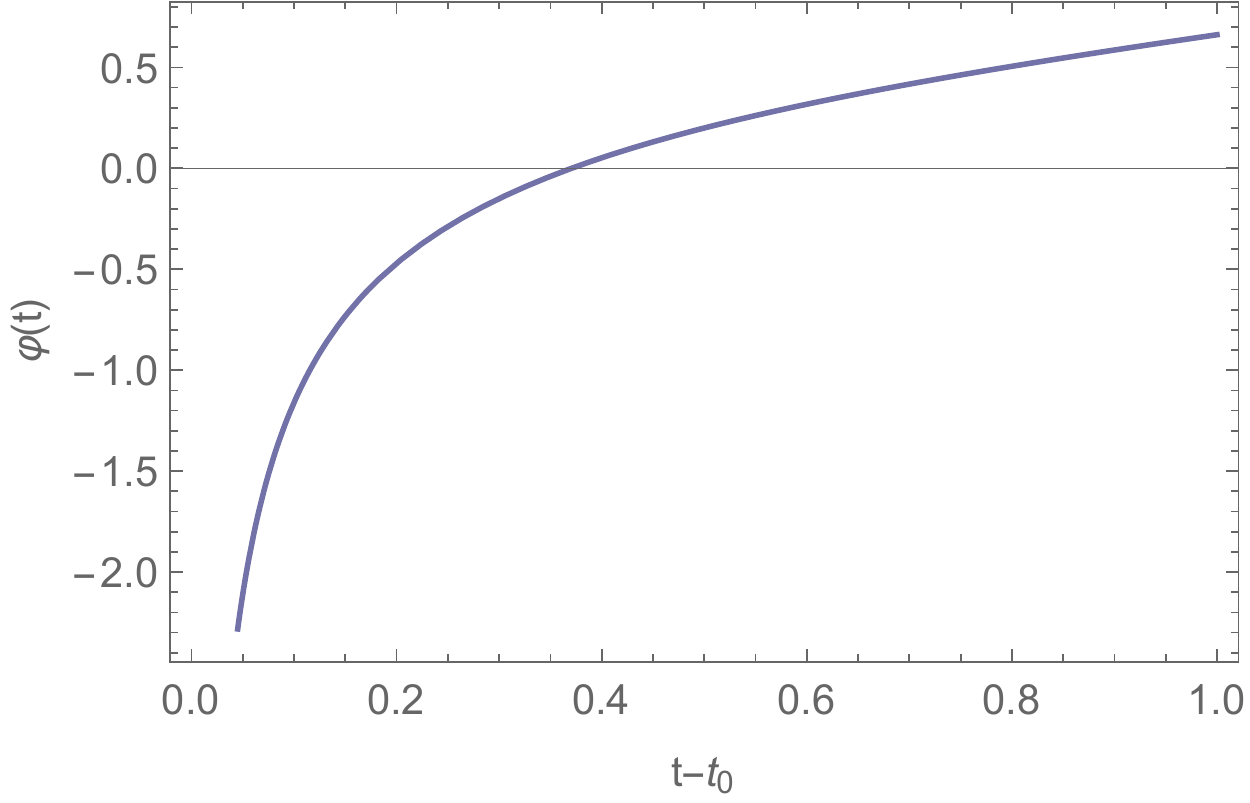}
\caption{Scalar field $\varphi(t)$}
\label{fig:cosphi6b}
\end{subfigure}
\caption{Solutions for the scalar field $\varphi\left(r\right)$ in Eq.~\eqref{cosphi6a} with $\kappa=\rho_0=a_0=\Lambda=\psi_0=\varphi_0=\varphi_1=1$, $t_0=w=0$ (left panel); and in Eq.~\eqref{cosphi6b} with $\kappa=\rho_0=V_0=\varphi_1=1$, $a_0=\sqrt{2}$, $\varphi_0=1/\sqrt{2}$, $t_0=w=0$ (right panel).}
\end{figure}

As examples, let us consider the previous specific cases for which Eq.\eqref{coskgphi6} still has an analytical solution. Consider the de-Sitter scale factor given in Eq.\eqref{cosscale1}, for which we have chosen a potential of the form in Eq.\eqref{cospot1}, and we obtained a solution for $\psi$ as in Eq.\eqref{cospsi1}. Inserting these results into Eq.\eqref{coskgphi6} and integrating we obtain
\begin{equation}\label{cosphi6a}
\varphi\left(t\right)=-\frac{\kappa^2\rho_0\left(a_0e^{\sqrt{\Lambda} t}\right)^{-3w}e^{-3\sqrt{\Lambda} t}}{3a_0^3{\Lambda}\left(4+3w\right)}+\psi_0e^{-6\sqrt{\Lambda} t_0}-\frac{2}{7}\psi_0e^{-6\sqrt{\Lambda} t}-2\psi_0e^{-3\sqrt{\Lambda}\left(t+t_0\right)}+\varphi_0e^{-4\sqrt{\Lambda} t}+\varphi_1e^{\sqrt{\Lambda} t},
\end{equation}
where $\varphi_0$ and $\varphi_1$ are constants of integration. This solution is plotted in Fig.\eqref{fig:cosphi6a}. This solution is complete since all variables are known.

Consider now the solution obtained via the simplification of the potential equation given in Eq.\eqref{cosconst2}, from which we obtained a scale factor that behaves as Eq.\eqref{cosscale2}, a potential of the form $V=V_0$, and a solution for the scalar field $\psi$ given in Eq.\eqref{cospsi2}. In this case, the solution for Eq.\eqref{coskgphi6} becomes
\begin{equation}\label{cosphi6b}
\varphi\left(t\right)=\frac{2}{15} V_0 \left(t-t_0\right)^2-\frac{\varphi_0}{\sqrt{ t-t_0}}+\varphi_1-\frac{4\kappa ^2 \rho_0 \left(t-t_0\right) }{3 a_0^2 (3 w-2)} \left(a_0 \sqrt{ t-t_0}\right)^{-\left(3w+1\right)},
\end{equation}
where $\varphi_0$ and $\varphi_1$ are constants of integration. This solution is plotted in Fig.\eqref{fig:cosphi6b}. This solution is complete because all variables have been determined.

\section{Forms of the $f\left(R,\mathcal R\right)$}

It is useful to obtain the form of the function $f\left(R,\mathcal{R}\right)$ in all the previous cases as the scalar-tensor form of the theory is only a derived form and used to provide an easier platform to perform the calculations. To  obtain $f\left(R,\mathcal{R}\right)$, we are going to use the definition of the potential given by Eq.~\eqref{ghmpgstpotential}. Replacing the definitions of the scalar fields $\varphi=f_R$ and $\psi=-f_\mathcal R$ into this equation, we obtain
\begin{equation}\label{cosdifpot}
V\left(f_R,f_\mathcal{R}\right)=-f\left(R,\mathcal{R}\right)+f_R R+f_\mathcal{R} \mathcal{R},
\end{equation}
where the subscripts $R$ and $\mathcal{R}$ represent derivatives with respect to these variables, respectively. Eq. \eqref{cosdifpot} is a partial differential equation for $f\left(R,\mathcal R\right)$. Due to its partial nature, the solutions are not unique and we shall obtain a family of solutions for each case.

As we have seen in the previous section, the solutions for the scalar field can have zeros and assume a negative sign. While this does not compromise the physical meaning of the solution in the generalized hybrid gravity picture (in this picture the scalar fields are not real matter fields) it can lead to further constraints on the form of the action. One might require that  the derivatives of $f$, which correspond to the scalar fields by $\varphi=f_R$ and $\psi=-f_\mathcal R$, are never zero. In this case the solution we have found will only be meaningful in the intervals in which the scalar field have a fixed sign.

\subsection{Constant potential}\label{Sec:coslinpot}

Let us first consider the cases where the potential obtained was a constant, i.e., $V=V_0$. This happened in all the solutions obtained by simpifying the potential equation with $V_\varphi=0$ and choosing $b\left(\psi\right)=V_0$. In this case, Eq.\eqref{cosdifpot} becomes
\begin{equation}\label{cosdifpot1}
f\left(R,\mathcal{R}\right)=f_R R+f_\mathcal{R} \mathcal{R}-V_0.
\end{equation}
The general solution of Eq.\eqref{cosdifpot1} can be written as 
\begin{equation}\label{cosdifpots1}
f\left(R,\mathcal{R}\right)=-V_0+g\left(\frac{\mathcal{R}}{R}\right)R+h\left(\frac{R}{\mathcal R}\right)\mathcal R,
\end{equation}
where $g$ and $h$ are arbitrary functions. However, not all forms of the function $g$ can be used to obtain the correct solutions. Consider for example the particular cases $g\left(x\right)=e^x$ and $g\left(x\right)=x^\gamma$, for some constant $\gamma$. Inserting these functions $g$ into Eq.\eqref{cosdifpots1}, inserting the result into Eq.\eqref{cosdifpot}, and using the definitions of the scalar fields $\varphi=f_R$ and $\psi=-f_\mathcal R$, we recover $V=V_0$. This happens because these two forms of the function $g$ allow us to invert the relations $\varphi\left(R,\mathcal R\right)$ and $\psi\left(R,\mathcal R\right)$ to obtain $R\left(\varphi,\psi\right)$ and $\mathcal R\left(\varphi,\psi\right)$ and thus the equivalence between the scalar-tensor and the geometrical representations is guaranteed. Other forms of the function $g$ might prevent this invertibility and thus, despite being compatible with Eq.\eqref{cosdifpot1}, one can not say immediatly that the theory will have the solutions reconstructed.

\subsection{Linear potential}

Consider now the cases where the potential was linear in $\varphi$ and $\psi$ in the form $V=V_0\left(\varphi-\psi\right)$. This form was obtained in two of the particular cases studied in the previous sections: the de-Sitter solution, for which we had $V_0=12\Lambda$, and the second simplified potential equation case, where $V_0=-3\Omega^2$. In these situations, Eq.\eqref{cosdifpot} becomes
\begin{equation}\label{cosdifpot2}
f\left(R,\mathcal{R}\right)=f_R \left(R-V_0\right)+f_\mathcal{R}\left(\mathcal{R}-V_0\right).
\end{equation} 
To solve this PDE, we define two new variables $\bar{R}=R-V_0$ and $\bar{\mathcal R}=\mathcal R-V_0$. Since $V_0$ is constant, the derivatives $f_R=f_{\bar{R}}$ and $f_\mathcal R=f_{\bar{\mathcal R}}$ remain of the same form, and Eq.\eqref{cosdifpot2} can be written in the form $f\left(\bar{R},\bar{\mathcal{R}}\right)=f_{\bar{R}}\bar{R}+f_{\bar{\mathcal{R}}}\mathcal{R}$, which is very similar to the linear potential case expressed in Eq.\eqref{cosdifpot1} but without the constant $V_0$. The general solution for this equation is then $f\left(\bar{R},\bar{\mathcal{R}}\right)=g\left({\bar{\mathcal{R}}}/{\bar{R}}\right)\bar{R}$, where $g$ is an arbitrary function, and replacing back the variables $\bar{R}$ and $\bar{\mathcal R}$ to $R$ and $\mathcal R$ one obtains
\begin{equation}\label{cosdifpots2}
f\left(R,\mathcal{R}\right)=g\left(\frac{\mathcal{R}-V_0}{R-V_0}\right)\left(R-V_0\right)+h\left(\frac{R-V_0}{\mathcal R-V_0}\right)\left(\mathcal R-V_0\right).
\end{equation}
Again, one must be careful to choose a specific form of the function $g$ for which the relations $\varphi\left(R,\mathcal R\right)$ and $\psi\left(R,\mathcal R\right)$ can be inverted to obtain $R\left(\varphi,\psi\right)$ and $\mathcal R\left(\varphi,\psi\right)$. A form $g\left(x\right)=e^x$ allows that invertibility but, unlike the previous case, a power-law of the form $g\left(x\right)=x^\gamma$ does not, and these forms shall thus be discarded as the equivalence between the two representations of the theory is no longer guaranteed and they may not possess the solutions reconstructed.

\subsection{Quadratic potential}

Finally, in the simplified scalar field equations case, the potential obtained was quadratic in $\varphi$ and $\psi$ in the form $V=V_0\left(\varphi-\psi\right)^2$. This was the form of the potential for which the potential-depending terms in Eqs.\eqref{cosfvkgphi} and \eqref{cosfvkgpsi} cancel each other. Inserting this potential in Eq.\eqref{cosdifpot} yields
\begin{equation}\label{cosdifpot3}
f\left(R,\mathcal{R}\right)=f_R R+f_\mathcal{R} \mathcal{R}-V_0\left(f_R+f_\mathcal{R}\right)^2.
\end{equation}
This equation is a specific form of the Clairaut equation in two variables, and one can obtain the so-called general solution of the Clairaut equation, of the form
\begin{equation}\label{cosdifpots3a}
f\left(R,\mathcal{R}\right)=RC_1+\mathcal{R}C_2-V_0\left(C_1+C_2\right)^2,
\end{equation}
where $C_1$ and $C_2$ are constants of integration. However, this form of the function $f$ presents a few problems. Note that from Eq.\eqref{cosdifpots3a}, we have $f_{RR}=f_{\mathcal R\mathcal R}=f_{R\mathcal R}=0$, from which one sees that the determinant of the matrix $\mathcal M$ in Eq.\eqref{ghmpgstmatrix} vanishes, and thus the equivalence between the geometrical and the scalar tensor representations is not guaranteed. We thus need a different solution for Eq.\eqref{cosdifpot3} that preserves $\det\mathcal M\neq 0$. To find this solution, let us take the derivatives of Eq.\eqref{cosdifpot3} with respect to $R$ and $\mathcal R$ and define the function $g\left(f_R,f_\mathcal R\right)=-V_0\left(f_R+f_\mathcal{R}\right)^2$. We thus obtain the system
\begin{eqnarray}
\left(R+g^{(1,0)}\right)f_{RR}+\left(\mathcal R+g^{(0,1)}\right)f_{R\mathcal R}=0,\nonumber \\
\left(\mathcal R+g^{(0,1)}\right)f_{\mathcal R\mathcal R}+\left(R+g^{(1,0)}\right)f_{R\mathcal R}=0,
\end{eqnarray}
where $g^{(i,j)}$ represents the derivative of $g\left(f_R,f_\mathcal R\right)$ $i$ times with respect to $f_R$ and $j$ times with respect to $\mathcal R$. This system of equation can be rewritten in terms of a matrix system of the form
\begin{equation}\label{cosdifpot3mat}
\mathcal A \textbf{v}=
\begin{bmatrix}
f_{RR} & f_{R\mathcal R} \\[0.8em]
f_{\mathcal R R} & f_{\mathcal R\mathcal R} 
\end{bmatrix}
\begin{bmatrix}
R+g^{(1,0)} \\[0.8em]
\mathcal{R}+g^{(0,1)}
\end{bmatrix}
=0,
\end{equation}
which has a unique solution if the determinant of the matrix $\mathcal A$ is non-zero, i.e. $\det\mathcal A\neq 0$. Comparing Eq.\eqref{cosdifpot3mat} with Eq.\eqref{ghmpgstmatrix}, one verifies that the matrices $\mathcal M$ and $\mathcal A$ are the same for the unique solution $R=\alpha$ and $\mathcal R=\beta$. This implies that guaranteeing that $\det\mathcal A\neq 0$ preserves the equivalence between the geometrical and the scalar-tensor representations of the theory. Also, the unique solution for $R$ and $\mathcal R$ is then $R=-g^{(1,0)}$ and $\mathcal R=-g^{(0,1)}$. In our particular case, we have $g^{(1,0)}=g^{(0,1)}=-2V_0\left(f_R+f_\mathcal R\right)$, from which we imply that $R=\mathcal R$. For this condition to be consistent with the solution obtained, one needs to verify that Eq.\eqref{ghmpgrelricsca} is also satisfied. Inserting the metric in Eq.\eqref{metricflrw}, the scale factor in Eq.\eqref{cosscale4} and the scalar field $\psi$ from Eq.\eqref{cospsi4} into Eq.\eqref{ghmpgrelricsca} and making $R=\mathcal R$ cancel out, one verifies that this equation is only satisfied if the integration constant $\psi_0=0$ vanishes. This represents an extra constraint one must verify for the solutions obtained to be consistent with the definition of the scalar-tensor representation.

Finally, to obtain the solution for the function $f\left(R,\mathcal R\right)$, we sum the two conditions $R=-g^{(1,0)}$ and $\mathcal R=-g^{(0,1)}$ to obtain an equation of the form 
\begin{equation}\label{cosdifpot3b}
R+\mathcal R=4V_0\left(f_R+f_\mathcal R\right).
\end{equation}
Defining two new variables $Z=R+\mathcal R$ and $W=R-\mathcal R$, one can rewrite Eq.\eqref{cosdifpot3b} in the form $Z-8V_0f_Z=0$, where the subscript $Z$ denotes a derivative with respect to $Z$, with $f_Z=\frac{1}{2}\left(f_R+f_\mathcal R\right)$. This equation can be integrated directly over $Z$, giving rise to an integration function $C\left(W\right)$, which vanishes identically since $R=\mathcal R$ and thus $W=0$ in our case. Transforming back the variable $Z$ to $R$ and $\mathcal R$ one obtains finally
\begin{equation}\label{cosdifpots3b}
f\left(R,\mathcal R\right)=\frac{\left(R+\mathcal R\right)^2}{16V_0}+C\left(R-\mathcal R\right).
\end{equation}
The invertibility of $\varphi\left(R,\mathcal R\right)$ and $\psi\left(R,\mathcal R\right)$ to obtain $R\left(\varphi,\psi\right)$ and $\mathcal R\left(\varphi,\psi\right)$ is already guaranteed from the condition $\det\mathcal A\neq 0$, and Eq.\eqref{cosdifpot3} is satisfied as long as $C\left(0\right)=0$.

\section{Conclusions}

In this chapter, we devised a number of reconstruction strategies to obtain  new exact cosmological solutions in the context of  the generalized hybrid metric-Palatini gravity. This theory can be recast in a scalar-tensor theory with two scalar fields whose solutions can be computed analytically for a convenient choice of the potential or the scale factor. Using the new methods we obtained a number of physically interesting solutions including  power-law and exponential  scale factors.

These solutions reveal some crucial differences not only with respect to GR but also to $f(R)$ gravity. For example it becomes evident that a given behavior of the scale factors in vacuum has always a counterpart in the presence of a perfect fluid, as shown in Sec.~\ref{sec:cosmat}. 

The impact in terms of the interpretation of the cosmological dark phenomenology via this class of theories is clear: a negligible matter distribution still gives rise to expansion laws which are compatible with the observations without requiring the total baryonic matter to be a relevant percentage of the total energy density. This is particularly important in the case of the de Sitter solution given in Sec.~\ref{sec:cosflat}, because it implies that in these theories inflation could start even if matter is not negligible, and also that matter is not central to obtain the classical $t^{2/3}$ Friedmann solution. 

Another interesting aspect of the solutions we found concern the behavior of vacuum cosmology and the role of the spatial curvature. In the first case we obtained that the generalized hybrid metric-Palatini theory can have a surprisingly complex behavior in the flat vacuum case. This has important consequences in terms, for example, of the far future of the cosmological models and the meaning of the cosmic no hair theorem in this framework. 

Like in any higher order model, in the generalized hybrid metric-Palatini theory the spatial curvature has  an important role. Even small deviations from perfect flatness are able to change dramatically the evolution of the cosmology. The result of Sec.~\ref{sec:coscurv}  gives us a glimpse of these differences and their magnitude, showing that vacuum closed models will collapse, whereas open ones expand forever both with a relatively simple expansion law. 

It is also interesting to note that our results imply that generalized hybrid metric-Palatini gravity can generate a de-Sitter evolution without the appearance of an explicit cosmological constant in the action. Instead, the presence of such a constant leads, throughout our method, to solutions which are radiation-like. This result gives us information on the properties of effective dark fluids related to the non Hilbert-Einstein terms present in this theory. It also implies,  by the cosmological no-hair theorem, that these solutions should be unstable. The stability of the other solutions we have found cannot be obtained with the same ease. This is a limitation of the reconstruction in general. It allows to obtain some exact solution but it does not offer information on their stability, which has to be investigated with different approaches. 

Our results suggest that the hybrid theory space is a priori large and requires further investigation not only in terms of the exploration of the solution space, but also in terms of the stability of these solutions. A final thought should be spent on the role of our results for the testability of this class of theory. It is clear that the solutions we have found can be used to perform a number of  cosmological tests, e.g., the ones based on distances, for example. However, a more complete use of the accuracy of the current observations, such as cosmic microwave background anisotropies, requires a full analysis of the cosmological perturbations. 
\cleardoublepage

\chapter{Cosmological phase-space of the generalized hybrid metric-Palatini gravity}
\label{chapter:chapter4}

In this chapter, we study the cosmological phase space of the generalized hybrid metric-Palatini gravity theory using the dynamical system approach. We formulate the propagation equations of the suitable dimentionless variables that describe FLRW universes as an autonomous system. The fixed points are obtained for four different forms of the function $f\left(R,\mathcal R\right)$ and the behaviour of the scale factor is computed. We show that due to the structure of the system, no global attractors can be present and also that two different classes of solutions for the scale factor exist. In addition, using a redefinition of the dynamic variables, we also compute solutions for static universes.

\section{Introduction}

The cosmology of generalized hybrid metric-Palatini theories can be efficiently analysed using the dynamical systems approach. A dynamical system is a system described by a set of variables that change over time obeying differential equations involving time derivatives. This approach allows one to predict the future behavior of the system by solving these equations either analytical or through numerical integrations. In the particular case of cosmology, this method consists in the definition of a set of specific variables by which the cosmological equations can be converted into an autonomous system of first order differential equations. The analysis characteristic of the phase space of this system can then offer some semi-quantitative information on the evolution of the cosmology. The application of phase space analysis to cosmology has a long history and a recent review on the application of these techniques to a variety of cosmological models can be found in \cite{bahamonde1}.

The  first phase space analysis of the scalar tensor representation of hybrid metric-Palatini theories  was performed in detail in \cite{capozziello1}. However both the definition of the scalar field and the one of the dynamical system variables, which correspond to a rearrangement of the degrees of freedom of the theory might hide some features of this class of theories both at the level of the phase space and of the space of solutions. In \cite{carloni1}, instead, the phase space of this theory was analysed without introducing scalar fields. In the same way, the first phase space analysis of the generalized hybrid metric-Palatini gravity was also performed in the scalar-tensor representation of the theory \cite{tamanini1}.

The objective of this work is to perform a similar (i.e. scalar field free) dynamical systems analysis of the cosmology of hybrid metric-Palatini  which precludes from the scalar-tensor representation. We will analyze four different models, find the respective fixed points and study their stability. We will find that most of the fixed points are saddle points, but in some specific cases it is possible to find attractors and repellers in the phase space.  

\section{Basic equations}

In this section we shall derive the main equations on which we shall be applying the dynamical system formalism. We will be working with the geometrical representation of the theory, i.e., the field equations are given by Eq.\eqref{ghmpgfield}, and to simplify the notation we choose units for which $\kappa^2=1$. Let us now define the Palatini-Einstein tensor $\mathcal G_{ab}$ and the functions $E\left(R,\mathcal R\right)$ and $F\left(R,\mathcal R\right)$ as
\begin{equation}
\mathcal{G}_{ab}\equiv \mathcal{R}_{ab}-\frac{1}{2}\mathcal{R}g_{ab},
\end{equation}
\begin{equation}
E\left(R,\mathcal R\right)=\frac{\partial f}{\partial R}; \ \ \ \ \ F\left(R,\mathcal R\right)=\frac{\partial f}{\partial \mathcal{R}},
\end{equation}
respectively. Using these definitions and the definition of the Einstein tensor $G_{ab}$, the field equations given in Eq.\eqref{ghmpgfield} can be written in the more convenient form
\begin{equation}\label{dynfield}
EG_{ab}+F\mathcal{G}_{ab}-\frac{1}{2}g_{ab}\left[f\left(R,\cal{R}\right)-ER-F\mathcal{R}\right]-\left(\nabla_a\nabla_b-g_{ab}\Box\right)E= T_{ab}.
\end{equation}

We shall be working with functions $f$ that satisfy the Schwartz theorem, and therefore their crossed derivatives are equal, which is also true for the functions $E$ and $F$. This feature imposes the following constraints on the derivatives of the functions $E$ and $F$:
\begin{equation}
E_\mathcal R=F_R,\ \ \ F_{\mathcal R R}=F_{R\mathcal R}=E_{\mathcal R\mathcal R},\ \ \ E_{R\mathcal R}=E_{\mathcal R R}=F_{RR}
\end{equation}
where the subscripts $R$ and $\mathcal R$ denote derivatives with respect to $R$ and $\mathcal R$, respectively. Eqs.\eqref{dynfield} are in principle of order four in the metric $g_{ab}$. However, there are functions $f$ for which these field equations contain only terms of order two. This happens if the following conditions are satisfied:
\begin{eqnarray}\label{dynreduce}
\nonumber &F_R^2-F_\mathcal R E_R =0,\\
&F_R^2F_{\mathcal R\mathcal R}-2F_RF_\mathcal RF_{R\mathcal R}+F_\mathcal R^2F_{RR}=0,\\
\nonumber &F_\mathcal R^3E_{RR}-3F_\mathcal R^2F_RF_{RR}+3F_\mathcal RF_R^2F_{R\mathcal R}-F_R^3F_{\mathcal R\mathcal R}=0.
\end{eqnarray}
A class of functions that satisfies the conditions given in Eq.\eqref{dynreduce} is 
\begin{equation}
f\left(R,\mathcal R\right)=f_0+\mathcal{R}g\left(\frac{R}{\mathcal{R}}\right)+Rh\left(\frac{\mathcal R}{R}\right),
\end{equation}
for some constant $f_0$. In the following we will examine in detail a member of this class of functions.

Since we are interested in exploring the cosmological phase space of the theory, we shall consider the FLRW spacetime with the spatial curvature parameter $k$, where $k$ can assume three values, $k=-1,0,1$. In the usual spherical coordinates $(t,r,\theta,\phi)$ the line element can be written as Eq.\eqref{metricflrw}, where $a(t)$ is the scale factor. We shall also define an auxiliary variable as $A=\sqrt{F}a\left(t\right)$, and a new time variable $\tau=\sqrt{F}t$, so that
\begin{equation}
X^{\dagger}=\frac{\dot X}{\sqrt{F}},
\end{equation}
where the symbol $\dagger$ denotes a derivative with respect to $\tau$ and a dot denotes a derivative with respect to $t$. With these considerations, we write the modified Hubble parameter $\mathcal H$ and the Palatini scalar curvature $\mathcal R$ as
\begin{equation}\label{dynhubble}
\mathcal{H}=\frac{\dot A}{A}=\frac{\dot a}{a}+\frac{\dot F}{2F},
\end{equation}
\begin{equation}\label{dynpalatini}
\mathcal{R}=6F\left[\frac{A^{\dagger\dagger}}{A}+\left(\frac{A^\dagger}{A}\right)^2+\frac{k}{A^2}\right]=6\left(\mathcal{\dot H}+\mathcal{H}^2+\mathcal{H}H+\frac{k}{a^2}\right).
\end{equation}

With these definitions and considering the stress-energy tensor of an isotropic perfect fluid given by $T^a_b=\text{diag}\left(-\rho,p,p,p\right)$, where $\rho$ is the energy density and $p$ is the isotropic pressure, we can write the two independent components of Eq.\eqref{dynfield}, the Friedmann equation and the Raychaudhuri equation, for this system in the forms
\begin{equation}\label{dynfried}
\left(\frac{\dot a}{a}\right)^2+\frac{k}{a^2}\left(1+\frac{F}{E}\right)+\frac{F}{E}\mathcal{H}^2+\frac{1}{6E}\left(f-ER-F\mathcal{R}\right)+\frac{\dot a \dot E}{aE}-\frac{\rho}{3E}=0,
\end{equation}
\begin{equation}\label{dynraych}
\frac{\ddot a}{a}-\frac{F}{E}\left(\frac{k}{a^2}+\mathcal{H}^2\right)+\frac{1}{6E}\left(f-RE\right)+\frac{1}{6E}\left(\rho+3p\right)+\frac{\dot a \dot E}{2aE}+\frac{\ddot E}{2E}=0.
\end{equation}

Note that in defining these equations we have divided by $E$. This operation will introduce a divergence when $E=0$. Such divergence will become relevant in the follwing. We also impose an equation of state of the form
\begin{equation}\label{dyneos}
p=w\rho
\end{equation}
for which the equation of conservation of the stress energy tensor given by $\nabla_aT^{ab}=0$ becomes
\begin{equation}\label{dyntabcons}
\dot\rho+3\frac{\dot a}{a}\left(1+w\right)\rho=0
\end{equation}
The four equations from Eq.\eqref{dynfried} to Eq.\eqref{dyntabcons} constitute a system for the variables $p$, $\rho$, $a$, $k$ and $f$. In the following, when we consider specific examples, we shall be selecting particular forms of the function $f$ in order to close the system. Note that, however, as explained in chapter \ref{chapter:chapter3}, the conversation equation for matter given in Eq.\eqref{dyntabcons} is not independent from the two cosmological equations given in Eqs.\eqref{dynfried} and \eqref{dynraych}, and thus we shall not use it directly. For more details on how these three equations are not independent, see Sec.\ref{Etabequiv}.

\section{Dynamical system approach}

In dealing with dynamical systems, one must study the dimensional structure of the theory considered, because the number of dynamical variables and equations needed to describe the system will depend on the number of dimensional constants present in the theory. Therefore, we introduce a new variable $R_0$ such that the products $R/R_0$ and $\mathcal R/R_0$ are dimensionless. In adition, we also introduce dimensionless parameters in the form of barred greek letters which will represent the product between the coupling constant of the additional invariants and a power of $R_0$. With these considerations, we can write Eq. \eqref{ghmpgaction} as
\begin{equation}
S=\int_\Omega\sqrt{-g}f\left(\frac{R}{R_0},\frac{\mathcal{R}}{R_0},\bar\alpha,...\right)d^4x+S_m(g_{ab}, \chi),
\end{equation}
where $\chi$ collectively denotes the matter fields, the function $f$ retains the same properties as in the action of Eq. \eqref{ghmpgaction} and $R_0$ is assumed to be non-negative. The advantage of this formalism is that instead of needing one dynamical variable for each dimensional constant, we only need a dynamical variable related to $R_0$, since the barred constants become dimensionless.

\subsection{Dynamical variables and equations}

Before proceeding, note that the cosmological equations Eqs. \eqref{dynfried} and \eqref{dynraych} depend non-trivially in time derivatives of the functions $F$ and $E$. These functions can be taken as general functions of $R$ and $\mathcal R$, so that their time derivatives can be written as functions of time derivatives of the curvature scalars, which are themselves functions of time. We therefore must compute the time derivatives of $R$ and $\mathcal R$ to proceed. To do so, we first define the dimentionless time variable $N=\log \left(a/a_0\right)$, where $a_0$ is some constant with dimensions of length, to guarantee that the argument of the logarithm is dimensionless. So, for any quantity $X$ we have
\begin{equation}
X'=\frac{\dot X}{H},
\end{equation}
where the prime $'$ denotes a derivative with respect to $N$ and $H$ is the Hubble parameter defined in Eq.\eqref{defhubble}, and we define new cosmological parameters $q, j,$ and $s$ as
\begin{equation}\label{dyncospar}
q=\frac{H'}{H},\ \ \ \ \ j=\frac{H''}{H},\ \ \ \ \ s=\frac{H'''}{H}.
\end{equation}
Using the previous definitions, the Ricci tensor $R$ and its derivatives with respect to $t$ become
\begin{equation}\label{dynricci}
R=6\left[\left(q+2\right)H^2+\frac{k}{a^2}\right],
\end{equation}
\begin{equation}\label{dyndricci}
\dot R=6H\left\{\left[j+q\left(q+4\right)\right]H^2-\frac{2k}{a^2}\right\},
\end{equation}
\begin{equation}\label{dynddricci}
\ddot R=6H^2\left\{\left[s+4j\left(1+q\right)+q^2\left(q+8\right)\right]H^2+2\left(2-q\right)\frac{k}{a^2}\right\}.
\end{equation}

To find expressions for the derivatives of $\mathcal{R}$ we compute the total derivarive of $F$ with respect to $t$ given by $\dot F=F_R \dot R+F_\mathcal R \dot{\mathcal R}$ and then use Eqs. \eqref{dynhubble} and \eqref{dynpalatini} to solve with respect to $\mathcal {\dot R}$. We obtain:
\begin{equation}\label{dyndpalatini}
\mathcal{\dot R}=\frac{1}{F_\mathcal{R}}\left[\left(\mathcal H-H\right)2F-F_R \dot R\right],
\end{equation}
\begin{eqnarray}
&&\mathcal{\ddot R}=-\frac{1}{F_\mathcal R^2}\left(F_{\mathcal R R}\dot R+F_{\mathcal R \mathcal R}\mathcal{\dot R}\right)\left[\left(\mathcal H-H\right)2F-F_R\dot R\right]+\frac{2F}{F_R}\left(\frac{\mathcal R}{6}-\mathcal H^2-\mathcal H H-\frac{k}{a^2}-qH^2\right)+\nonumber \\
&&+\frac{1}{F_\mathcal R}\left[2\left(\mathcal H-H\right)\left(F_R \dot R + F_\mathcal{R} \mathcal{\dot R}\right)-F_R\ddot R-\dot R\left(F_{RR}\dot R+F_{R\mathcal R}\mathcal{\dot R}\right)\right],\label{dynddpalatini}
\end{eqnarray}
where $\dot R$, $\ddot R$ and $\mathcal{\dot R}$ are computed in Eqs. \eqref{dyndricci}, \eqref{dynddricci} and \eqref{dyndpalatini}, respectively. These results completely determine the forms of the first and second time derivatives of $F$ and $E$.

Let us now define a set of dynamical dimensionless variables. We choose to work with the following variables:
\begin{equation}\label{dynvariables}
K=\frac{k}{a^2H^2},\ \ \ X=\frac{\mathcal H}{H},\ \ \ Y=\frac{R}{6H^2}, \ \ \ Z=\frac{\mathcal R}{6H^2}, \ \ \ J=j, \ \ \ Q=q,\ \ \ \Omega=\frac{\rho}{3H^2E},\ \ \ A=\frac{R_0}{6H^2}.
\end{equation}
The Jacobian $J$ of this definition of variables can be written in the form
\begin{equation}\label{dynjacobian}
J=\frac{1}{108a^2H^9E},
\end{equation}
which means that it has a different form for each choice of the function $f$. In order to guarantee that the variables in Eq. \eqref{dynvariables} cover the entire phase space of the cosmological equations i.e. they constitute a global set of coordinates for it, $J$ must always be finite and non zero, i.e., regular. When the Jacobian is not regular the definition in Eq. \eqref{dynvariables} is not invertible and therefore there can be features of the field equations which are not preserved in the phase space of Eq. \eqref{dynvariables} and features of the phase space of Eq. \eqref{dynvariables} which are spurious, including fixed points. From Eq.\eqref{dynjacobian} it is evident that a regular Jacobian corresponds to $E\neq\{0,\infty\}$. The case $E=\infty\, (J=0)$ corresponds to a true singularity in both Eqs. \eqref{dynfried} and \eqref{dynraych}, and Eqs. \eqref{dynfield}.  The case $E=0\, (J=\infty)$, instead, corresponds to a singularity for Eqs. \eqref{dynfried} and \eqref{dynraych} but not for Eqs. \eqref{dynfield}. This implies that the solutions of Eqs. \eqref{dynfield} associated to $E=0$ will not be represented in the phase space.  

In the following the fixed points for which $J=0/\infty$  will not be included in our analysis unless they are attractors in the phase space. The only exception to this choice will be the fixed points representing static universe solutions which we will consider in Sec. \ref{sec:dynstatic}. We will see that these points have $E=0$, but it is easy to prove via Eqs.\eqref{dynfield} that they  represent true solutions for the field equations.

To simplify the forms of the dynamical equations, it is also useful to define a set of auxiliary dimensionless functions associated with the derivatives of $F$ and $E$. We thus define the following functions:
\begin{eqnarray}
&&\textbf{A}=\frac{F}{E},\ \ \ \textbf{B}=\frac{f}{6EH^2},\ \ \ \textbf{C}=\frac{F_R}{F_\mathcal R},\ \ \ \textbf{D}=\frac{F}{3H^2F_\mathcal R},\ \ \ \textbf{E}=\frac{3H^2F_{R\mathcal R}}{F_\mathcal R},\nonumber \\
&&\textbf{F}=\frac{3H^2F_{RR}}{F_\mathcal R},\ \ \ \textbf{G}=\frac{3H^2F_{\mathcal R\mathcal R}}{F_\mathcal R},\ \ \ \textbf{H}=\frac{3H^2E_{RR}}{F_\mathcal R},\ \ \ \textbf{I}=\frac{E_R}{F_\mathcal R}.\label{dynfunctions}
\end{eqnarray}

Using the definitions of the dynamical variables given in Eq.\eqref{dynvariables} and the definitions of the dynamical functions given in Eq.\eqref{dynfunctions}, we can rewrite the cosmological equations in Eqs.\eqref{dynfried} and \eqref{dynraych} in the forms
\begin{equation}\label{dynfriedvar}
1-Y+\textbf{B}+K+\textbf{A}\left[K+X^2+2\textbf{C}\left(X-1\right)-Z\right]+\frac{2\textbf{A}}{\textbf{D}}\left(\textbf{I}-\textbf{C}^2\right)\left[J+Q\left(Q+4\right)-2K\right]-\Omega=0.
\end{equation}
\begin{eqnarray}
&&1+Q-Y+\textbf{B}+\frac{1+3w}{2}\Omega+\textbf{A}\left\{-\left(K+X^2\right)+\textbf{C}\left[Z-\left(X^2+1\right)-K-Q\right]+\right.\nonumber\\
&&\left.+\frac{\textbf{I}-\textbf{C}^2}{\textbf{D}}\left[J\left(5+4Q\right)+Q\left(4+9Q+Q^2\right)+2K\left(1-Q\right)+s\right]+\right.\nonumber\\
&&\left.+2\left(X-1\right)\left[\left(\textbf{E}\textbf{D}-\textbf{G}\textbf{C}\textbf{D}+\textbf{C}\right)\left(X-1\right)+2\left(J+Q\left(Q+4\right)-2K\right)\left(\textbf{F}+\textbf{G}\textbf{C}^2-2\textbf{E}\textbf{C}\right)\right]+\right.\nonumber\\
&&\left.+\frac{2}{\textbf{D}}\left[J+Q\left(Q+4\right)-2K\right]^2\left(\textbf{H}-3\textbf{C}\textbf{F}+3\textbf{C}^2\textbf{E}-\textbf{C}^3\textbf{G}\right)\right\}=0,\label{dynraychvar}
\end{eqnarray}
respectively, and also rewrite the definitions of $R$ and $\mathcal R$ given by Eqs. \eqref{dynricci} and \eqref{dynpalatini} in the forms
\begin{equation}\label{dynriccivar}
Y=K+Q+2,
\end{equation}
\begin{equation}\label{dynpalatinivar}
Z=\frac{\mathcal{\dot H}}{H^2}+X\left(X+1\right)+K,
\end{equation}
respectively. The dynamical system of equations for the variables in Eq.\eqref{dynvariables} is obtained from the derivatives of the variables with respect to the dimensionless time variable $N$, which must vanish in the fixed points. These equations become
\begin{eqnarray}
K'&=&-2K\left(Q+1\right),\nonumber \\
X'&=&Z-X\left(X+1+Q\right)-K, \nonumber \\
Y'&=&J+Q\left(Q+4\right)-2K-2YQ, \nonumber \\
Z'&=&\textbf{D}\left(X-1\right)+\textbf{C}\left[2K-J-Q\left(Q+4\right)\right]-2ZQ,\label{dynsystem} \\
Q'&=&J-Q^2, \nonumber \\
J'&=&s-QJ,\nonumber \\
\Omega'&=&-\Omega\left\{3\left(1+3w\right)+2Q+2\textbf{A}\left[\textbf{C}\left(X-1\right)+\frac{\textbf{I}-\textbf{C}^2}{\textbf{D}}\left(J+Q\left(Q+4\right)-2K\right)\right]\right\},\nonumber \\
A'&=&-2AQ,\nonumber
\end{eqnarray}
where the cosmological parameter $s$ can be obtained from Eq.\eqref{dynraychvar} and is given by
\begin{eqnarray}
s=\frac{H'''}{H}&=&\frac{\textbf{D}}{\textbf{I}-\textbf{C}^2}\left\{-\frac{1}{\textbf{A}}\left(1+Q-Y+\textbf{B}+\frac{1+3w}{2}\Omega\right)+\left(K+X^2\right)-\textbf{C}\left[Z-\left(X^2+1\right)-K-Q\right]\right.-\nonumber \\
&&\left.-2\left(X-1\right)\left[\left(\textbf{E}\textbf{D}-\textbf{G}\textbf{C}\textbf{D}+\textbf{C}\right)\left(X-1\right)+2\left(J+Q\left(Q+4\right)-2K\right)\left(\textbf{F}+\textbf{G}\textbf{C}^2-2\textbf{E}\textbf{C}\right)\right]-\right. \label{dynsolpars}\\
&&\left.-\frac{2}{\textbf{D}}\left[J+Q\left(Q+4\right)-2K\right]^2\left(\textbf{H}-3\textbf{C}\textbf{F}+3\textbf{C}^2\textbf{E}-\textbf{C}^3\textbf{G}\right)\right\}-\nonumber \\
&&-J\left(5+4Q\right)-Q\left(4+9Q+Q^2\right)-2K\left(1-Q\right).\nonumber
\end{eqnarray}
Finally, Eqs.\eqref{dynfriedvar} and \eqref{dynriccivar} take the role of two constraints for the variables in Eq.\eqref{dynvariables} and allow us to eliminate two variables from the system in Eq.\eqref{dynsystem}. For simplicity, we choose to eliminate $Q$ and $J$, leaving a simplified system of the form
\begin{eqnarray}
K'&=&2K\left(K-Y+1\right),\nonumber \\
X'&=&Z-X\left(X+Y-1\right)+K\left(X-1\right),\nonumber \\
Y'&=&2Y\left(2+K-Y\right)+\frac{\textbf{D}}{2\textbf{A}\left(\textbf{C}^2-\textbf{I}\right)}\left\{1+\textbf{B}+K-Y+\textbf{A}\left[K+2\textbf{C}\left(X-1\right)+X^2-Z\right]-\Omega\right\}, \nonumber \\
Z'&=&\frac{1}{2\textbf{A}\left(\textbf{C}^2-\textbf{I}\right)}\left\{4\textbf{A}\textbf{C}^2\left(2+K-Y\right)-\right.\label{dynsystemsim} \\
&&\left.-2\textbf{A}\textbf{I}\left[\textbf{D}\left(X-1\right)+2Z\left(2+K-Y\right)\right]-\textbf{C}\textbf{D}\left[1+\textbf{B}+K-Y-\Omega+\textbf{A}\left(K+X^2-Z\right)\right]\right\},\nonumber\\
\Omega'&=&-\Omega\left[-2+3w-\textbf{B}-3\left(K-Y\right)-\textbf{A}\left(K+X^2-Z\right)+\Omega\right]\nonumber \\
A'&=&2A\left(2+K-Y\right).\nonumber
\end{eqnarray}
In the above system we have  implemented the constraints given in Eqs. \eqref{dynfriedvar} and \eqref{dynriccivar} to keep the equation to a manageable size. The implementation of the constraints introduces some non trivial structural changes in the system, like the cancelations of the divergences. In the following  we will use the above equations to explore the phase spaces of specific models with a given form of the function $f$.  

\subsection{Solution associated to a fixed point}

\begin{figure}
\centering 
\begin{subfigure}{0.48\textwidth}
\includegraphics[scale=0.65]{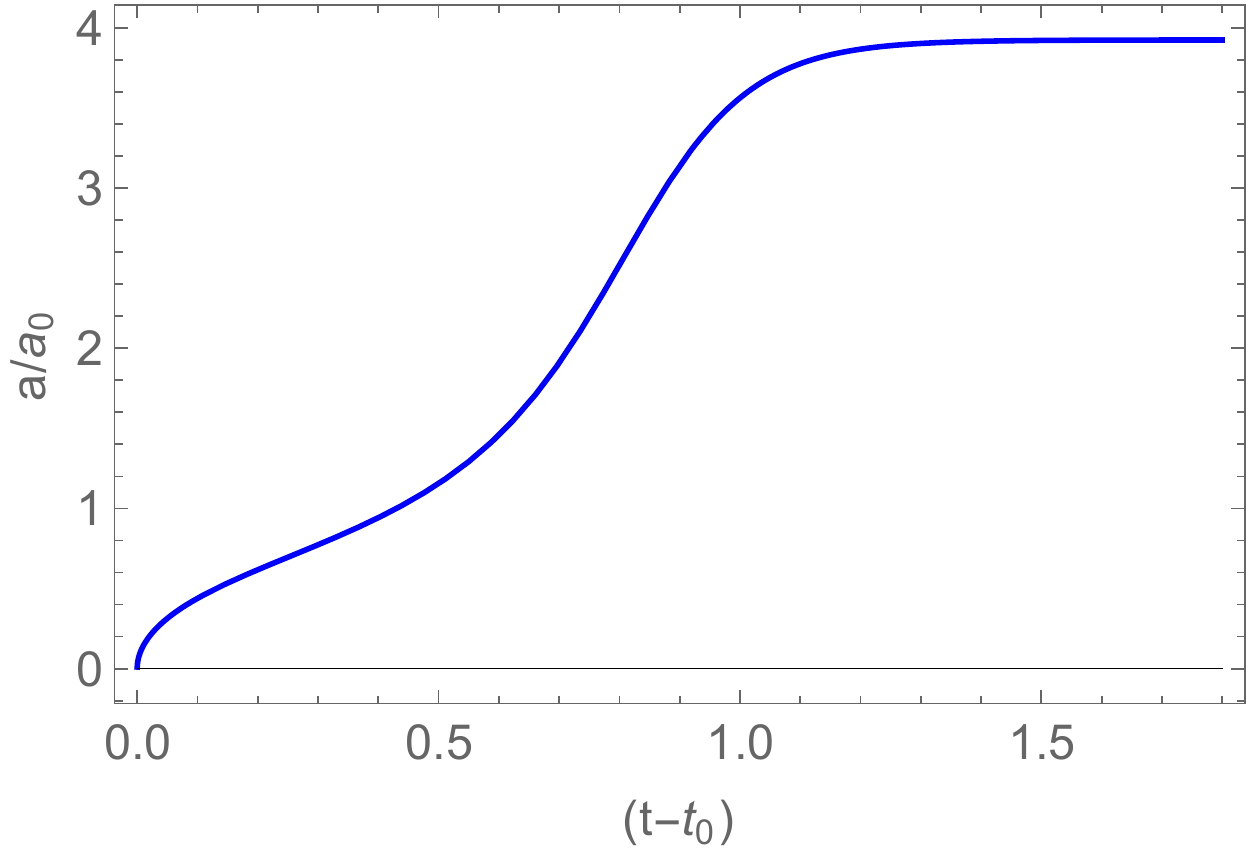}
\caption{Scale factor $a\left(t\right)$ for $s\neq 0$}
\label{fig:dynscale1}
\end{subfigure}
\ \ \ \ \ 
\begin{subfigure}{0.48\textwidth}
\includegraphics[scale=0.65]{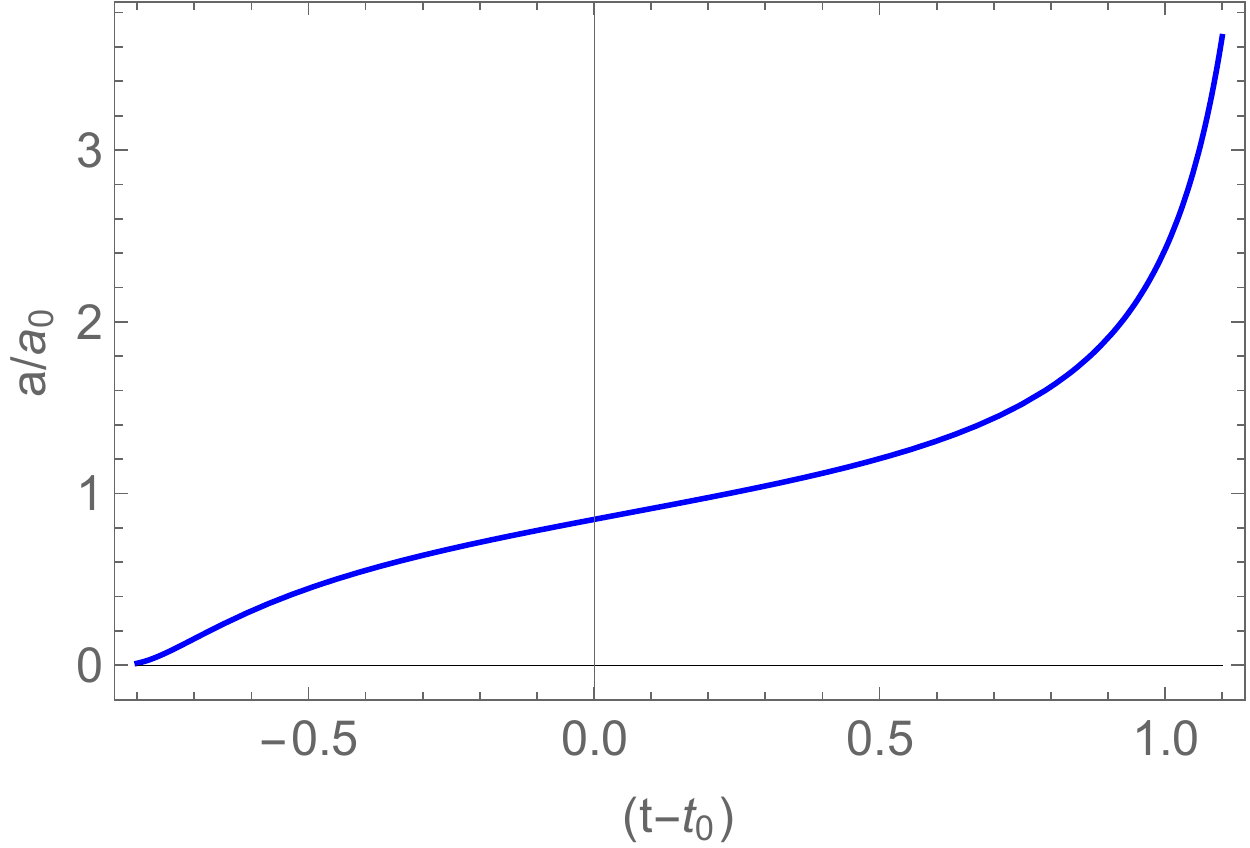}
\caption{Scale factor $a\left(t\right)$ for $s=0$}
\label{fig:dynscale2}
\end{subfigure}
\caption{Solutions for the scale factor $a\left(t\right)$ in Eq.~\eqref{dynsolhubble} with $H_0=H_1=H_2=1$, and $p=2$, $a_0=0.01$ (left panel) or $p=0$, $a_0=1$ (right panel)}
\end{figure}

The solution for a specific fixed point can be obtained by computing the value of $s$ using the values of the dynamic variables and functions at that given fixed point in Eq.\eqref{dynsolpars}. Then, Eq.\eqref{dyncospar} becomes a differential equation for the scale factor $a$ which has two possible forms depending on $s$:
\begin{eqnarray}\label{dynsolhubble}
&&\frac{\dot a}{a}=H_0+H_1\log a+H_2\left(\log a\right)^2, \ \ \ \ \ s=0;\nonumber\\
&&\frac{\dot a}{a}=H_0a^{-p}+a^{\frac{p}{2}}\left[H_1\cos\left(\frac{p\sqrt{3}}{2}\log a\right)+H_2\sin\left(\frac{p\sqrt{3}}{2}\log a\right)\right],\ \ \ \ \ s\neq 0,
\end{eqnarray}
where $p=-\sqrt[3]{s}$ and $H_i$ are constants of integration. The equation for $s=0$ can be solved analitically and the result is
\begin{equation}\label{dynsolscale}
a\left(t\right)=a_0\exp\left\{\frac{\sqrt{4H_2H_0-H_1^2}}{2H_2}\tan\left[\frac{1}{2}\left(t-t_0\right)\sqrt{4H_2H_0-H_1^2}\right]-\frac{H_1}{2H_2}\right\},
\end{equation}
where $a_0$ is a constant of integration. The two types of solution given in Eqs.\eqref{dynsolhubble} have a crucial difference: when $s\neq0$  the time evolution of the solution will approach a constant value of the scale factor, see Fig.\ref{fig:dynscale1}, whereas when $s=0$ a finite time singularity can appear, see Fig.\ref{fig:dynscale2}. The solution for $s=0$ depends on three constants $H_i$ and depending on the values of these constants, the asymptotic character of an orbit approaching a fixed point with $s=0$ can be a constant or a finite-time singularity. In particular, for $H_1^2=4H_2H_0$, we have $a\left(t\right)=a_0\exp\left(-2H_1/H_2\right)$, which is a constant value. The presence of attractors with this character therefore might be a sign of a potential instability of the model for a certain set of initial conditions.

Note also that these two solutions for the scale factor correspond to the solutions we would find if our initial condition corresponds to the fixed point itself, and the solution would be valid from the inital time $t=t_0$ up to the divergence time (in the $s=0$ case) or to infinity (in the $s\neq 0$ case). If our initial conditions do not correspond to a fixed point, then it will orbit through the cosmological phase space and more complicated solutions for the scale factor can arise.

\section{Analysis of specific cases}

In this section we apply the general method presented above to a number of forms of the function $f$. Some of the models are chosen for their simplicity and the analogy with some interesting $f(R)$-gravity theories. Others, like the one of Sec.\ref{sec:dynmodel4}, are chosen for the special form assumed by their field equations and the connection with the results obtained in chapter \ref{chapter:chapter3} for the specific forms of the function $f$ for linear and constant potentials.

\subsection{$R^\alpha \mathcal R^\beta$ gravity}

Let us start by considering the following form for the action
\begin{equation}\label{dynmodel1}
S=\int\sqrt{-g} \gamma \frac{R^\alpha\mathcal R^\beta}{R_0^{\alpha+\beta}} d^4x + S_m,
\end{equation}
where $\gamma$ and $R_0$ are constants. Note that in this case $\gamma$ can be factored out of the action without loss of generality by defining $\bar{S}_m=\gamma^{-1}{S}_m$. As a consequence, there will be no need for the variable $A$ associated to the constant $R_0$, i.e., this is a degenerate case, much in the same way of the case $f(R)=R^n$ in \cite{carloni1}. The Jacobian given in Eq.\eqref{dynjacobian} for this case can be written in terms of the dynamic variables and parameters as
\begin{equation}\label{dynjacobian1}
J=\frac{Y^{1-\alpha}Z^{-\beta}}{\alpha 2^{1+\alpha+\beta}3^{2+\alpha+\beta}H^{7+2\left(\alpha+\beta\right)}a^2}.
\end{equation}
For this Jacobian to be finite, we must exclude the value $\alpha=0$ from the analysis, and also constraint our results for the fixed points to have values for the variables $Y$ and $Z$ different from zero. The dynamical functions in Eq. \eqref{dynfunctions} in this case become
\begin{eqnarray}
&&\textbf{A}=\frac{\beta Y}{\alpha Z},\ \ \ \textbf{B}=\frac{Y}{\alpha},\ \ \ \textbf{C}=\frac{\alpha Z}{\left(\beta-1\right)Y},\ \ \ \textbf{D}=\frac{2Z}{\beta-1},\ \ \ \textbf{E}=\frac{\alpha}{2Y},\nonumber \\
&&\textbf{F}=\frac{\alpha\left(\alpha-1\right)Z}{2\left(\beta-1\right)Y^2},\ \ \ \textbf{G}=\frac{\beta-2}{2Z},\ \ \ \textbf{H}=\frac{\alpha\left(\alpha-1\right)\left(\alpha-2\right)Z^2}{2\beta\left(\beta-1\right)Y^3},\ \ \ \textbf{I}=\frac{\alpha\left(\alpha-1\right)Z^2}{\beta\left(\beta-1\right)Y^2},\label{dynfunctions1}
\end{eqnarray}
and, once the constraints  \eqref{dynfriedvar} and \eqref{dynriccivar} are implemented, the dynamical system from Eq. \eqref{dynsystemsim} becomes 
\begin{eqnarray}
K'&=&2K\left(K-Y+1\right),\nonumber \\
X'&=&Z-X\left(X+Y-1\right)+K\left(X-1\right),\nonumber \\
Y'&=&Y\left\{2\left(2+K-Y\right)+\frac{\left(\beta-1\right)}{\alpha+\beta-1}\left\{1+Y\left(\frac{1}{\alpha}-1\right)+\right.\right.\nonumber\\
&&\left.\left.+K+\frac{\beta Y}{\alpha Z}\left[K+2\frac{\alpha Z}{\left(\beta-1\right)Y}\left(X-1\right)+X^2-Z\right]-\Omega\right\}\right\}, \nonumber\\
Z'&=&\frac{\left(\beta-1\right)}{2\left(\alpha+\beta-1\right)}\left\{4\alpha\beta \left(2+K-Y\right)-2\left(\alpha-1\right)\left[\frac{2Z}{\beta-1}\left(X-1\right)+2Z\left(2+K-Y\right)\right]-\right.\label{dynsystem1} \\
&&\left.-\frac{2\alpha Z}{\left(\beta-1\right)}\left[1+Y\left(\frac{1}{\alpha}-1\right)+K-\Omega+\frac{\beta Y}{\alpha Z}\left(K+X^2-Z\right)\right]\right\},\nonumber\\
\Omega'&=&-\Omega\left[-2+3w-\frac{Y}{\alpha}-3\left(K-Y\right)-\frac{\beta Y}{\alpha Z}\left(K+X^2-Z\right)+\Omega\right]\nonumber 
\end{eqnarray}
Eqs. \eqref{dynsystem1} present divergences for specific values of the parameters  for $\alpha=0$ or $\beta=1$ and for any $\alpha+\beta=1$, which implies that our formulation is not valid for these cases. Indeed, when this is the case the functions in Eq. \eqref{dynfunctions1} are divergent and the analysis should be performed starting again from the cosmological Eqs. \eqref{dynfried} and \eqref{dynraych}. The dynamical system also presents some divergences for  $Y=0$ and $Z=0$, these are due to the very structure of the  gravitation field equations for this choice of the action. Because of these singularities the dynamical system is not $C(1)$ in the entire phase space and one can use the standard analysis tool of the phase space only when $Y, Z\neq0$. We will pursue this kind of analysis here.

The system also presents the $K=0$ and $\Omega=0$ invariant submanifolds together with the invariant submanifold $Z=0$. The presence of the latter submanifolds, allows to solve partially the problem about the singularities in the phase space. Indeed  the presence of the $Z=0$  submanifold implies that no orbit will cross this surface. However the issue remain for the $Y=0$ hypersurface. The presence of this submanifold also prevents the presence of a global attractor  for this case. Such attractor should have  $Z=0$, $K=0$ and $\Omega=0$ and therefore would correspond to a singular state for the theory. This feature also allows to discriminate sets of initial conditions and of parameters values which will lead to a give time-asymptotic state for the system.

\begin{figure}[h!]
\centering
\includegraphics[scale=0.6]{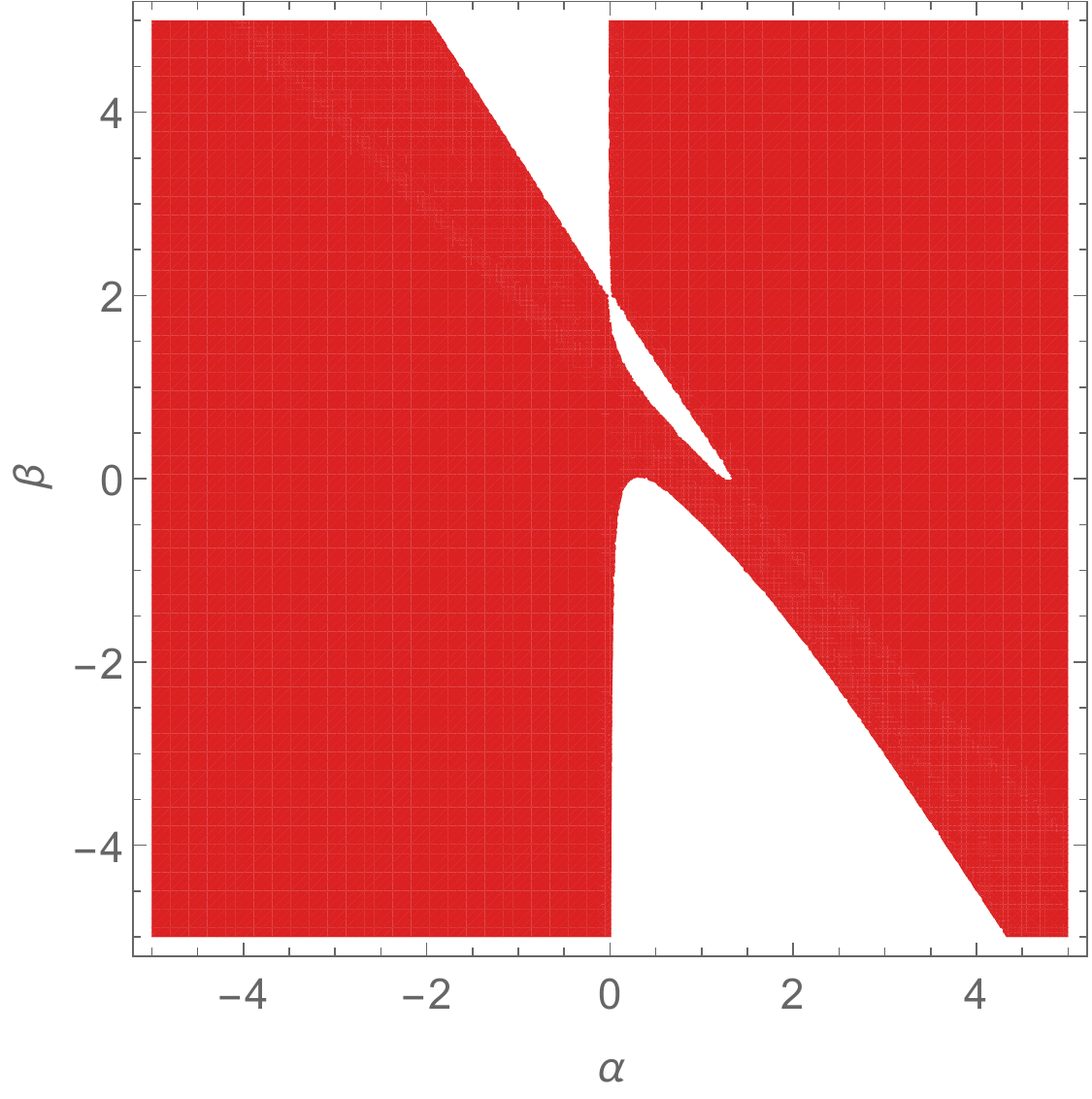}
\caption{Region of the space $\{\alpha,\beta\}$ where the fixed points $\mathcal E_{\pm}$ are defined, for the system give by Eq. \eqref{dynsystem1}.}
\label{fig:dynregion}
\end{figure}

The fixed points of Eq. \eqref{dynsystem1} are at most ten, of which four of them have $Y=0$ and are unstable. Therefore they will be excluded by our analysis. The fixed points $\mathcal E_{\pm}$ are only defined in a specific region of the parameters $\alpha$ and $\beta$ where the coordinates are real. This region is shown in Fig. \ref{fig:dynregion}. Using the constraints from Eqs.\eqref{dynfriedvar} and \eqref{dynriccivar} we also realise that the same happens for  points $\mathcal B$ and $\mathcal C$. 

The stability of the fixed points varies non trivially with the parameters $\alpha, \beta$ and $w$, and it is very difficult to present in a compact way all the general results. By inspection, however, we verify that that only points $\mathcal B$ and $\mathcal E_{\pm}$ can be (local) attractors in the phase space, whereas the other points are always unstable. Point $\mathcal B$ corresponds to a solution of the type shown in Eq. \eqref{dynsolscale} and therefore can lead to a singularity at finite time. Instead, points ~$\mathcal E_{\pm}$ represent a solution approaching a constant scale factor. 

In Table \ref{tab:dynfixed1} we show the most general form of the fixed points for the system in Eq.\eqref{dynsystem1}. The solution for the parameter $s$ in the fixed point $\mathcal E_\pm$ can not be represented in an easy way because of its complexity. The  same happens for the additional conditions arising from the constraints \eqref{dynfriedvar} and \eqref{dynriccivar}. In Tables \ref{tab:dynfixed1a}
 and \ref{tab:dynfixed1b} one can find an explicit analysis of the specific cases $\alpha=1$, $\beta=3$ and $w=1$, and for $\alpha=-1$, $\beta=3$ and $w=0$.

\begin{table}
\centering
\begin{tabular}{c c c c c}
\hline
Point & Coordinates & Existance & Stability & Parameter s \\ \hline
\multirow{7}{*}{$\mathcal A$}
							 &$K=2\alpha^2+2\alpha\left(\beta-1\right)-1$&&&\\
                             &$X=2-\alpha-\beta$&&&\\
                             &$Y=2\alpha\left(\alpha+\beta-1\right)$&&&\\
                             &$Z=\left(\alpha+\beta-1\right)\left[\beta+3\left(\alpha-1\right)\right]$&$3\alpha+\beta\neq 3$&Saddle&$-1$\\
                             &$Q=-1$&&&\\
                             &$J=1$&&&\\
                             &$\Omega=0$&&&\\ \hline
\multirow{7}{*}{$\mathcal B$}
							 &$K=0$&&&\\
                             &$X=1$&&&\\
                             &$Y=\frac{2\alpha}{2\alpha+\beta-2}$&$\alpha+\beta=2$&Saddle&\\
                             &$Z=2$&$2\alpha+\beta\neq 2$&or&0\\
                             &$Q=0$&&Attractor&\\
                             &$J=0$&&&\\
                             &$\Omega=0$&&&\\ \hline
\multirow{7}{*}{$\mathcal C$}
							 &$K=0$&&&\\
                             &$X=\frac{1}{\alpha+\beta}$&$\alpha+3\alpha^3+8\alpha^2\beta+7\alpha\beta^2+$&&\\
                             &$Y=\frac{2\left(\alpha+\beta\right)-1}{\alpha+\beta}$&$+2\beta\left(1+\beta^2\right)^2-5\left(\alpha+\beta\right)^2=0$&&\\
                             &$Z=\frac{1}{\alpha+\beta}$&$\alpha\neq 0$&Saddle&$\frac{-1}{\left(\alpha+\beta\right)^3}$\\
                             &$Q=-\frac{1}{\alpha+\beta}$&$\alpha+\beta\neq 0$&&\\
                             &$J=\frac{1}{\left(\alpha+\beta\right)^2}$&$\alpha+\beta\neq \frac{1}{2}$&&\\
                             &$\Omega=0$&&&\\ \hline
\multirow{7}{*}{$\mathcal D$}
							 &$K=0$&&&\\
                             &$X=\frac{\left(\alpha+\beta\right)\left(3w+1\right)-3\left(w+1\right)}{2\left(\alpha+\beta\right)}$&$w\neq \left\{\frac{1}{3},0\right\}$&&\\
                             &$Y=\frac{4\left(\alpha+\beta\right)-3\left(1+3w\right)}{\alpha+\beta}$&$\alpha+\beta\neq 0$&&\\
                             &$Z=\frac{\left(3w-1\right)\left[\left(\alpha+\beta\right)\left(3w+1\right)-3\left(w+1\right)\right]}{4\left(\alpha+\beta\right)}$&$4\left(\alpha+\beta\right)\neq 3\left(1+w\right)$&Saddle&$-\frac{27}{8}\left(\frac{1+w}{\alpha+\beta}\right)^3$\\
                             &$Q=-\frac{3\left(w+1\right)}{2\left(\alpha+\beta\right)}$&$\left(\alpha+\beta\right)\left(3w+1\right)\neq 3\left(1+w\right)$&&\\
                             &$J=\frac{9\left(w+1\right)^2}{4\left(\alpha+\beta\right)^2}$&&&\\
                             &$\Omega=W\left(\alpha,\beta,w\right)$&&&\\ \hline
\multirow{7}{*}{$\mathcal E_{\pm}$}
							 &$K=0$&&&\\
                             &$X=\frac{\beta+\alpha\left(\alpha+\beta+2\right)-\left[2\pm f\left(\alpha,\beta\right)\right]}{2\alpha\left[2\left(\alpha+\beta\right)-1\right]}$&$\alpha+\beta\neq 0$&&\\
                             &$Y=2-\frac{1\pm f\left(\alpha,\beta\right)}{2\alpha\left(\alpha+\beta-1\right)}+\frac{3\left(1-\alpha\right)\pm 2f\left(\alpha,\beta\right)}{2\alpha\left[2\left(\alpha+\beta\right)-1\right]}$&$\alpha+\beta\neq 1$&Saddle&\\
                             &$Z=g_\pm\left(\alpha,\beta\right)$&$+$&or&NA\\
                             &$Q=\frac{\beta-\left[2\pm f\left(\alpha,\beta\right)\right]+\alpha\left[4-3\left(\alpha+\beta\right)\right]}{2\alpha\left(\alpha+\beta-1\right)\left[2\left(\alpha+\beta\right)-1\right]}$&aditional&Attractor&\\
                             &$J=h_\pm\left(\alpha,\beta\right)$&conditions&&\\
                             &$\Omega=0$&&&\\ \hline
\multicolumn{5}{c}{} \\
\multicolumn{5}{c}{$W\left(\alpha,\beta,w\right)=\frac{1}{2\left(3w-1\right)}\left\{\frac{8\beta}{\alpha-\frac{2\left[11+3w+\alpha\left(-4+9w+9w^2\right)\right]}{\alpha}}-\frac{9\left(1+w\right)^2}{\left(\alpha+\beta\right)^2}+\frac{3\left(1+w\right)\left[4+\left(3+9w\right)\alpha\right]}{\alpha\left(\alpha+\beta\right)}\right\}$} \\ 
\multicolumn{5}{c}{} \\
\multicolumn{5}{c}{$f\left(\alpha,\beta\right)=\sqrt{\left[2+\alpha\left(5\alpha-8\right)\right]^2+2\beta\left[\left(\alpha-1\right)\alpha\left(33\alpha-38\right)-2\right]+\beta^2\left[1+\alpha\left(57\alpha-62\right)\right]+16\alpha\beta^3}$} \\ 
\multicolumn{5}{c}{} \\
\multicolumn{5}{c}{$g_\pm\left(\alpha,\beta\right)=\frac{1}{8\alpha^2}\left\{1-12\alpha+15\alpha^2+8\alpha\beta-\frac{4\left[1\pm f\left(\alpha,\beta\right)\right]\left(\alpha-1\right)}{\alpha+\beta-1}+\frac{3\left(\alpha-1\right)\left[3\left(1-\alpha\right)\pm2f\left(\alpha,\beta\right)\right]}{\left[2\left(\alpha+\beta\right)-1\right]^2}+\frac{3\left(\alpha-2\right)\left(\alpha+1\right)\pm f\left(\alpha,\beta\right)\left(\alpha-5\right)}{2\left(\alpha+\beta\right)-1}\right\}$} \\ 
\multicolumn{5}{c}{} \\
\multicolumn{5}{c}{$h_\pm\left(\alpha,\beta\right)=\frac{17\alpha^4+\left(\beta-2\right)^2+\alpha^3\left(42\beta-52\right)\pm f\left(\alpha,\beta\right)\left[2-\beta+\alpha\left(3\left(\alpha+\beta\right)-4\right)\right]+2\alpha\left(\beta-2\right)\left[6+\beta\left(4\beta-9\right)\right]+\alpha^2\left[56+\beta\left(33\beta-86\right)\right]}{2\alpha^2\left(\alpha+\beta-1\right)^2\left[2\left(\alpha+\beta\right)-2\right]^2}$} \\ 
\multicolumn{5}{c}{} \\ \hline
\end{tabular}
\caption{Fixed points for the system given by Eq. \eqref{dynsystem1}}
\label{tab:dynfixed1}
\end{table}


\begin{table}
\centering
\begin{footnotesize}
\begin{tabular}{c c c c c c c c c c}
\hline
& $K$ & $X$ & $Y$ & $Z$ & $Q$ & $J$ & $\Omega$ & Stability & Parameter $s$ \\ \hline
$\mathcal A$ & 5 & $-2$ & 6 & 9 & $-1$ & 1 & 0 & Saddle & $-1$ \\ 
$\mathcal D$ & 0 & $-\frac{5}{4}$ & $\frac{5}{4}$ & $\frac{5}{4}$ & $-\frac{3}{4}$ & $\frac{9}{16}$ & $-\frac{41}{16}$ & Saddle &  $-\frac{27}{63}$ \\
$\mathcal E_-$ & 0 & $\frac{1}{14}\left(7-\sqrt{385}\right)$ & $\frac{1}{42}\left(77-\sqrt{385}\right)$ & $\frac{23}{7}-\sqrt{\frac{55}{7}}$ & $\frac{1}{42}\left(-7-\sqrt{385}\right)$ & $\frac{1}{126}\left(31+\sqrt{385}\right)$ & 0 & Saddle & $\frac{-301-19\sqrt{385}}{2646}$ \\ 
$\mathcal E_+$ & 0 & $\frac{1}{14}\left(7+\sqrt{385}\right)$ & $\frac{1}{42}\left(77+\sqrt{385}\right)$ & $\frac{23}{7}+\sqrt{\frac{55}{7}}$ & $\frac{1}{42}\left(-7+\sqrt{385}\right)$ & $\frac{1}{126}\left(31-\sqrt{385}\right)$ & 0 & Attractor & $\frac{-301+19\sqrt{385}}{2646}$ \\ 
\end{tabular}
\end{footnotesize}
\caption{Fixed points for the system given by Eq. \eqref{dynsystem1} in the specific case $\alpha=1$, $\beta=3$, $w=1$.}
\label{tab:dynfixed1a}
\end{table}

\begin{table}
\centering
\begin{tabular}{c c c c c c c c c c}
\hline
& $K$ & $X$ & $Y$ & $Z$ & $Q$ & $J$ & $\Omega$ & Stability & Parameter $s$ \\ \hline
$\mathcal A$ & $-3$ & 0 & -2 & -3 & $-1$ & 1 & 0 & Saddle & $-\frac{1}{27}$ \\ 
$\mathcal B$ & 0 & 1 & 2 & 2 & 0 & 0 & 0 & Attractor & 0 \\ 
$\mathcal D$ & 0 & $\frac{1}{4}$ & $\frac{5}{4}$ & $\frac{1}{8}$ & -$\frac{3}{4}$ & $\frac{9}{16}$ & $-\frac{9}{8}$ & Saddle &  $-\frac{27}{64}$ \\ 
\end{tabular}
\caption{Fixed points for the system given by Eq. \eqref{dynsystem1} specific case $\alpha=-1$, $\beta=-3$, $w=0$.}
\label{tab:dynfixed1b}
\end{table}

\subsection{$\alpha R^n+\beta \mathcal R^m$ gravity}

In this section we consider an action of the form
\begin{equation}\label{dynmodel2}
S=\int\sqrt{-g}\left[\bar\alpha \left(\frac{R}{R_0}\right)^n+\bar\beta\left(\frac{\mathcal R}{R_0}\right)^m\right] d^4x + S_m,
\end{equation}
where $R_0$ is a constant with the same units as $R$. Since multiplying the action by a constant does not affect the resultant equations of motion, we can take $\bar\alpha$ out of the action and write
\begin{equation}
S=\int\sqrt{-g}\left[\left(\frac{R}{R_0}\right)^n+\gamma\left(\frac{\mathcal R}{R_0}\right)^m\right] d^4x + \bar{S}_m,
\end{equation}
where $\bar{S}_m=\alpha^{-1}{S}_m$ and $\gamma=\bar\alpha/\bar\beta$ is a parameter that allows us to select which of the two terms is dominant. For $\gamma\ll 1$ we have a dominant $f\left(R\right)$ term, and for $\gamma\gg 1$ we have a dominant $f\left(\mathcal R\right)$ term. The Jacobian from Eq. \eqref{dynjacobian} for this case can be written in terms of the dynamic variables and parameters as
\begin{equation}\label{dynjacobian2}
J=\frac{A^n Y^{1-n}}{18na^2H^7}.
\end{equation}
For this Jacobian to be finite, we must exclude the value $n=0$ from the analysis, and also constraint our results for the fixed points to have values for the variables $Y$ and $A$ different from zero. The dynamical functions in Eq. \eqref{dynfunctions} become
\begin{eqnarray}
&&\textbf{A}=\frac{\gamma mZ^{m-1}A^{n-m}}{ n Y^{n-1}},\ \ \ \textbf{B}=\frac{Y}{n}+\frac{\gamma Z^mA^{n-m}}{n Y^{n-1}},\ \ \ \textbf{D}=\frac{2Z}{m-1},\ \ \ \textbf{G}=\frac{m-2}{2Z},\nonumber \\
&&\textbf{H}=\frac{n \left(n-1\right)\left(n-2\right)Y^{n-3}A^{m-n}}{2\gamma m \left(m-1\right)Z^{n-2}},\ \ \ \textbf{I}=\frac{n \left(n-1\right)Y^{n-2}A^{m-n}}{\gamma m \left(m-1\right)Z^{m-2}},\ \ \ \textbf{C}=\textbf{E}=\textbf{F}=0,\label{dynfunctions2}
\end{eqnarray}
and, using the constraints from Eqs. \eqref{dynfriedvar} and \eqref{dynriccivar}, the dynamical system from Eq. \eqref{dynsystemsim} becomes
\begin{eqnarray}
K'&=&2K\left(K-Y+1\right),\nonumber \\
X'&=&Z-X\left(X+Y-1\right)+K\left(X-1\right),\nonumber \\
Y'&=&Y\left\{2\left(2+K-Y\right)+\frac{1}{1-n}\left[1+\frac{n\left(n-1\right)Y^{n-2}Z^{2-m}}{m\left(m-1\right)\gamma A^{n-m}}+\right.\right.\nonumber\\
&&\left.\left.+K-Y+\frac{m\gamma Y^{1-n}Z^{m-1}}{n A^{m-n}}\left(K+X^2-Z\right)-\Omega\right]\right\}, \nonumber\\
Z'&=&2Z\left(\frac{X-1}{m-1}+2+K-Y\right),\label{dynsystem2}\\
\Omega'&=&-\Omega\left[-2+3w-\frac{Y}{n}\left(1+\frac{\gamma Z^m}{Y^n A^{m-n}}\right)-3\left(K-Y\right)-\frac{\gamma m Z^{m-1}}{n Y^{n-1} A^{m-n}}\left(K+X^2-Z\right)+\Omega\right]\nonumber \\
A'&=&2A\left(2+K-Y\right).\nonumber
\end{eqnarray}

Eqs. \eqref{dynsystem2} also present divergences for specific values of the parameters, in this case $m=0, m=1,$ and $n=1$, which implies again that our formulation is not valid for these cases. Like it was stated in the previous section, when this is the case the functions in Eq. \eqref{dynfunctions2} are divergent and the analysis should be performed starting again from the cosmological Eqs. \eqref{dynfried} and \eqref{dynraych}. The dynamical system also presents the usual divergences for $Y=0$ and $Z=0$, which are due to the very structure of the gravitation field equations. Because of these singularities the dynamical system is not $C(1)$ for the whole phase space and one can only use the usual dynamical system formalism analysis of the phase space when $Y, Z\neq0$.

Just like the previous case, the system given in Eq. \eqref{dynsystem2} also presents the  $K=0$ and $\Omega=0$ invariant submanifolds together with the invariant submanifolds $Z=0$ and $A=0$. The presence of these submanifolds allows to solve partially the problem about the singularities in the phase space. The presence of the $Z=0$ submanifold implies once again that no orbit can cross this surface. On the other hand, the presence of the $Y=0$ submanifold would prevent the presence of a global attractor for this case, but as we will see this model does not present any attractors. Knowing this, it is possible again to analyze discrete sets of initial conditions and parameters and verify the time-asymptotic state for the system.

Much in the same way of the $f(R)-$gravity analogue of this model, the Jacobian $J$ vanishes for all but one of the eighteen fixed points of this model.  Since one can prove that all of these fixed points correspond to singular states of the field equations and that they are unstable, we will ignore them. Therefore the theory has only one relevant fixed point which we will call $\mathcal B$.  As from Table \ref{tab:dynfixed2} the existence of  $\mathcal B$ depends on the values of the parameters $n$, $m$ and $\gamma$ and the point is always a saddle point. If $n\neq m$ it only exists if $\left(m-n\right)$ is an odd number. Instead, for $n=m$, the equation for the variable $A$ decouples from the rest of the system. However, the equation for $A$ can be considered as an extra constraint for the system which carries a memory of the properties of the complete system, such as divergences for $A=0$ and specific values of $\gamma$ that allow the existence of point $\mathcal B$. More specifcally, if $n=m=2$, $\mathcal B$ exists for any value of $\gamma$, whereas for $n=m\neq 2$ then $\mathcal B$ exists only for $\gamma=-1$.  

For any value of the parameters, $\mathcal B$ is associated to $s=0$ and therefore it corresponds to a solution of the type shown in Eq. \eqref{dynsolscale}. This implies that the theory can incur in a singularity at finite time.

\begin{table}
\centering
\begin{tabular}{c c c c c}
\hline
Set & Coordinates & Existance & Stability & Parameter s \\ \hline
\multirow{9}{*}{$\mathcal B$}
							 &$K=0$&&&\\
                             &$X=1$&&&\\
                             &$Y=2$&$n\neq\{0,1\}$&&\\
                             &$Z=2$&&&\\
                             &$Q=0$&If $n=m\neq 2$, then $\gamma=-1$&Saddle&0\\
                             &$J=0$&&&\\
                             &$\Omega=0$&If $n\neq m$, then $n-m=$ odd&&\\
                             &$A\neq 0$, if $n=m$&&&\\ 
                             &$A=-2\left(\frac{n-2}{\gamma\left(m-2\right)}\right)^\frac{1}{n-m}$, if $n\neq m$&&&\\ \hline
\end{tabular}
\caption{Fixed points for the system given by Eq. \eqref{dynsystem2}.}
\label{tab:dynfixed2}
\end{table}

\subsection{$\exp\left(R/\mathcal R\right)$ gravity}

In this section we consider an action of the form
\begin{equation}\label{dynmodel3}
S=\int\sqrt{-g}R_0\exp\left(\frac{R}{\mathcal R}\right) d^4x + \bar{S}_m,
\end{equation}
where $\bar{S}_m=\bar{\alpha}^{-1}{S}_m$ is the dimensionless matter action used in the previous sections. The motivation to test a model of the form of Eq. \eqref{dynmodel3} is that in chapter \ref{chapter:chapter3} this form of the action was one of the solutions for the function $f\left(R,\mathcal R\right)$ for which the geometrical and the scalar-tensor formulations of the theory were equivalent. The Jacobian from Eq. \eqref{dynjacobian} for this case can be written in terms of the dynamic variables and parameters as
\begin{equation}\label{dynjacobian3}
J=\frac{e^{\frac{Y}{Z}}Z}{18a^2H^7}.
\end{equation}
For this Jacobian to be finite, we must constraint our results for the fixed points to have values for the variable $Z$ different from zero. The dynamical functions in Eq. \eqref{dynfunctions} become in this case
\begin{eqnarray}
&&\textbf{A}=-\frac{Y}{Z},\ \ \ \textbf{B}=Z,\ \ \ \textbf{C}=-\frac{Z\left(Y+Z\right)}{Y\left(Y+2Z\right)},\ \ \ \textbf{D}=-\frac{2Z^2}{Y+2Z},\ \ \ \textbf{E}=\frac{Y^2+4YZ+2Z^2}{2Y^2Z+4YZ^2},\nonumber \\
&&\textbf{F}=-\frac{1}{2Y},\ \ \ \textbf{G}=-\frac{Y^2+6YZ+6Z^2}{2Z^2\left(Y+2Z\right)},\ \ \ \textbf{H}=\frac{Z}{2Y^2+4YZ},\ \ \ \textbf{I}=\frac{Z^2}{Y^2+2YZ},\label{dynfunctions3}
\end{eqnarray}
and the dynamical system from Eq. \eqref{dynsystemsim} becomes
\begin{eqnarray}
K'&=&2K\left(K-Y+1\right),\nonumber \\
X'&=&Z-X\left(X+Y-1\right)+K\left(X-1\right),\nonumber \\
Y'&=&Y\left\{2\left(2+K-Y\right)+\left(2+\frac{Y}{Z}\right)\left\{1+Z+K-Y-\right.\right.\nonumber\\
&&-\left.\left.\frac{Y}{Z}\left[K-\frac{2Z\left(Y+Z\right)}{Y\left(Y+2Z\right)}\left(X-1\right)+X^2-Z\right]-\Omega\right\}\right\},\nonumber \\
Z'&=&\frac{2\left(Y+Z\right)^2}{Z^2}\left(2+K-Y\right)-\frac{2Y}{Z}\left(Y+2Z\right)\left[-\frac{Z\left(X-1\right)}{Y+2Z}+2+K-Y\right]+\label{dynsystem3}\\
&&+\left(Y+Z\right)\left[1+Z+K-Y-\Omega-\frac{Y}{Z}\left(K+X^2-Z\right)\right],\nonumber\\
\Omega'&=&-\Omega\left[-2+3w-Z-3\left(K-Y\right)+\frac{Y}{Z}\left(K+X^2-Z\right)+\Omega\right]\nonumber \\
A'&=&2A\left(2+K-Y\right).\nonumber
\end{eqnarray}
where we have used the constraints in Eqs. \eqref{dynfriedvar} and \eqref{dynriccivar}. Eqs. \eqref{dynsystem3} present divergences for specific values of $Y$ and $Z$.  These divergences occur for $Y=0$, $Z=0$, and $Y+2Z=0$, and are due to the  very structure of the  gravitation field equations. The reincidence of the divergences $Y=0$ and $Z=0$ for different choices of the action seems to indicate that there should be few forms of the action for which one can find regularity in these values. These singularities imply again that the dynamical system is not $C(1)$ for the whole phase space and one can use the standard dynamical system analysis of the phase space only when $Y, Z\neq0$ and $Y\neq -2Z$. 

The system also presents the usual $K=0$ and $\Omega=0$ invariant submanifolds together with the invariant submanifold $Z=0$. As stated in the previous sections, the presence of the latter submanifolds allows to solve partially the problem about the singularities in the phase space. The presence of the $Z=0$ submanifold implies again that no orbit will cross this surface, whereas the issue remains for the $Y=0$ hypersurface. The presence of global attractors is also prevented in this case due to the existance of this submanifold. Such attractor should have  $Z=0$, $K=0$ and $\Omega=0$, which would correspond to a singular state for the theory. We can once again use this information to discriminate sets of initial conditions and of parameters values to analyze the time-asymptotic state for the system.

The system given by Eq. \eqref{dynsystem3} presents at most three fixed points, which are shown in Table \ref{tab:dynfixed3} with their stability and associated solution. The fixed point $\mathcal E_-$ has $s\neq 0$ and thus this system may approach a constant scale factor.

\begin{table}
\centering
\begin{tabular}{c c c c c}
\hline
Point & Coordinates & Stability & Parameter s \\ \hline
\multirow{8}{*}{$\mathcal A$}
							 &$K=-6$&&\\
                             &$X=2$&&\\
                             &$Y=-5$&&\\
                             &$Z=-2$&Saddle&$-1$\\
                             &$Q=-1$&&\\
                             &$J=1$&&\\
                             &$\Omega=0$&&\\ \hline
\multirow{8}{*}{$\mathcal E_{\pm}$}
							 &$K=0$&&\\
                             &$X=-\frac{1}{2}\left(5\pm \sqrt{33}\right)$&&\\
                             &$Y=\frac{1}{2}\left(11\pm\sqrt{33}\right)$&$\mathcal E_+$: Saddle&\\
                             &$Z=-\left(5\pm\sqrt{33}\right)$&&$\frac{1}{2}\left(259\pm 45\sqrt{33}\right)$\\
                             &$Q=\frac{1}{2}\left(7\pm\sqrt{33}\right)$&$\mathcal E_-$: Attractor&\\
                             &$J=\frac{1}{2}\left(41\pm\sqrt{33}\right)$&&\\
                             &$\Omega=0$&&\\ \hline
\end{tabular}
\caption{Fixed points for the system given by Eq. \eqref{dynsystem3}.}
\label{tab:dynfixed3}
\end{table}

\subsection{$R\exp\left(\mathcal R/ R\right)$ gravity}\label{sec:dynmodel4}

To finalize, in this section we consider an action of the form
\begin{equation}\label{dynmodel4}
S=\int\sqrt{-g}R\exp\left(\frac{\mathcal R}{R}\right) d^4x + \bar{S}_m,
\end{equation}
where also in this case we have set $\bar{S}_m=\bar{\alpha}^{-1}{S}_m$. This particular form of $f$ not only corresponds to one of the particular cases obtained in chapter \ref{chapter:chapter3} but also satisfies Eq. \eqref{dynreduce} and therefore the field equations are effectively of order two. In terms of the dynamical functions in Eq. \eqref{dynfunctions}, the relations in Eqs. \eqref{dynreduce} read
\begin{eqnarray}
& \textbf{C}^2-\textbf{I}=0,\\
& \textbf{F}+\textbf{G}\textbf{C}^2-2\textbf{E}\textbf{C}=0,\\
& \textbf{H}-3\textbf{C}\textbf{F}+3\textbf{C}^2\textbf{E}-\textbf{C}^3\textbf{G}=0,
\end{eqnarray}
and the cosmological Eqs. \eqref{dynfried} and \eqref{dynraych} can be written, respectively
\begin{equation}\label{dynfriedred}
\frac{1}{Y-Z}\left[Y\left(1+2K+X^2-\Omega\right)+Z\left(1-Z-2X+\Omega\right)\right]=0,
\end{equation}
\begin{eqnarray}\label{dynraychred}
\frac{1}{Y-Z}\left\{Y\left[2\left(Q-K+Z-1\right)+2X\left(4-3X\right)+\left(1+3w\right)\Omega\right]-\right.&&\\
\left.-Z\left[2\left(2-K+Z\right)+2X\left(X-4\right)+\left(1+3w\right)\right]\right\}&=&0.\nonumber
\end{eqnarray}
At this point, using Eq. \eqref{dynriccivar}, we can write $Y$, $Z$ in terms of $X, K, \Omega$ and
substituting in Eqs. \eqref{dynsystem} we obtain:
\begin{eqnarray}
K'&=&-2K\left(1+Q\right),\nonumber \\
X'&=&-K-\left(1+Q\right)X-X^2+Z,\label{dynsystem4}\\
\Omega'&=&\frac{\Omega}{Y-Z}\left[Z\left(1+3w+2Q+2X\right)-Y\left(3+3w+2Q\right)\right],\nonumber
\end{eqnarray}
where $Y=Y(X, K, \Omega)$, $Z=Z(X, K, \Omega)$ and $Q=Q(X, K, \Omega)$ have not been fully substituted for sake of simplicity. The Jacobian in Eq. \eqref{dynjacobian} for this case can be written in terms of the dynamic variables and parameters as
\begin{equation}
J=\frac{Ye^{-\frac{Z}{Y}}}{108a^2H^9\left(Y-Z\right)}.
\end{equation}
For this Jacobian to be regular, we must exclude the fixed points that have values of $Y=Z$ or $Y=0$, which also represent divergences for the system in Eq. \eqref{dynsystem4} and the very gravitation field equations for this choice of the action. Because of these singularities the dynamical system Eqs.\eqref{dynsystem4} is not $C(1)$ in the entire phase space and one can use the standard analysis tool of the phase space only when $Y\neq Z$ and $Z\neq 0$.

Eqs. \eqref{dynsystem4} also presents the usual $K=0$ and $\Omega=0$ invariant submanifolds together with the invariant submanifold $Z=0$, which is not explicit but still exists. The presence of this last submanifold allows to solve partially the problem about the singularities in the phase space. The analysis is the same as before: the presence of the $Z=0$  submanifold implies that no orbit will cross this surface, which also prevents the presence of a global attractor  for this case. Such attractor should have  $Z=0$, $K=0$ and $\Omega=0$ and therefore would correspond to a singular state for the theory. We can therefore discriminate sets of initial conditions and of parameters values which give rise to a given time-asymptotic state for the system. 

The system Eq. \eqref{dynsystem4} presents at most three fixed points, which are shown in Table \ref{tab:dynfixed4} with their stability and associated solution. These points are all non hyperbolic, i.e., linear analysis cannot be used to ascertain their stability. A standard tool for the analysis for this type of points is the analysis of the central manifold. The method consists in rewriting the system Eqs.~\eqref{dynsystem4} in terms of new variables $(U_1, U_2, U_3)$ in the form
\begin{eqnarray}
U_1'=AU_1+F_1\left(U_1,U_2,U_3\right),\nonumber \\
U_2'=BU_2+F_2\left(U_1,U_2,U_3\right),\\
U_3'=CU_3+F_3\left(U_1,U_2,U_3\right),\nonumber \\
\end{eqnarray}
where $A$, $B$, and $C$ are constants and the functions $F_i$ respect the conditions $F_i\left(0,0,0\right)=0$ and $\frac{\partial F_i}{\partial U_j}\left(0,0,0\right)=0$, for $\{i,j\}=1,2,3$. Supposing that the real part of the quantity $A$ vanished, the variables $U_2$ and $U_3$ can be written as
\begin{eqnarray}
U_2=h_2\left(U_1\right),\nonumber \\
U_3=h_3\left(U_1\right),
\end{eqnarray}
and the centre manifold can be defied by the equations
\begin{eqnarray}
h_2'\left(U_1\right)\left[A U_1+F_1\left(U_1,h_2\left(U_1\right),h_3\left(U_1\right)\right)\right]-Bh_2\left(U_1\right)-F_2\left(U_1,h_2\left(U_1\right),h_3\left(U_1\right)\right)=0,\nonumber \\
h_3'\left(U_1\right)\left[A U_1+F_1\left(U_1,h_2\left(U_1\right),h_3\left(U_1\right)\right)\right]-Ch_3\left(U_1\right)-F_3\left(U_1,h_2\left(U_1\right),h_3\left(U_1\right)\right)=0.
\end{eqnarray}
which can be solved by series. The stability of the non hyperbolic point will be then determined by the structure of the equation
\begin{eqnarray}\label{dynmanifold}
U_1'=AU_1+F_1\left(U_1,h_2(U_1),h_3(U_1)\right).
\end{eqnarray}

Let us now analyze the stability of the three fixed points obtained for this model. For point $\mathcal A$, the variable transformation is 
\begin{equation}
U_1=K,\ \ \ \ \ U_2=X-\frac{1}{2},\ \ \ \ \ U_3=\Omega,
\end{equation}
and Eq.\eqref{dynmanifold} takes the form
\begin{equation}
U_1'=\frac{8}{5}U_1^2+\mathcal O\left(U_1^3\right)
\end{equation}
from which we see that $U_1'>0$ for any value of $U_1$ near zero, which implies that this point is a saddle. For point $\mathcal B$ the variable transformation is 
\begin{equation}
U_1=K+1,\ \ \ \ \ U_2=X-\frac{1}{2}\left(1-3w\right),\ \ \ \ \ U_3=\Omega+1+3w,
\end{equation}
and Eq.\eqref{dynmanifold} takes the form
\begin{equation}
U_1'=2U_1^2+\mathcal O\left(U_1^3\right),
\end{equation}
which implies again that this point is a saddle. Finally, for point $\mathcal C$ the variable transformation is 
\begin{equation}
U_1=K,\ \ \ \ \ U_2=X+\frac{3}{2}\left(w-1\right),\ \ \ \ \ U_3=\Omega-2+3w,
\end{equation}
and Eq.\eqref{dynmanifold} takes the form
\begin{equation}
U_1'=2\left[\frac{1}{3\left(1-w\right)}+\frac{1}{9w-5}\right]U_1^2+\mathcal O\left(U_1^3\right).
\end{equation}
For $0<w<1$ also this point is a saddle. The solutions associated to the fixed points can be found by the relation
\begin{equation}
\frac{\dot{H}}{H^2}=q=Q
\end{equation}
and using $Q=Q(X, K, \Omega)$ obtained by Eqs.\eqref{dynfriedred} and \eqref{dynraychred} and evaluated at the fixed point. In general we have  
\begin{eqnarray}
a\left(t\right)&=&a_0\exp\left(H_0t\right),\ \ \ \ \ q_*=0,\nonumber \\
a\left(t\right)&=&a_0\left(t-t_0\right)^{-\frac{1}{q}},\ \ \ \ \ q_*\neq 0\label{dynsolscalered},
\end{eqnarray}
where $H_0$, $a_0$ and $t_0$ are constants of integration and $q_*$ is the value of $Q$ at the fixed point. Notice that the fixed points are characterised by only two different values of $q_*$, i.e. $-1$ and $-2$. Using Eq.\eqref{dynsolscalered}, we verify that the solution for $q=-1$ corresponds to a linearly growing scale factor, whereas the solution for $q=-2$ corresponds to a solution for the scale factor that grows with $\sqrt{t}$.

One of the motivations behind the choice to analyze an action of the form of Eq.~\eqref{dynmodel3} was the comparison with the results obtained in chapter \ref{chapter:chapter3}. In this work some forms of the function $f\left(R,\mathcal R\right)$, including the one of Eq.~\eqref{dynmodel3},  were obtained by reconstruction from a given cosmological solution. The phase space  analysis we have performed allow to understand the stability of such solutions, which was impossible to obtain by reconstruction. In particular, point $\mathcal A$ corresponds to the solution found in chapter \ref{chapter:chapter3} i.e. a flat ($K=0$) vacuum ($\Omega=0$) universe with $a\left(t\right)$ proportional to $\sqrt{t}$ and we determined that such solution is unstable. This shows that we can use the dynamical system to determine the stability of the fixed point obtained in our previous work even of these results were obtained using a non trivial redefinition of the action. It should be stressed, however, that it is not necessarily true that an exact solution found for the cosmological equation of a given theory corresponds to a fixed point of our phase space.  For example,  in chapter \ref{chapter:chapter3} an exact non flat vacuum solution was found in $f=\mathcal{R}\exp\left(R/\mathcal R\right)$ which does not correspond to a fixed point of the phase space. However, in general the phase space analysis is useful to understand in a deeper way not only the stability of the solution, but also the consequences of the reorganisation of the degree of freedom that is often employed to analyse this class of theories.

\begin{table}
\centering
\begin{tabular}{c c c c c}
\hline
Point & Coordinates & Stability & Parameter q \\ \hline
\multirow{6}{*}{$\mathcal A$}
							 &$K=0$&&\\
                             &$X=\frac{1}{2}$&&\\
                             &$Y=0$&Saddle&$-2$\\
                             &$Z=-\frac{1}{4}$&&\\
                             &$\Omega=0$&&\\ \hline
\multirow{6}{*}{$\mathcal B$}
							 &$K=-1$&&\\
                             &$X=\frac{1}{2}\left(1-3w\right)$&&\\
                             &$Y=0$&Saddle&$-1$\\
                             &$Z=\frac{3}{4}\left(1+3w\right)\left(w-1\right)$&&\\
                             &$\Omega=-\left(3w+1\right)$&&\\ \hline
\multirow{6}{*}{$\mathcal C$}
							 &$K=0$&&\\
                             &$X=-\frac{3}{2}\left(w-1\right)$&&\\
                             &$Y=0$&Saddle&$-2$\\
                             &$Z=\frac{3}{4}\left(3w-1\right)\left(w-1\right)$&&\\
                             &$\Omega=2-3w$&&\\ \hline
\end{tabular}
\caption{Fixed points for the system given by Eq. \eqref{dynsystem4}.}
\label{tab:dynfixed4}
\end{table}

\section{Stability of static universes}\label{sec:dynstatic}

The variables we have defined in the previous section (see Eq.\eqref{dynvariables}) are efficient in determining the fixed points corresponding to a finite value of the quantities they represent. However, since variables have a term $H$ or $H^2$ in the denominator, our setting excludes an interesting case which is connected with the existence of solutions characterised by  $H=0$, i.e., the static Universes. In particular, fixed points (if any) associated to this kind of solution would be at the infinite boundary of the phase space. In order to look for solutions with $H=0$ one has, therefore, to investigate the asymptotics of the dynamical system. 

There are many approaches that can be adopted for this purpose. One could employ, for example, stereographic projections by which the infinite boundary is mapped to a finite radius sphere. In the following we will use a different strategy which allow to analyse the stability of a static universe without having to explore the entire asymptotics. More specifically, we will redefine all the variables and functions in such a way to bring the static fixed point in the finite part of the phase space. As said, the exploration of this extended phase space, is clearly not a complete analysis of the asymptotic, but it will allow an easier analysis of the stability of these solutions.

We start by redefining the cosmological parameters as:
\begin{equation}\label{dyncosparsta}
\bar q=\frac{H'}{\left(H+c\right)}\ \ \ \ \ \bar j=\frac{H''}{\left(H+c\right)}\ \ \ \ \ \bar s=\frac{H'''}{\left(H+c\right)},
\end{equation}
where $c$ is an arbitrary constant with units of $H$. In the same philosophy, the dynamic variables can be defined as:
\begin{eqnarray}
&&\bar K=\frac{k}{a^2\left(H+c\right)^2},\ \ \ \bar X=\frac{\mathcal H}{\left(H+c\right)},\ \ \ \bar Y=\frac{R}{6\left(H+c\right)^2}, \ \ \ \bar Z=\frac{\mathcal R}{6\left(H+c\right)^2}, \nonumber \\ 
&&\ \bar J=\bar j, \ \ \ \bar Q=\bar q,\ \ \ \bar\Omega=\frac{\rho}{3\left(H+c\right)E},\ \ \ \bar A=\frac{R_0}{6\left(H+c\right)^2}, \ \ \ T=\frac{H}{\left(H+c\right)}.\label{dynvariablessta}
\end{eqnarray}
The Jacobian $J$ of this definition of variables can be written in the form
\begin{equation}\label{dynjacobiansta}
J=\frac{1}{108a^2\left(H+c\right)^9E}.
\end{equation}
This means that for each specific model, the constraints that arise from imposing that the Jacobian must be finite and different from zero are the same as in the analysis of the previous section. From this point on we shall drop the bars to simplify the notation. The evolution equations become in this case:
\begin{eqnarray}
K'&=&-2K\left(Q+T\right)\nonumber \\
X'&=&Z-X\left(Q+X+T\right)-K\nonumber \\
Y'&=&J-2KT+Q\left(4T-2Y+Q\right)\nonumber \\
Z'&=&-\textbf{C}\left(J+Q^2-2KT+4QT\right)+\textbf{D}T^2\left(X-T\right)-2QZ\\
Q'&=&J-Q^2 \nonumber \\
J'&=&s-JQ \nonumber \\
\Omega'&=&\frac{\Omega}{\textbf{D}T}\left\{-\textbf{D}T^2\left[2Q+3T\left(1+w\right)\right]+2\textbf{A}\left[\left(\textbf{C}^2-\textbf{I}\right)\left(J+Q^2-2KT+4QT\right)+\textbf{C}\textbf{D}T^2\left(T-X\right)\right]\right\}  \nonumber
\end{eqnarray}
and the Friedmann equation from Eq.\eqref{dynfried} becomes
\begin{eqnarray}
\textbf{A}\left[2\left(\textbf{I}-\textbf{C}^2\right)\left(J+Q^2-2KT-4QT\right)+2\textbf{D}\textbf{C}T^2\left(X-T\right)+\textbf{D}T\left(K+X^2-Z\right)\right]+&&\nonumber\\
+\textbf{D}T\left[K+\left(1+\textbf{B}\right)T^2-Y-\Omega\right]&=&0
\end{eqnarray}
The Raychaudhuri equation given in Eq.\eqref{dynraych} is too long to be reported here but can be computed easily. Using this new system of equations we can investigate the extended phase space. As we will see, we will be able to  find  all the previously discovered fixed points as points with $T=1$ plus extra sets of fixed points with $H=0$ which will have $T=0$. As an example of how to apply the new dynamical system approach, we will perform the analisys for the models from Eqs. \eqref{dynmodel1}, \eqref{dynmodel2}, and \eqref{dynmodel3}. In all the three cases, there is only one fixed point with H=0, which corresponds to the origin, i.e. 
\begin{equation}
\mathcal O =\{K=0,X=0,Y=0,Z=0,Q=0,J=0,\Omega=0,T=0\}.
\end{equation}
This result is somewhat expected. The fact that we are looking for fixed points with $H=0$ implies directly that $T=0$. Then, both $K=0$ and $\Omega=0$ are invariant submanifolds. Now, if $H=0$, then $a$ is a constant and $\dot a=\ddot a=0$, from which $Q=J=0$. Since $k=0$ from $K=0$, then we also have $R=0$ and therefore $Y=0$. This been said, the only values of both $Z$ and $X$ that make $X'=0$ and $Z'=0$ using the results explained in this paragraph are $Z=0$ and $X=0$, and the fixed point is the origin.

\paragraph*{Model $R^\alpha\mathcal R^\beta$:}
For the model from Eq. \eqref{dynmodel1} the fixed point $\mathcal O$ is always unstable, but might correspond to a saddle point or to a repeller depending on the parameters $\alpha$ and $\beta$. In fact, If $1-\beta<\alpha<0$ or $-\beta>\alpha>0$, this point corresponds to a repeller. Any other combination of the parameters gives rise to a saddle point.

\paragraph*{Model $R^n+\mathcal R^m$:}
For the model from Eq. \eqref{dynmodel2}, note that we have one extra variable $A$ which also vanishes in this calculation. This implies that not all values of $m$ and $n$ are allowed, since the power of $A$ must be positive for the system to converge. Despite that, the analysis of the stability in this case reveals that, for all the combinations of the parameters $m$, $n$ and $\gamma$ for which the fixed point exists, it is always a saddle point.

\paragraph*{Model $\exp\left(\frac{R}{\mathcal R}\right)$:}
For the model from Eq. \eqref{dynmodel3}, the analysis of the stability of the point $\mathcal O$ reveals that the fixed point is unstable, since the eigenvalues associated to it are all either positive or zero. However, to verify if the fixed point is a saddle point or a repeller, one would have to make use of the central manifold theorem again. For our purposes however, it is enough for us to note that, since all the other eigenvalues of the point are of alternate sign, we can conclude directly that the point is unstable.

\paragraph*{Model $R\exp\left(\frac{\mathcal R}{R}\right)$:}

For the model from Eq.\eqref{dynmodel4}, there are no fixed points corresponding to static universe solutions, i.e., point $\mathcal O$ does not exist in this model. 

\section{Conclusions}

In this chapter we have applied the methods of dynamical systems to analyze the structure of the phase space of the generalized hybrid metric-Palatini gravity in a cosmological frame. Using the symmetries of the theory in $R$ and $\mathcal R$ we obtained the cosmological equations and, defining the appropriate dynamical variables and functions, we derived a closed system of dynamical equations that allows us to study the phase space of different forms of the function $f\left(R,\mathcal R\right)$.  We studied four different models of the function $f$: namely Eqs. \eqref{dynmodel1}, \eqref{dynmodel2}, \eqref{dynmodel3} and \eqref{dynmodel4}.

The structure of the phase space is similar in all the studied cases. In none of the particular cases there were global attractors due to the fact that one of the invariant submanifolds present in all the cases, $Z=0$, corresponds to a singularity in the phase space, and therefore a global attractor which would have to be in the intersection of all the invariant submanifolds would correspond to a singular state of the theory. However, the presence of these submanifolds allows us to discriminate sets of initial conditions and predict the time asymptotic states of the theory.

Except for the particular case shown in Eq. \eqref{dynmodel3}, all the theories we analysed feature a fixed point that we denoted by $\mathcal B$. This fixed point stands in the intersection of two of the invariant submanifold, $\Omega=0$ and $K=0$, with a positive value $Z=2$. In the model of Eq. \eqref{dynmodel2} $\mathcal B$ is always unstable, but in the model from Eq. \eqref{dynmodel1}  $\mathcal B$ is an attractor for some particular values of $\alpha$ and $\beta$ (see table \ref{tab:dynfixed1}). This means that all the orbits starting with a positive value of $Z$ and  with a value of $Y$ with the same sign as the one arising from the particular choice of parameters in $\mathcal B$, can eventually reach this fixed point. Note that it is even possible to chose sets of parameters such that $\mathcal B$ is the only finite attractor for the system (see table \ref{tab:dynfixed1b}). The solution associated to point $\mathcal B$ is characterised by $s=0$ and contains three constants of integration $H_i$. Depending of the values of these constants it can have two different types asymptotic limit: a constant or a finite type singularity. The value of the constants $H_i$ and therefore the possibility of the occurrence of the singularity depends on observational constraints on higher order cosmological parameters (e.g. jerk, snap, etc.). This suggests that models for which $\mathcal B$ is an attractor generalized hybrid-metric Palatini theories, like many $f(R)$-gravity models, can incur in finite time singularities.

The other possible attractors in the phase space of the theories we have analysed are the fixed points $\mathcal E_\pm$. They appear in the case of Eqs. \eqref{dynmodel1} and \eqref{dynmodel3}. Contrary to the fixed point $\mathcal B$, the fixed points $\mathcal E_\pm$ have always a solution $s\neq 0$ which asymptotically tend to a constant scale factor. These fixed points also lie in the intersection of the $\Omega=0$ and $K=0$ invariant submanifolds, thus we expect that some of the orbits should reach this fixed point. For the model in Eq. \eqref{dynmodel1} we have shown that it is possible to choose sets of parameters such that $\mathcal E_+$ is the only finite attractor for the system, see table \ref{tab:dynfixed1a}. On the other hand, for the model in Eq. \eqref{dynmodel3} $\mathcal E_-$ is always the only finite attractor of the system and all the orbits starting from a positive value of $Y$ and a negative value of $Z$ might reach this fixed point.

For the specific case shown in Eq. \eqref{dynmodel4}, the system of dynamical equations becomes much simpler since only two variables are needed to fully describe the phase space and the solutions. However, the study of the stability becomes more complicated because the fixed points are non-hyperbolic and their stability analysis requires the use of central manifolds. The behaviour of the solutions reduces to simple power-laws or exponentials.

Although the expected scale factor behaviors can not be obtained due to the higher order terms in the field equations, some of the fixed points obtained in this analysis can be mapped to the well-known fixed points in GR. The cosmological phase space of FLRW universes in GR presents at most two fixed points, the Milner universe, given by $k=-1$ and $\Omega=0$, and the flat Friedmann-Lemaître universe, with $k=0$ and $\Omega=1$. From Tab.\ref{tab:dynfixed1}, we verify that in the limit to GR which corresponds to $\alpha=1$ and $\beta=0$, the fixed point $\mathcal A$ recovers $k=-1$ and $\Omega=0$, and the fixed point $\mathcal D$ recovers $k=0$ and $\Omega\neq 0$, with a value that is not necessarily $\Omega=1$ due to our different definition of $\Omega$.  

Our analysis connects directly with the results of chapter \ref{chapter:chapter3} in which some pairs function-exact solution were found via reconstruction method. One of the limitation of the reconstruction technique was the impossibility to understand the stability of of the solution obtained. The phase space analysis gives us a tool to determine this stability. In particular, we determined that the solution found in chapter \ref{chapter:chapter3} is actually unstable.

Since the variables Eqs.\eqref{dynvariables} have $H$ in the denominator, they cannot be used directly to investigate the presence of static ($H=0$) fixed points, as they would be located at the asymptotic boundary of the phase space. In order to study static universe solutions, we generalised Eqs. \eqref{dynvariables} in such a way to move possible static fixed points  to the finite part of the phase space. This is different from a complete asymptotical analysis, but it allows to obtain information on static universes in an easier way. All the theories we have considered  with Eq. \eqref{dynvariables} turn out to present a static fixed point which is always unstable. Therefore, as in GR, in these models the static universe is always unstable. However, differently from GR, the solution associate to this points are spatially flat and empty (i.e. no cosmological constant). Such peculiar forms of static universes  are the result of the action of the non trivial geometrical terms appearing in the field equations.

The existence of unstable static solutions in the phase space points to the existence in the context of generalised hybrid metric-Palatini gravity to phenomena such as bounces, turning points, and loitering phases, which are represented by the orbits bouncing against the static fixed points. These open the way to a series of scenarios which could be interesting to investigate further.

Finally, note that for the particular case in Eq. \eqref{dynmodel2}, no finite nor asymptotic attractors were found. One explanation for this result is that the orbits in the phase space do not tend asymptotically to a given solution but could instead be closed upon themselves. A structure similar to this one occurs for example in the frictionless pendulum, where the orbits are closed and there are no attractors in the phase space. This structure could indicate that the solutions represented by orbits in the phase space actually correspond to cyclic universes. 
\cleardoublepage

\chapter{Junction conditions and thin shells in the generalized hybrid metric-Palatini gravity}
\label{chapter:chapter5}

In this chapter we study the junction conditions of a recently proposed modified theory of gravity called generalized hybrid metric-Palatini gravity, which arises as a natural generalization of the successful hybrid metric-Palatini gravity. We compute the junction conditions for a smooth matching of two regions of the spacetime and also the conditions for a matching with a thin-shell of matter at the separation hypersurface. The dynamically equivalent scalar-tensor representation of the theory is also derived and the equivalent junction conditions in this representation are also computed for a smooth matching with and without a thin-shell. These junction conditions are then used in two specific examples, the matching of a Minkowski spacetime to a Schwarzschild spacetime with a thin shell separating the two spacetimes, for which it is shown that, unlike in GR, the matching can only be performed at a specific value of the shell radius that corresponds to the Buchdahl limit $R=9M/4$; and also a matching between the Minkowski spacetime and the Schwarzschild spacetime using a thick shell of constant density perfect fluid, for which it is shown that a smooth matching between the fluid and the Schwarzschild exterior is possible if we constrain the surface of the shell to the light ring $R=3M$, and all the energy conditions are satisfied for the whole spacetime.

\section{Introduction}

In the pursuit of finding solutions for the EFE in GR, sometimes the following situation arises: a hypersurface separates the whole spacetime into two regions which are described by two different metric tensors expressed in two different coordinate systems. In these situations, it is natural to ask which conditions must the two metric tensors satisfy in order for the two regions to be matched smoothly at the hypersurface. These conditions are called the junction conditions.

The junction conditions in GR have been deduced long ago \cite{darmois1,israel1}, and imply that the induced metric across the hypersurface that separates the two spacetime regions must be continuous, and also that the extrinsic curvature must be continuous. To obtain these conditions, a standard way consists in writing the general metric $g_{ab}$ for the whole spacetime as a sum of distribution functions defined only in each of the two subspaces, via the use od the Heaviside distribution function $\Theta\left(l\right)$. One then computes the geometrical quantities that depend on the metric, namely the Christoffel symbols $\Gamma^c_{ab}$, the Riemann tensor $R^a_{bcd}$, the Ricci tensor $R_{ab}$ and the Ricci scalar $R$, insert these results into the field equations, and verify which conditions must be satisfied by these quantities in order to avoid the presence of divergences. These conditions have been used to derive new solutions for the EFEs, such as constant density stars with an exterior schwarzschild, the Openheimer-Snyder stellar collapse \cite{oppenheimer1}, and the matching between FLRW spacetimes with Vaidya (and consequently, Schwarzschild) exteriors \cite{senovilla1}. 

If the extrinsic curvature is not continuous, then the matching between the two regions of spacetime can still be done but implies the existance of a thin-shell of matter at the junction radius\cite{israel1,lanczos1,lanczos2}.  The thermodynamics of these shells has also been studied \cite{martinez1} and the shell's entropy has been computed in various specific cases such as rotating shells \cite{lemos2,lemos3} and electrically charged shells\cite{lemos4,lemos5}. Colisions of spacetimes with two shells have also been studied with numerical approaches \cite{brito2}.

In the context of modified gravity, each modified theory of gravity will have its own junction conditions, which must be deduced from the modified field equations and the equations of motion of the extra fields, if any. In particular, the junction conditions have been deduced for $f\left(R\right)$ theories of gravity with \cite{vignolo1} and without torsion \cite{senovilla2,deruelle1}, scalar-tensor theories \cite{barrabes1,suffern1}, and also Gauss-Bonet gravity \cite{davis2}. The objective of this chapter is to deduce the junction conditions for smooth matching and matching with thin shells in the generalized hybrid metric-Palatini gravity and provide examples of application.

\subsection{Notation and assumptions}

Let us now introduce a few considerations about the formalism used in this chapter. Let $\Sigma$ be a hypersurface that separates the spacetime $\mathcal V$ into two regions, $\mathcal V^+$ and $\mathcal V^-$. Let us consider that the metric $g_{ab}^+$, expressed in coordinates $x^a_+$, is the metric in region $\mathcal V^+$ and the metric $g_{ab}^-$, expressed in coordinates $x^a_-$, is the metric in region $\mathcal V^-$, where the latin indeces run from $0$ to $3$. Let us assume that a set of coordinates $y^\alpha$ can be defined in both sides of $\Sigma$, where greek indeces run from $0$ to $2$. The projection vectors from the 4-dimensional regions $\mathcal V^\pm$ to the 3-dimensional hypersurface $\Sigma$ are $e^a_\alpha=\partial x^a/\partial y^\alpha$. We define $n^a$ to be the unit normal vector on $\Sigma$ pointing in the direction from $\mathcal V^-$ to $\mathcal V^+$. Let $l$ denote the proper distance or time along the geodesics perpendicular to $\Sigma$ and choose $l$ to be zero at $\Sigma$, negative in the region $\mathcal V^-$, and positive in the region $\mathcal V^+$. The displacement from $\Sigma$ along the geodesics parametrized by $l$ is $dx^a=n^adl$, and $n_a=\epsilon \partial_a l$, where $\epsilon$ is either $1$ or $-1$ when $n^a$ is a spacelike or timelike vector, respectively, i.e. $n^an_a=\epsilon$. 

We shall be working using distribution functions. For any quantity $X$, we define $X=X^+\Theta\left(l\right)+X^-\Theta\left(-l\right)$, where the indeces $\pm$ indicate that the quantity $X^\pm$ is the value of the quantity $X$ in the region $\mathcal V^\pm$, and $\Theta\left(l\right)$ is the Heaviside distribution function, with $\delta\left(l\right)=\partial_l\Theta\left(l\right)$ the Dirac distribution function. We also denote $\left[X\right]=X^+|_\Sigma-X^-|_\Sigma$ as the jump of $X$ across $\Sigma$, which implies by definition that $\left[n^a\right]=\left[e^a_\alpha\right]=0$.

\section{Geometrical representation}

Let us first deduce the junction conditions in the geometrical representation of the theory. The equations of motion in this representation are given by the field equations in Eq.\eqref{ghmpgfield} and the relation between $R_{ab}$ and $\mathcal R_{ab}$ in Eq.\eqref{ghmpgrelricten}. To simplify the analysis, we shall use Eq.\eqref{ghmpgrelricten} to cancel the terms depending on $\mathcal R_{ab}$ in Eq.\eqref{ghmpgfield}, leading to 
\begin{equation}\label{jcgeofield}
\left(R_{ab}-\nabla_a\nabla_b+g_{ab}\Box\right)\left(f_R+f_\mathcal R\right)-\frac{3}{2}g_{ab}\Box f_\mathcal R+\frac{3}{2f_\mathcal R}\partial_af_\mathcal R\partial_bf_\mathcal R-\frac{1}{2}g_{ab}f=8\pi T_{ab}.
\end{equation}
Another thing we are going to need is an explicit form of the derivatives of the function $f$ written in terms of the derivatives of $R$ and $\mathcal R$. Considering that $f$ is a function of two variables $R$ and $\mathcal R$, then we can write the partial derivatives $\partial_a f_X$, and the covariant derivatives $\nabla_a\nabla_bf_X$, with $X$ being either $R$ or $\mathcal R$, as
\begin{equation}\label{jcgeodf}
\partial_a f_X= f_{XR}\partial_aR+f_{X\mathcal R}\partial_a\mathcal R,
\end{equation}
\begin{equation}
\nabla_a\nabla_af_X=f_{XR}\nabla_a\nabla_bR+f_{X\mathcal R}\nabla_a\nabla_b\mathcal R+f_{XRR}\partial_aR\partial_bR+f_{X\mathcal R\mathcal R}\partial_a\mathcal R\partial_b\mathcal R+
2f_{XR\mathcal R}\partial_{(a}R\partial_{b)}\mathcal R.
\end{equation}
These results, along with $\Box=\nabla^c\nabla_c$, allow us to expand the terms with derivatives of $f_R$ or $f_\mathcal R$ in Eq.\eqref{jcgeofield} and write them as derivatives of either $R$ or $\mathcal R$. We shall not write the resultant equation due to its size.

\subsection{Smooth matching}

In this subsection we shall derive the junction conditions of the geometrical representation of the generalized hybrid metric-Palatini gravity. First of all, note that in the field equations given by Eq.\eqref{jcgeofield} there is a term dependent on $R_{ab}$. Similarly to the GR case, when we write the metric $g_{ab}$ in terms of distribution functions,
\begin{equation}
g_{ab}=g_{ab}^+\Theta\left(l\right)+g_{ab}^-\Theta\left(-l\right),
\end{equation}
this term is going to contribute with the two usual junction conditions in GR, i.e. the induced metric on $\Sigma$, which is $h_{\alpha\beta}=g_{ab}e^a_\alpha e^b_\beta$ must be continuous, and the extrinsic curvature $K_{\alpha\beta}$ must also be continuous. The first of these conditions comes from the fact that when one takes the derivative of the metric $g_{ab}$, written in the distribution formalism, with respect to $x^\gamma$, there will be terms depending on $\delta\left(l\right)$. When one computes the Christoffel symbols, these terms must vanish because otherwise we would have products of the form $\Theta\left(l\right)\delta\left(l\right)$ which are not distributions and the formalism would cease to be valid. On the other hand, the second junction condition assures that no $\delta\left(l\right)$ terms are present in the stress-energy tensor $T_{ab}$ in the field equations. Note that this condition is not mandatory because it does not give rise to terms of the form $\Theta\left(l\right)\delta\left(l\right)$, and therefore if this condition is violated we can still perform the matching with a thin-shell of matter at the hypersurface $\Sigma$, which we shall see in the next section. These two conditions, written in terms of the jumps of $h_{\alpha\beta}$ and $K_{\alpha\beta}$ become
\begin{equation}\label{jcgeocondgab}
\left[h_{\alpha\beta}\right]=\left[K_{\alpha\beta}\right]=0.
\end{equation}
Now, notice that Eq.\eqref{jcgeofield} depends directly on the function $f$, which can be any general function of $R$ and $\mathcal R$. This means that, in general, there will be terms in $f$ that are power-laws or products of $R$ and $\mathcal R$. When we write these terms as distribution functions, this implies that $R$ and $\mathcal R$ must not depend directly on the Dirac $\delta\left(l\right)$ function, to avoid the presence of singular factors in these terms of the form $\delta\left(l\right)$, or terms of the form $\Theta\left(l\right)\delta\left(l\right)$ which are not distributions. In general, the Ricci scalar can be written in terms of distribution functions as as
\begin{equation}\label{jcgeodistricci}
R=R^+\Theta\left(l\right)+R^-\Theta\left(-l\right)+A\delta\left(l\right),
\end{equation}
where $A=-2\epsilon\left[K\right]$ is a scalar related to the jump of the trace of the extrinsic curvature. Since $A=0$ is mandatory to avoid the presence of the $\delta\left(l\right)$ terms, then we have an extra junction condition which must be verified, which is
\begin{equation}\label{jcgeocondK}
\left[K\right]=0.
\end{equation}
Note that this condition is automatically verified if Eq.\eqref{jcgeocondgab} holds. However, if we consider the matching with a thin shell, then we have to impose that Eq.\eqref{jcgeocondK} is verified even when Eq.\eqref{jcgeocondgab} is not. Since $R_{ab}$ and $\mathcal R_{ab}$ are the Ricci tensors of metrics conformally related to each other, then imposing the conditions in Eqs.\eqref{jcgeocondgab} and \eqref{jcgeocondK} to the metric $g_{ab}$, automatically solves the same conditions for any metric conformally related to $g_{ab}$. We then argue that $\mathcal R=g^{ab}\mathcal R_{ab}$ can be written in terms of distribution functions as
\begin{equation}\label{jcgeodistpalatini}
\mathcal R=\mathcal R^+\Theta\left(l\right)+\mathcal R^-\Theta\left(-l\right).
\end{equation}
Computing the partial derivatives of $R$ and $\mathcal R$ expressed in Eqs.\eqref{jcgeodistricci} and \eqref{jcgeodistpalatini} leads to
\begin{eqnarray}
&&\partial_a R=\partial_aR^+\Theta\left(l\right)+\partial_aR^-\Theta\left(-l\right)+\epsilon\delta\left(l\right)\left[R\right]n_a,\nonumber \\
&&\partial_a \mathcal R=\partial_a\mathcal R^+\Theta\left(l\right)+\partial_a\mathcal R^-\Theta\left(-l\right)+\epsilon\delta\left(l\right)\left[\mathcal R\right]n_a.\label{jcgeodistdricci}
\end{eqnarray}
In the field equations in Eq.\eqref{jcgeofield}, we can see that due to the existence of the term $\partial_af_\mathcal R\partial_bf_\mathcal R$, there will be terms depending on products of these derivatives, such as $\partial^cR\partial_cR$, and the same for $\mathcal R$. Given the results in Eq.\eqref{jcgeodistdricci}, these products would depend on $\delta\left(l\right)^2$, which are singular terms, or on $\Theta\left(l\right)\delta\left(l\right)$, which are not distributions. Therefore, to avoid the presence of these terms we obtain the junction conditions for $R$ and $\mathcal R$ as
\begin{equation}\label{jcgeocondricci}
\left[R\right]=\left[\mathcal R\right]=0,
\end{equation}
i.e., $R$ and $\mathcal R$ must be continuous across the hypersurface $\Sigma$. Therefore, let us denote the values of $R$ and $\mathcal R$ at $\Sigma$ as $R_\Sigma$ and $\mathcal R_\Sigma$, respectively. Using Eq.\eqref{jcgeocondricci}, the terms with first derivatives of $R$ and $\mathcal R$ become regular. On the other hand, the second order terms can be written as
\begin{equation}
\nabla_a\nabla_bR=\nabla_a\nabla_bR_+\Theta\left(l\right)+\nabla_a\nabla_bR_-\Theta\left(-l\right)+\epsilon\delta\left(l\right)n_a\left[\partial_bR\right],
\end{equation}
\begin{equation}
\Box R=\Box R_+\Theta\left(l\right)+\Box R_-\Theta\left(-l\right)+\epsilon\delta\left(l\right)n^a\left[\partial_aR\right],
\end{equation}
and the same for $\mathcal R$. Again, we want to avoid the presence of the singular terms depending on $\delta\left(l\right)$, which implies the second junction condition for $R$ and $\mathcal R$ as
\begin{equation}\label{jcgeoconddricci}
\left[\partial_cR\right]=\left[\partial_c\mathcal R\right]=0,
\end{equation}
i.e., the first derivatives of $R$ and $\mathcal R$ are continuous, or $R$ and $\mathcal R$ cross the hypersurface $\Sigma$ smoothly. The same conditions as in Eq.\eqref{jcgeoconddricci} are also obtained if we apply the distribution formalism to the trace of Eq.\eqref{ghmpgrelricten}. If these conditions are verified, along with Eqs.\eqref{jcgeocondgab},\eqref{jcgeocondK} and \eqref{jcgeocondricci}, then we can match the two metrics $g_{ab}^\pm$ smoothly on $\Sigma$. Note that the second order terms in $R$ and $\mathcal R$, unlike the first order terms, are not multiplied by each other, and therefore terms of the form $\Theta\left(l\right)\delta\left(l\right)$ will not appear in the field equations even if Eq.\eqref{jcgeoconddricci} is not satisfied. Thus, this condition is not mandatory, and if not satisfied will give rise to a thin shell of matter, which we shall discuss now.

\subsection{Matching with thin shells}

Let us now verify which of the conditions derived in the previous section must be verified in order to do the matching featuring a thin shell of matter at the hypersurface $\Sigma$. To do so, we shall write the stress-energy tensor $T_{ab}$ as a distribution function of the form
\begin{equation}\label{jcgeodisttab}
T_{ab}=T_{ab}^+\Theta\left(l\right)+T_{ab}^-\Theta\left(-l\right)+\delta\left(l\right)S_{ab},
\end{equation}
where $S_{ab}$ is the 4-dimensional stress-energy tensor of the thin shell, which can be written as a 3-dimensional tensor at $\Sigma$ as
\begin{equation}
S_{ab}=S_{\alpha\beta}e^\alpha_a e^\beta_b.
\end{equation}
Now, let us keep all the functions in the previous section that give rise to terms proportional to $\delta\left(l\right)$ but not to terms of the form $\Theta\left(l\right)\delta\left(l\right)$, i.e., we shall keep $\left[K_{\alpha\beta}\right]\neq 0$, $\left[\partial_a R\right]\neq 0$, and $\left[\partial_a\mathcal R\right]\neq 0$. With these considerations, the $\delta\left(l\right)$ terms in the field equations Eq.\eqref{jcgeofield} at the hypersurface $\Sigma$ can be written as
\begin{equation}\label{jcgeosabaux}
8\pi S_{\alpha\beta}=-\left(f_R+f_\mathcal R\right)\epsilon\left[K_{\alpha\beta}\right]+h_{\alpha\beta}\left[\left(f_{RR}-\frac{1}{2}f_{\mathcal R R}\right)\epsilon n^c\left[\partial_cR\right]+\left(f_{R\mathcal R}-\frac{1}{2}f_{\mathcal R\mathcal R}\right)\epsilon n^c\left[\partial_c\mathcal R\right]\right],
\end{equation}
where $f$ and its derivatives are evaluated considering $R=R_\Sigma$ and $\mathcal R=\mathcal R_\Sigma$. In this step, we used the property $n_ae^a_\alpha=0$. The jumps of the derivatives of $R$ and $\mathcal R$ are not independent of each other, and the relationship between them can be obtained from the trace of Eq.\eqref{ghmpgrelricten} written in terms of the distribution functions, which in this case becomes
\begin{equation}\label{jcgeodricciaux}
f_{\mathcal R R}n^a\left[\partial_aR\right]+f_{\mathcal R\mathcal R}n^a\left[\partial_a\mathcal R\right]=0.
\end{equation}
Inserting Eq.\eqref{jcgeodricciaux} into Eq.\eqref{jcgeosabaux} and raising one index using the inverse induced metric $h^{\alpha\beta}$, yields finally the equation that allows to compute the stress-energy tensor of the thin shell as
\begin{equation}\label{jcgeosab}
8\pi S_\alpha^\beta=\epsilon\delta_\alpha^\beta n^c\left[\partial_cR\right]\left(f_{RR}-\frac{f_{\mathcal R R}^2}{f_{\mathcal R\mathcal R}}\right)-\left(f_R+f_\mathcal R\right)\epsilon\left[K_\alpha^\beta\right],
\end{equation}
where $\delta_\alpha^\beta=h^{\beta\gamma}h_{\gamma\alpha}$ is the identity matrix. In this representation, $S_\alpha^\beta$ is a diagonal matrix of the form $S_\alpha^\beta=\text{diag}\left(-\sigma,p,p\right)$ where $\sigma$ is the surface density of the thin shell and $p$ is the transverse pressure of the thin shell. Using this representation and noticing that, since $\left[K\right]=0$, the angular components of $\left[K_\alpha^\beta\right]$ are the same and $\left[K_0^0\right]=-2\left[K_\theta^\theta\right]$, then we obtain the surface density and the pressure of the thin shell as
\begin{equation}\label{jcgeosigma}
\sigma=\frac{\epsilon}{8\pi}\left[\left(f_R+f_\mathcal R\right)\left[K_0^0\right]-n^c\left[\partial_cR\right]\left(f_{RR}-\frac{f_{\mathcal R R}^2}{f_{\mathcal R\mathcal R}}\right)\right],
\end{equation}
\begin{equation}\label{jcgeopressure}
p=\frac{\epsilon}{8\pi}\left[n^c\left[\partial_cR\right]\left(f_{RR}-\frac{f_{\mathcal R R}^2}{f_{\mathcal R\mathcal R}}\right)+\frac{1}{2}\left(f_R+f_\mathcal R\right)\left[K_0^0\right]\right].
\end{equation}
If the derivatives of the Ricci scalar $R$ perpendicular to $\Sigma$ and the extrinsic curvature $K_\alpha^\beta$ are continuous across $\Sigma$, then $\sigma=p=0$ and we recover a smooth junction. Note that by Eq.\eqref{jcgeodricciaux}, the jumps of the normal derivatives of $R$ and $\mathcal R$ are not independent and this extra condition must be verified in order for the matching to be done.

\section{Scalar-tensor representation}

Let us now turn to the scalar-tensor representation of the theory. In this case, the equations of motion are given by the field equations in Eq.\eqref{ghmpgstfield} and the modified Klein-Gordon equations given by Eqs.\eqref{ghmpgstkgphi} and \eqref{ghmpgstkgpsi}. It shall be clear after this section that the scalar-tensor formalism is easier to work with than the geometrical formalism when it comes to junction conditions and thin shells, and then in the upcoming sections we shall provide examples to shoe that explicitly.

\subsection{Smooth matching}

We shall now derive the junction conditions for the scalar-tensor representation of the generalized hybrid metric-Palatini gravity. First of all, note that in the field equations given by Eq.\eqref{ghmpgstfield}, there is only one term that depends on derivatives of the metric $g_{ab}$, which is the Einstein tensor $G_{ab}$. This implies again that the junction conditions for the metric are going to be the same as in GR, i.e., the induced metric on $\Sigma$, which is $h_{\alpha\beta}=g_{ab}e^a_\alpha e^b_\beta$, must be continuous, and the extrinsic curvature $K_{\alpha\beta}$ must also be continuous, for the same reasons explained in the geometrical representation of the theory. Again, note that the second junction condition for the metric is not mandatory because it does not give rise to terms of the form $\Theta\left(l\right)\delta\left(l\right)$, and therefore if this condition is violated we can still perform the matching with a thin-shell of matter at the hypersurface $\Sigma$. These two conditions can be written in terms of the jumps of $h_{\alpha\beta}$ and $K_{\alpha\beta}$ as
\begin{equation}\label{jcscacondgab}
\left[h_{\alpha\beta}\right]=\left[K_{\alpha\beta}\right]=0.
\end{equation}
We are then left with the junction conditions for the scalar fields. To deduce these conditions, let us write both scalar fiels generically as $\phi=\{\varphi,\psi\}$, and write them as distribution functions of the form
\begin{equation}\label{jcscadistscalar}
\phi=\phi_+\Theta\left(l\right)+\phi_-\Theta\left(-l\right),
\end{equation}
where $\phi$ represents either $\varphi$ or $\psi$. Computing the derivatives of the scalar fields in this distribution representation yields
\begin{equation}\label{jcscadistdscalar}
\partial_a\phi=\partial_a\phi_+\Theta\left(l\right)+\partial_a\phi_.\Theta\left(-l\right)+\epsilon\delta\left(l\right)\left[\phi\right]n_a=0.
\end{equation}
Note that in the field equations given by Eq.\eqref{ghmpgstfield}, there are terms that depend on products of derivatives of the scalar field, such as $\partial^c\phi\partial_c\phi$. Given the result in Eq.\eqref{jcscadistdscalar}, these products would have terms depending $\delta\left(l\right)^2$, which are divergent, or terms of the form $\Theta\left(l\right)\delta\left(l\right)$, which are not distribution functions. Therefore, to avoid the presence of these terms, the first junction condition for the scalar fields implies that
\begin{equation}\label{jcscacondscalar}
\left[\varphi\right]=\left[\psi\right]=0,
\end{equation}
i.e., the scalar fields must be continuous across the hypersurface $\Sigma$. Then, let us define the value of the scalar fields at $\Sigma$ to be $\varphi_\Sigma$ and $\psi_\Sigma$. Now, using Eq.\eqref{jcscacondscalar}, we verify that the terms $\partial^c\phi\partial_c\phi$ and $\partial_a\phi\partial_b\phi$ in Eq.\eqref{ghmpgstfield} become regular. On the other hand, the second order terms in the scalar fields become
\begin{equation}
\nabla_a\nabla_b\phi=\nabla_a\nabla_b\phi_+\Theta\left(l\right)+\nabla_a\nabla_b\phi_-\Theta\left(-l\right)+\epsilon\delta\left(l\right)n_a\left[\partial_b\phi\right],
\end{equation}
\begin{equation}
\Box\phi=\Box\phi_+\Theta\left(l\right)+\Box\phi_-\Theta\left(-l\right)+\epsilon\delta\left(l\right)n^a\left[\partial_a\phi\right].
\end{equation}
To avoid the presence of the divergent terms $\delta\left(l\right)$ in the field equations,  second junction condition for the scalar fields must be then
\begin{equation}\label{jcscaconddscalar}
\left[\partial_c\varphi\right]=\left[\partial_c\psi\right]=0,
\end{equation}
i.e., the first derivatives of the scalar fields are continuous, or the scalar fields cross the hypersurface $\Sigma$ smoothly. The same conditions are obtained when we apply the distribution formalism to the scalar field equations given by Eqs. \eqref{ghmpgstkgphi} and \eqref{ghmpgstkgpsi}. If these conditions are verified, along with the junction conditions for the metric in Eq.\eqref{jcscacondgab} and the first junction condition for the scalar fields written in Eq.\eqref{jcscacondscalar}, then we can match the two metrics $g_{ab}^\pm$ smoothly on $\Sigma$. Note that Eq.\eqref{jcscaconddscalar} is not mandatory since the second order terms in the scalar fields, unlike the first order ones, are not multiplied by each other, and thus no terms of the form $\Theta\left(l\right)\delta\left(l\right)$ arise. This means that if Eq.\eqref{jcscaconddscalar} is not satisfied, the matching is still possible with the help of a thin shell of matter at the hypersurface $\Sigma$, which we study now.

\subsection{Matching with thin shells}

Let us now verify which of the conditions derived in the previous section must be verified in order to do the matching featuring a thin shell of matter at the hypersurface $\Sigma$. To do so, we shall write the stress-energy tensor $T_{ab}$ as a distribution function of the form
\begin{equation}\label{jcscadisttab}
T_{ab}=T_{ab}^+\Theta\left(l\right)+T_{ab}^-\Theta\left(-l\right)+\delta\left(l\right)S_{ab},
\end{equation}
where $S_{ab}$ is the 4-dimensional stress-energy tensor of the thin shell, which can be written as a 3-dimensional tensor at $\Sigma$ as
\begin{equation}
S_{ab}=S_{\alpha\beta}e^\alpha_a e^\beta_b.
\end{equation}
Now, let us keep all the functions in the previous section that give rise to terms proportional to $\delta\left(l\right)$ but not to terms of the form $\Theta\left(l\right)\delta\left(l\right)$, i.e., we shall keep $\left[K_{\alpha\beta}\right]\neq 0$, $\left[K\right]\neq 0$, $\left[\partial_a \varphi\right]\neq 0$, and $\left[\partial_a\psi\right]\neq 0$. Using these considerations, the $\delta\left(l\right)$ factors of the modified field equations given by Eq.\eqref{ghmpgstfield} at the hypersurface $\Sigma$ can be written as
\begin{equation}\label{jcscasabaux}
8\pi S_{\alpha\beta}=\epsilon h_{\alpha\beta}n^c\left(\left[\partial_c\varphi\right]-\left[\partial_c\psi\right]\right)-\left(\varphi_\Sigma-\psi_\Sigma\right)\epsilon\left(\left[K_{\alpha\beta}\right]-\left[K\right]h_{\alpha\beta}\right),
\end{equation}
where $K=K^\alpha_\alpha$ is the trace of the extrinsic curvature. In this calculation, we used the fact that $n_ae^a_\alpha=0$. On the other hand, the $\delta\left(l\right)$ factors of the scalar field equations given by Eqs.\eqref{ghmpgstkgpsi} and \eqref{ghmpgstkgphi} become, respectively
\begin{equation}\label{jcscakgpsi}
n^a\left[\partial_a\psi\right]=0,
\end{equation}
\begin{equation}\label{jcscakgphi}
\epsilon n^a\left[\partial_a\varphi\right]=\frac{8\pi}{3}S,
\end{equation}
where $S=S^\alpha_\alpha$ is the trace of the stress-energy tensor of the thin shell. Eqs.\eqref{jcscakgpsi} and \eqref{jcscakgphi} are the junction conditions for the scalar fields $\psi$ and $\varphi$ when we allow for a thin shell to exist at $\Sigma$. Inserting these results into Eq.\eqref{jcscasabaux} we obtain
\begin{equation}\label{jcscasabaux2}
8\pi\left(S_{\alpha\beta}-\frac{1}{3}h_{\alpha\beta} S\right)=-\epsilon\left(\varphi_\Sigma-\psi_\Sigma\right)\left(\left[K_{\alpha\beta}\right]-h_{\alpha\beta}\left[K\right]\right).
\end{equation}
Tracing Eq.\eqref{jcscasabaux2} with the inverse induced metric $h^{\alpha\beta}$ cancels the stress-energy tensor terms and we recover the result obtained in the geometrical representation of the theory
\begin{equation}\label{jcscacondK}
\left[K\right]=0.
\end{equation}
This result implies that the extrinsic curvature does not need to be continuous at the hypersurface $\Sigma$ like it is needed for the smooth matching, but it must at least have a continuous trace across $\Sigma$, as expected from the geometrical representation. Inserting this result into Eq.\eqref{jcscasabaux2} and raising one of the indeces using the inverse induced metric $h^{\alpha\beta}$, yields finally the condition to compute the stress-energy tensor of the thin shell
\begin{equation}\label{jcscasab}
8\pi\left(S_\alpha^\beta-\frac{1}{3}\delta_\alpha^\beta S\right)=-\epsilon\left(\varphi_\Sigma-\psi_\Sigma\right)\left[K_\alpha^\beta\right].
\end{equation}
where $\delta_\alpha^\beta=h^{\alpha\gamma}h_{\gamma\beta}$ is the identity matrix. In this representation, the tensor $S_\alpha^\beta$ is a diagonal matrix which can be written as $S_\alpha^\beta=\text{diag}\left(-\sigma,p,p\right)$ where $\sigma$ is the surface density of the thin shell and $p$ is the transverse pressure of the thin shell. Using this representation and noticing that, since $\left[K\right]=0$, the angular components of $\left[K_\alpha^\beta\right]$ are the same and $\left[K_0^0\right]=-2\left[K_\theta^\theta\right]$, then Eqs.\eqref{jcscasab} and \eqref{jcscakgphi} in the form
\begin{equation}
\frac{16\pi}{3}\left(\sigma+p\right)=\epsilon\left(\varphi_\Sigma-\psi_\Sigma\right)\left[K_0^0\right],
\end{equation}
\begin{equation}
\epsilon n^a\left[\partial_a\varphi\right]=\frac{8\pi}{3}\left(2p-\sigma\right),
\end{equation}
become a system of two equations for the two unknowns $\sigma$ and $p$. We can thus obtain the density and transverse pressure of the thin shell in the final forms
\begin{equation}\label{jcscasigma}
\sigma=\frac{\epsilon}{8\pi}\left[\left(\varphi_\Sigma-\psi_\Sigma\right)\left[K_0^0\right]-n^a\left[\partial_a\varphi\right]\right]
\end{equation}
\begin{equation}\label{jcscapressure}
p=\frac{\epsilon}{8\pi}\left[n^a\left[\partial_a\varphi\right]+\frac{1}{2}\left(\varphi_\Sigma-\psi_\Sigma\right)\left[K_0^0\right]\right].
\end{equation}
If the derivatives of the scalar field $\varphi$ perpendicular to $\Sigma$ and the extrinsic curvature $K_{\alpha\beta}$ are continuous across $\Sigma$, then $\sigma=p=0$ and we recover the smooth junction. Note that this result is equivalent to the one obtained in the geometrical representation of the theory: when we map back $\varphi=f_R$, use Eq.\eqref{jcgeodf} to expand the partial derivatives of $f_R$, and use Eq.\eqref{jcgeoconddricci} to relate the partial derivatives of $R$ and $\mathcal R$, we recover Eqs.\eqref{jcgeosigma} and \eqref{jcgeopressure}.

\section{Example: Minkowski to Schwarzschild}

The simplest possible application for the junction conditions with thin shells is the matching between a Minkowski spacetime to a Schwarzschild spacetime, as done in \cite{martinez1}. In this section, we start by showing how to do the same analysis in the generalized hybrid metric-Palatini gravity in both the geometrical and the scalar-tensor representations of the theory, thus emphasizing the equivalence between the two formalisms. We then extend our analysis to a more complicated spacetime in which the matching between the interior Minkowski solutions and the exterior Schwarzschild solution is performed via the use of thick shell of perfect fluid with a constant density. For simplicity, this latter example is done in the scalar-tensor representation of the theory.

\subsection{With a thin shell}

\subsubsection{Matching in the geometrical representation}

In this section we shall provide a simple example of how to use the junction conditions derived above to match two different spacetimes with a thin shell in between. To do so, let us consider a very simple form of the function $f\left(R,\mathcal R\right)$ given by
\begin{equation}\label{jcex1function}
f\left(R,\mathcal R\right)=g\left(R\right)+\mathcal R h\left(\frac{R}{R_0}\right)
\end{equation}
where $g\left(R\right)$ and $h\left(R/R_0\right)$ are well behaved functions of their arguments, and $R_0$ is a constant with units of $R$. For this specific choice of the function $f$, we can write the derivatives $f_R$ and $f_\mathcal R$ as
\begin{equation}\label{jcex1dfunction}
f_R=g'\left(R\right)+h'\left(\bar R\right)\bar{\mathcal R},\ \ \ \ \ \ \ f_\mathcal R=h\left(\bar R\right),
\end{equation}
where $\bar{\mathcal R}=\mathcal R/R_0$ and $\bar{R}=R/R_0$ are now adimentional variables. Now, Eq.\eqref{ghmpgrelricten} becomes an equation for $\mathcal R_{ab}$ as a function of $R_{ab}$ and $R$ as
\begin{equation}\label{jcex1relricten}
\mathcal R_{ab}=R_{ab}-\frac{1}{h\left(\bar R\right)}\left(\nabla_a\nabla_b+\frac{1}{2}g_{ab}\Box\right)h\left(\bar R\right)+\frac{3}{2 h\left(\bar R\right)^2}\partial_ah\left(\bar R\right)\partial_bh\left(\bar R\right).
\end{equation}
Notice  now the importance of our specific choice of the function $f$ in Eq.\eqref{jcex1function}. This choice allows us to write $\mathcal R_{ab}$ as a function of $R_{ab}$ and $R$ only, implying that we can use Eq.\eqref{jcex1relricten} and its trace, and use Eqs.\eqref{jcex1function} and \eqref{jcex1dfunction} to cancel the terms depending on $\mathcal R_{ab}$ and $\mathcal R$ in the modified field equation given in Eq.\eqref{ghmpgfield} and obtain an equation that only depends on the metric $g_{ab}$ and its derivatives (we shall not write this equation due to its size). This is not possible in general. For any choice of the function $f$ for which $f_\mathcal R$ depends on $\mathcal R$, Eq.\eqref{ghmpgrelricten} becomes a partial differential equation for $\mathcal R$ and the problem is much more complicated. 

Consider now the line elements of the Minkowski and Schwarzschild spacetimes given in spherical coordinates $\left(t,r,\theta,\phi\right)$ by
\begin{equation}\label{metricminkowski}
ds^2=-dt^2+dr^2+r^2d\Omega^2,
\end{equation}
\begin{equation}\label{metricschwarzschild}
ds^2=-\left(1-\frac{2M}{r}\right)dt^2+\frac{1}{1-\frac{2M}{r}}dr^2+r^2d\Omega^2,
\end{equation}
respectively, where $M$ is the Schwarzschild mass, and $d\Omega^2=d\theta^2+\sin^2\theta d\phi^2$. Both these spacetimes are solutions of the modified field equations in vacuum, i.e. with $T_{ab}=0$. Also, both solutions have $R=0$ and consequently $\mathcal R=0$ by Eq.\eqref{jcex1relricten}. Thus, for these spacetimes the junction conditions $\left[R\right]=\left[\mathcal R\right]=0$ and $\left[\partial_cR\right]=\left[\partial_c\mathcal R\right]=0$ are automatically satisfied.

The junction condition for the jump of the trace of the extrinsic curvature, $\left[K\right]=0$ is given in this case by the equation
\begin{equation}\label{jcex1condK}
\left[K\right]=-\frac{2}{r}-\frac{3M-2r}{r^2\sqrt{1-\frac{2M}{r}}}=0
\end{equation}
Solving Eq.\eqref{jcex1condK} for $r$ one obtains $r=9M/4\equiv r_\Sigma$, which corresponds to the buchdahl limit for compactness of a fluid star. This is the main difference between GR and the generalized hybrid metric-Palatini gravity and it is also a feature of other simpler theories like $f\left(R\right)$ and the non-generalized version of the hybrid metric-Palatini gravity. In GR, the matching between these two spacetimes could be performed for any value of the radial coordinate $r>2M$ outside the event horizon of the Schwarzschild solution. However, in here we are forced to perform the matching at a specific value of $r$ to satisfy the extra $\left[K\right]=0$ junction condition.

Notice that the jump in the extrinsic curvature $\left[K_{ab}\right]$ is never zero, and thus we need a thin shell at $r=r_\Sigma$ to perform the matching. Let us now study the stress-energy tensor of this thin shell. From Eqs.\eqref{jcgeosigma} and \eqref{jcgeopressure} we obtain for our specific case
\begin{equation}
\sigma=2p=\frac{1}{8\pi}\left\{\left[g'\left(R\right)+h\left(\bar R\right)\right]\left[K_0^0\right]\right\},
\end{equation}
where we have used $\epsilon=1$ since $n^a$ points in the radial direction and thus is a spacelike vector. Also, in this case $\left[K_0^0\right]$ must be computed at $r=r_\Sigma$ from which we obtain 
\begin{equation}
\left[K_0^0\right]|_{r=r_\Sigma}=\frac{16}{27M}>0.
\end{equation}
Now we just have to specify the forms of the functions $g$ and $h$. Choosing $g\left(R\right)=R$ and $h\left(R/R_0\right)=1+R/R_0$, for which the function $f$ has the very simple form $f\left(R,\mathcal R\right)=R+\mathcal R+\left(R\mathcal R\right)/R_0$. We obtain then, for our solutions with $R=\mathcal R=0$ that
\begin{equation}\label{jcex1sab}
\sigma=2p=\frac{4}{27\pi M}.
\end{equation}
For this particular choice of the function $f$, all the energy conditions are satified at the shell.

\subsubsection{Matching in the scalar-tensor representation}

To emphasize the equivalence between the geometrical and the scalar-tensor representations of the theory, let us now perform the calculations in the scalar-tensor representation. Using the same choice of the function $f$ given by $f\left(R,\mathcal R\right)=R+\mathcal R+\left(R\mathcal R\right)/R_0$, the scalar fields $\varphi$ and $\psi$ can be written as invertible functions of $R$ and $\mathcal R$ as
\begin{equation}\label{jcex1phi}
\varphi=1+\frac{\mathcal R}{R_0} \Leftrightarrow \mathcal R=R_0\left(\varphi-1\right),
\end{equation}
\begin{equation}\label{jcex1psi}
-\psi=1+\frac{R}{R_0} \Leftrightarrow~R=-R_0\left(\psi+1\right),
\end{equation}
respectively. This invertibility allows us to find the form of the potential $V\left(\varphi,\psi\right)$ associated with the specific choice of the function $f$ by Eq.\eqref{ghmpgstpotential}, from which we obtain
\begin{equation}\label{jcex1pot}
V\left(\varphi,\psi\right)=V_0\left(\varphi-1\right)\left(\psi+1\right),
\end{equation}
where $V_0=-3R_0$ is a constant. Plugging the metrics for the Minkowski and Schwarzschild spacetimes given in Eqs.\eqref{metricminkowski} and \eqref{metricschwarzschild} respectively, and the potential in Eq.\eqref{jcex1pot} into the modified field equations for the scalar-tensor representation from Eq.\eqref{ghmpgstfield}, yields a PDE for the fields $\varphi$ and $\psi$. This equation, along with the two scalar field equations given by Eqs.\eqref{ghmpgstkgphi} and \eqref{ghmpgstkgpsi}, is a set of three PDEs for the two scalar fields $\varphi$ and $\psi$. A solution for these equations is $\varphi=1$ and $\psi=-1$, constants. The same result can be obtained directly from Eqs.\eqref{jcex1phi} and \eqref{jcex1psi} by setting $R=\mathcal R=0$.

Since the scalar fields are constant through the two spacetimes, the junction conditions $\left[\varphi\right]=\left[\psi\right]=0$ and $\left[\partial_c\varphi\right]=\left[\partial_c\psi\right]=0$ are automatically verified. Also, since the metrics are the same, then the condition $\left[K\right]=0$ also yields the constraint $r=r_\Sigma=9M/4$, and thus $\left[K_0^0\right]=16/\left(27M\right)$ at this hypersurface. The density and transverse pressure of the thin shell are then given by Eqs.\eqref{jcscasigma} and \eqref{jcscapressure} as
\begin{equation}
\sigma=2p=\frac{1}{8\pi}\left\{\left(\varphi-\psi\right)\left[K_0^0\right]\right\}=\frac{4}{27\pi M},
\end{equation}
in agreement with Eq.\eqref{jcex1sab}, as expected. Again, all the energy conditions are satisfied at the shell.

\subsection{With a perfect fluid thick shell}

\subsubsection{Solution for the constant density perfect fluid shell}

In this section we aim to build a spacetime that consists of an interior Minkowski spacetime, a shell of perfect fluid with a finite thickness, and an exterior Schwarzschild spacetime. We shall be working in the scalar-tensor representation of the theory assuming $V\left(\varphi,\psi\right)=0$ to simplify the problem. This choice of the potential, using Eq.\eqref{ghmpgstpotential}, can be shown to be equivalent to chosing a function $f$ of the form $f\left(R,\mathcal R\right)=R g\left(\mathcal R/R\right)+\mathcal R h\left(R/\mathcal R\right)$, for some well-behaved functions $g$ and $h$ (see Sec.\ref{Sec:coslinpot} where we considered $V=V_0$ constant and take the limit $V_0=0$). In this representation, it can be shown that if $\varphi^{(i,e)}=\varphi^{(i,e)}_0$ and $\psi^{(i,e)}=\psi^{(i,e)}_0\neq\varphi^{(i,e)}_0$, for some constants $\varphi_0^{(i,e)}$ and $\psi_0^{(i,e)}$, where the indeces ${(i,e)}$ denote the interior and exterior regions of the spacetime respectively, then the Minkownski (interior) and Schwarzschild (exterior) spacetimes described by the line elements
\begin{equation}\label{metricminkowskiint}
ds^2_{(i)}=-dt^2+dr^2+r^2d\Omega^2,
\end{equation}
\begin{equation}\label{metricschwarzschildext}
ds^2_{(e)}=-\left(1-\frac{2M}{r}\right)dt^2+\frac{1}{1-\frac{2M}{r}}dr^2+r^2d\Omega^2,
\end{equation}
respectively, are solutions of the modified field equations and scalar field equations given by Eqs.\eqref{ghmpgstfield} to \eqref{ghmpgstkgphi} in vacuum i.e. with $T_{ab}=0$. Thus, we only have to find a solution for the thick shell of matter in between these two spacetimes. We shall use no index for the variables in this region to distinguish between the interior and exterior regions. To do so, let us assume a general form of a spherically symmetric and static metric of the form
\begin{equation}\label{metricspherical}
ds^2=-e^{\zeta\left(r\right)}dt^2+\left[1-\frac{2m\left(r\right)}{r}\right]dr^2+r^2d\Omega^2,
\end{equation}
where $\zeta\left(r\right)$ is called the redshift function and $2m\left(r\right)=b\left(r\right)$, with $b\left(r\right)$ the shape function, all assumed to be functions of the radial coordinate only to preserve the staticity of the solution. We shall also assume that the distribution of matter can be described by an anisotropic perfect fluid, i.e. $T_a^b=\text{diag}\left(-\rho,p_r,p_t,p_t\right)$, where $\rho\left(r\right)$ is the energy density, $p_r\left(r\right)$ is the radial pressure, and $p_t\left(r\right)$ is the transverse pressure, all functions of the radial coordinate only. Inserting Eq.\eqref{metricspherical} and the ansatz for $T_{ab}$ into Eqs.\eqref{ghmpgstfield} to \eqref{ghmpgstkgphi} yields a set of five independent equations which read, after rearrangements:
\begin{equation}\label{jcex2fieldrho}
\kappa^2\rho=\frac{2m'}{r^2}\left(\varphi-\psi\right)-\frac{3}{4}\left(1-\frac{2m}{r}\right)\frac{\psi'^2}{\psi}+\left(\varphi'-\psi'\right)\left(\frac{3m}{r}+m'-2\right)-\left(r-2m\right)\left(\varphi''-\psi''\right)
\end{equation}
\begin{equation}\label{jcex2fieldpr}
\kappa^2 p_r=\frac{\varphi-\psi}{r}\left[\left(1-\frac{2m}{r}\right)\zeta'-\frac{2m}{r^2}\right]+\left[\left(\varphi'-\psi'\right)\left(\frac{2}{r}+\frac{\zeta'}{2}\right)-\frac{3}{4}\frac{\psi'^2}{\psi}\right]\left(1-\frac{2m}{r}\right),
\end{equation}
\begin{eqnarray}
&&\kappa^2 p_t=\frac{\varphi-\psi}{2}\left[\left(1-\frac{2m}{r}\right)\left(\zeta''+\frac{\zeta'^2}{2}\right)+\frac{\zeta'}{r}\left(1-m'-\frac{m}{r}\right)-\frac{2}{r^2}\left(m'-\frac{m}{r}\right)\right]+\label{jcex2fieldpt}\\
&&+\left(1-\frac{2m}{r}\right)\left[\left(\varphi''-\psi''\right)+\frac{3}{4}\frac{\psi'^2}{\psi}\right]+\left(\varphi'-\psi'\right)\left[\frac{\zeta'}{2}\left(1-\frac{2m}{r}\right)+\frac{1-m'}{r}-\frac{m}{r^2}\right],\nonumber
\end{eqnarray}
\begin{equation}\label{jcex2kgpsi}
\psi''-\frac{\psi'^2}{2\psi}+\frac{\psi'}{2r}\left[4+r\zeta'-\frac{2\left(m'r-m\right)}{r-2m}\right]=0,
\end{equation}
\begin{equation}\label{jcex2kgphi}
\left(\zeta''+\frac{\zeta'^2}{2}\right)\left(1-\frac{2m}{r}\right)-\frac{4m'}{r^2}-\frac{\zeta'}{r}\left(\frac{3m}{r}+m'-2\right)=0,
\end{equation}
where a prime denotes a derivative with respect to $r$. Eq.\eqref{jcex2kgphi} is obtained by subtracting Eq.\eqref{ghmpgstkgpsi} from Eq.\eqref{ghmpgstkgphi}, using Eqs.\eqref{jcex2fieldrho} to \eqref{jcex2fieldpt} to cancel the terms depending on $\rho$, $p_r$ and $p_t$, and assuming $\varphi\neq\psi$. We now have a set of five independent equations for seven independent variables, namely $\rho$, $p_r$, $p_t$, $\varphi$, $\psi$, $m$ and $\zeta$. This means that we can impose two constraints to determine the system. The first constraint we impose is simply that $\psi=\psi_0$ is a constant, which imediatly verifies Eq.\eqref{jcex2kgpsi}. Then, we choose to impose a constraint on the function $m$ of the form
\begin{equation}\label{jcex2mass}
m\left(r\right)=M\left(\frac{r}{R_\Sigma^{(e)}}\right)^\alpha,
\end{equation}
where $M$ is a constant with units of mass that will correspond to the Schwarzschild mass of the exterior solution, $R_\Sigma^{(e)}$ is a constant with units of $r$ that, upon matching with the exterior Schwarzschild spacetime, will correspond to the radius of the outer surface of the thick shell, and $\alpha$ is an exponent that we set to $\alpha=1$ to guarantee the existance of analytical solutions. A more realistic solution could be analysed by setting $\alpha=3$ and solving the equations numerically. It can be seen that for $r=0$ we obtain $m=0$, i.e. there is no mass inside a sphere of radius zero, and also that $m\left(r=R_\Sigma^{(e)}\right)=M$, which is in agreement with the mass of the exterior Schwarzschild spacetime.

Inserting Eq.\eqref{jcex2mass} into Eq.\eqref{jcex2kgphi} yields a second order ODE for the function $\zeta\left(r\right)$ which can be solved analytically for the general solution
\begin{equation}\label{jcex2zeta}
\zeta\left(r\right)=\zeta_0-\left(1+\beta\right)\log\left(\frac{r}{M}\right)+2\log\left[\left(\frac{r}{M}\right)^\beta+\zeta_1\right],
\end{equation}
\begin{equation}
\beta=\sqrt{\frac{1+6\bar{M}}{1-2\bar{M}}},\ \ \ \ \ \ \bar{M}=\frac{M}{R_\Sigma^{(e)}}\nonumber,
\end{equation}
and $\zeta_i$ are constants of integration. In this case, we set $\zeta_1=0$ and leave $\zeta_0$ as a free parameter. This parameter will be useful later when we have to impose the continuity of the metric at $R_\Sigma^{(e)}$. Now we are interested in finding solutions with constant density. To do so, we insert Eqs.\eqref{jcex2zeta} and \eqref{jcex2mass}, along with the constraint $\psi=\psi_0$ into Eq.\eqref{jcex2fieldrho} and search for solutions for $\varphi$ that guarantee that the right hand side (and consequently $\rho$) are constant. The solution for $\varphi$ is extremely long so we shall not write it explicitly. Inserting these solutions for $m$, $\zeta$, $\varphi$ and $\psi$ into Eqs.\eqref{jcex2fieldpr} and \eqref{jcex2fieldpt} yields the results for both $p_r$ and $p_t$ and our solution is complete. We do not show the solutions explicitly due to their size, but the results are plotted in Fig.\ref{fig:jcex2tshell}.

\subsubsection{Exterior matching with a Schwarzschild spametime}

Let us now turn to the matching between the thick shell and the exterior Schwarzschild spacetime. In this case, a smooth matching between these two spacetimes is possible. Since we have chosen $\psi=\psi_0$ to be a constant, the junction conditions $\left[\psi\right]=0$ and $\left[\partial_c\psi\right]=0$ are automatically satisfied as long as we choose $\psi_0=\psi_0^{(e)}$.

For the matching to be smooth we have to impose that $\left[\varphi\right]=0$ and $\left[\partial_c\varphi\right]=0$. The first of these two matching conditions can be satisfied simply by setting $\varphi_0^{(e)}=\varphi\left(R_\Sigma^{(e)}\right)$. The second junction condition can be analysed by differentiating the solution for $\varphi$ in the thick shell and equaling it to zero. This solution depends on two integration constants, namely $\varphi_1$ and $\varphi_2$, and also on $\bar M$. 

It is also natural to search for solutions in which the pressure vanishes at the surface, i.e. $p_r\left(R_\Sigma^{(e)}\right)=p_t\left(R_\Sigma^{(e)}\right)=0$. The solutions for $p_r$ and $p_t$ also depend on $\varphi_1$, $\varphi_2$, and $\bar M$. We therefore can write three equations:
\begin{equation}
p_r\left(R_\Sigma^{(e)}\right)=0,\ \ \ \ \ p_t\left(R_\Sigma^{(e)}\right)=0,\ \ \ \ \ \partial_r\varphi\left(R_\Sigma^{(e)}\right)=0,
\end{equation}
and solve them for the three constants $\varphi_1$, $\varphi_2$ and $\bar M$. Doing so, we obtain $\bar M=1/3$, $\varphi_1=54\pi M^4\rho_0$ and $\varphi_2=40\pi M\rho_0$. These results guarantee that the junction condition $\left[\partial_c\varphi\right]=0$ is satisfied. Note that $\bar M=1/3$ implies that $R_\Sigma^{(e)}=3M$, that is, for the matching to be smooth and also for the pressures to vanish at the surface of the shell, one has to perform the matching exactly at the radius of the light ring. Using the obtained values for $\varphi_1$, $\varphi_2$, and $\bar M$, the scalar field $\varphi$ in the shell can then be written as
\begin{equation}\label{jcex2solphi}
\varphi\left(r\right)=\psi_0+\left(\frac{54M^4}{r^2}+40Mr-6r^2\right)\pi\rho_0.
\end{equation}
At the surface, we have $\varphi\left(3M\right)=72M^2\pi\rho_0=\varphi_0^{(e)}$, and we guarantee that $\left[\varphi\right]=0$ is satisfied.

Finally, one has to verify that the metric $g_{ab}$ and the extrinsic curvature $K_{ab}$ are continuous at the surface of the shell. Inserting $\bar M=1/3$ into Eq.\eqref{jcex2zeta}, one verifies that $g_{tt}$ of the shell can be written as
\begin{equation}
g_{tt}=-e^{\zeta_0}\left(\frac{r}{M}\right)^{\beta-1}=-e^{\zeta_0}\left(\frac{r}{M}\right)^{2}.
\end{equation}
Imposing continuity of this result with the exterior metric at the surface $r=R_\Sigma^{(e)}$ yields the equation
\begin{equation}
e^{\zeta_0}\left(\frac{R_\Sigma^{(e)}}{M}\right)^{2}=1-\frac{2M}{R_\Sigma^{(e)}}.
\end{equation}
But $R_\Sigma^{(e)}=3M$, from which we obtain $\zeta_0=-\log\left(27\right)$. These values of the parameters $\zeta_0$, $\varphi_0^{(e)}$, $\varphi_1$, $\varphi_2$ and $\bar M$ guarantee that the matching between the fluid shell and the exterior Schwarzschild solution is smooth.

\subsubsection{Interior matching with a Minkowski spametime}

The fluid shell solution features some problems for small values of the radial coordinate $r$. First of all, in the limit $r\to 0$ the redshift function $\zeta$ diverges and $g_{tt}=0$, which corresponds to an horizon. Also, in this limit both the pressures $p_r$ and $p_t$ diverge, and even far from the limit there are regions where $p_r$ becomes negative and the energy conditions are violated. Therefore, for our solution to have physical significance, we have to perform a matching with a regular interior solution, that we choose to be Minkowski. Let us assume then that the interior metric can be written as
\begin{equation}\label{jcex2metricint}
g_{ab}^{(i)}=\Omega\eta_{ab},
\end{equation}
where $\eta_{ab}$ is the Minkowski metric and $\Omega$ is a constant conformal factor. The metric $g_{ab}^{(i)}$ is a solution of the modified field equations when $V=0$, $\varphi^{(i)}=\varphi_0^{(i)}$ and $\psi^{(i)}=\psi_0^{(i)}$ for some constants $\varphi_0^{(i)}$ and $\psi_0^{(i)}$. The junction conditions $\left[\psi\right]=0$ and $\left[\partial_c\psi\right]=0$ are then automatically satisfied if we choose $\psi_0^{(i)}=\psi_0$ of the thin shell, which is still a free parameter. 

Using the metrics in Eqs.\eqref{metricspherical} and \eqref{jcex2metricint}, one verifies that the junction condition $\left[K\right]=0$ imposes a constraint on the value of $\Omega$, which is $\Omega=4/3$, rather than a constraint on the coordinate $r$. In this case, the condition is satisfied for any $r$ and we can choose to perform the matching at any $r=R_\Sigma^{(i)}$ that we find appropriate.

\begin{figure}
\centering
\includegraphics[scale=0.6]{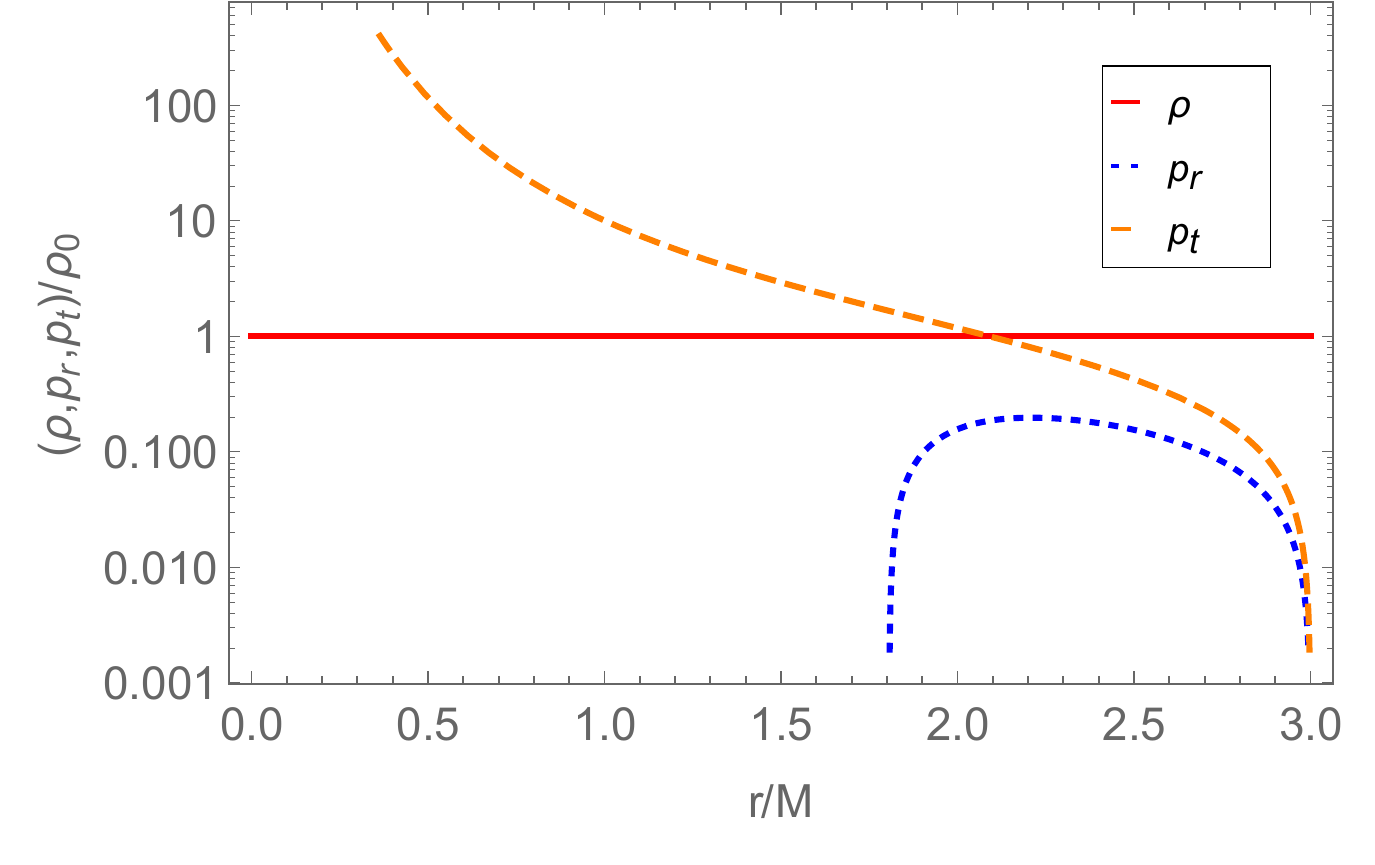}
\caption{Density $\rho$, radial pressure $p_r$ and transverse pressure $p_t$ of the thick shell solution in Eq.\eqref{metricspherical}, normalized to $\rho_0$, with $\zeta_0=-\log\left(27\right)$, $\bar M=1/3$, $\varphi_1=54\pi M^4\rho_0$ and $\varphi_2=40\pi M\rho_0$.}
\label{fig:jcex2tshell}
\end{figure}

To find a suitable radius to perform the matching, let us consider the validity of the energy conditions. As can be seen from Fig.\ref{fig:jcex2tshell}, $\rho$ is always positive as long as we choose $\rho_0>0$, which we can because it is still a free parameter. Then, it can be shown that for $r\gtrsim 2.08787M$ we have $\rho>0$, $\rho+p_r>0$, $\rho+p_t>0$, $\rho>|p_r|$, $\rho>|p_t|$, and $\rho+p_r+2p_t>0$, and all the energy conditions are satisfied in this region. Also, in this region we have both $p_r>0$ and $p_t>0$, which are important conditions to the physical significance of the solution. We shall thus perform the matching for some $r=R_\Sigma^{(i)}$ between $\sim 2.08787M$ and $3M$.

We can now compute the surface density and pressure of the thin shell using Eqs.\eqref{jcscasigma} and \eqref{jcscapressure}, where we use $\left[\partial_c\varphi\right]=-\partial_r\varphi$, where $\partial_r\varphi$ is the derivative of Eq.\eqref{jcex2solphi}, and $\left[K_0^0\right]=1/\left(\sqrt{3}r\right)$ from which we obtain
\begin{equation}
\sigma=\frac{\rho_0}{12r^3}\left[27\left(\sqrt{3}-6\right)M^4+20\left(3+\sqrt{3}\right)Mr^3-3\left(6+\sqrt{3}\right)r^4\right],
\end{equation}
\begin{equation}
p=\frac{\rho_0}{24r^3}\left[27\left(12+\sqrt{3}\right)M^4+20\left(\sqrt{3}-6\right)Mr^3-3\left(\sqrt{3}- 12\right)\right],
\end{equation}
respectively, where we considered again $\epsilon=1$ because $n^a$ points in the radial direction and thus is a spacelike vector. For $\rho_0>0$, we verify that in the region $\sim 2.08787M<r<3M$ we have $\sigma>0$, $\sigma+p>0$, $\sigma>|p|$ and $\sigma+2p>0$, from which we conclude that all the energy conditions are verified no matter the value of $R_\Sigma^{(i)}$ that we choose in this region, and we can perform the matching anywhere. To simplify, let us choose $R_\Sigma^{(i)}=2.5M$ from which we obtain $\sigma\sim 2.43963M\rho_0$ and $p\sim0.640817M\rho_0$. 

The only step left to complete the matching is to verify the junction condition $\left[\varphi\right]=0$. Using Eq.\eqref{jcex2solphi} we verify that $\varphi\left(2.5M\right)=\left(3557/50\right)M^2\pi\rho_0+\psi_0$. Since the constant $\varphi_0^{(i)}$ of the interior solution is a free parameter, we can choose $\varphi_0^{(i)}=\varphi\left(2.5M\right)$ and the matching is complete. The final solution still has three free parameters, namely $M$, $\rho_0$ and $\psi_0$ that can be chosen to be anything. Ideally, one should choose $\psi_0<\varphi$ everywhere, to guarantee that the coupling $\left(\varphi-\psi\right)$ in the field equations is positive.

\section{Conclusions}

In this chapter we have deduced the junction conditions for a smooth matching with and without a thin shell of matter for both the geometrical and the scalar-tensor representation of the generalized hybrid metric-Palatini gravity. As expected, the presence of extra fields and the generalizarion of the action to $f\left(R,\mathcal R\right)$ leads to new junction conditions in both representations, which implies that matched solutions in GR may not necessarily be solutions in this theory, e.g., self-gravitating thin shells of matter are contrained to have a specific radius $R=9M/4$, unlike in GR where they can have any radius $R>2M$ for a given mass $M$.

In the geometrical representation of the theory, we obtained some expected results, namely the fact that the trace of the extrinsic curvature and the Ricci scalar must be continuous for a smooth matching, and also that the normal derivative of the Ricci scalar might or not be continuous, giving rise to the existance of a thin shell of matter if not. These results are the same as obtained for $f\left(R\right)$ theories. Given the conformal relation between $R$ and $\mathcal R$, it was also expected that the continuity of $R$ implied the continuity of $\mathcal R$. Also, this conformal relation implies, in the case of a matching with a thin shell, that the jump of the derivatives of $R$ and $\mathcal R$ does not contribute independently to the stress energy tensor of the shell. In fact, the two quantities must have a specific relation, expressed in Eq.~\eqref{jcgeodricciaux}.

Some other expected results also appeared in the scalar-tensor representation of the theory. Just like previously obtained for the Brans-Dicke theory of gravity, we verified that the scalar fields and their normal derivatives must be continuous for the matching to be smooth. However, since only the scalar field $\varphi$ is coupled to matter, we verified that only the derivative of $\varphi$ is allowed to be discontinuous in order to contribute to the stress energy of the thin shell. Also, unlike the Brans-Dicke case, the trace of the extrinsic curvature must also be continuous, which emphasizes the relation between the geometrical and the scalar-tensor representation of the theory.

Although the equivalence between the two representations can be obvious in some results such as the continuity of the trace of the extrinsic curvature or the direct implication between the continuity of the scalar fields to the continuity of $R$ and $\mathcal R$, the equivalence of the remaining results is not so obvious. In the results for the density and transverse pressure of the thin shell, we verify that a transformation $\varphi=f_R$ in the scalar-tensor results yields the same expression as obtained for the geometrical representation. Also, the relation between the derivatives of $R$ and $\mathcal R$ given by Eq.~\eqref{jcgeodricciaux} can also be found in the scalar tensor representation, although not so obviously, through Eq.~\eqref{jcscakgpsi}, since a replacement $\psi=f_\mathcal R$ here leads to the correct relation. We may then conclude that, as expected, the junction conditions in both representations are equivalent.

The importance and usefulness of these conditions is explicit where we successfully deduced a three-region toy model solution to the modified field equations representing a constant density shell of perfect fluid with a finite thickness. The use of the junction conditions allowed us to satisfy all the energy conditions for the whole space time, and also avoid problems in the metric such as horizons. Other more realistic solutions could also be obtained numerically by choosing $\alpha=3$ in the ansatz for $m\left(r\right)$, which is out of the scope of this thesis.

Another interesting system studied in GR using the junction conditions is the Oppenheimer-Snyder collapse of a FLRW dust sphere into a black-hole. Unfortunately, similarly to the $f\left(R\right)$ case, this solution can not be obtained in the GHMPG theory due to the fact that the Ricci scalar $R$ must be continuous across the matching hypersurface and this solution presents two different values of $R$ for the inside and the outside solutions. Finding a solution for a collapsing star in this theory is still an open problem and will be adressed in a future work. 
\cleardoublepage

\chapter{Wormhole solutions in the generalized hybrid metric-Palatini gravity}
\label{chapter:chapter6}

Wormhole solutions in a generalized hybrid metric-Palatini matter theory are found.  The
solutions are worked out in the scalar-tensor representation of the theory.  The main interest in the solutions found is that the matter field obeys the null energy condition (NEC) everywhere, including the throat and up to infinity, so that there is no need for exotic matter. The wormhole geometry with its flaring out at the throat is supported by the higher-order curvature terms, or equivalently, by the two fundamental scalar fields, which either way can be interpreted as a gravitational fluid. Thus, in this theory, in building a wormhole, it is possible to exchange the  exoticity of matter by the exoticity of the gravitational sector. The specific wormhole displayed, built to obey the matter NEC from the throat to infinity, has three regions, namely, an interior region containing the throat, a thin shell of matter, and a vacuum Schwarzschild anti-de Sitter (AdS) exterior region.  For hybrid metric-Palatini matter theories this wormhole solution is the first where the NEC for the matter fields is verified for the entire spacetime keeping the solution under asymptotic control, in this case asymptotically AdS. 

\section{Introduction}

Within general relativity, wormholes were found as exact solutions connecting two different asymptotically flat regions of spacetime \cite{morris1,visser1,lobo1} as well as two different asympotically de Sitter (dS) or anti-de Sitter (AdS) regions \cite{lemos1}.  The fundamental ingredient in wormhole physics is the existence of a throat satisfying a flaring-out condition. In general relativity this geometric condition entails the violation of the null energy condition (NEC). This NEC states that $T_{ab}k^a k^b \geq 0$, where $T_{ab}$ is the matter stress-energy tensor and $k^a$ is any null vector. Matter that violates the NEC is denoted as exotic matter. Wormhole solutions have been also found in other theories, the most relevant being the Brans-Dicke theory \cite{agnese1,nandi1}.  In these works the NEC for the matter is also violated.  However due to its nature, it is important and useful to minimize its usage.

In fact, in the context of modified theories of gravity, it has been shown in principle that normal matter may thread the wormhole throat, and it is the higher-order curvature terms, which may be interpreted
as a gravitational fluid, that support these nonstandard wormhole geometries.  Indeed, in \cite{lobo2} it was shown explicitly that in $f(R)$ theories wormhole throats can be theoretically constructed without the presence of exotic matter, in \cite{garcia1} nonminimal couplings were used to build such wormholes, and in \cite{harko2} generic modified gravities were used also with that aim in mind, i.e., the wormhole throats are sustained by the fundamental fields presented in the modified gravity alone.  This type of solutions were also found in Brans-Dicke theories \cite{anchordoqui1}, and in a hybrid metric-Palatini gravitational theory \cite{capozziello5}.  It is our aim to find wormhole solutions whose matter obeys the NEC not only at the throat but everywhere in a generalized hybrid metric-Palatini gravity.

In this chapter, our aim is to find static and spherically symmetric wormholes solutions in the generalized $f(R,{\cal R})$ hybrid metric-Palatini matter theory in which the matter satisfies the NEC everywhere, from the throat to infinity, so there is no need for exotic matter. This fills a gap in the literature as most of the work that has been done in this area has been aimed at finding solutions where the NEC is satisfied solely at the wormhole throat paying no attention to the other regions. Also, we find that matter also satisfies the weak energy condition (WEC) which states that $T_{ab}u^a u^b \geq 0$, where $u^a$ is any timelike vector, for the entire spacetime except for the needed thin-shell of matter.

\section{Wormhole ansatz and equations}

In this section we shall derive the equations and conditions for our wormhole solution to be not only traversable but also to not violate the NEC. We start by imposing that the wormhole solutions are described by a static and spherically symmetric metric which is diagonal in the usual spherical $\left(t,r,\theta,\phi\right)$ coordinates and whose line element has the form
\begin{equation}\label{metricwormhole}
ds^2=-e^{\zeta\left(r\right)}dt^2+\left[1-\frac{b\left(r\right)}{r}\right]^{-1}dr^2+r^2\left( d\theta^2+\sin^2\theta d\phi^2\right),
\end{equation}
where  $\zeta\left(r\right)$ is known as the redshift function and $b\left(r\right)$ is the shape function. The shape function $b(r)$ should obey the boundary condition $b\left(r_0\right)=r_0$, where $r_0$ is defined as the radius of the wormhole throat, and also the flaring-out condition discussed in the upcoming sections. We further assume that matter is described by an anisotropic perfect fluid, i.e., the stress-energy tensor is given by $T^a_b=\text{diag}\left(-\rho, p_r, p_t, p_t\right)$, from which $T=-\rho+p_r+2p_t$ where $\rho\left(r\right)$ is the energy density, $p_r\left(r\right)$ is the radial pressure, and $p_t\left(r\right)$ is the radial pressure, all assumed to be functions of $r$ only to preserve the staticity of the solution. We shall be working  in the scalar-tensor representation of the theory, for which the field equations are given by Eq.\eqref{ghmpgstfield} and the scalar field equations are given by Eqs.\eqref{ghmpgstkgphi} and \eqref{ghmpgstkgpsi}, where the scalar fields $\varphi\left(r\right)$ and $\psi\left(r\right)$ are also assumed to be functions of $r$ only.

Under these assumptions, Eq.\eqref{ghmpgstfield} has three independent components given by
\begin{equation}\label{wormfieldrho}
\left(\varphi-\psi\right)\frac{b'}{r^2}-\frac{\left(\varphi-\psi\right)'}{2r}\left(1-b'\right)-\frac{V}{2}-\left(1-\frac{b}{r}\right)\left[\left(\varphi-\psi\right)''+\frac{3\left(\varphi-\psi\right)'}{2r}+\frac{3\psi'^2}{4\psi}\right]=\kappa^2\rho,
\end{equation}
\begin{equation}\label{wormfieldpr}
\left(\varphi-\psi\right)\left[\frac{\zeta'}{r}\left(1-\frac{b}{r}\right)-\frac{b}{r^3}\right]+\left(1-\frac{b}{r}\right)\left[-\frac{3\psi'^2}{4\psi}+\frac{2\left(\varphi-\psi\right)'}{r}+\frac{\zeta'\left(\varphi-\psi\right)'}{2}\right]+\frac{V}{2}=\kappa^2p_r,
\end{equation}
\begin{eqnarray}
\left(\varphi-\psi\right)\Bigg[\left(1-\frac{b}{r}\right)\left(\frac{\zeta''}{2}+\frac{\zeta'^2}{4}+\frac{\zeta'}{2r}\right)+\frac{b-rb'}{2r^3}\left(1+\frac{\zeta'r}{2}\right)\Bigg]+\frac{\varphi'}{2r}\left(1-b'\right)+\frac{V}{2}+\nonumber \\
+\left(1-\frac{b}{r}\right)\left[\left(\varphi-\psi\right)''+\frac{\zeta'\left(\varphi-\psi\right)'}{2}+\frac{3\psi'^2}{4\psi}+\frac{\left(\varphi-\psi\right)'}{2r}\right]=\kappa^2p_t\label{wormfieldpt}
\end{eqnarray}
respectively, where a prime $'$ denotes a derivative with respect to the radial coordinate $r$. These equations will work as equations for the matter fields $\rho, p_r$ and $p_t$. An equation for the scalar field $\psi$ can be obtained directly from Eq.\eqref{ghmpgstkgpsi}, which becomes
\begin{equation}\label{wormkgpsi}
\left(1-\frac{b}{r}\right)\left(\psi''+\frac{\zeta'\psi'}{2}+\frac{3\psi'}{2r}-\frac{\psi'^2}{2\psi}\right)+\frac{\psi'}{2r}\left(1-b'\right)-\frac{\psi}{3}\left(V_\varphi+V_\psi\right)=0.
\end{equation}
The equation for the scalar field $\varphi$ given by Eq.\eqref{ghmpgstkgphi} is extremely complicated for it not only depends on both scalar fields $\varphi$ and $\psi$ but also on the matter fields $\rho, p_r$ and $p_t$ via the trace $T$. To find a more suitable equation to work with, we use Eqs.\eqref{wormfieldrho} to \eqref{wormfieldpt} to write $T$ as a function of the scalar fields $\varphi$ and $\psi$, and the metric functions $\zeta$ and $b$. We insert the result into Eq.\eqref{ghmpgstkgphi} and use Eq.\eqref{wormkgpsi} to cancel the terms depending on $\psi''$ (or equivalently, into the difference between Eqs.\eqref{ghmpgstkgphi} and \eqref{ghmpgstkgpsi}). The result is an equation that only depends on $\varphi$ via the potential as
\begin{equation}\label{wormkgphi}
\left(1-\frac{b}{r}\right)\left(\zeta''+\frac{\zeta'^2}{2}+\frac{2\zeta'}{r}\right)-\frac{b+rb'}{r^3}+\frac{b-rb'}{r^3}\left(1+\frac{r\zeta'}{2}\right)+V_\varphi=0.
\end{equation}

We thus have a system of five independent equations given by Eqs.\eqref{wormfieldrho} to \eqref{wormkgphi} for eight independent variables, namely the scalar fields $\varphi$ and $\psi$, the matter fields $\rho$, $p_r$ and $p_t$, the metric functions $\zeta$ and $b$, and the potential $V$. This means that one can impose three extra constraints to the problem to close the system and obtain an unique solution.

\subsection{Null energy condition}

The main aim in our wormhole construction is that throughout the wormhole solution the matter must obey the NEC. In pure general relativity, it is known that at the wormhole throat, the NEC for the matter fields is violated in order to allow for the wormhole flare out \cite{visser1}. In the scalar representation of the generalized hybrid metric-Palatini gravity that we are studying, the gravitational field is complemented by the two other fundamental gravitational fields, the scalar fields $\varphi$ and $\psi$, whereas the matter fields in this theory are still encoded in the stress-energy tensor $T_{ab}$. Thus, to allow for a mandatory flare out, the matter $T_{ab}$ can obey the NEC, as long as some appropriate combination of $T_{ab}$ with the other fundamental fields does not obey the NEC. 

To build such a wormhole might be a difficult task as the NEC can be obeyed in a certain region and then be violated at some other region.  Our strategy is then the following.  The most critical radius is the wormhole throat radius.  So we impose that the NEC for the matter fields is obeyed at the throat and its vicinity. We then look for a solution in the vicinity of the throat. When the solution starts to violate the NEC above a certain $r$ away from the throat, we cut the solution there. We then join it to a vacuum solution for larger radii and build a thin shell solution at a junction radius up to which the matter fields obey the NEC.

To be specific, we start by finding wormhole solutions for which the matter threading the throat is normal matter, i.e., the matter stress-energy tensor $T_{ab}k^ak^b$, for two null vectors $k^a$ and $k^b$, must obey the NEC, i.e., $T_{ab}k^ak^b\geq 0$. This NEC is to be obeyed at the throat and its vicinity. Noting that ${T_a}^b$ is of the form $T^a_b=\text{diag}\left(-\rho, p_r, p_t, p_t\right)$, taking into account Eqs.\eqref{wormfieldrho} to \eqref{wormfieldpt}, and choosing in the frame of Eq.\eqref{metricwormhole} a null vector $k^a$ of the form $k^a=\left(1,1,0,0\right)$, one obtains from $T_{ab}k^ak^b \geq 0$ that
\begin{equation}
\rho+p_r\geq 0.
\label{wormnecpr}
\end{equation}
For a null vector $k^a$ of the form $k^a=\left(1,0,1,0\right)$, we find that $T_{ab}k^ak^b\geq 0$ means
\begin{equation}
\rho+p_t \geq 0.
\label{wormnecpt}
\end{equation}
It might happen that the weak energy condition (WEC) for the matter alone is also be obeyed at the wormhole throat, although in our wormhole construction we do not impose it. To verify whether the WEC
is verified or not, we choose a timelike vector $u^a$ of the form $u^a=\left(1,0,0,0\right)$, and see if $T_{ab}u^au^b\geq 0$. For a perfect fluid this becomes
\begin{equation}
\rho\geq 0.
\label{wormwec}
\end{equation}
Thus the matter fields that build the wormhole we are constructing must obey the NEC Eqs.\eqref{wormnecpr} and \eqref{wormnecpt}, and might, but not necessarily, obey the WEC Eq.\eqref{wormwec}.

\subsection{Flaring-out condition}

The shape function $b(r)$ obeys two boundary conditions. The first one already metioned is $b\left(r_0\right)=r_0$, where $r_0$ is the radius of the wormhole throat.  Secondly, the fundamental condition in wormhole physics is that the throat flares out which is translated by the condition at the throat $(b-b'r)/b^2>0$. This imposes that at the throat we have \cite{morris1,visser1}
\begin{equation}\label{wormfoc}
b'\left(r_0\right)<1,
\end{equation}
Which is known as the flaring-out condition. This condition is general. Indeed it is a condition on the Einstein tensor $G_{ab}$. In general relativity, since the field equation is $G_{ab}=\kappa^2T_{ab}$, it is also a  condition on the matter stress-energy tensor $T_{ab}$ and it directly implies that $T_{ab}$ violates the NEC. However, in the scalar field representation of the generalized hybrid metric-Palatini theory there are two extra gravitational fundamental fields $\varphi$ and $\psi$. So, the flaring out
condition on $G_{ab}$, in the hybrid generalization, translates into a condition on a combination of the matter stress-energy $T_{ab}$ with an appropriate tensor built out of $\varphi$, $\psi$, and geometric tensorial quantities. Let us see this in detail. Note that Eq.~\eqref{ghmpgstfield} can be written in the form
\begin{equation}
\left(\varphi-\psi\right)\left(G_{ab}+H_{ab}\right)=\kappa^2T_{ab},
\end{equation}
\begin{equation}
H_{ab}=\frac{1}{\varphi-\psi}[(\Box\varphi-\Box\psi+\frac{1}{2}V+\frac{3}{4\psi}\partial^c\psi\partial_c\psi)g_{ab}-\frac{3}{2\psi}\partial_a\psi\partial_b\psi-\nabla_a\nabla_b\varphi+\nabla_a\nabla_b\psi].
\end{equation}
We can define for clarity an effective stress-energy tensor $T_{ab}^{(eff)}$ as $T_{ab}^{(eff)}=\frac{T_{ab}}{\varphi-\psi}-\frac{H_{ab}}{\kappa^2}$. Our aim, defined from the start, is that the NEC on the matter $T_{ab}$ is obeyed everywhere, including the wormhole throat. But, since for wormhole construction one needs to flare out at the throat then the NEC on $T_{ab}^{(eff)}$ has to be violated there. This means that the contraction of the effective stress-energy tensor with two null vectors, $k^a$ and $k^b$ say, must be negative, i.e., $T_{ab}^{\rm eff} \,k^ak^b<0$, at the throat. From the definition of $T_{ab}^{\rm eff}$ above, this condition can be converted into $T_{ab}k^ak^b-g H_{ab}\,k^ak^b<0$, where $g(\varphi,\psi)\equiv\frac{\varphi-\psi}{\kappa^2}$. So it gives $T_{ab}k^ak^b<g H_{ab}\,k^ak^b$. We assume that $g>0$ and write thus this condition as $T_{ab}k^ak^b < g H_{ab}k^ak^b$, at the throat. Noting that $T_{ab}$ is of the form $T^a_b=\text{diag}\left(-\rho, p_r, p_t, p_t\right)$, taking into account Eqs.\eqref{wormfieldrho} to \eqref{wormfieldpt} and choosing in the frame of Eq.\eqref{metricwormhole} a null vector $k^a$ of the form $k^a=\left(1,1,0,0\right)$, we find that $T_{ab}k^ak^b < g H_{ab}k^ak^b$
at $r_0$, the throat, is then
\begin{eqnarray}
\kappa^2\left(\rho+p_r\right) &<&\frac{Vr}{r-b}-e^\zeta\left(1-\frac{b}{r}\right)\left(\frac{Vr}{r-b}+\frac{3}{4}\frac{\psi'^2}{\psi}+\frac{3}{2}\frac{\varphi'-\psi'}{r}+\varphi''-\psi''\right)+\nonumber\\
&+&\frac{\varphi'-\psi'}{2r}\left[4+r\zeta'+e^\zeta\left(b'-1\right)\right]-\frac{3}{4}\frac{\psi'^2}{\psi}.
\end{eqnarray}
On the other hand, for a null vector $k^a$ of the form $k^a=\left(1,0,1,0\right)$, we find that $T_{ab}k^ak^b < g H_{ab}k^ak^b$ takes the form
\begin{eqnarray}
\kappa^2\left(\rho+p_t\right)&<&\left(1-e^\zeta\right)\left[\frac{V}{2}+\left(1-\frac{b}{r}\right)\left(\frac{3}{4}\frac{\psi'^2}{\psi}+\psi''-\varphi''\right)+\frac{\varphi'-\psi'}{2r}\left(1-b'\right)\right]+\nonumber\\
&+&\frac{\varphi'-\psi'}{2r}\left(1-\frac{b}{r}\right)\left(r\zeta'-3e^\zeta\right).
\end{eqnarray}

The violation of the NEC implies the violation of the WEC at the wormhole throat \cite{visser1}. We choose a timelike vector $u^a$ of the form $u^a=\left(1,0,0,0\right)$, and we find that $T_{ab}u^au^b < g H_{ab}u^au^b$ becomes then
\begin{equation}
\kappa^2\rho <\left(1-\frac{b}{r}\right)\left(\frac{3}{4}\frac{\psi'^2}{\psi}+\frac{3}{2}\frac{\varphi'-\psi'}{r}+\varphi''-\psi''\right)+\frac{1-b'}{2r}\left(\varphi'-\psi'\right)+\frac{V}{2}.
\end{equation}
These inequalities at the wormhole throat, coming out of $T_{ab}^{(eff)}$ to obey the flaring out condition, and the NEC and WEC on $T_{ab}$, Eqs.\eqref{wormnecpr} and \eqref{wormnecpt} are not incompatible at all. It is thus possible in principle to find wormhole solutions in the generalized hybrid theory obeying the matter NEC.

\section{Wormhole regions}

In this section we shall build an analytical wormhole solution patch by patch. We start by investigating the interior region near the throat and we shall verify that although the NEC and the WEC are satisfied at the throat radius, there is a larger radius from which these conditions are violated up to infinity. We thus continue our construction by deriving an exterior spherically symmetric vacuum solution for which all the energy conditions are automatically satisfied. Finally, we apply the junction conditions derived in chapter \ref{chapter:chapter5} to match the interior solution to the exterior solution in a region where the NEC and WEC are still satisfied in the interior solution. To do so, one needs a thin shell of perfect fluid at the junction radius, and we show that this thin shell can be fine-tuned to satisfy the NEC as well, thus guaranteeing that the NEC is valid for the whole spacetime.

\subsection{Interior solution}

The system composed by Eqs.\eqref{wormfieldrho} to \eqref{wormkgphi} is a system of five equations to eight variables, and thus we can impose three constraints to close the system and obtain an unique solution. We decide to impose constraints on the redshift function $\zeta\left(r\right)$, to guarantee that this function is finite and thus no horizons are present; to the shape function $b\left(r\right)$ to guarantee that the flaring-out condition in Eq.\eqref{wormfoc} is always satisfied; and on the potential $V\left(\varphi,\psi\right)$ to simplify the scalar field equations in the same line of thought as employed in chapter \ref{chapter:chapter3}.

A quite general class of redshift $\zeta\left(r\right)$ and shape $b\left(r\right)$ metric functions in Eq.\eqref{metricwormhole} that verify the traversability conditions for a wormhole is
\begin{equation}\label{wormredshift}
\zeta\left(r\right)=\zeta_0\left(\frac{r_0}{r}\right)^\alpha\exp\left(-\gamma\frac{r-r_0}{r_0}\right),
\end{equation}
\begin{equation}\label{wormshape}
b\left(r\right)= r_0\left(\frac{r_0}{r}\right)^\beta\exp\left(-\delta\frac{r-r_0}{r_0}\right),
\end{equation}
where $\zeta_0$ is a dimensionless constant, $r_0$ is the wormhole throat radius, and $\alpha$, $\gamma$, $\beta$, and $\delta$, are free exponents. We also assume that $\gamma>0$ so that the redshift function $\zeta(r)$ given in Eq.\eqref{wormredshift} is finite and nonzero for every value of $r$ between $r_0$ and infinity, and $\zeta_0$ is the value of $\zeta$ at $r_0$. We further assume that $\delta\geq0$, so that the shape function $b(r)$ given in Eq.\eqref{wormshape} yields $b\left(r=r_0\right)=r_0$ and $b'\left(r_0\right)<1$, as it should for a traversable wormhole solution, see Eq.\eqref{wormfoc}. We consider further a power-law potential of the form 
\begin{equation}\label{wormpotential}
V\left(\varphi,\psi\right)=V_0\left(\varphi-\psi\right)^\eta,
\end{equation}
where $V_0$ is a constant and $\eta$ a free exponent. With these choices of Eqs.\eqref{wormredshift} to \eqref{wormpotential} and for various combinations of the parameters $\alpha, \beta, \gamma, \delta$. and
$\eta$ one can then find specific equations for $\varphi$ and $\psi$ from Eqs.\eqref{wormkgphi} and \eqref{wormkgpsi}, respectively, and try so solve them either analytically or numerically to obtain the wormhole solutions.

\begin{figure}
\centering 
\begin{subfigure}{0.48\textwidth}
\includegraphics[scale=0.55]{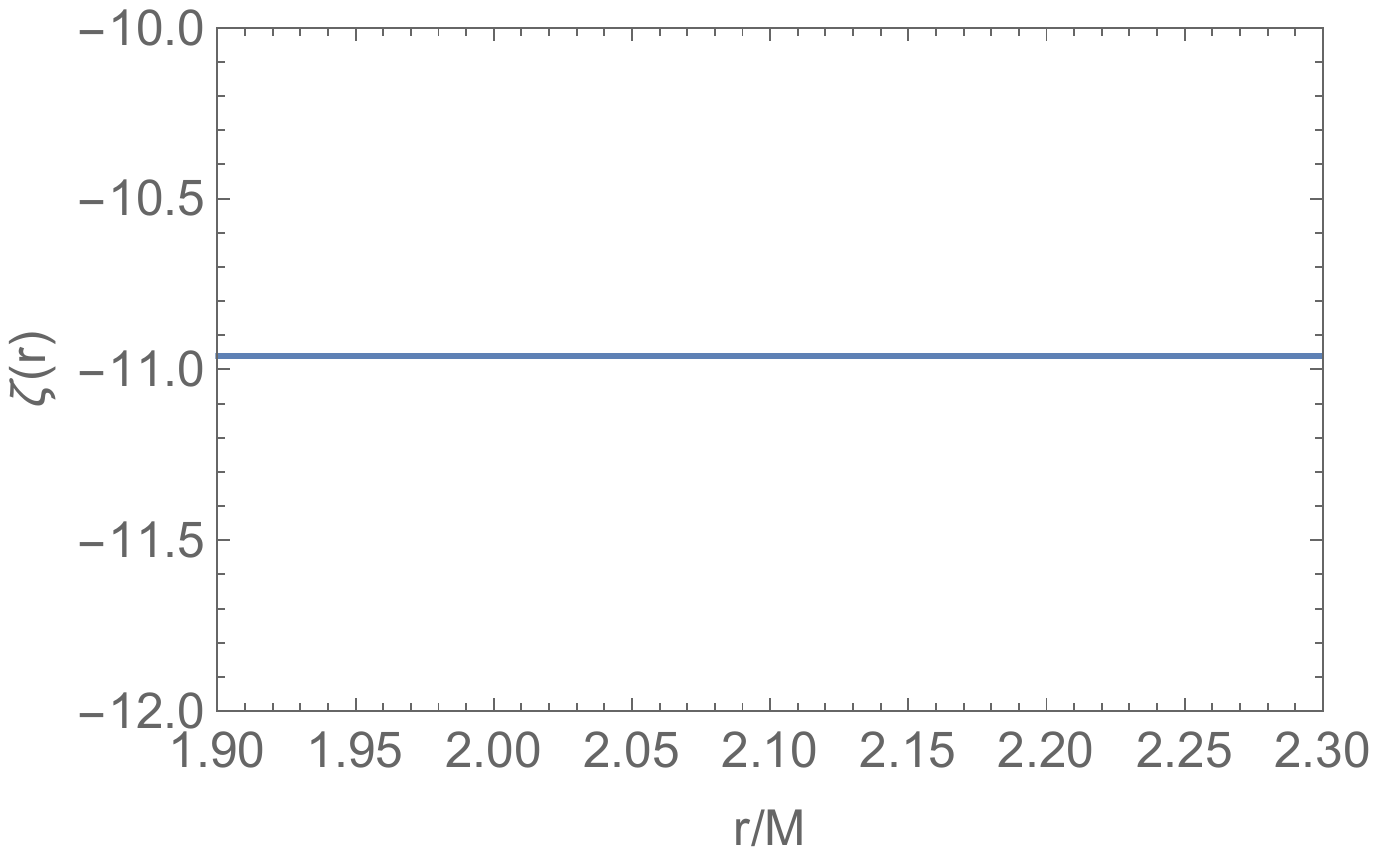}
\caption{Redshift function $\zeta\left(r\right)$}
\label{fig:wormredshift1}
\end{subfigure}
\ \ \ \ \ 
\begin{subfigure}{0.48\textwidth}
\includegraphics[scale=0.55]{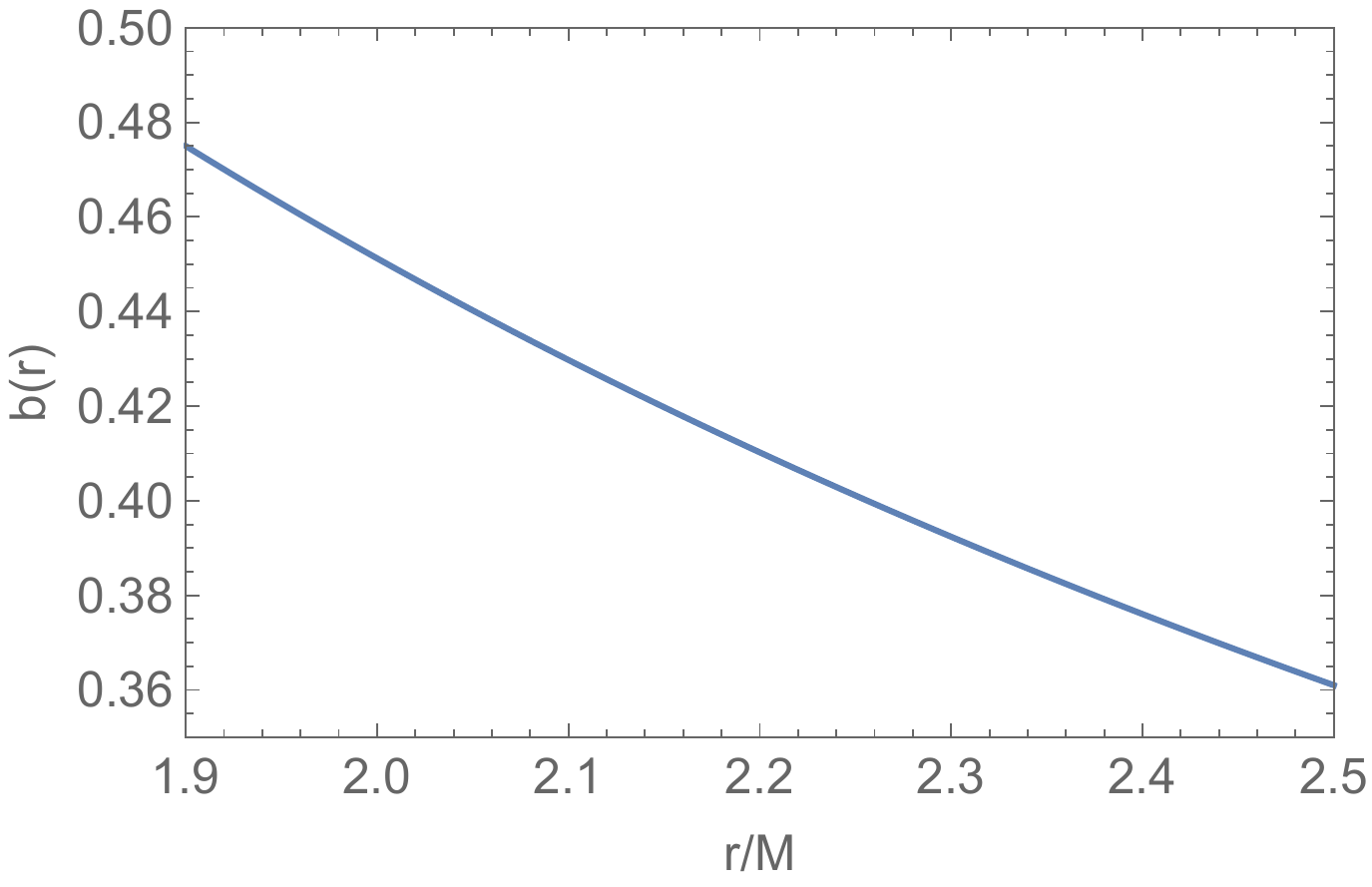}
\caption{Shape function $b\left(r\right)$}
\label{fig:wormshape1}
\end{subfigure}
\ \\
\ \\
\ \\
\begin{subfigure}{0.48\textwidth}
\includegraphics[scale=0.55]{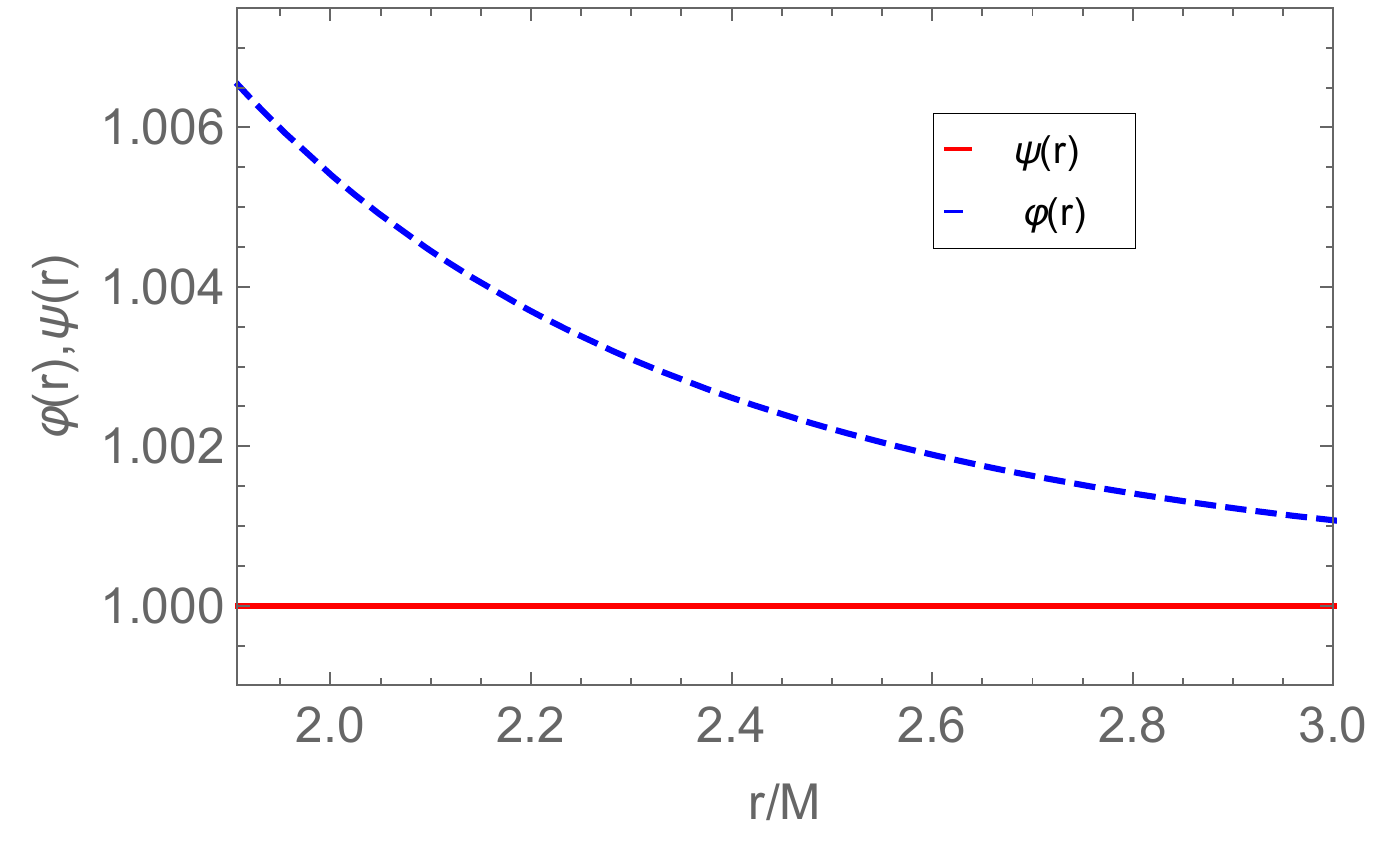}
\caption{Scalar fields $\varphi$ and $\psi(t)$}
\label{fig:wormfields1}
\end{subfigure}
\ \ \ \ \
\begin{subfigure}{0.48\textwidth}
\includegraphics[scale=0.55]{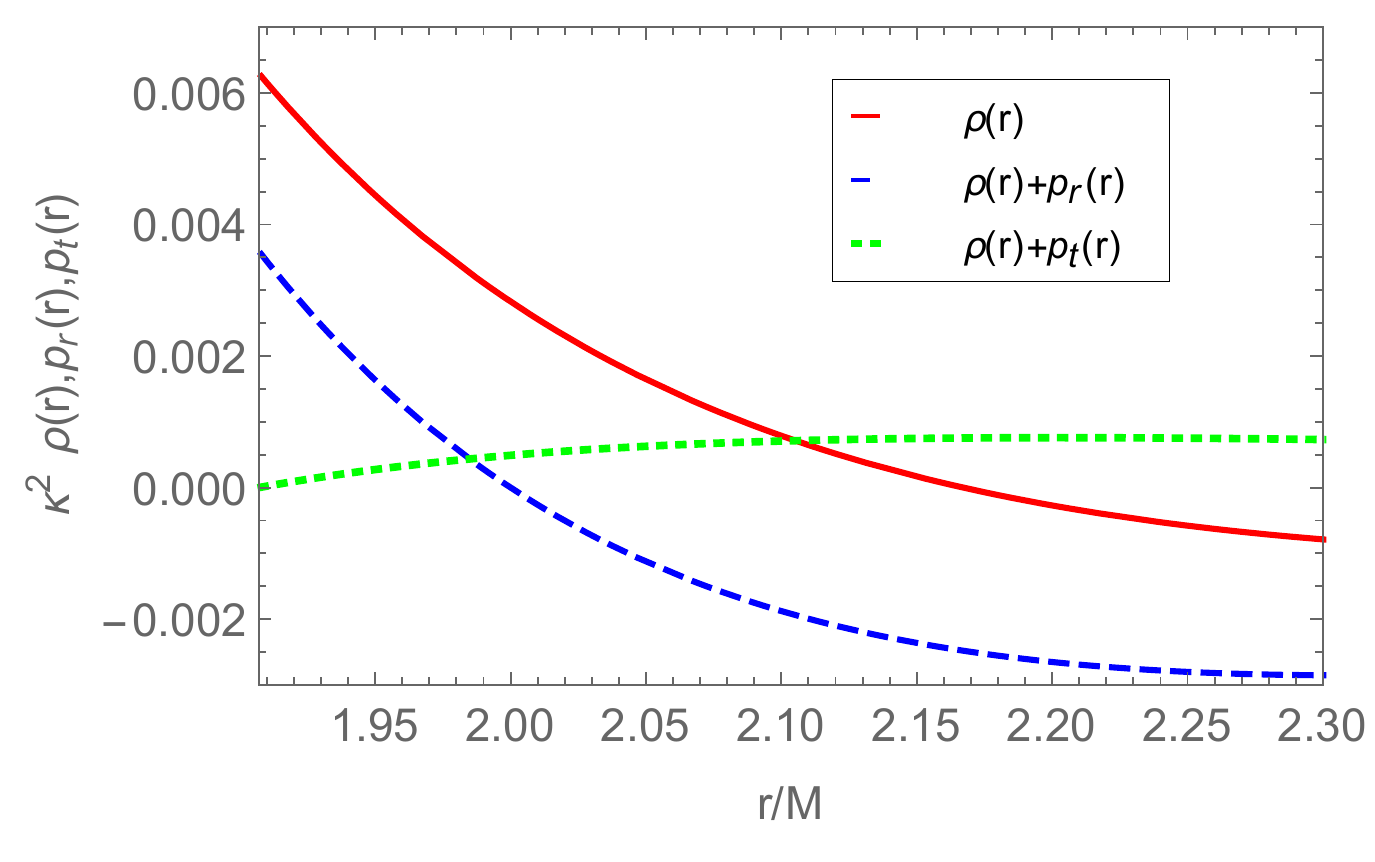}
\caption{Matter fields $\rho\left(r\right)$, $p_r\left(r\right)$ and $p_t\left(r\right)$}
\label{fig:wormmatter1}
\end{subfigure}
\caption{Metric fields $\zeta(r)$ and $b(r)$, scalar fields $\varphi(r)$ and $\psi(r)$, and matter fields
$\rho$, $p_r$ and $p_t$, with $\zeta_0=-10.96$, $M=1$, $r_0=2\sqrt{10/11}=1.90693$, $\psi_0=1$, $\psi_1=0$, $V_0=-42$, for the interior wormhole solution}
\label{fig:wormsolint}
\end{figure}

Here we are interested in finding solutions that do not violate the matter NEC anywhere, Eqs.\eqref{wormnecpr} and \eqref{wormnecpt}, in particular at the throat, $\rho\left(r_0\right)+p_r\left(r_0\right)>0$ and $\rho\left(r_0\right)+p_t\left(r_0\right)>0$, and possibly at its neighborhood. As we will see, an analytical solution that respects these conditions is obtained for $\alpha=\gamma=\delta=0$, and $\beta=1$. For these choices, Eqs.\eqref{wormredshift} and \eqref{wormshape} yield 
\begin{equation}\label{wormredshift1}
\zeta\left(r\right)=\zeta_0,
\end{equation}
\begin{equation}\label{wormshape1}
b\left(r\right)=\frac{r_0^2}{r},
\end{equation}
respectively. These solutions are plotted in Figs.\ref{fig:wormredshift1} and \ref{fig:wormshape1}. Thus, the line element Eq.\eqref{metricwormhole} can be written for the interior wormhole region containing the throat as
\begin{equation}\label{worminterior}
ds^2=-e^{\zeta_0}dt^2+\left[1-\frac{r_0^2}{r^2}\right]^{-1}dr^2+r^2\left( d\theta^2+\sin^2\theta d\phi^2\right).
\end{equation}
Note that one can absorb the factor $e^{\zeta_0}$ in the time coordinate by performing a redefinition of the form $d\bar t=e^{\zeta_0/2}dt$. Choosing $\eta=2$ Eq.\eqref{wormpotential} yields
\begin{equation}\label{wormpotential1}
V\left(\varphi,\psi\right)=V_0\left(\varphi-\psi\right)^2.
\end{equation}
Inserting our assumptions for the redshift function, the shape function and the potential given respectively in Eqs.\eqref{wormredshift1}, \eqref{wormshape1} and \eqref{wormpotential1}, into the scalar field equations given by Eqs.\eqref{wormkgpsi} and \eqref{wormkgphi} leads to 
\begin{equation}\label{wormkgpsi1}
\psi''-\frac{\psi'^2}{2\psi}+\frac{\psi'}{r}\left[\frac{2-\left(\frac{r_0}{r}\right)^2}{1-\left(\frac{r_0}{r}\right)^2}\right]=0,
\end{equation}
\begin{equation}\label{wormkgphi1}
r_0^2+r^4V_0\left(\varphi-\psi\right)=0,
\end{equation}
respectively. The equation Eq.\eqref{wormkgpsi1} can be directly integrated over the radial coordinate $r$ to obtain a solution for $\psi$. On the other hand, note that Eq.\eqref{wormkgphi1} is a linear relation between $\varphi$ and $\psi$, from which we see that as soon as we obtain a solution for $\psi$ we automatically have a solution for $\varphi$ with no integrations needed. Doing so, the solutions for the scalar fields $\psi$ and $\varphi$ become then
\begin{equation}\label{wormsolpsi}
\psi\left(r\right)=\left[\sqrt{\psi_0}+\sqrt{\psi_1} \arctan\left(\sqrt{\frac{r^2}{r_0^2}-1}\right)\right]^2,
\end{equation}
\begin{equation}\label{wormsolphi}
\varphi\left(r\right)=\left[\sqrt{\psi_0}+\sqrt{\psi_1 }\arctan\left(\sqrt{\frac{r^2}{r_0^2}-1}\right)\right]^2-\frac{r_0^2}{r^4V_0},
\end{equation}
respectively. These scalar fields are plotted in Fig.\eqref{fig:wormfields1}. It is also possible to simplify the problem even further and obtain non-trivial solutions. Let us choose for example $\psi_1=0$ in such a way that $\psi$ becomes a constant and the solutions for $\psi$ and $\varphi$ become finally
\begin{equation}\label{wormsolpsi1}
\psi(r)=\psi_0,
\end{equation}
\begin{equation}\label{wormsolphi1}
\varphi(r)=\psi_0-\frac{r_0^2}{r^4V_0},
\end{equation}
respectively. This simplification was actually mandatory. Recall the analysis done in chapter \ref{chapter:chapter3} on which we computed the forms of the functions $f\left(R,\mathcal R\right)$ that correspond to specific forms of the potential $V\left(\varphi,\psi\right)$. Back then, we showed that for a quadratic potential of the form in Eq.\eqref{wormpotential1} we must have $R=\mathcal R$. Imposing this constraint, the metric in Eq.\eqref{worminterior} and the scalar field in Eq.\eqref{wormsolpsi} into the relation between $R$ and $\mathcal R$ from Eq.\eqref{ghmpgrelricsca} one verifies that for this relation to be satisfied we must have $\psi_1=0$. From Eqs.\eqref{wormsolpsi1} and \eqref{wormsolphi1} we see that both scalar fields are finite at the throat $r=r_0$, for values of $r$ in its vicinity, and indeed for all $r$ up to infinity. 

We are now in conditions to find the solutions for the matter fields and study in which cases they obey the matter NEC. Inserting the solutions for the metric fields from Eqs.\eqref{wormredshift1} and \eqref{wormshape1}, the potential form Eq.\eqref{wormpotential1} and the scalar fields from Eqs.\eqref{wormsolpsi1} and \eqref{wormsolphi1} into the field equations given in Eqs.\eqref{wormfieldrho} to \eqref{wormfieldpt}, we obtain the energy density, the radial pressure, and the tangential pressure as
\begin{equation}\label{wormwec1}
\rho=\frac{r_0^2}{2r^6V_0\kappa^2}\left(24-31\frac{r_0^2}{r^2}\right),
\end{equation}
\begin{equation}
p_r=\frac{r_0^2}{2r^6V_0\kappa^2}\left(16-13\frac{r_0^2}{r^2}\right),
\end{equation}
\begin{equation}
p_t=\frac{r_0^2}{2r^6V_0\kappa^2}\left(-32+39\frac{r_0^2}{r^2}\right),
\end{equation}
respectively. It is more useful to combine $\rho$ with $p_r$ and $p_t$ to obtain expressions directly connected to the NEC. These combinations give
\begin{equation}\label{wormnecpr1}
\rho+p_r=\frac{2r_0^2}{r^6V_0\kappa^2}\left(10-11\frac{r_0^2}{r^2}\right),
\end{equation}
\begin{equation}\label{wormnecpt1}
\rho+p_t=\frac{4r_0^2}{r^6V_0\kappa^2}\left(-1+\frac{r_0^2}{r^2}\right).
\end{equation}
The matter fields from Eqs.\eqref{wormwec1}, \eqref{wormnecpr1} and \eqref{wormnecpt1} are plotted in Fig.\ref{fig:wormmatter1}. At the throat, noting that $r_0>0$ and $\kappa^2>0$, from Eq.\eqref{wormnecpr1} we verify that we need $V_0<0$ to preserve the NEC. The quantity $\rho+p_r$ is then positive up to a radius $r$ given by $r=\sqrt{11/10}r_0$. On the other hand, the quantity $\rho+p_t$ vanishes at the throat and is positive for every value of $r$. Thus the quantity $\rho+p_r$ sets the violation of the NEC. We then impose that the interior solution is valid within the following range of $r$:
\begin{equation}\label{worminteriorrange}
r\leq\sqrt{\frac{11}{10}}r_0,
\end{equation}
so the interior solution must stop at most at $r=\sqrt{\frac{11}{10}}r_0$. From Eq.\eqref{wormwec1} we see that in this range of $r$ the WEC given by $\rho\geq0$, is also obeyed. The interior solution with its metric fields $\zeta(r)$ and $b(r)$, scalar fields $\varphi(r)$ and $\psi(r)$, and matter fields $\rho$,
$\rho+p_r$, and $\rho+p_t$ are displayed in Fig.\ref{fig:wormsolint} for adjusted values of the free parameters. Both the scalar fields are finite for all $r$ and converge to the same constant at infinity, which implies $\varphi-\psi\to 0$ at infinity. Relative to the NEC, note that $\rho+p_r$ is positive at $r=r_0$ and changes sign at a finite value of $r$, $r=2M$, which is precisely $\sqrt{11/10}r_0$ for $r_0/M=2\sqrt{10/11}\sim 1.90693$. On the other hand, $\rho+p_t$ is positive for every value of $r$.  Note that $\rho$ is also positive at $r=r_0$ and changes sign at a higher value of $r$, $r=2.16M$, after which the WEC is violated.  Thus, the NEC and WEC are satisfied at the wormhole throat but are violated somewhere further up.  In order to maintain the validity of the NEC, the interior solution must at most stop at $r=2M$.

\subsection{Exterior solution}

To guarantee that the complete solution does obey the NEC for any value of $r$, we need to match at some $r_\Sigma<\sqrt{\frac{11}{10}}\,r_0$ the interior solution obtained in the previous section to an external vacuum spherically symmetric solution. To do so, one has to derive a vacuum exterior solution under asymptotic control, and then use the junction conditions of this theory to perform the matching with some thin shell at the boundary between the interior and exterior regions. Afterwards, we still have to ensure that the matter of this thin shell obeys the matter NEC, so that the complete spacetime obeys this condition.

To find an exterior vacuum solution, we start by setting the stress-energy tensor to zero i.e. $T_{ab}=0$, such that $\rho=p_r=p_t=0$. In addition, we choose the scalar fields to be constant in this exterior solution, i.e. $\varphi\left(r\right)=\varphi_e$ and $\psi\left(r\right)=\psi_e$, with $\varphi_e$ and $\psi_e$ constants and $\varphi_2\neq\psi_e$, where the subscript $e$ stands for "exterior". The junction of these assumptions for the exterior solution and the solutions obtained for the interior solution are plotted in Fig.\ref{fig:wormmatter2} for the matter fields and Fig.\ref{fig:wormfields2} for the scalar fields.

For continuity we choose the potential to be $V=V_0\left(\varphi_e-\psi_e\right)^2$. Note that from Eq.\eqref{ghmpgstpotential} it can be seen that choosing a particular form of the potential $V$ we are also setting a particular solution for the function $f$. This means that both the interior and the exterior spacetimes must be solutions of the field equations with the same form of the potential $V$, because otherwise they would not be solutions of the same form of the function $f\left(R,\mathcal R\right)$. These choices imply that the field equation Eq.\eqref{ghmpgstfield} can be
written as
\begin{equation}
G_{ab}+\frac{V_0}{2}\left(\varphi_e-\psi_e\right)g_{ab}=0.
\end{equation}
Considering that in the second term the multiplicative factors of the metric $g_{ab}$ are constant, we see that the field equation is of the same form as the Einstein field equations (see Eq.\eqref{efe}) in vacuum with a cosmological constant given by $\frac{V_0}{2}\left(\varphi_e-\psi_e\right)$. Thus, the Schwarzschild solution with a cosmological constant of general relativity is a vacuum solution of the
generalized hybrid theory in the particular form we are studying. This class of solutions is also known as
the Kottler solution, as well as  Schwarzschild-de-Sitter solution if the constant cosmological term is positive or Schwarzschild-Anti-de-Sitter solution if the constant cosmological term is negative. The metric fields $\zeta(r)$ and $b(r)$ for the exterior region outside some radius $r_\Sigma$ are then
\begin{equation}\label{wormredshift2}
e^{\zeta(r)}=\left[1-\frac{2M}{r}-\frac{V_0\left(\varphi_e-\psi_e\right)r^2}{6}\right]e^{\zeta_e}, 
\end{equation}
\begin{equation}\label{wormshape2}
b\left(r\right)=2M+\frac{V_0\left(\varphi_e-\psi_e\right) r^3}{6},
\end{equation}
respectively, where $M$ is a constant of integration and represents the mass, and the factor $e^{\zeta_e}$ is also a constant that we introduce for convenience. These solutions are plotted in Figs.\ref{fig:wormredshift2} and \ref{fig:wormshape2}, respectively. The line element of the generalized field equations given by Eq.\eqref{ghmpgstfield} in the case where the scalar fields are constant is then
\begin{equation}\label{metrickottler}
ds^2=-\left[{1-\frac{2M}{r}-\frac{V_0\left(\varphi_e-\psi_e\right) r^2}{6}}\right]e^{\zeta_e}dt^2+\left[1-\frac{2M}{r}-\frac{V_0\left(\varphi_e-\psi_e\right)r^2}{6}\right]^{-1}dr^2+r^2d\Omega^2,
\end{equation}
The sign of the term $\frac{V_0}{2}\left(\varphi_e-\psi_e\right)$ determines whether the solution is Schwarzschild-dS or Schwarzschild-AdS. This sign will be determined by the matching surface and the imposition that the NEC holds everywhere. Wormholes in dS and AdS spacetimes in general relativity were treated in \cite{lemos1}. The exterior solution is now complete since all variables are known.

\subsection{Thin shell}

To match the interior to the exterior solution we need the junction conditions for the generalized hybrid metric-Palatini gravity.  The deduction of the junction equations has been performed in chapter \ref{chapter:chapter5}.  There are seven junction conditions in this theory that must be respected in order to perform the matching, two of which imply the existence of a thin shell of matter at the junction radius $r_\Sigma$. For the reader's convenience, we shall rewrite these conditions here, which are
\begin{eqnarray}
&&\left[h_{\alpha\beta}\right]=0,\label{wormjchab}\\
&&\left[K\right]=0,\label{wormjck}\\
&&\left[\varphi\right]=0,\label{wormjcphi}\\
&&\left[\psi\right]=0,\label{wormjcpsi}\\
&&n^a\left[\partial_a\psi\right]=0, \label{wormjcdpsi}\\
&&n^a\left[\partial_a\varphi\right]=\frac{\kappa^2}{3}S,\label{wormjcdphi}\\
&&S_\alpha^\beta-\frac{1}{3}\delta_\alpha^\beta S=-\frac{\left(\varphi_\Sigma-\psi_\Sigma\right)}{\kappa^2} \left[K_\alpha^\beta\right],\label{wormjcsab}
\end{eqnarray}
where $h_{\alpha\beta}$ is the induced metric at the junction hypersurface $\Sigma$, with Greek indeces standing for $0$ and $2,3$, the brackets $\left[X\right]$ denote the jump of a quantity $X$ across $\Sigma$, $n^a$ is the unit normal vector to $\Sigma$, $K=K_\alpha^\alpha$ is the trace of the extrinsic curvature $K_{\alpha\beta}$ of the surface $\Sigma$, $S=S_\alpha^\alpha$ is the trace of the stress-energy tensor $S_{\alpha\beta}$ of the thin shell, $\delta_\alpha^\beta$ is the Kronecker delta, and the subscripts $\Sigma$ indicate the value computed at the hypersurface radius.	

Now, the matter NEC given in Eqs.\eqref{wormnecpr} and \eqref{wormnecpt} should also apply to the shell, since we want this condition to be valid throughout the whole spacetime. Let us write the stress-energy tensor of the thin shell as $S_\alpha^\beta=\text{diag}\left(-\sigma,p,p\right)$, such that $S=-\sigma+2p$, where $\sigma$ is the surface energy density and $p$ is the transverse pressure of the thin shell. When applied for the particular case of the thin shell the matter NEC becomes
\begin{equation}\label{wormnecshell}
\sigma+p\geq0
\end{equation}

The condition Eq.\eqref{wormjchab} gives, on using Eqs.\eqref{wormredshift1} and \eqref{wormredshift2}, that we must choose
\begin{equation}\label{wormrelredshift}
e^{\zeta_e}=\frac{e^{\zeta_0}}{1-\frac{2M}{r_\Sigma}-\frac{V_0\left(\varphi_e-\psi_e\right) r_\Sigma^2}{6}}.
\end{equation}
This factor $e^{\zeta_e}$ has been chosen to guarantee that the time coordinate $t$ for the interior region, Eq.\eqref{metricwormhole}, is the same as the coordinate $t$ for the exterior, Eq.\eqref{metrickottler}. The angular part of the metrics Eq.~\eqref{metricwormhole} and Eq.\eqref{metrickottler} are continuous. Thus, the line element at the surface $\Sigma$ and outside it is
\begin{equation}\label{metrickottler2}
ds^2=-\left(\frac{1-\frac{2M}{r}-\frac{V_0\left(\varphi_e-\psi_e\right) r^2}{6}}{1-\frac{2M}{r_\Sigma}-\frac{V_0\left(\varphi_e-\psi_e\right)r_\Sigma^2}{6}}\right)e^{\zeta_0}dt^2+\left(1-\frac{2M}{r}-\frac{V_0\left(\varphi_e-\psi_e\right)r^2}{6}\right)^{-1}dr^2+r^2d\Omega^2,
\end{equation}\label{kottler2}
where we are using that the $\varphi$ and $\psi$ are continous at the matching.

The conditions Eq.\eqref{wormjcphi} and \eqref{wormjcpsi} mean that the scalar fields must be given by 
\begin{equation}\label{wormsolphi2}
\varphi_e=\varphi_\Sigma=\psi_0-\frac{r_0^2}{r_\Sigma^4V_0},
\end{equation}
\begin{equation}\label{wormsolpsi2}
\psi_e=\psi_\Sigma=\psi_0,
\end{equation}
where we have used the solutions for the scalar fields given in Eqs.\eqref{wormsolphi1} and \eqref{wormsolpsi1}. Whenever it occurs we keep the notation $\varphi_e$. The condition Eq.\eqref{wormjcdpsi} is automatically satisfied since the scalar field $\psi$ is constant throught the whole spacetime. 

Since we have choosen the scalar fields to be constant in the exterior, then Eq.\eqref{wormjcdphi} features only a contribution from the interior derivative of $\varphi$ and the resultant condition is
\begin{equation}\label{wormtshell}
\frac{4r_0^2}{r_\Sigma^5V_0}=-\frac{\kappa^2}{3}S.
\end{equation}

The condition Eq.~\eqref{wormjcsab} can be used to detemine the surface energy density $\sigma$ and the surface transverse pressure $p$. Note that $\sigma$ is given by $S_0^0=-\sigma$ and $p$  is given by $S_1^1=S_2^2=p$. Then, from Eq.\eqref{wormjcsab} we have 
\begin{equation}\label{wormshellrho}
\sigma=\frac{4r_0^2}{\kappa^2V_0r_\Sigma^5}\left(1-\frac14\, r_\Sigma\left[K_0^0\right]\right),
\end{equation}
\begin{equation}\label{wormshellp}
p=-\frac{4r_0^2}{\kappa^2V_0r_\Sigma^5}\left(1+\frac18\, r_\Sigma\left[K_0^0\right]\right),
\end{equation}
where $\left[K_0^0\right]$ can be obtained from the interior solution in Eq.\eqref{metricwormhole}, the exterior solution in Eq.\eqref{metrickottler2}, and the conditions from Eqs.\eqref{wormsolphi2} to \eqref{wormtshell}, which becomes
\begin{equation}\label{wormkab}
\left[K_0^0\right]=\frac{r_0\zeta_0}{2r_\Sigma^2}\sqrt{1-\frac{r_0^2}{r_\Sigma^2}}+\frac{\frac{r_0^2}{r_\Sigma}-6M}{6r_\Sigma^2\sqrt{1-\frac{2M}{r_\Sigma}+\frac{r_0^2}{6r_\Sigma^2}}}
\end{equation}

Finally, the condition Eq.~\eqref{wormjck} means that the matching has to be performed at the radius $r_\Sigma$ such that the jump in the trace of the extrinsic curvature is zero. Upon using Eqs.\eqref{wormsolphi2} to \eqref{wormtshell}, we obtain 
\begin{equation}\label{wormk}
\left[K\right]=\frac{1}{r_\Sigma}\left[\frac{r_0\zeta_0}{2r_\Sigma}\sqrt{1-\left(\frac{r_0}{r_\Sigma}\right)^2}-\sqrt{1-\left(\frac{r_0}{r_\Sigma}\right)^4}-\sqrt{1-\left(\frac{r_0}{r_\Sigma}\right)^5}+\frac{2-\frac{3M}{r_\Sigma}+\frac{r_0^2}{2r_\Sigma^2}}{\sqrt{1-\frac{2M}{r_\Sigma}+\frac{r_0^2}{6r_\Sigma^2}}}\right]=0.
\end{equation}

The condition Eq.~\eqref{wormnecshell} is the matter NEC that should be valid on the shell. In this way
we can keep the validity of the matter NEC throughout the whole spacetime. Since the relevant quantity is $\sigma+p$, we use Eqs.\eqref{wormshellrho} and \eqref{wormshellp} to write
\begin{equation}
\sigma+p=-\frac{3r_0^2}{2\kappa^2V_0r_\Sigma^4}\left[K_0^0\right].
\end{equation}

Finding a combination of parameters that fulfills all the junction conditions plus the matter NEC everywhere is a problem that requires some fine-tuning. We now turn to this.

\section{Full wormhole solution}

Before proceeding, let us note that the solution scales with the exterior mass $M$ appearing in
Eq.\eqref{metrickottler} and so all quantities can be normalized to it. The parameter $V_0$, as we already argued, must be negative, $V_0<0$, so that the NEC is verified at the throat and its vicinity. In this case  the WEC is also verified at the throat and its vicinity.

The junction condition $\left[K\right]=0$, see Eq.\eqref{wormk}, can be achieved only for certain values of the radius $r\equiv r_\Sigma$.  We certainly need to guarantee that the radius $r_\Sigma$ at which $\left[K\right]=0$ happens is inside the region $r<\sqrt{11/10}\ r_0$, so that the interior solution does not violate the NEC. Manipulation of the parameters of the interior solution shows that these conditions can be achieved by changing the parameter $r_0$.

Once we have set values for $r_0$ and $r_\Sigma$, Eq.\eqref{wormk} automatically sets the value of the parameter $\zeta_0$ and hence, Eq.\eqref{wormkab} sets the value of $\left[K_0^0\right]$. Then, by Eqs.\eqref{wormshellrho} and \eqref{wormshellp}, we see that choosing a value of the parameter $V_0$ determines the values of $\sigma$ and $p$. However, since we already argued that $V_0$ must be negative in order for the NEC to be verified at the throat, then the signs of $\sigma$ and $p$ are determined even without specifying an exact value for $V_0$. Moreover, since $V_0<0$, the NEC at the shell, Eq.\eqref{wormnecshell}, is satisfied if $\left[K_0^0\right]\geq0$.

Having all the parameters determined, one has in general to verify that $r_\Sigma$ is greater that the gravitational radius $r_g$ of the solution. If this were not the case then there would be a horizon and the solution would be invalid. In our solution, one has $r_g=2M\left[1+2V_0(\varphi_e-\psi_e)M^2/3\right]$, see Eq.\eqref{metrickottler}. However, using Eqs.\eqref{wormsolphi2} and \eqref{wormsolpsi2}, one can write the gravitational radius as 
\begin{equation}
r_g=2M\left(1-\frac{2M^2r_0^2}{3r_\Sigma^4}\right),
\end{equation}
which is smaller than $r_\Sigma$ for any value of $r_\Sigma$ between $r_0$ and $\sqrt{\frac{11}{10}}r_0$. This implies that, in the range of solutions that we are interested in, this step is automatically satisfied.

We are now in a position to specify a set of parameters for which the matter NEC is obeyed in the whole spacetime, i.e., in the interior,  on the shell, and in the exterior, the latter being a trivial case since there is no matter in this region. We give an example and consider a wormhole with mass $M$. Note that the conclusions are the same for any combination of $r_0$ and $r_\Sigma$ within the allowed region for these two parameters. A concrete convenient example is to have a  wormhole interior region for which the radius $r_0$ of the throat has the value $r_0=2M\sqrt{10/11}$. Then we can set $r_\Sigma=2M$ and have the matter NEC satisfied in the interior. With these values of $r_0$ and $r_\Sigma$, Eq.\eqref{wormk} sets the value of $\zeta_0$ to $\zeta_0=-10.96$. From Eq.\eqref{wormkab} we get $\left[K_0^0\right]= 0.049$, and using Eqs.\eqref{wormshellrho} and \eqref{wormshellp} we compute the values of the stress-energy tensor at the shell, which become $\kappa^2\sigma= 0.44/V_0$ and $\kappa^2 p= -0.46/V_0$, respectively, and hence $\kappa^2\left(\sigma+p\right)= -0.02/V_0$. Since $V_0<0$ we see that $\sigma+p>0$ and thus the NEC is satisfied at the shell. However, we have $\sigma<0$, and the  WEC does not hold on the shell but holds everywhere else. 

\begin{figure}
\centering 
\begin{subfigure}{0.48\textwidth}
\includegraphics[scale=0.55]{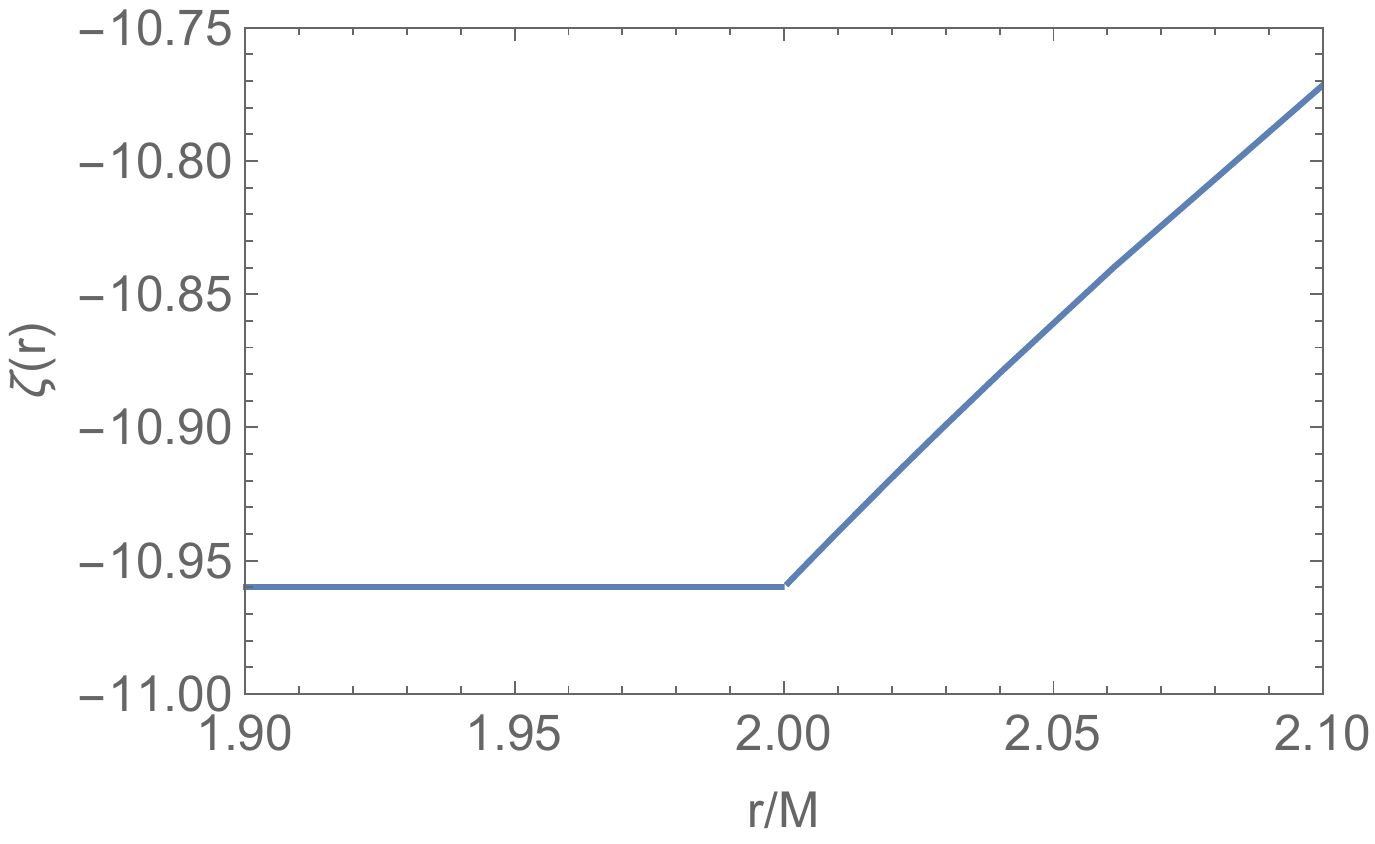}
\caption{Redshift function $\zeta\left(r\right)$}
\label{fig:wormredshift2}
\end{subfigure}
\ \ \ \ \ 
\begin{subfigure}{0.48\textwidth}
\includegraphics[scale=0.55]{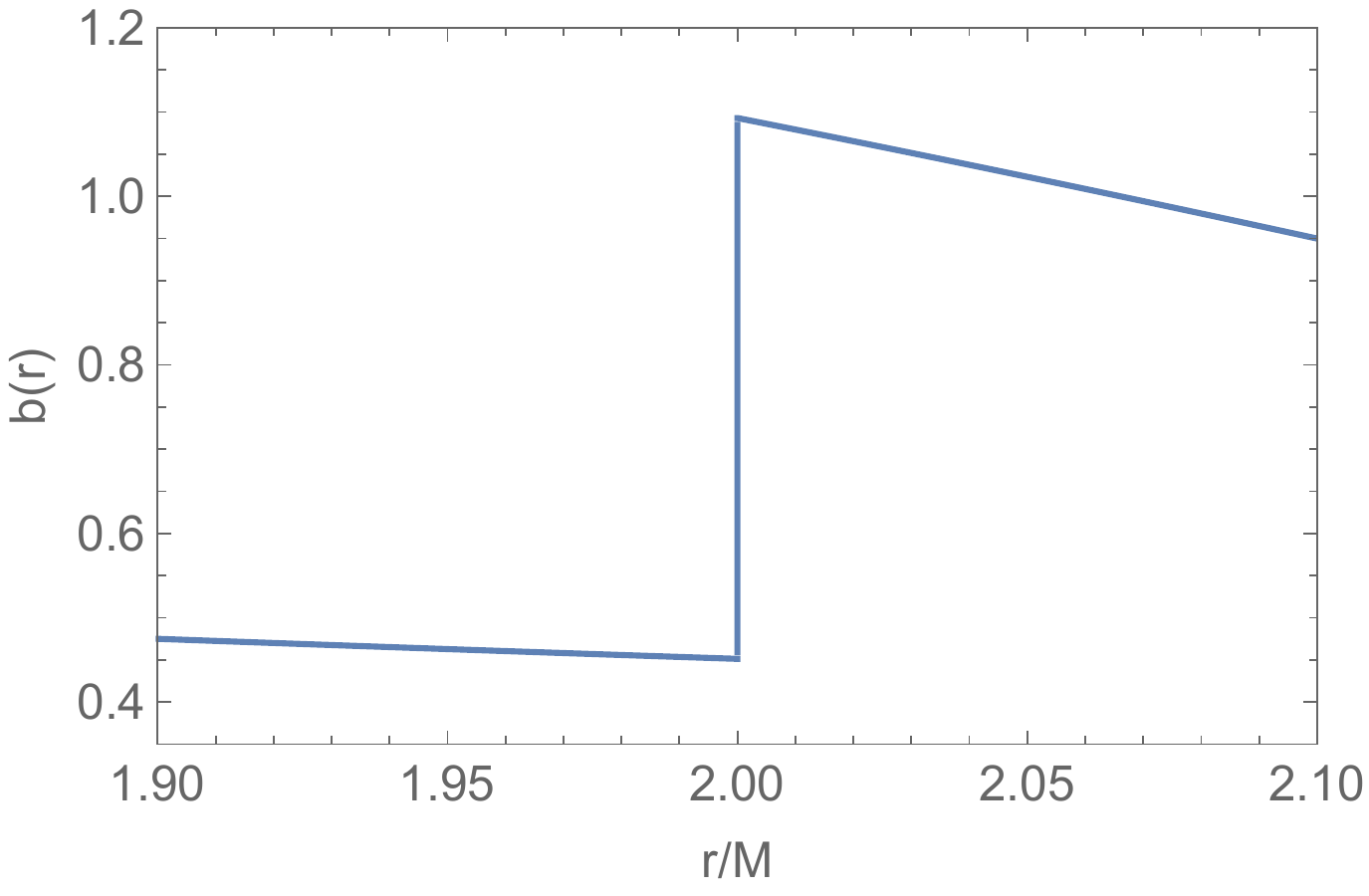}
\caption{Shape function $b\left(r\right)$}
\label{fig:wormshape2}
\end{subfigure}
\ \\
\ \\
\ \\
\begin{subfigure}{0.48\textwidth}
\includegraphics[scale=0.55]{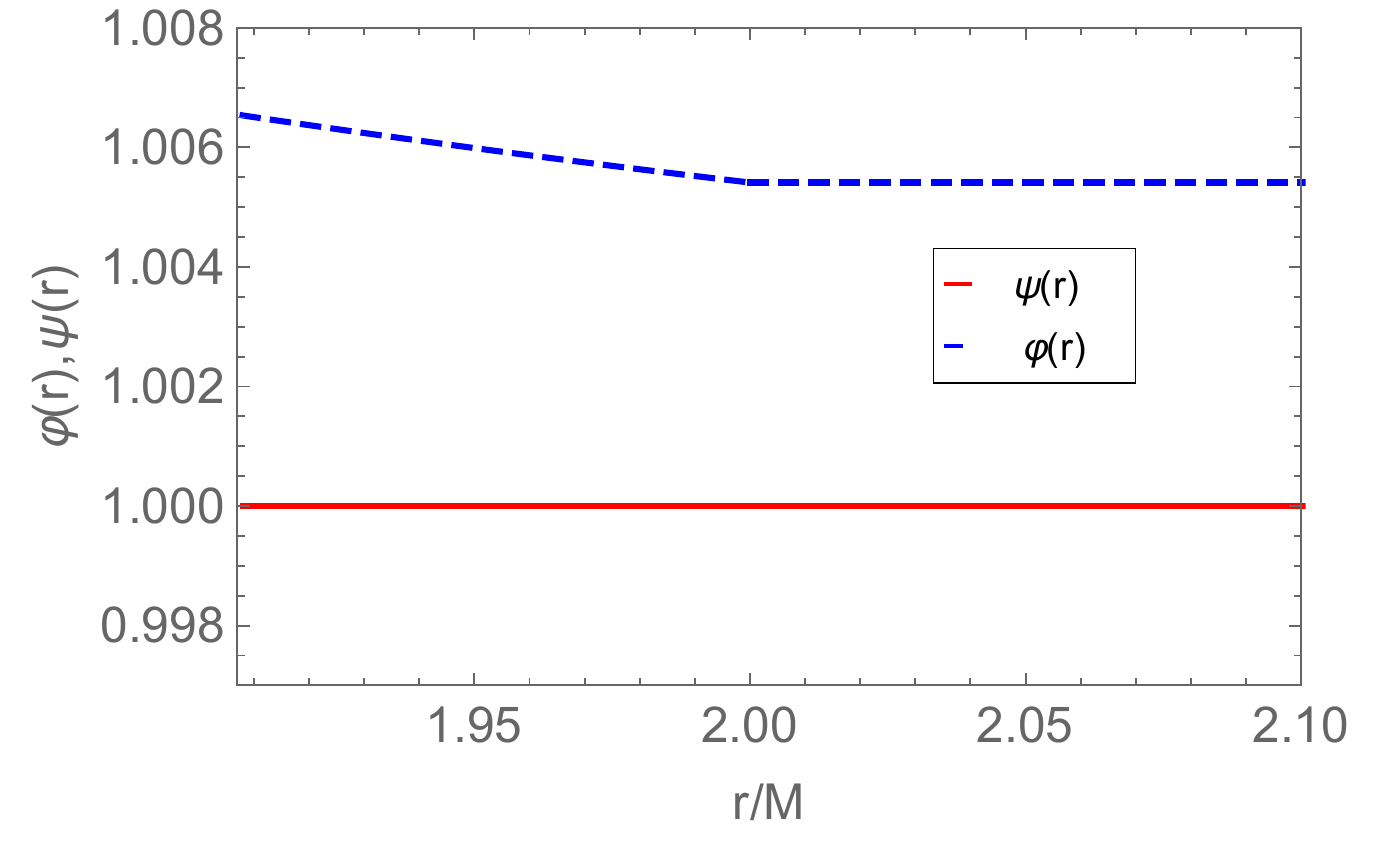}
\caption{Scalar fields $\varphi$ and $\psi(t)$}
\label{fig:wormfields2}
\end{subfigure}
\ \ \ \ \
\begin{subfigure}{0.48\textwidth}
\includegraphics[scale=0.55]{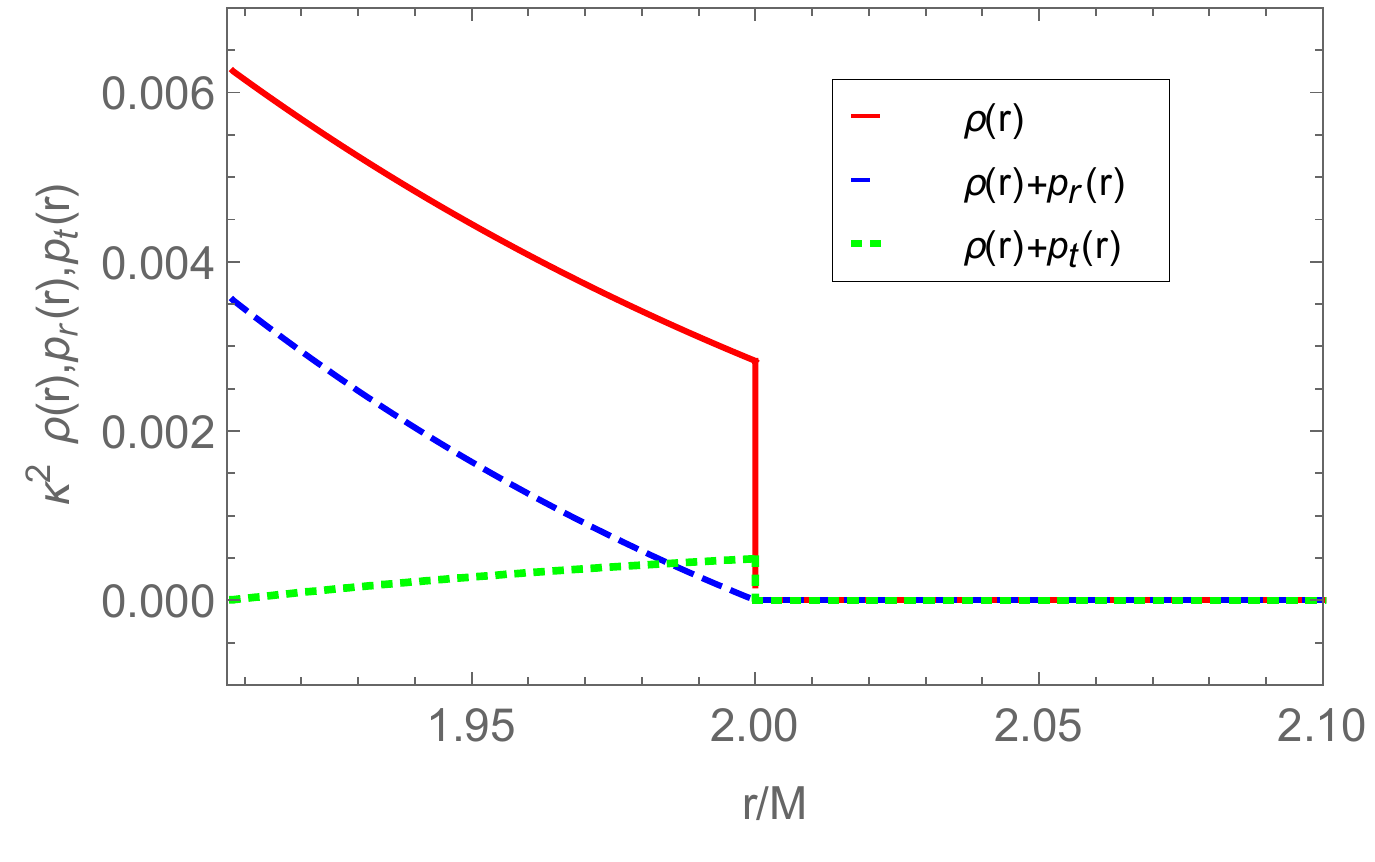}
\caption{Matter fields $\rho\left(r\right)$, $p_r\left(r\right)$ and $p_t\left(r\right)$}
\label{fig:wormmatter2}
\end{subfigure}
\caption{Metric fields $\zeta(r)$ and $b(r)$, scalar fields $\varphi(r)$ and $\psi(r)$, and matter fields
$\rho$, $p_r$ and $p_t$, with $\zeta_0=-10.96$, $M=1$, $r_0=2\sqrt{10/11}=1.90693$, $\psi_0=1$, $\psi_1=0$, $V_0=-42$, for the full wormhole solution}
\label{fig:wormsoltot}
\end{figure}

\begin{figure}
\centering
\includegraphics[scale=0.45]{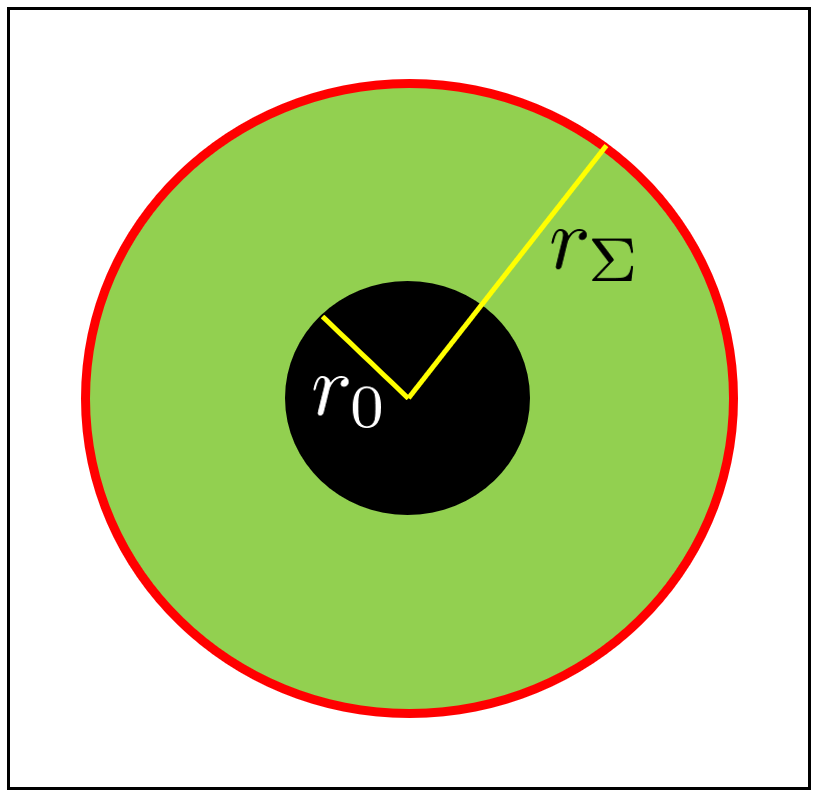}
\caption{Pictorical diagram of the full wormhole solutions, not to scale. The colors represent the throat (black), the perfect fluid (green), the thin shell (red), and vacuum (white)}
\label{fig:wormhole}
\end{figure}

This completes our full wormhole solution. Fig.\ref{fig:wormsoltot} displays a solution for this specific choice of the parameters. The metric fields $\zeta(r)$ and $b(r)$ are asymptotically AdS, and a thin shell of matter is perceptibly present at the matching surface $r_\Sigma=2$ (more properly at $r_\Sigma=2M$ in our solution, and here we put $M=1$), thus outside the gravitational radius of the solution.  Both the scalar fields are finite for all $r$. Relative to the NEC, note that $\rho+p_r$ is positive at $r=r_0$ and up to $r=2$ and for $r>2$ is zero. The quantity $\rho+p_t$ is zero at $r=r_0$, positive for every other value of $r<2$, and for $r>2$ is zero again. At the shell $r=2$. This wormhole solution obeys the matter NEC everywhere in conformity with our aim. A pictorical diagram is also provided in Fig.\ref{fig:wormhole}.

Other wormhole solutions in this generalized hybrid theory with other choices of parameters, within the
ranges stated above, that obey the matter NEC everywhere can be found along the lines we have presented. 

\section{Conclusions}

In this chapter, we found traversable asymptotically AdS wormhole solutions that obey the NEC everywhere in the scalar-tensor representation of the generalized hybrid metric-Palatini gravity theory, so there is no need for exotic matter. The solution was obtained solely via the use of analytical methods, and thus it is the simplest possible solution one can find. More complicated solutions could be obtained by considering more general forms of the redshift function, the shape function, and the potential, and solving the system numerically.

The interior wormhole solution obtained verifies the NEC near and at the throat of the wormhole.  The matching of the interior solution to an exterior vacuum solution yields a thin shell respecting also the NEC. Finding a combination of parameters that allow for the interior and the exterior solution to be
matched without violating the NEC in the interior solution and at the thin shell is a problem that requires fine-tuning, but can be acomplished. We presented a specific combination of parameters with which it is possible to build the full wormhole solution and found that it is asymptotically AdS. Most of the work that has been done in  hybrid metric-Palatini gravity theories has been aimed to find a solution
where the NEC is satisfied solely at the throat of the wormhole. Within these teories our solution is the first where the NEC is verified for the entire spacetime.

There are two main interesting conclusions. On one hand, the existence of these solutions is in agreement with the idea that traversable wormholes supported by extra fundamental gravitational fields, here in
the form of scalar fields, can exist without the need of exotic matter. On the other hand, the rather contrived construction necessary to build a spacetime complete wormhole solution may indicate that
there are not many such solutions around in this class of theories, at least in what comes to analytical solutions.

It is also worth to mention that in this work we have never forced the conservation equation for the stress-energy tensor $\nabla_aT^{ab}=0$ to be verified. In fact, if one verifies if this equation is satisfied \textit{a posteriori}, one gets to the conclusion that it is not. However, since $\nabla_aG^{ab}=0$ by construction, then $\nabla_aT^{ab}_{(eff)}=0$ which implies that $\nabla_aT^{ab}=-\nabla_aT^{ab}_{\varphi\psi}$, where $T_{ab}^{\varphi\psi}$ is the contribution to the effective stress-energy tensor of the scalar-field dependent terms. This implies that the perfect fluid and the scalar fields are interacting with each other and, more specifically, there is energy being transfered from the perfect fluids to the scalar-fields and vice-versa in such a way guaranteeing that, despite $T_{ab}$ not being conserved, $T_{ab}^{(eff)}$ is conserved and we are not violating the laws of energy conservation.  
\cleardoublepage

\chapter{Black-hole perturbations in the generalized hybrid metric-Palatini gravity}
\label{chapter:chapter7}

In this chapter, we study black-hole perturbations in the geometrical representation of the generalized hybrid metric-Palatini gravity. We start by verifying which are the most general conditions on the function $f\left(R,\mathcal R\right)$ that allow for the Kerr solution in GR to be also a solution of this theory. We then perturb the metric tensor, which consequently imposes a perturbation in both the Ricci and Palatini scalar curvatures. To first order in the perturbations, the equations of motion, namely the field equations and the equation that relates the Ricci and the Palatini curvatures can be rewritten in terms of a 4th order wave equation for the perturbation $\delta R$ which can be factorized into two 2nd order massive wave equations for the same variable. The usual ansatz and separation methods are applied and stability bounds on the effective mass of the Ricci perturbation are obtained. These stability regimes are studied case by case and specific forms of the function $f\left(R,\mathcal R\right)$ that allow for stable solutions to exist are obtained.

\section{Introduction}

In an astrophysical context, black holes are known to arise from the collapse of very massive stars \cite{kippenhahn1}. These resulting astrophysical black holes are known to be rotating, due to angular momentum conservation from the initial star, and neutral due to the acretion of matter \cite{gibbons1} and the surrounding plasma \cite{blandford1}. In General Relativity (GR), one can describe these objects using the Kerr metric \cite{kerr1}. To study the stability of these objects, one often uses the methods of perturbation theory, in which small deviations (perturbations) to the background metric are considered and their evolution in time is studied. Since the deviations are arbitrarily small, it is usual to keep only the first order perturbative terms in the equations. Depending on the obtained behavior for the perturbation over time, one can conclude if the background metric is stable (the perturbation fades away) or unstable (the perturbation grows exponentially).
As physically realistic objects, Kerr black holes must be stable against exterior perturbations. These perturbations have been studied in the context of GR for scalar, vectorial and tensorial perturbations. For massless perturbations, the Kerr black hole was shown to be stable. However, for massive scalar, vectorial and tensor perturbations, the confinement of superradiant modes will lead to an amplification of the perturbation ad infinitum, giving rise to instabilities such as the black hole bomb \cite{press1}. The study of these instabilities allows one to impose constraints on the masses of ultralight degrees of freedom such as the photon \cite{pani1} and the graviton \cite{brito1}, constraints on dark matter candidates\cite{arvanitaki1}, and also to study the existence of black hole solutions that violate the so-called no-hair theorems \cite{herdeiro1}, being an evidence of the importance of these studies not only in an astrophysical context but also with implications to particle physics and high energy physics.

Black hole perturbations have also been studied in a wide variety of modified theories of gravity. For instance, the stability of the $f(R)$ (Schwarzschild) black hole was investigated in the scalar-tensor theory by introducing two auxiliary scalars \cite{myung1}. It was shown that the linearized curvature scalar becomes a scalaron, so that the linearized equations are second order, which are the same equations for the massive Brans-Dicke theory. Furthermore, the authors proved that the $f(R)$ black hole solution is stable against the external perturbations if the scalaron does not have a tachyonic mass. The authors extended their analysis to the stability of the $f(R)$-AdS (Schwarzschild-AdS) black hole obtained from $f(R)$ gravity \cite{moon1}, and concluded that stable solutions against the external perturbations exist if the scalaron is once again free from the tachyon. The stability of the Schwarzschild black hole was also analysed in several extensions of $f(R)$ gravity \cite{moon2,myung2,myung3}.

Relative to the Kerr solution, it has been shown to be unstable in the case of $f(R)$ gravity \cite{myung4,myung5}, due to the fact that the perturbation equation for the massive spin-0 graviton in this theory, or equivalently the perturbed Ricci scalar after imposing the Lorentz gauge, is analogous to the Klein-Gordon equation for a massive scalar field in GR which has been intensively studied. The objective of this chapter is to show that the Kerr solution exists in the generalized hybrid metric-Palatini gravity and to show that for certain choices of the function $f\left(R,\mathcal R\right)$ it can be stable.

\section{Perturbations of solutions with $R_{ab}=0$}

In this section, we are interested in obtaining the equations that describe metric perturbations in the generalized hybrid metric-Palatini gravity. More specifically, we are interested in atrophysical black-hole solutions, which are very well described by the Kerr metric, a solution to the EFE wirh $R_{ab}=0$. We start by computing a general form of the function $f\left(R,\mathcal R\right)$ for which solutions with $R_{ab}=0$ in GR are also solutions of the generalized hybrid metric-Palatini theory and afterwards we compute the linearized equations of motion fo the massive scalar degree of freedom of the theory.

\subsection{General conditions}

In this section we assume a general form for the function $f\left(R,\mathcal R\right)$ that guarantees that general relativity solutions with $R_{ab}=0$, such as the Schwarzschild and Kerr solutions, are also solutions of the GHMP theory. To do so, let us assume at first two very general conditions on the function $f$: First, consider that the function $f$ is analytical in both $R$ and $\mathcal R$ around a point $\left\{R_0,\mathcal R_0\right\}$, where $R_0$ and $\mathcal R_0$ are constants, and therefore can be expanded in a Taylor series of the form
\begin{equation}\label{bhfunction}
f\left(R,\mathcal R\right)=\sum_{\left\{n,m\right\}=0}^\infty\frac{\partial^{\left(n+m\right)}f\left(R_0,\mathcal R_0\right)}{\partial^n R\ \partial^m\mathcal R}\frac{\left(R-R_0\right)^n}{n!}\frac{\left(\mathcal R-\mathcal R_0\right)^m}{m!}.
\end{equation}
The second assumption consists in imposing that the function $f$ has a zero in the point where we perform the Taylor's series expansion, that is $f\left(R_0,\mathcal R_0\right)=0$. In general, if both $R$ and $\mathcal R$ happen to be constants, this can be achieved by performing the series expansion around the values of these constants. 

Let us now show that for a function $f$ that satisfies these conditions, then it is always possible for a general relativity solution with $R=0$ to be also a solution in the GHMP gravity. To start with, denote generically $X$ as meaning $R$, $\cal R$, or any combination of the form $R\cal R$, and so on, and
$f_X$ the derivative of $f$ in relation to $X$. Then, the derivatives of the functions $f_X$ with respect to the coordinates $x^a$ can be written as functions of the derivatives of $R$ and $\mathcal R$ by
making use of the chain rule, from which we obtain 
\begin{equation}\label{bhderiv1}
\partial_a f_X=f_{XR}\partial_a R + f_{X\mathcal R}\partial_a\mathcal R,
\end{equation}
for the partial derivatives, which also allow us to write the terms $\nabla_a\nabla_b f_X$ and $\Box f_X$ as functions of $R$ and $\mathcal R$ as
\begin{equation}\label{bhderiv2}
\nabla_a\nabla_bf_X=f_{XRR}\nabla_aR\nabla_bR+f_{X\mathcal R\mathcal R}\nabla_a\mathcal R\nabla_b\mathcal R +2f_{XR\mathcal R}\nabla_{(a}R\nabla_{b)}\mathcal R+f_{XR}\nabla_a\nabla_bR+f_{X\mathcal R}\nabla_a\nabla_b\mathcal R,
\end{equation}
where indexes in parenthesis are symmetrized, and $\Box f_X=g^{ab}\nabla_a\nabla_bf_X$. Now, let us first use Eq. \eqref{ghmpgrelricten} to eliminate the term $\mathcal R_{ab}$ in Eq. \eqref{ghmpgfield}, from which we get
\begin{equation}\label{bhfield}
\left(f_R+f_\mathcal R\right)R_{ab}-\left(\nabla_a\nabla_b+\frac{1}{2}g_{ab}\Box\right)f_\mathcal R+\frac{3}{2f_\mathcal R}\partial_a f_\mathcal R\partial_b f_\mathcal R-\frac{1}{2}g_{ab}f-\left(\nabla_a\nabla_b-g_{ab}\Box\right)f_R=\kappa^2 T_{ab}.
\end{equation}
With the assumption $R_{ab}=0$, one has $T_{ab}=0$ and $R=0$. Expansions above should be inserted here but we shall not write the results due to their lengthy character. Using $f\left(R_0,\mathcal R_0\right)=0$, Eq.\eqref{bhfield} becomes a partial differential equation for $\mathcal R$ that in principle can not be solved until we choose a particular form for the function $f$. However, notice that if $\mathcal R=\mathcal R_0$, where $\mathcal R_0$ is a constant, then Eq.\eqref{bhfield} is identically zero upon using Eqs.\eqref{bhderiv1} and \eqref{bhderiv2} with the assumption in Eq.\eqref{bhfunction} guaranteeing that all the terms in Eq.\eqref{bhfield} are finite at $R=0$ and $\mathcal R=\mathcal R_0$. Using then Eq.\eqref{ghmpgrelricsca} assuming $R=0$ and using $\mathcal R=\mathcal R_0$ from the previous equation, we obtain directly that $\mathcal R_0=0$. Thus, solutions of GR with $R_{ab}=0$ are also solutions of the generalized hybrid metric-Palatini gravity with $\mathcal R=0$. This result is coherent with the fact that we have chosen a specific value for both $R$ and $\mathcal R$ in the previous paragraph, which implies that the conformal factor between the metrics $g_{ab}$ and $h_{ab}$, given by $f_\mathcal R$, is constant, the two metrics thus have the same Ricci tensor, and so $\mathcal R=g^{ab}\mathcal R_{ab}=g^{ab}R_{ab}=R$. Note that the field equation and the relation between the scalar curvatures are both partial differential equations, and therefore their solutions are not unique. We choose this particular solution because it allows us to perform the following analysis without specifying a form for the function $f$ besides the two assumptions already made. 

\subsection{Linearized equations}

We now consider a perturbation $\delta g_{ab}$ in the background metric $\bar g_{aB}$, such that the new metric can be written as
\begin{equation}\label{bhperturbgab}
g_{ab}=\bar{g}_{ab}+\epsilon\delta g_{ab}, 
\end{equation}
where $\epsilon$ is a small parameter, and the bar represents the unperturbed quantities. This perturbation in the metric induces a perturbation in the Ricci tensor of the form $R_{ab}=\bar{R}_{ab}+\epsilon\delta R_{ab}$ and Ricci scalar as $R=\bar{R}+\epsilon\delta R$. Due to the fact that the metric $h_{ab}$ is conformally related to $g_{ab}$, it also induces a perturbation in the Palatini curvature tensor as $\mathcal R_{ab}=\bar{\mathcal R}_{ab}+\epsilon\delta\mathcal R_{ab}$ and scalar as $\mathcal R=\bar{\mathcal R}+\epsilon\delta\mathcal R$. The perturbation $\delta R_{ab}$ and $\delta R$ can be written in terms of $\delta g_{ab}$ and $\delta g$ as
\begin{equation}\label{bhperturbrab}
\delta R_{ab}=\frac{1}{2}\left(2\nabla^c\nabla_{(a}\delta g_{b)c}-\Box\delta g_{ab}-\nabla_a\nabla_b\delta  g\right)
\end{equation}
\begin{equation}\label{bhperturbric}
\delta R=\nabla_a\nabla_b\delta g^{ab}-\Box\delta g=-\frac{1}{2}\Box\delta g.
\end{equation}
Since the unperturbed quantities $\bar{R}$ and $\bar{\mathcal R}$ vanish in the solutions we are considering, then the expansion of the function $f$ given in Eq. \eqref{bhfunction} and its derivatives $f_X$, to first order in $\epsilon$, yields
\begin{equation}\label{bhperturbfunction}
f_X=\bar f_X+\epsilon \left(\bar f_{XR}\delta R+ \bar f_{X\mathcal R}\delta \mathcal R\right).
\end{equation}
This equation can also be deduced directly from the expansion in Eq.\eqref{bhfunction} by $f\left(R,\mathcal R\right)=f\left(\bar R+\delta R,\bar{\mathcal R}+\delta\mathcal R\right)$. Note that the barred functions are constants because they represent the coefficients of the Taylor expansion of the unperturbed function $f$, and therefore they can be taken out of the derivative operators unchanged, e.g. $\partial_a f=\epsilon\left(\bar f_R \partial_a \delta R+\bar f_\mathcal R\partial_a\delta\mathcal R\right)$. To simplify the notation, from now on we shall drop the bars and any term containing the function $f$ is to be considered as a constant, unless stated otherwise.

The equations of  motion, Eqs.~\eqref{ghmpgfield} and \eqref{ghmpgrelricsca} then become, in vacuum and to first order in $\epsilon$:
\begin{equation}\label{bhperturbeom1}
f_R\delta R_{ab}+f_{\mathcal R}\delta\mathcal R_{ab}-\frac{1}{2}\bar g_{ab}\left(f_R\delta R+f_{\mathcal R}\delta \mathcal R\right)-\left(\nabla_a\nabla_b-g_{ab}\Box\right)\left(f_{R R}\delta R+f_{R \mathcal R}\delta\mathcal R\right)=0,
\end{equation}
\begin{equation}\label{bhperturbeom2}
\delta\mathcal R_{ab}=\delta R_{ab}-\frac{1}{f_{\mathcal R}}\left(\nabla_a\nabla_b+\frac{1}{2}\bar g_{ab}\Box\right)\left(f_{\mathcal R R}\delta R+f_{\mathcal R \mathcal R}\delta\mathcal R\right),
\end{equation}
respectively. These equations are fourth-order equations in the metric perturbation $\delta g_{ab}$, which are very difficult to handle. However, a simpler system of equations for $\delta R$ and $\delta\mathcal R$ can be obtained by taking the trace of Eqs.~\eqref{bhperturbeom1} and \eqref{bhperturbeom2}. This approach is well motivated: similarly to the $f\left(R\right)$ theories of gravity, the generalized hybrid metric-Palatini theory presents three degrees of freedom without ghosts, two for massless spin-2 gravitons, and one for a massive spin-0 scalar graviton (one might think that there are two scalar degrees of freedom corresponding to both $f_R$ and $f_{\mathcal R}$, but these actually correspond to the same degree of freedom due to their conformal relation expressed by the trace of Eq.~\eqref{ghmpgrelricsca}). The scalar degree of freedom is described by the trace $\delta g$. Using the Lorenz gauge, i.e., $\nabla_b\delta g^{ab}=(1/2)\nabla^a\delta g$ Eq.~\eqref{bhperturbric} turns into
\begin{equation}
\delta R=\nabla_a\nabla_b\delta g^{ab}-\Box\delta g=-\frac{1}{2}\Box\delta g.
\end{equation}.
So, under this gauge, the perturbation $\delta R$ is directly related to $\delta g$ which represents the massive spin-0 degree of freedom of the theory. We restrict ourselves to the study of the massive scalar degree of freedom of the theory by the analysis of the perturbation $\delta R$.  Thus, we will study stability against scalar mode perturbations. 

To obtain an equation for the perturbation in the Ricci scalar $\delta R$ we shall work with the trace of Eq.\eqref{ghmpgfield} and with Eq.\eqref{ghmpgrelricsca}. Writing these equations in terms of the expansions of the functions $f_X$ to first order in $\epsilon$ given by Eq. \eqref{bhperturbfunction} yields then
\begin{equation}\label{bhsystem1}
f_R\delta R+f_\mathcal R\delta\mathcal R-3f_{RR}\Box\delta R-3f_{R\mathcal R}\Box\delta\mathcal R=0,
\end{equation}
\begin{equation}\label{bhsystem2}
\delta\mathcal R=\delta R-\frac{3}{f_\mathcal R}\left(f_{\mathcal R\mathcal R}\Box\delta\mathcal R+f_{\mathcal R R}\Box\delta R\right),
\end{equation}
respectively, where we used $T_{ab}=0$ and $\bar f=f\left(0,0\right)=0$. Note that the perturbations $\delta R$ and $\delta\mathcal R$ cannot be equal. If they are, then one of the equations above immediately sets $f\left(R,\mathcal R\right)=f\left(R-\mathcal R\right)$, and thus the perturbations cancel completely in the other equation and we obtain an identity. This is not a feature of the first order expansion, for it can be shown that for any order in $\epsilon$ that we choose, if $f\left(R,\mathcal R\right)=f\left(R-\mathcal R\right)$ then the perturbations cancel identically in these two equations.

Both Eqs. \eqref{bhsystem1} and \eqref{bhsystem2} can be rewritten in the form $\left(\Box+a_1\right)\delta R=a_2\left(\Box+a_3\right)\delta\mathcal R$, where $a_i$ are constants that depend only on the values of $\bar f_X$ and that are different for both equations. To obtain an equation that depends only on $\delta R$, we proceed as follows: First, we solve Eq.\eqref{bhsystem2} with respect to $\Box\delta\mathcal R$ and we replace it into Eq.\eqref{bhsystem1} to obtain an equation of the form $\left(\Box+b_1\right)\delta R=b_2\delta\mathcal R$, where $b_i$ are constants; then we solve Eq.\eqref{bhsystem2} with respect to $\Box\delta R$ and insert the result into Eq.\eqref{bhsystem1} to obtain an equation of the form $\left(\Box+c_1\right)\delta \mathcal R=c_2\delta R$, where $c_i$ are constants; and finally we use the first of these two equations to replace the term depending on $\delta\mathcal R$ in the second equation. The resultant equation is
\begin{equation}\label{bhbox2ricci}
\Box^2\delta R+A\Box\delta R+B\delta R=0,
\end{equation}
where the constants $A$ and $B$ are given in terms of the functions $\bar f_X$ as
\begin{eqnarray}
A&=&\frac{f_Rf_{\mathcal R\mathcal R}-2f_\mathcal Rf_{\mathcal R R}-f_\mathcal Rf_{RR}}{3\left(f_{\mathcal R R}^2-f_{RR}f_{\mathcal R\mathcal R}\right)},\label{bhparA} \\
B&=&\frac{f_\mathcal R\left(f_\mathcal R+f_R\right)}{9\left(f_{\mathcal R R}^2-f_{RR}f_{\mathcal R\mathcal R}\right)}.\label{bhparB}
\end{eqnarray}
Note that Eq.\eqref{bhbox2ricci} is a 4th order equation on the perturbation $\delta R$, unlike what happens in $f\left(R\right)$ or with the Klein-Gordon equation for theories with a scalar field minimally coupled to GR, and therefore could be very difficult to solve. However, since $A$ and $B$ are constants, it is possible to factorize Eq.\eqref{bhbox2ricci} into
\begin{equation}\label{bhscalareq}
\left(\Box-\mu_+^2\right)\left(\Box-\mu_-^2\right)\delta R=0,
\end{equation}
where the constants $\mu_\pm^2$ can be expressed in terms of the constants $A$ and $B$ as
\begin{equation}\label{bhscalarmass}
\mu_\pm^2=-\frac{1}{2}\left(A\pm\sqrt{A^2-4B}\right).
\end{equation}
Note that due to the fact that $\mu_\pm^2$ are constants, the terms $\left(\Box-\mu_\pm^2\right)$ can commute in Eq.\eqref{bhscalareq}, and so we can reduce this equation to a set of two equations of the form $\left(\Box-\mu_\pm^2\right)\delta R=0$, which are of the form of a Klein-Gordon equation for a scalar field where the constants $\mu_\pm^2$ take the role of the field's mass. 

\section{Perturbations of Kerr BHs}

Let us now restrict our study to the Kerr metric in the generalized-hybrid metric-Palatini gravity. The Kerr metric is given by, in Boyer-Lindquist coordinates by
\begin{equation}\label{metrickerr}
ds^2=-\left(1-\frac{2Mr}{r\rho}\right)dt^2+\frac{\rho^2}{\Delta}dr^2+\rho^2d\theta^2-\frac{4Mra\sin^2\theta}{\rho^2}dtd\phi+\left(r^2+a^2+\frac{2Mra^2\sin^2\theta}{\rho^2}\right)d\phi^2,
\end{equation}
\begin{equation}
\Delta=r^2+a^2-2Mr,\ \ \ \ \  \rho^2=r^2+a^2\cos^2\theta,\ \ \ \ \  a=\frac{J}{M},
\end{equation}
where $M$ is the black hole mass and $J$ is the black hole angular momentum. Equations of the form of Eq.\eqref{bhscalareq} have been extensively studied and are known to be separable for the Schwarzschild and Kerr metrics.

\subsection{Separability and quasi-bound states}

To study the separability of the equations of motion, we first note that Eq.\eqref{bhscalareq} is a 4th order PDE for $\delta R$, and should therefore have four linearly independent solutions. Each of the equations $\left(\Box-\mu_\pm^2\right)\delta R=0$ has two solutions corresponding to ingoing and outgoing waves. To verify this, we use the usual procedure to deal with these kind of equations. We start by chosing an ansatz of the form
\begin{equation}\label{bhansatz}
\delta R=\psi\left(r\right)S\left(\theta\right)\exp\left(-i\omega t+im\phi\right), 
\end{equation}
where $\psi\left(r\right)$ is the radial wavefunction, $\omega$ is the wave angular frequency, $m$ is the azimuthal number, and $S\left(\theta\right)$ are the scalar spheroidal harmonics defined by
\begin{equation}\label{bhharmonics}
\left[\lambda-m^2+a^2\left(\omega^2-\mu_\pm^2\right)\cos^2\theta\right]\sin^2\theta S\left(\theta\right)+\sin\theta\partial_\theta\left[\sin\theta\partial_\theta S\left(\theta\right)\right]=0,
\end{equation}
where $\lambda=l\left(l+1\right)+\mathcal O\left(c\right)$, with $c=a^2\left(\omega^2-\mu_\pm^2\right)$ is the separability constant, and $l$ is the angular momentum number. Note that for the study of superradiant instabilities, one is interested in studying the existance of quasi bound-states (QBS), which occur for $\omega^2\sim\mu_\pm^2$, and therefore $c\sim 0$ and the spheroidal harmonics can be approximated by the spherical harmonics, with constant of separation $\lambda=l\left(l+1\right)$. 

Using this ansatz, we can separate each of the factors $\left(\Box-\mu_\pm^2\right)\delta R=0$ into a radial and an angular equations. The angular equation is exactly Eq.\eqref{bhharmonics}. To find a more suitable way to write the radial equation, it is useful to redefine the radial wave function and the radial coordinate. Let us define the so-called tortoise coordinate $r_*$ and the new radial wavefunction $u\left(r\right)$ as
\begin{equation}\label{bhwave}
\frac{dr}{dr_*}=\frac{\Delta}{r^2+a^2},\ \ \ \ \  u\left(r\right)=\sqrt{r^2+a^2}\psi\left(r\right),
\end{equation}
so that the new radial equation can be written in the form of a wave equation in the presence of a potential barrier as
\begin{equation}
\frac{d^2u}{dr_*^2}+\left[\omega^2-V\left(r\right)\right]u=0,
\end{equation}
where the potential $V\left(r\right)$ is given by
\begin{equation}
V\left(r\right)=\frac{\Delta}{r^2+a^2}\left[\frac{\Delta+\Delta'r}{\left(r^2+a^2\right)^2}-\frac{3r^2\Delta}{\left(r^2+a^2\right)^3}+\frac{1}{r^2+a^2}\left(\mu_\pm^2r^2-\omega^2a^2+\frac{4Mram\omega}{\Delta}-\frac{m^2a^2}{\Delta}+\lambda\right)\right].
\end{equation}
Eq.\eqref{bhwave} admits two solutions, one ingoing wave and one outgoing wave. Since the potential is a very complicated function of the radial coordinate, we can not find an analytical solution that holds for the entire space. Therefore, this equation is usually solved numerically by imposing appropriate boundary conditions at the horizon, where $r=M+\sqrt{M^2-a^2}\equiv r_+$ and $r_*=-\infty$, and at infinity, where $r=r_*=+\infty$. At these boundaries, the potential takes the forms
\begin{equation}\label{bhpotential}
V\left(r_+\right)=\omega^2-\left(\omega-m\Omega\right)^2,\ \ \ \ \  V\left(\infty\right)=\mu_\pm^2,
\end{equation}
where $\Omega=\frac{a}{2Mr_+}$ is the angular speed of the event horizon. Since the horizon functions as a one directional membrane, we want our boundary condition at the horizon to be given by a purely ingoing wave. On the other hand, at infinity, we want our solution to decay exponentially to give rise to a quasi-bound state. In general, these boundary conditions will have the forms $u\left(r\right)=A_i e^{i\omega_ir_*}+B_ie^{-i\omega_ir_*}$, where the subscript $i$ can be either $h$ or $\infty$ for boundary conditions at the horizon or at infinity, respectively. Using Eq. \eqref{bhpotential} one verifies that $\omega_h=\omega-m\Omega$ and $\omega_\infty=\sqrt{\omega^2-\mu_\pm^2}$, which implies that for the bound states to form one needs not only $\omega^2\sim\mu_\pm^2$ but also $\mu_\pm^2>\omega^2$, and therefore our boundary conditions become
\begin{eqnarray}
u\left(r\to r_+\right)=B_he^{-i\left(\omega-m\Omega\right)r_*},\nonumber \\
u\left(r\to\infty\right)=A_\infty e^{-\sqrt{\mu_\pm^2-\omega^2}r_*},\label{bhboundary}
\end{eqnarray}
respectively.

Computing the quasi-bound states consists of integrating the radial Eq. \eqref{bhwave} subjected to the boundary conditions in Eq. \eqref{bhboundary} and computing the roots of $\omega$. These roots will be of the form $\omega=\omega_R+i\omega_I$. As can be seen from Eq. \eqref{bhansatz}, if $\omega_I<0$ the perturbation decays exponentially with time, but if $\omega_I>0$ the wavefunction grows exponentially and can no longer be considered a perturbation.  These modes have been extensively studied and so we shall skip this study here and jump directly into the stability analysis.

\subsection{Perturbations in the scalar-tensor representation}

The objective of this section is to show that one can perform the perturbative analysis in the equivalent scalar-tensor representation of the theory and that the perturbation equations and the results are the same. Consider the field equations in the scalar-tensor representation given by Eq.\eqref{ghmpgstfield} and the relation between $R_{ab}$ and $\mathcal R_{ab}$ given by Eq.\eqref{ghmpgrelricten}. Using Eq.\eqref{ghmpgrelricten} and its trace to cancel the terms $\mathcal R_{ab}$ and $\mathcal R$ in Eq.\eqref{ghmpgstfield}, and tracing the result, one verifies that one of the possible ways for a solution in GR with $R=0$ to be a solution for this representation of the g.h.m.P.gravity is to impose that $V=0$ and also that both scalar fields $\varphi$ and $\psi$ are constants. Note that the trace of the field equation is a PDE for $\varphi$ and $\psi$, like in it was a PDE for $\mathcal R$ in the previous sections, and therefore these solutions are not unique. We choose constant scalar fields as solutions because this is equivalent to setting $\mathcal R=\mathcal R_0$ for some constant $\mathcal R_0$, and we recover the results of the geometrical representation. Then, using $\psi=\psi_0$ for some constant $\psi_0$ in Eq.\eqref{ghmpgrelricsca} one verifies that $\mathcal R=R=0$, which is the same result we obtained before. The constraint $V=0$, for solutions with $\mathcal R=R=0$ is equivalent to the constraint $f\left(0,0\right)=0$ that we imposed in the previous sections. On the other hand, constraining $\varphi$ and $\psi$ to be constants is equivalent to constraining $f_R$ and $f_\mathcal R$ to be constants in the geometrical representation, which is exactly what happens for $\mathcal R=R=0$, and thus these results are coherent with the ones from the geometrical representation.

Now, let us perturb the metric $g_{ab}$ in the same way we did in Eq.\eqref{bhperturbgab}. This will again impose a perturbation in both $R$ and $\mathcal R$ of the forms $R=\bar R+\epsilon\delta R$ and $\mathcal R=\bar{\mathcal R}+\epsilon\delta\mathcal R$, plus additional perturbations on the scalar fields of the forms
\begin{equation}
\varphi=\bar\varphi+\epsilon\delta\varphi\nonumber, \ \ \ \ \ \psi=\bar\psi+\epsilon\delta\psi
\end{equation}
From the definitions of the scalar fields in terms of the auxiliary fields $\alpha$ and $\beta$ from Eq.\eqref{ghmpgstauxaction} and using the fact that $\alpha=R$ and $\beta=\mathcal R$ for the equivalence between the two representations of the theory to hold, we can rewrite the perturbations in the scalar fields as
\begin{equation}
\delta \varphi=\frac{\partial^2f}{\partial\alpha^2}\delta\alpha+\frac{\partial^2f}{\partial\alpha\partial\beta}\delta\beta=\bar f_{RR}\delta R+\bar f_{R\mathcal R}\delta\mathcal R,
\end{equation}
\begin{equation}
\delta \psi=\frac{\partial^2f}{\partial\beta^2}\delta\beta+\frac{\partial^2f}{\partial\beta\partial\alpha}\delta\alpha=\bar f_{\mathcal R\mathcal R}\delta \mathcal R+\bar f_{\mathcal R R}\delta R.
\end{equation}
Inserting these perturbations into the traces of Eqs.\eqref{ghmpgstfield} and \eqref{ghmpgrelricten} and keeping only the terms in leading order in $\epsilon$ yields again the same equations as Eqs. \eqref{bhsystem1} and \eqref{bhsystem2}, and the procedure is the same as it was done in the geometrical representation. We therefore conclude that the analysis of metric perturbations in both representations of the theory is equivalent, as anticipated.

\section{Stability regimes}

As explained, each of the terms $\left(\Box-\mu_\pm^2\right)\delta R=0$ gives rise to a set of two different solutions, an ingoing and an outgoing waves. Since these terms commute in the full equation given by Eq. \eqref{bhscalareq}, then the complete solution for this equation is given by a linear combination of the two sets of solutions for each of the $\mu_\pm^2$'s. Since Eq. \eqref{bhscalareq} is a fourth-order equation, then these four solutions represent all the possible solutions for the equation. As the masses $\mu_\pm$ are different in general, the two sets of solutions will form quasi-bound states for different ranges of the angular frequency $\omega$. However, if one of the two sets of solutions is unstable, then the entire solution will also be unstable, even if the other set is stable. 

Two conditions are needed for these instabilities to occur: the superradiant condition given by $\omega<m\Omega$ must be satisfied, and a potential well between the ergoregion and the boundstate potential barriers must exist, which can be translated into a condition for $\omega$ as $\mu_\pm^2/2<\omega^2<\mu_\pm^2$\cite{hod1}. There are then two different ways for one to avoid the existance of instabilities in this theory. The first way is to consider the masses $\mu_\pm>\sqrt{2}m\Omega$, such that these two conditions cannot be verified simultaneously. Quasi-bound states will form for certain frequencies $\omega$, but the superradiant condition will never be satisfied for the same frequencies. The second way to avoid instabilities is to consider massless perturbations, $\mu_\pm^2=0$, such that quasi-bound states never form even if $\omega$ satisfies the superradiant condition. Note that we can also achieve stability for a combination of the two cases above, i.e., one of the masses might vanish and the other be in the region $\mu>\sqrt{2}m\Omega$.

\subsubsection{The case $\mu_\pm^2=0$}

Let us start by studying the particular case where the masses $\mu_\pm^2$ vanish, which implies that quasi-bound states can never form and hence no instabilities can occur. From Eqs. \eqref{bhbox2ricci} and \eqref{bhscalarmass}, we verify that if both $A$ and $B$ vanish, then Eq. \eqref{bhscalareq} becomes simply $\Box^2\delta R=0$. If we can find a form of the function $f\left(R,\mathcal R\right)$ such that both $A$ and $B$ vanish, then the Kerr solution will always be stable in this theory. To guarantee that none of the equations of motion diverge, we need to verify that all the first and second derivatives of $f$, i.e. $f_R$, $f_\mathcal R$, $f_{RR}$, $f_{\mathcal R\mathcal R}$, $f_{R\mathcal R}$, must be finite. On the other hand, the factors $A$ and $B$, given by Eqs. \eqref{bhparA} and \eqref{bhparB}, will vanish if the following conditions are satisfied:
\begin{eqnarray}
&&f_{R\mathcal R}^2-f_{RR}f_{\mathcal R\mathcal R}\neq 0,\nonumber\\
&&f_Rf_{\mathcal R\mathcal R}-2f_\mathcal Rf_{R\mathcal R}-f_\mathcal Rf_{RR}=0,\nonumber\\
&&f_R+f_\mathcal R=0.
\end{eqnarray}
Note that these conditions must be satisfied at $R=\mathcal R=0$. There are many different functions $f$ that verify these conditions. The simplest class of functions $f$ that satisfy these conditions is
\begin{equation}\label{bhsolf1}
f\left(R,\mathcal R\right)=\left(a_1+a_2R+a_3\mathcal R\right)\left(R-\mathcal R\right)
\end{equation}
where $a_i\neq 0$ are constants that must satisfy the constraints $a_2\neq -a_3$. Any higher order form of the function $f\left(R,\mathcal R\right)$ obtained from Eq. \eqref{bhsolf1} by adding terms such as $R^3$ or $R^2\mathcal R$ will also have stable solutions because all these extra terms vanish when we set $R=0$ and $\mathcal R=0$ in Eqs. \eqref{bhparA} and \eqref{bhparB}.

\subsubsection{The case $\mu_+=0$ with $\mu_->\sqrt{2}m\Omega$}

We now want one of the $\mu_\pm^2$ to vanish and the other to be large enough to satisfy the condition $\mu_\pm>\sqrt{2}m\Omega$. However, this is not possible for both masses. As can be seen from Eq. \eqref{bhscalarmass}, the only way for $\mu_-=0$ is to have $B=0, A>0$, but in this region we have $\mu_+<0$, which gives rise to tachyonic instabilities. We therefore can only set $\mu_+=0$ by choosing $B=0, A<0$, and in this region we have $\mu_->0$ and can choose the function $f$ in such a way that $\mu_->\sqrt{2}m\Omega$.

Let us then study the case where $\mu_+=0$ and $\mu_->\sqrt{2}m\Omega$. As before, in order to avoid divergences in the equations of motion we have to verify that all the first and second derivatives of $f$, i.e. $f_R$, $f_\mathcal R$, $f_{RR}$, $f_{\mathcal R\mathcal R}$, $f_{R\mathcal R}$, must be finite, and the extra constraints on the function $f$ such that $B=0$ and $A\neq 0$ are
\begin{eqnarray}
&&f_{R\mathcal R}^2-f_{RR}f_{\mathcal R\mathcal R}\neq 0,\nonumber\\
&&f_Rf_{\mathcal R\mathcal R}-2f_\mathcal Rf_{R\mathcal R}-f_\mathcal Rf_{RR}\neq 0,\nonumber\\
&&f_R+f_\mathcal R=0.
\end{eqnarray}
Again, these conditions must be satisfied at $R=\mathcal R=0$. A simple class of functions $f$ that satisfies these constraints is
\begin{equation}\label{bhsolf2}
f\left(R,\mathcal R\right)=a_1\left(R-\mathcal R\right)+a_2R^2+a_3\mathcal R^2+a_4R\mathcal R,
\end{equation}
where $a_i\neq 0$ are constants. This form of the function $f$ implies, by Eqs. \eqref{bhparA}, \eqref{bhparB} and \eqref{bhscalarmass}, that $A$ and $\mu_+$ can be written in terms of the constants $a_i$ as
\begin{equation}\label{bhsolmass2}
\mu_-=-A=\frac{2a_1\left(a_2+a_3+a_4\right)}{12 a_2a_3-3a_4^2}>\sqrt{2}m\Omega,
\end{equation}
where the inequality guarantees that the solutions are stable. For us to obtain a finite $A$, both the numerator and the denominator of Eq. \eqref{bhsolmass2} must be $\neq 0$. 

Now, let us try to find a specific combination of the constants $a_i$ such that the Eq. \eqref{bhsolmass2} is satisfied. There are many combinations of $a_i$ that work, but let us take for example the case where $a_3$ and $a_4$ are set and verify if there is a value of $a_2$ that solves the problem. Note that the denominator of Eq.~\eqref{bhparA} diverges to $+\infty$ when we take the limit $a_2\to a_4^2/\left(4
a_3\right)^+$. Also, if we choose both $a_3>0$ and $a_4>0$, then $a_2>0$ and the numerator of Eq.~\eqref{bhparA} is positive in this limit. This implies that we can always choose a finite value of
$a_2\gtrsim a_4^2/\left(4 a_3\right)$ arbitrarily close to $a_4^2/\left(4 a_3\right)$ such that for any $m$ and $\Omega$ the condition $\mu_->\sqrt{2}m\Omega$ is always satisfied. Note that $m$ does not have an upper limit, but it has been shown that superradiant instabilities are exponentially suppressed for larger values of $l$ and $m$, and so for $m \gg 1$ the effects of these instabilities is negligible.

Again, any higher order form of the function $f\left(R,\mathcal R\right)$ will also have stable solutions because all these extra terms vanish when we set $R=0$ and $\mathcal R=0$ in Eqs. \eqref{bhparA} and \eqref{bhparB}.

\subsubsection{The case $\mu_\pm>\sqrt{2}m\Omega$}

Finally, we turn to the case where both masses are $\mu_\pm>\sqrt{2}m\Omega$ and the superradiant instability condition is not verified even if boundstates are formed. In this case we need both $A$ and $B$ to be finite. Let us now analyze the regions of the phase space of $A$ and $B$ that allow for these solutions to exist. From Eq. \eqref{bhscalarmass}, we can see that there are three regions of the phase space that must be excluded: 1. Region $4B>A^2$, because both $\mu_\pm^2$ are complex; 2. Region $B<0$, because $\mu_+$ is always negative; 3. Region $B>0$ and $A>0$, because both $\mu_\pm^2$ are negative. We are then constrained to work in the region defined by the three conditions $B>0$, $A<0$ and $A^2>4B$.

To approach this problem, let us first study the structure of the $\mu_\pm^2$ as functions of $A$ and $B$ in Eq. \eqref{bhscalarmass}. In the limit of large $\mu_\pm^2$, we must have $A\to -\infty$ like in the previous case. However, in this limit, the quantity inside the square root is going to depend on how $B$ is proportional to $A$. If $B\propto |A|^n$ with $n < 2$, then in this limit $\sqrt{A^2-4B}\to |A|$ and $\mu_+ \to 0$, and we recover the previous case. On the other hand, if $B\propto |A|^n$ with $n>2$, then in this limit we eventually break the relation $A^2>4B$ and the $\mu_\pm^2$ become complex. Therefore, we need a behaviour or $B$ of the form $B=CA^2$ for some constant $C$. Then, the constraints $B>0$ and $A^2>4B$ imply that the constant $C$ must be somewhere in the region $0<C<1/4$. Inserting this form of $B$ into Eq. \eqref{bhscalarmass} leads to the relation 
\begin{equation}
\mu_\pm^2=-(1/2)A\left(1\mp\sqrt{1-4C}\right)>0,
\end{equation}
where the inequality arises since we know that $A<0$ and $C<1/4$. At this point, if we can choose a specific form of the function $f$ such that we can make $A$ arbitrarily large, our problem is solved. Also note that in the limit $C=0$ we recover the previous case where $\mu_+=0$, and in the limit $C=1/4$ we obtain $\mu_+=\mu_-$.

Consider now the most general form of the function $f$ that avoids divergences in the equations of motion, i.e. a function $f$ for which $f_R$, $f_\mathcal R$, $f_{RR}$, $f_{\mathcal R\mathcal R}$ and $f_{R\mathcal R}$ are finite:
\begin{equation}
f\left(R,\mathcal R\right)=a_1 R+a_2\mathcal R+a_3R^2+a_4\mathcal R^2+a_5R\mathcal R,
\end{equation}
where $a_i\neq 0$ are constants. Again, any higher order form of the function $f\left(R,\mathcal R\right)$ will also work because the extra terms vanish for $R=\mathcal R=0$. For this particular choice of $f$, we see by Eq. \eqref{bhparA} that $A$ diverges to $-\infty$ if the denominator is positive and we take the limit $a_3\to a_5^2/\left(4 a_4\right)^+$, or if the denominator is negative and we take the limit $a_3\to a_5^2/\left(4 a_4\right)^-$. However, this happening is not enough to conclude that we can make $\mu_\pm^2$ arbitrarily large. We also need to verify that $B=CA^2$, with $0<C<1/4$, i.e. from Eq. \eqref{bhparA} we must have
\begin{equation}\label{bhconstraintC}
f_\mathcal R\left(f_\mathcal R+f_R\right)\left(f_{\mathcal R R}^2-f_{RR}f_{\mathcal R\mathcal R}\right)=C\big(f_Rf_{\mathcal R\mathcal R}-2f_\mathcal Rf_{\mathcal R R}-f_\mathcal Rf_{RR}\big)^2,
\end{equation}
\begin{equation}
f_{\mathcal R R}^2-f_{RR}f_{\mathcal R\mathcal R}\neq 0.
\end{equation}
Finding the most general combinations of $a_i$ for which the function $f$ satisfy these constraints is a fine-tuning problem. To solve this problem we proceed as follows. First write $a_3=a_5^2/\left(4
a_4\right)+\epsilon$, for some $\epsilon\gtrsim 0$ that must be finite but we can make it arbitrarily small. We also need to have $a_2\neq a_1$ to guarantee that $B\neq 0$, so we write $a_2=-\left(a_6+1\right)a_1$, where $a_6\neq 0$. Inserting these considerations into Eq.~\eqref{bhconstraintC}, we verify that $C$ is positive in the small $\epsilon$ limit only if the condition $a_4a_6\left(1+a_6\right)<0$. This happens in the regimes: $-1<a_6<0$ with $a_4>0$, $a_6<-1$ with $a_4<0$ or $a_6>0$ with $a_4<0$. As an example, let us consider $a_4=1$ and $a_6=-1/2$, which corresponds to
the maximum of the polynomial $-a_6\left(a_6+1\right)$. Finally, we have to guarantee that $A<0$ in this regime. Inserting these results into Eq.~\eqref{bhparA} we verify that $A<0$ requires that the quantities
$\left(a_5+1\right)$ and $a_1$ have the same sign. For simplicity, let us take $a_1=a_5=1$. Note that other choices for the values of the parameters $a_i$ could also be chosen following the same reasoning. In
here, our aim is simply to provide an example of a combination that works. We are thus left with:
\begin{equation}
A=-\frac{13+4\epsilon}{48\epsilon},\quad B=\frac{1}{144\epsilon},\quad C=\frac{16\epsilon}{\left(13+4\epsilon\right)^2}.
\end{equation}
From these results, we verify that for any $\epsilon>0$ we have $B>0$, $0<C<1/4$, and also that in the limit $\epsilon\to 0$ we have $A\to -\infty$ and thus $\mu^2_\pm\to +\infty$. As a verification, note also
that these results preserve the equality $B=CA^2$. We can thus consider $\epsilon$ arbitrarily small and force $\mu^2_\pm>\sqrt{2}m\Omega$ for any $m$ and $\Omega$. Again, we note that $m$ does not have an upper bound, but superradiant instabilities are exponentially supressed for large values of $m$ and we can neglect their effects.

\section{Conclusions}

We have shown that it is always possible to choose a specific value for $\mathcal R$, namely $\mathcal R=\mathcal R_0=R$ for some solution in GR with constant $R=R_0$ such that this solution is also a solution for the GHMP gravity for any form of the function $f$ that satisfies two very general conditions: $f$ must be analytical in the point $\{R_0,\mathcal R_0\}$, and $f$ must have a zero in the same point i.e. $f\left(R_0,\mathcal R_0\right)=0$. This result is in agreement with the fact that for constant $R$ and $\mathcal R$, the conformal factor between the metric $g_{ab}$ and $h_{ab}$, which is given by $f_\mathcal R$, is constant and therefore both metrics $g_{ab}$ and $h_{ab}$ must have the same Ricci tensor. Note that when we set a value for $R$, Eq. \eqref{ghmpgrelricsca} becomes a PDE for $\mathcal R$, and thus the solution is not unique and other forms of $\mathcal R$ might also allow for the Kerr solution. However, one would have to specify a form of the function $f$ to continue the analysis, and the aim of keeping $f$ as general as possible would not be fulfilled.

When we perturb the metric tensor, the equation that describes the perturbation in the Ricci scalar is a 4th order massive wave equation with two different masses. Since the Ricci perturbation depends on 2nd order derivatives of the metric perturbation, we would be confronted with a 6th order differential equation for $\delta g_{ab}$, with a very complicated and untreatable form. However, using the Lorentz gauge, one can reduce this equation to 4th order equation for the massive spin-0 degree of freedom. This 4th order equation can be factorized into two commutative 2nd order equations of the form of the Klein-Gordon equation for a massive scalar field in GR. One can then apply the usual separation methods to expand the perturbation in spheroidal harmonics and a radial wavefunction and the usual numerical integration techniques to compute the quasi-bound state frequencies.

The stability of the Kerr metric against superradiant instabilities is dictated by two phenomena: the existence of superradiant amplification in the regime $\omega<m\Omega$, and the confinement of the wave in a potential well in the regime $\mu^2/2<\omega^2<\mu^2$. We have shown that it is possible to select specific well-behaved forms of the function $f$ such that these two conditions are not satisfied simultaneously for any value of the angular frequency $\omega$, more specifically by forcing the masses $\mu_\pm^2$ to vanish or to be large enough for $\mu_\pm>\sqrt{2}m\Omega$. Also, since the masses only depend on the values of $f$ and its derivatives at $R=\mathcal R=0$, any higher order term on $R$ and $\mathcal R$ up to infinity can be added to the function $f$ leaving these results unaffected, being then coherent with the two general constraints we imposed on the function $f$ to begin with.

As an azimuthal number, $m$ does not have an upper bound, and so one could argue that for any constant value of $\mu_\pm^2$, there is always a value of $m$ such that $\mu_\pm<\sqrt{2}m\Omega$. However, it has been shown that superradiant instabilities are exponentially supressed for larger values of $m$. This implies that we can consider an upper bound on $m$ for which the instability timescale is greater than the age of the universe, say $m_{max}$, and only after choose an appropriate value of $\mu_\pm^2$ that satisfies the inequality $\mu_\pm>\sqrt{2}m_{max}\Omega$. This guarantees that even if the instabilities occur, their effects would not be seen in the present universe. 
\cleardoublepage

\chapter{Cosmological phase-space of higher-order $f\left(R,\Box R\right)$ gravity}
\label{chapter:chapter4.5}

\def\ua{^{\alpha}}  
\def\ub{^{\beta}}
\def\da{_{\alpha}}
\def\db{_{\beta}}
\def\ug{^{\gamma}}
\def\dg{_{\gamma}}
\def\l{\mathcal L}
\def\g{\mathds G}
\def\fo{\mathcal F}
\def\uamu{^{\alpha\mu}}
\def\uanu{^{\alpha\nu}}
\def\uab{^{\alpha\beta}}
\def\dab{_{\alpha\beta}}
\def\dabgd{_{\alpha\beta\gamma\delta}}
\def\uabgd{^{\alpha\beta\gamma\delta}}
\def\udeab{^{;\alpha\beta}}
\def\ddeab{_{;\alpha\beta}}
\def\ddemunu{_{;\mu\nu}}
\def\udemunu{^{;\mu\nu}}
\def\ddemu{_{;\mu}}  \def\udemu{^{;\mu}}
\def\ddenu{_{;\nu}}  \def\udenu{^{;\nu}}
\def\ddea{_{;\alpha}}  \def\udea{^{;\alpha}}
\def\ddeb{_{;\beta}}  \def\udeb{^{;\beta}}
\def\ad{\dot{a}}
\def\add{\ddot{a}}
\def\vol{\int d^4x\,\sqrt{-g}} 
\def\grav{\frac{1}{16 \pi G}}
\def\half{\frac{1}{2}}
\def\gu{g^{ab}}
\def\gd{g_{ab}}
\def\naba{\nabla_{\alpha}}
\def\nabb{\nabla_{\beta}}
\def\pmu{\partial_{\mu}}
\def\pnu{\partial_{\nu}}
\def\pa{\partial}
\def \dsQ {{\mathds Q}}
\def \dsJ {{\mathds J}}
\def \dsS {{\mathds S}}
\def \A {{\mathbb A}}
\def \K {{\mathbb K}}
\def \R {{\mathbb R}}
\def \Q {{\mathbb Q}}
\def \S {{\mathbb S}}

In chapter \ref{chapter:chapter3} we have studied the cosmological phase space of the generalized hybrid metric-Palatini gravity where the action is a function of the form $f\left(R,\mathcal R\right)$. The equations of motion for this sort of action are effectively 4th order in the derivatives of the metric $g_{ab}$, see chapter \ref{chapter:chapter2}. However, more complicated actions of hybrid metric-Palatini gravity theories can be obtained by including higher order terms such as $\Box \mathcal R$, for which the field equations are effectively of order 6th or higher. The equations of motion for such a theory are extremely complicated to work with, even in the dynamical system formalism. As an introductory step to these higher order metric-Palatini theories, we provide an anaysis of simpler higher order theories of gravity where the action is a function of the form $f\left(R,\Box R\right)$, and leave the analysis of the higher order hybrid theories for a future work. The phase space of these cosmology reveals that higher-order terms can have a dramatic influence on the evolution of the cosmology, avoiding the onset of finite time singularities. We also confirm and extend some of results which were obtained in the past for this class of theories.

\section{Introduction}

General relativity (GR) deals with second-order differential equations for the metric $g_{ab}$. Higher-order modifications of the gravitational interaction have been for a long time the focus of intense investigation. They have been proposed for a number of reasons including the first attempts of unification of gravitation and other fundamental interactions. Nowadays, the main reason why one considers this kind of extension in GR is of quantum origin. Studies on the renormalisation of the stress-energy tensor of quantum fields in the framework of a semi classical approach to GR, i.e., what we call quantum field theory in curved spacetime, show that such corrections are needed to take into account the differences between the gravitation of quantum fields and the gravitation of classical fluids \cite{buchbinder1,birrell1}. 

With the introduction of the paradigm of inflation and the requirement of a field able to drive it, it was natural, although not obvious, to consider these quantum corrections as the engine of the inflationary mechanism. Starobinski \cite{starobinski1} was able to show explicitly in the case of fourth-order corrections to GR that this was indeed the case: quantum corrections could induce an inflationary phase.  Such result should not be surprising. Fourth-order gravity carries an additional scalar degree of freedom and this scalar degree of freedom can drive an inflationary phase. In the following years other researchers \cite{amendola1,gottlober1,wands1} tried to look at the behaviour of sixth-order corrections, to see if in this case one could obtain a richer inflationary phase and more specifically a cosmology with multiple inflationary phases. However it turned out that this is not the case: in spite of the presence of an additional scalar degree of freedom, multiple inflationary phases were not possible. The reason behind this result is still largely unknown.

The possibility of dark energy as a candidate to explain the late-time cosmic acceleration offered yet another application for the additional degree of freedom of higher-order gravity. Like in the case of inflation, this perspective offered an elegant way to explain
dark energy: higher-order corrections were a geometrical way to interpret the mysterious new component of the universe \cite{capozziello1,sotiriou1,capozziello6,nojiri2}. Here an important point should be stressed: differently from the standard perturbative investigation of a physical system, in the case of higher-order gravity, the behaviour of the new theory cannot be deduced as a small perturbation of the original second-order one. The reason is that, since the equations of motion switch order, the dynamics of the perturbed system are completely different from the non-perturbed one whatever the (non zero) value of the smallness parameter. For this reason, the properties of higher-order gravity cannot be deduced from their lower-order counterpart, even if the higher-order terms are suppressed by a small coupling constant. This fact calls for a complete reanalysis of the phenomenology of these theories. Such study should be performed with tools designed specifically for this task, which therefore contain no hidden assumptions or priors which might compromise the final result. One of these tools, which will be used in the following is the so called Dynamical System Approach (DSA), already explored in chapter \ref{chapter:chapter3}.

Among the many unresolved issues that are known to affect higher-order theories, it is worth to mention briefly the so-called Ostrogradski theorem \cite{woodard1}. The theorem shows that for a generic system with an higher-order Lagrangian, there exist a conserved quantity $H$ corresponding to time shift invariance. When this quantity is interpreted as an Hamiltonian, by the definition of a suitable Legendre transformation, it can be shown that such Hamiltonian, not being limited from below, leads to the presence of undesirable features of the theory upon quantisation, whereas the classical behaviour, which includes classical cosmology, has no problem. In view of this conclusion higher-order theories, with the notable exception of $f(R)$ gravity, are deemed as unphysical. The most important issue for this work is then, why bother with higher-order gravity? We can give two arguments. The first is that, as mentioned above, the higher-order terms we will consider are terms of a series of corrections arising in a renormalisation procedure. In this perspective, therefore, there is no requirement that the truncated series had the same convergence property of its sum. A typical example is the Taylor series of $\sin(x)$. The truncated series is not bound, whereas its full sum is. In the same way the truncation of the original semiclassical model that gives rise to a higher-order theory might be fundamentally flawed on the quantum point of view. The problem only arises if one chooses the {\em complete} theory of quantum gravity to be given by a $n-$order truncation.  The second is that a study of the behaviour of the truncation allows an understanding of the interplay between the different terms of the development and in particular if and how the pathologies of the theory at a certain order are changed by the therms of higher-order. This  on one hand allows to give statements on the validity of the procedure of renormalisation in quantum field theory in curved spacetime and, on the other hand, it is  interesting in the context of the cosmology of fourth-order gravity, as it is known that this class of theories can present a number of issues, i.e., scale factor can evolve towards a singularity at finite time (see chapter \ref{chapter:chapter3}) which is independent form the Ostrogradski instability. An analysis of the higher-order theories can therefore shed light on the real nature of these pathologies.

In this work we propose a DSA able to give a description of the dynamics of cosmological models based on a subclass of $2n+4$-order gravity represented by the Lagrangian density $\mathcal L=f(R, \Box R)$. We show that the higher-order terms in these theories act in an unexpected way on the cosmology, being dominant and preventing the appearance of finite time singularities. The calculations involved in this task are formidable so we will give the full expression only when strictly necessary.

\section{Basic equations}

Let us start with the general action for a relativistic theory of order six of the form
\begin{equation}\label{hogaction}
S=\int\sqrt{-g}f\left(R,\Box R\right)d^4x+S_m,
\end{equation}
where $g$ is the metric determinant of the metric $g_{ab}$, $f$ is a generic function of the Ricci scalar $R$ and of its d'Alembertian $\Box R$, and $S_m$ is the matter action. This theory is in general of order eight in the derivatives of the metric. Since we consider the boundary terms as irrelevant, integrating by parts leads to a series  of important simplifications in the theory above. First, it is important to note that, not differently from the case of the Einstein-Hilbert action, if $f$ is linear in $\Box R$ the theory is only of order six. In fact, we can use this line of reasoning to obtain a second important observation on Eq. \eqref{hogaction}: any non linear term in $\Box R$ appearing in $f$ can always be recast as an higher-order term. Thus, for example, $(\Box R)^2$  can be written as
\begin{equation}
(\Box R)^2= \nabla_\mu \left(\nabla^\mu R \Box R-R\nabla^\mu \Box R\right)+ R \Box^2 R\,.
\end{equation}
So terms of the type $(\Box R)^n$ can be converted into terms of order $2n+4$ of the
form $R\Box^n R$. In general, therefore, there exists a subset  of theories of gravity of order $2n+4$ which is represented by action in Eq. \eqref{hogaction} and can be written as \cite{gottlober1,wands1}
\begin{equation}\label{hogactionsimp}
S=\int \sqrt{-g} \left[f_0(R)+ \sum_{i=1}^{n} a_i(\Box R)^i\right]d^4x+S_m,
\end{equation}
where the $a_i$ are coefficients in the expansion.

For our purposes, these considerations will be crucial. They will allow to simplify greatly the equations for theories of order six and  to generate a minimal extension of the approach to treat  any theory of order $2n+4$ that can be written in the form in Eq. \eqref{hogactionsimp}. In the following we will start  describing the general theory and then, using  the considerations above, we will present explictly a dynamical systems formalism for Lagrangians of order six, showing that the same formalism can be easily extended to theories of generic order $2n+4$ with the form in Eq. \eqref{hogactionsimp}.

Variation of the action given by Eq. \eqref{hogactionsimp} with respect to the metric tensor $g_{ab}$ gives the gravitational field equations of the form
\begin{equation}
\mathds G G_{ab}=\frac{1}{2} g_{ab}\left(f-\mathds G R\right)+\nabla_a\nabla_b\mathds G-
g_{ab}\Box \mathds G -\frac{1}{2}g_{ab}\left(\nabla_c\mathcal F\nabla^cR+\mathcal F\Box R\right)+\nabla_{(a}\mathcal F\nabla_{b)}R+T_{ab},
\end{equation}
where $T_{ab}$ is the standard stress energy tensor and we define
\begin{equation}
\mathds G=\frac{\partial f}{\partial R}+\Box\mathcal F, \ \ \ \ \ \ \ \mathcal F=\frac{\partial f}{\partial \Box R}.
\end{equation}

In the Friedmann-Lemaître-Robertson-Walker metric given by Eq.\eqref{metricflrw} and assuming the matter component to be an isotropic perfect fluid, i.e., $T_a^b=\left(-\mu,p,p,p\right)$, where $\mu$ is the energy density and $p$ is the pressure of the fluid, we obtain the cosmological equations which are usually written as
\begin{equation}
\mathds G \left(H^2+\frac{k}{a^2}\right)=\frac{1}{6}\left(R \mathds G-f+\mathcal F\Box R+\dot{R}\dot{\mathcal F}\right)\\
-H\dot{\mathds G}+\frac{\mu}{3},
\label{hogfriedmann}
\end{equation}
\begin{equation}
\mathds G \left(\dot{H}+H^2\right)=-\frac{1}{6}\left(f-R\mathds G-{\mathcal F}\Box R\right)-\frac{1}{3}\dot{R}\dot{\mathcal F}-\frac{1}{2}\ddot{\mathds G}-\frac{1}{2}H\dot{\mathds G}-\frac{1}{6}(\mu+3 p),
\label{hograychaudhuri}
\end{equation}
where $k=-1,0,1$ is the spatial curvature, $H$ is the hubble parameter defined in Eq.\eqref{defhubble}, $a\left(t\right)$ is the scale factor, a dot $\dot{\ }$ denotes a derivative with respect to time, and the Ricci scalar $R$ and its D'Alembertian operator can be written as
\begin{eqnarray}
&&R=6\left[\dot{H}+2H^2+\frac{k}{a^2}\right],\label{hogdefR}\\
&&\Box R=-\ddot{R}-3H\dot{R}.\label{hogdefboxR}
\end{eqnarray}
With an abuse of terminology, we will sometimes refer to Eq.\eqref{hogfriedmann} as the  Friedmann equation and to Eq.\eqref{hograychaudhuri} as the Raychaudhuri equation. We introduce now  the logarithmic time $N=\log\frac{a}{a_0}$ and we also define a set of seven cosmological parameters as
\begin{equation} \label{hogcosmopar}
{\mathfrak q} =\frac{H^{(1)}_{N}}{H},\quad {\mathfrak j} =\frac{H^{(2)}_{N}}{H},\quad{\mathfrak s}=\frac{H^{(3)}_{N}}{H},\quad{\mathfrak s}_1=\frac{H^{(4)}_{N}}{H},\quad{\mathfrak s}_2=\frac{H^{(5)}_{N}}{H},\quad{\mathfrak s}_3=\frac{H^{(6)}_{N}}{H}\quad{\mathfrak s}_4=\frac{H^{(7)}_{N}}{H},
\end{equation}
where $H^{(i)}_{N}$ represent the $i$th-derivative of $H$ with respect to $N$. One can write these equations in terms of these variables, but  this is a long and rather tedious exercise which does not really add anything to the understanding of the problem. For this reason we will not show them here, giving directly the equations in terms of the dynamical variables in the following sections.

\section{Dynamical system approach to the 6th order case}

Let us start looking at the sixth-order case. Recalling the argument of the previous section, all theories of order six that have the form $f(R,\Box R)$ can be written without loss of generality as 
\begin{equation}\label{hog6oaction}
f= f_1(R)+ f_2(R) \Box R\,,
\end{equation}
where $f_1$ and $f_2$ are in general different funtions of $R$. Eq.\eqref{hog6oaction} has the immediate consequence that ${\mathfrak s}_3$ and ${\mathfrak s}_4$ are not present in the cosmological equations and the analysis of this classes of modes is greatly simplified.

In order to apply the scheme presented in chapter \ref{chapter:chapter4} the action will need to be written in a dimensionless way. We introduce therefore the constant $R_0$, with $R_0>0$, which has dimension of a length squared and we will consider the function $f$ of the type $f= R_0 \bar{f}(R_0^{-1} R, R_0^{-2}\Box R)$, for some function $\bar f$. This implies the definition of an auxiliary dynamical variable related to $R_0$. We then define the set of dynamical variables
\begin{eqnarray}\label{hogdynvar}
\mathbb{R}=\frac{R}{6 H^2},\quad \mathbb{B}=\frac{\Box R}{6H^4},\quad  \mathbb{K}=\frac{k}{a^2 H^2},\quad \Omega =\frac{\mu }{3H^2},\\
\mathbb{J}={\mathfrak j},\quad \mathbb{Q}={\mathfrak q},\quad \mathbb{S}={\mathfrak s},\quad \mathbb{S}_1={\mathfrak s}_1\nonumber
\end{eqnarray}
Note that in the above setting $\mathbb{A}$ and $\Omega $ are defined positive so that all fixed points with $\mathbb{A}<0$ or $\Omega<0$ should be excluded. The Jacobian of this variable definition reads
\begin{equation}\label{hog6ojacobian}
J=-\frac{1}{108 a^2 H^{32}},
\end{equation}
which implies that the variables are always regular if $H\neq 0$ and $a \neq 0$. The requirement to have a closed system of equations requires the introduction of the auxiliary quantities
\begin{eqnarray}\label{hogdynfun}
&& {\bf X}_1\left(\mathbb{A},\mathbb{R}\right)= \frac{f_1\left(\mathbb{A},\mathbb{R}\right)}{H^2},\nonumber\\ 
&& {\bf X}_2\left(\mathbb{A},\mathbb{R} \right)= H^2f_2\left(\mathbb{A},\mathbb{R}\right),\nonumber\\ 
&& {\bf Y}_1\left(\mathbb{A},\mathbb{R} \right)= f'_{1}\left(\mathbb{A},\mathbb{R}\right),\nonumber\\ 
&& {\bf Y}_2\left(\mathbb{A},\mathbb{R} \right)=H^4 f'_{2}\left(\mathbb{A},\mathbb{R}\right),\nonumber\\
&& {\bf Z}_1\left(\mathbb{A},\mathbb{R}\right)= H^2 f''_1\left(\mathbb{A},\mathbb{R}\right),\\ 
&& {\bf Z}_2\left(\mathbb{A},\mathbb{R} \right)=H^6 f''_2\left(\mathbb{A},\mathbb{R}\right),\nonumber\\ 
&& {\bf W}_1\left(\mathbb{A},\mathbb{R}\right)=H^4 f^{(3)}_1\left(\mathbb{A},\mathbb{R}\right),\nonumber\\ 
&& {\bf W}_2\left(\mathbb{A},\mathbb{R} \right)=H^8 f^{(3)}_2\left(\mathbb{A},\mathbb{R}\right),\nonumber\\ 
&& {\bf T}\left(\mathbb{A},\mathbb{R} \right)= H^{10}f^{(4)}_{2}\left(\mathbb{A},\mathbb{R}\right),\nonumber\\
\end{eqnarray}
where the prime represents the derivative with respect to the Ricci scalar $\mathbb{R}$. The dynamical equations can then be written as
\begin{eqnarray}
\frac{d\mathbb{R}}{dN}=&&\mathbb{J}+(\mathbb{K}-2) \mathbb{K}-(\mathbb{R}-2)^2 ,\nonumber\\ 
\frac{d\mathbb{B}}{dN}=&&\mathbb{B} (3 \mathbb{K}-3 \mathbb{R}+7)+\frac{1}{2}
   \left[\mathbb{J}+\mathbb{K}^2-2 \mathbb{K} (\mathbb{R}+1)+\mathbb{R}^2-4\right]^2\nonumber\\
   &&~~~+\frac{1}{\mathbf{Y}_2}\left\{\frac{\Omega }{12}+18 \left[\mathbb{J}+\mathbb{K}^2-2
   \mathbb{K} (\mathbb{R}+1)+\mathbb{R}^2-4\right]^3 \mathbf{W}_2\right.\nonumber\\
   &&~~~-\frac{1}{72} \mathbf{X}_1-\frac{1}{12} (\mathbb{K}-\mathbb{R}+1)
   \mathbf{Y}_1+\left(2-\frac{\mathbb{J}}{2}-\frac{\mathbb{K}^2}{2}+\mathbb{K}
   \mathbb{R}+\mathbb{K}-\frac{\mathbb{R}^2}{2}\right)
    \mathbf{Z}_1\nonumber\\
   &&~~~-3 \left[4 \mathbb{B}-(\mathbb{K}-\mathbb{R}-5)
   \left(\mathbb{J}+\mathbb{K}^2-2 \mathbb{K}
   (\mathbb{R}+1)+\mathbb{R}^2-4\right)\right] \times\nonumber\\
   &&~~~\left.+\left[\mathbb{J}+\mathbb{K}^2-2 \mathbb{K} (\mathbb{R}+1)+\mathbb{R}^2-4\right]
   \mathbf{Z}_2 \right\}\label{hogdynsys6o}\\
\frac{d\Omega}{dN}=&&\Omega  (1-3w+2 \mathbb{K}-2 \mathbb{R}),\nonumber\\ 
\frac{d\mathbb{J}}{dN}=&&\mathbb{J} (5 \mathbb{K}-5
   \mathbb{R}+3)+(\mathbb{K}-\mathbb{R}) \left[\mathbb{K}^2-\mathbb{K} (2 \mathbb{R}+7)+\mathbb{R}
   (\mathbb{R}+5)\right],\nonumber\\
   &&~~~-\mathbb{B}-22 \mathbb{K}+20 \mathbb{R}-12,\\
\frac{d\mathbb{K}}{dN}=&&2 \mathbb{K} (\mathbb{K}-\mathbb{R}+1),\nonumber\\
\frac{d\mathbb{A}}{dN}=&&2 \mathbb{A}(\mathbb{K}-\mathbb{R}+2)\nonumber .
\end{eqnarray}

To eliminate the equations for $\mathbb{S}_1$, $\mathbb{Q}$, $\mathbb{S}$ we have implemented in the equations above the Friedmann equation, Eq.\eqref{hogfriedmann}, and the following constraints coming from the definition of $R$ and $\Box R$ in Eqs.\eqref{hogdefR} and \eqref{hogdefboxR}
\begin{equation}
\mathbb{R}=\mathbb{K}+\mathbb{Q}+2,
\end{equation}
\begin{equation}
\mathbb{B}=-4 \mathbb{J} \mathbb{Q}-7 \mathbb{J}+2 \mathbb{K} \mathbb{Q}+2\mathbb{K}-\mathbb{Q}^3-11 \mathbb{Q}^2-12 \mathbb{Q}-\mathbb{S}.
\end{equation}

The solutions associated to the fixed points can be derived calculating $\mathfrak{s}_2$ from the modified Raychaudhuri equation, Eq.\eqref{hograychaudhuri}, in a fixed point and then solving the differential equation
\begin{equation}\label{hoggenequation}
\frac{1}{H}\frac{ d^5{H}}{d N^5}=\mathfrak{s}^*_2,
\end{equation}
where here, and in the following, the asterisk $*$ indicates the value of a variable in a fixed point. 
Eq. \eqref{hoggenequation} can be solved in general to give
\begin{equation} \label{hoggensolH}
H=\sum_{i=0}^{2}\exp \left(p\,\alpha_i N\right)\left[H_i \cos\left(\beta_i p N\right)+\bar{H}_i \sin\left(\beta_i p N\right)\right], 
\end{equation}
where $p=-\sqrt[5]{\mathfrak{s}^*_2}$, $H_i$ and $\bar{H}_i$ are integration constants and  $a_i$ and $b_i$ are given by
\begin{equation}
\begin{array}{ll}
\alpha_0=-1, & \beta_0=0,\\
\alpha_1=\frac{1}{4}(\sqrt{5}+1), &\beta_1=\sqrt{\frac{5}{8}-\frac{\sqrt{5}}{8}},\\
\alpha_2=\frac{1}{4}(1-\sqrt{5}), &\beta_2=\sqrt{\frac{5}{8}+\frac{\sqrt{5}}{8}},
\end{array}
\end{equation}
i.e. are connected with the fifth root of unity. We are obviously interested in real solutions, which can be derived by a suitable redefinition of the integration constants.  The solutions above are oscillating, however  they do not correspond to oscillating scale factors. Indeed the scale factor is given by the equation
\begin{equation} \label{hoggeneqscale}
\dot{a}=\sum_{i=0}^{2}a^{1+p\,\alpha_i}\left[H_i \cos\left(\beta_i p \ln a\right)+\bar{H}_i \sin\left(\beta_i p \ln a\right)\right].
\end{equation}
This equation can be solved numerically. When $p>0$ it is useful to look at the particular solution of the equation above setting $H_i=0$ for $i>1$. However, one should bear in mind that this solution is not necessarily the one that the equation approaches at late time, i.e., for large $a$. This is evident form the form of Eq.\eqref{hoggeneqscale}. In the following we will calculate the solutions of these equations numerically for each example. In the case $\mathfrak{s}^*_2=0$ the equation to solve becomes
\begin{equation}\label{hoggeneqscale0}
\dot{a}=a\sum_{i=0}^{4}H_i (\log a)^i,
\end{equation}
which is more easily analysed using the logarithmic time $N$. In this variable it can be written as
\begin{equation}\label{hoggeneqN}
\dot{N}=\sum_{i=0}^{4}H_i N^i.
\end{equation}
This equation can be solved by separation of variables and it has a solution that depends on the roots of the polynomial in $N$ on the right hand side. In particular, the scale factor can have a finite time singularity if any of the roots of the polynomial are complex, otherwise it evolves asymptotically towards a constant value of the scale factor, i.e., a static universe.  Therefore,  a fixed point with  $\mathfrak{s}_2=0$, will correspond to one of these two cosmic histories depending on the value of the constants $H_i$. Considering that Eq.\eqref{hoggeneqscale0} can be viewed as an approximation of the general integral of the cosmology, then the values of the constants $H_i$ should match the initial conditions of the orbit. This implies that the solution in the fixed point will depend on the initial condition of the orbit that reaches it.

In the following we will examine two specific examples. The first one will show the phase space of a theory in which only sixth-order terms are present other than the Einstein-Hilbert one. This example will clarify the action of these terms. The second one will contain also fourth-order terms, so that the interaction between sixth  and fourth-order corrections can be observed explicitly.

\subsection{$R+\gamma R \Box R$ gravity}\label{Sec:hog61}

Let us start our analysis of specific cases by considering a simple action where only the Einstein-Hilbert term and a linear sixth-order term $\Box R$ are present. This is an interesting example as it clarifies the interplay between these terms. In this case the action can be written as 
\begin{equation}
S=\int\sqrt{-g} \left(  R+ \gamma R_0^{-2} R \Box R\right) d^4x+S_m,
\end{equation}
which corresponds to the particular case where the functions $f_1\left(R\right)$ and $f_2\left(R\right)$ take the forms $f_1=R_0^{-1} R$ and $f_2= \gamma R_0^{-3} R$. Considering these forms, then the only non-zero auxiliary quantities in Eq.\eqref{hogdynfun} are
\begin{equation}
{\bf X}_1\left(\mathbb{A},\mathbb{R}\right)= 6 \mathbb{R},\qquad {\bf X}_2\left(\mathbb{A},\mathbb{R} \right)= \frac{6\gamma \mathbb{R}}{\mathbb{A}},\qquad {\bf Y}_1\left(\mathbb{A},\mathbb{R} \right)= 1,\qquad {\bf Y}_2\left(\mathbb{A},\mathbb{R} \right)= \frac{\gamma}{\mathbb{A}^2}.
\end{equation}
Ideally, we should also provide the specific forms of the Friedmann and Raychaudhuri equations given in Eqs.\eqref{hogfriedmann} and \eqref{hograychaudhuri}. However, the size of these equations is enormous, so we shall avoid to do so in all of the specific examples. Let us just remind that the Friedmann equations in Eq.\eqref{hogfriedmann} works as an extra constraint for us to eliminate one of the dynamical system equations, whereas the Raychaudhuri equation in Eq.\eqref{hograychaudhuri} is used to compute the behavior of the scale factor from the value of $\mathfrak{s}^*_2$, see Eq.\eqref{hoggenequation}. The  dynamical system provided in Eq.\eqref{hogdynsys6o} becomes then
\begin{eqnarray}
&&\frac{d\mathbb{R}}{dN}=\mathbb{J}+(\mathbb{K}-2) \mathbb{K}-(\mathbb{R}-2)^2 ,\nonumber\\ 
&&\frac{d\mathbb{B}}{dN}=\mathbb{B} (3 \mathbb{K}-3 \mathbb{R}+7)-\frac{\mathbb{A}^2 (\mathbb{K}-\Omega +1)}{12 \gamma }+\frac{1}{2} \left(\mathbb{J}+\mathbb{K}^2-2 \mathbb{K} (\mathbb{R}+1)+\mathbb{R}^2-4\right)^2,\nonumber\\
&&\frac{d\mathbb{J}}{dN}=\mathbb{J} [5 (\mathbb{K}-\mathbb{R})+3]-\mathbb{B}-22 \mathbb{K}+20 \mathbb{R}-12+(\mathbb{K}-\mathbb{R}) \left[\mathbb{K}^2-\mathbb{K} (2 \mathbb{R}+7)+\mathbb{R}(\mathbb{R}+5)\right],\nonumber\\
&&\frac{d\Omega}{dN} =\Omega  (1-3w+2 \mathbb{K}-2 \mathbb{R}),\label{hogdynsys6oa}\\ 
&&\frac{d\mathbb{K}}{dN}=2 \mathbb{K} (\mathbb{K}-\mathbb{R}+1),\nonumber\\
&&\frac{d\mathbb{A}}{dN}=2 \mathbb{A}(\mathbb{K}-\mathbb{R}+2).\nonumber
\end{eqnarray}
The system presents three invariant submanifolds $\Omega=0$, $\mathbb{K}=0$ and $\mathbb{A}=0$, therefore only points that belong to all of these three submanifolds can be global attractors. The fixed points of the system can be found in Table \ref{Tab:hogfixed1}, together with their associated solutions. Point $\mathcal{C}$ has a solution of the type in Eq. \eqref{hoggeneqscale0} and as such can indicate the occurrence of a finite time singularity.

\begin{table*}[h!]
\begin{center}
\caption{Fixed points for the system given by Eq.\eqref{hogdynsys6oa} } \label{Tab:hogfixed1}
\begin{tabular}{lllcc} \hline
Point & Coordinates  & Stability & Parameter ${\mathfrak s}_2$  \\
&$\{\mathbb{R},\mathbb{B},\mathbb{J},\Omega,\mathbb{K}, \mathbb{A}\}$ &  \\ \hline\\
$\mathcal{A}$ & $\left\{ 0,0,1,0, -1 , 0\right\}$  & Saddle & $-1$\\ \\
$\mathcal{B}$ & $\left\{ 0, 0, 4, 0, 0, 0\right\}$  & \makecell[c]{ Reppeler for $w < 1/3$ \\ Saddle for $w > 1/3$} & $-32$ \\ \\
$\mathcal{C}$ & $\left\{ 2, 0, 0,0, 0, 0\right\}$  & Saddle & $0$ \\ \\
$\mathcal{I}_1$ & $\left\{ a_\mathcal{I}^-, b_\mathcal{I}^-,c_\mathcal{I}^+,0,0,0\right\}$  & Saddle & ${\mathfrak s}_{2}^{\mathcal{I}_1}$  \\ \\
$\mathcal{I}_2$ & $\left\{ a_\mathcal{I}^+, b_\mathcal{I}^+,c_\mathcal{I}^-,0,0,0\right\}$  & Attractor  & ${\mathfrak s}_2^{\mathcal{I}_2}$  \\ \\
\hline\\
\multicolumn{5}{c}{\makecell[c]{$a_\mathcal{I}^\pm=\frac{1}{10}\left(16\pm\sqrt{46}\right) $\qquad $b_\mathcal{I}^\pm=-\frac{9}{250}\left(74\pm 9\sqrt{46}\right) $ \qquad $c_\mathcal{I}^\pm= \frac{1}{50}\left(31\pm4\sqrt{46}\right)$\\ \\ ${\mathfrak s}_{2}^{\mathcal{I}_1} =-\frac{1}{25000}\left(18196+2689\sqrt{46}\right)$\qquad ${\mathfrak s}_2^{\mathcal{I}_2}=-\frac{1}{25000}\left(18196-2689\sqrt{46}\right)$
 }}\\\\\hline\\
 \end{tabular}
  \end{center}
\end{table*}

The stability of fixed points $\mathcal{B}$ for $w\neq 1/3$, $\mathcal{I}_1$ and $\mathcal{I}_2$,  can be deduced by the Hartmann-Grobmann theorem and it is also shown in Table \ref{Tab:hogfixed1}. Points $\mathcal{B}$ and $\mathcal{I}_1$ are unstable, whereas $\mathcal{I}_2$ is an attractor. Indeed this point is a global attractor for the cosmology as it lays on the intersection of the three invariant submanifolds of the phase space. The remaining points $\mathcal{A}$, $\mathcal{B}$ for $w = 1/3$, and $\mathcal{C}$, are non hyperbolic, as they have a zero eigenvalue. Their stability can be analyzed via the central manifold theorem \cite{wiggins1}, as performed in chapter \ref{chapter:chapter3}. However, we can also evaluate the character of these points in a faster way. In fact, point $\mathcal{A}$ has eigenvalues $\{4,-2,2,2,0,-(1+3w)\}$, i.e., with alternate signs. Therefore, regardless of the behaviour of the central manifold, this point is in fact always a saddle. This implies that in some cases we can evaluate the stability of a non hyperbolic fixed point without analysing in detail the central manifold. Clearly this is insufficient if the aim is to characterise the exact behaviour of the flow in the phase space. However, since we are mainly interested in the attractors in the phase space, such less precise analysis will be sufficient here. 

\subsection{$R+\alpha R^3+R\Box R$ gravity}\label{Sec:hog62}

We now turn to the case where the action is given by the Einstein-Hilbert plus fourth- and 
sixth-order correction terms of the form $R^3$ and $R\Box R$, respectively, such that the interaction between these orders can be appreciated. Consider then the action
\begin{equation}
S=\int \sqrt{-g} \left[ R+ \alpha  R_0^{-2} R^{3}+ \gamma R_0^{-2} R \Box R\right]d^4x+S_m,
\end{equation}
which corresponds to the particular case where the functions $f_1\left(R\right)$ and $f_2\left(R\right)$ are given by the forms $f_1=R_0 R+\alpha  R_0^3 R^{3}$ and $f_2= R_0^{-3} R$, respectively. Under these considerations, the only non zero auxiliary quantities from Eq.\eqref{hogdynfun} are
\begin{eqnarray}
&&{\bf X}_1\left(\mathbb{A},\mathbb{R}\right)= 6\mathbb{R}+\frac{216\alpha \mathbb{R}^3}{\mathbb{A}^2},\qquad {\bf X}_2\left(\mathbb{A},\mathbb{R} \right)=\frac{6\gamma \mathbb{R}}{\mathbb{A}^2},\qquad {\bf Y}_1\left(\mathbb{A},\mathbb{R} \right)= 1+\frac{108\alpha \mathbb{R}^2}{\mathbb{A}^2},\nonumber\\
&&{\bf Y}_2\left(\mathbb{A},\mathbb{R} \right)=\frac{\gamma}{\mathbb{A}^2},\qquad {\bf Z}_1\left(\mathbb{A},\mathbb{R} \right) =\frac{36 \alpha  \mathbb{R}}{\mathbb{A}^2},\qquad {\bf W}_1\left(\mathbb{A},\mathbb{R} \right) =\frac{6 \alpha }{\mathbb{A}^2},
\end{eqnarray}
and the cosmological equations can be decoupled to give an explicit equation for $\mathbb{S}_1$ and $\mathbb{S}_2$. We do not show these equations due to their size. The  dynamical system given in Eq.\eqref{hogdynsys6o} becomes then
\begin{eqnarray}
\frac{d\mathbb{R}}{dN}=&&\mathbb{J}+(\mathbb{K}-2) \mathbb{K}-(\mathbb{R}-2)^2 ,\nonumber\\ 
\frac{d\mathbb{B}}{dN}=&&\mathbb{B} (3 \mathbb{K}-3 \mathbb{R}+7)^2+\mathbb{K} \left((8-2 \mathbb{J}) \mathbb{R}-2 \mathbb{J}-2\mathbb{R}^3-2 \mathbb{R}^2+8\right)+\mathbb{K}^2 \left(\mathbb{J}+3 \mathbb{R}^2+4\mathbb{R}-2\right)\nonumber\\
  &&-2\mathbb{K}^3 (\mathbb{R}-1)+\frac{\mathbb{K}^4}{2}+\frac{\mathbb{R}^4}{2}+(\mathbb{J}-4)
   \mathbb{R}^2+\frac{\mathbb{J}-4}{2}+\frac{A^2}{12 \gamma } (-\mathbb{K}+\Omega -1)\nonumber\\
  &&-\frac{\alpha }{\gamma } \left[18 (\mathbb{J}-4) \mathbb{R}+18 \mathbb{K}^2 \mathbb{R}-\mathbb{K}
   \left(27 \mathbb{R}^2+36 \mathbb{R}\right)+12 \mathbb{R}^3+9 \mathbb{R}^2\right]\label{hogdynsys6ob}\\
\frac{d\Omega}{dN}=&&\Omega  (1-3w+2 \mathbb{K}-2 \mathbb{R}),\nonumber\\ 
\frac{d\mathbb{J}}{dN}=&&-\mathbb{B}+\mathbb{J} (5
   \mathbb{K}-5 \mathbb{R}+3)+(\mathbb{K}-\mathbb{R}) \left(\mathbb{K}^2-\mathbb{K} (2 \mathbb{R}+7)+\mathbb{R}(\mathbb{R}+5)\right)-22 \mathbb{K}+20 \mathbb{R}-12,\nonumber\\
\frac{d\mathbb{K}}{dN}=&&2 \mathbb{K} (\mathbb{K}-\mathbb{R}+1),\nonumber\\
\frac{d\mathbb{A}}{dN}=&&2 \mathbb{A}(\mathbb{K}-\mathbb{R}+2)\nonumber .
\end{eqnarray}

The system above in Eq.\eqref{hogdynsys6ob} presents the same invariant submanifolds as the previous system in Eq.\eqref{hogdynsys6oa} and therefore we can draw the same conclusions for the existence of global attractors. Table \ref{Tab:hogfixed2} summarises the fixed points for this system with the associated solution and their stability. All  the solutions associated to the fixed points are characterized by $\mathfrak{s}_2\neq0$ with the exception of $\mathcal{C}$ which is characterised by the solution in Eq.\eqref{hoggeneqscale0}. 

\begin{table*}[h!]
\begin{center}
\caption{Fixed points for the system given by Eq.\eqref{hogdynsys6ob}.} \label{Tab:hogfixed2}
\begin{tabular}{llllll} \hline
Point & Coordinates  & Existence & Stability &  Parameter $\mathfrak{s}_2$ \\
&$\{\mathbb{R},\mathbb{B},\mathbb{J},\Omega,\mathbb{K}, \mathbb{A}\}$ &  &  \\ \hline\\
$\mathcal{A}$ & $\left\{ 0,0,1,0, -1 , 0\right\}$  & Always & Saddle & $-1$\\ \\
$\mathcal{B}$ & $\left\{ 0, 0, 4, 0, 0, 0\right\}$  & Always & Reppeler & $-32$\\ \\
$\mathcal{C}$ & $\left\{ 2, 0, 0,0, 0, 12\sqrt{\alpha}\right\}$ & $\alpha>0$ & Saddle & $0$\\ \\
$\mathcal{G}$ & $\left\{ 12+ \frac{2\gamma}{3\alpha}, 0, 1,0, 11+ \frac{2\gamma}{3\alpha}, 0\right\}$  & $\{\alpha,\gamma\}\neq0$ & Saddle & $\mathfrak{s}_\mathcal{G}$  \\ \\
$\mathcal{H}_1$ & $\left\{\mathbb{R}^*_1 , -6\mathbb{R}^*_1(\mathbb{R}^*_1-1)(\mathbb{R}^*_1-2),(\mathbb{R}^*_1-2)^2,0,0,0\right\}$  & Fig.~\ref{Fig:hogexistH} &   Fig.~\ref{Fig:hogstabR1} & $\sigma_1$  \\ \\
$\mathcal{H}_2$ & $\left\{\mathbb{R}^*_2 , -6\mathbb{R}^*_2(\mathbb{R}^*_2-1)(\mathbb{R}^*_2-2),(\mathbb{R}^*_2-2)^2,0,0,0\right\}$  & Fig.~\ref{Fig:hogexistH} &   Fig.~\ref{Fig:hogstabR2} & $\sigma_2$ \\ \\
$\mathcal{H}_3$ & $\left\{\mathbb{R}^*_3 , -6\mathbb{R}^*_3(\mathbb{R}^*_3-1)(\mathbb{R}^*_3-2),(\mathbb{R}^*_3-2)^2,0,0,0\right\}$  & Fig.~\ref{Fig:hogexistH} &   Saddle & $\sigma_3$  \\ \\
\hline \\ 
\multicolumn{5}{l}{$\mathfrak{s}_\mathcal{G}=-577-5184\frac{\alpha}{\gamma}-11\frac{\gamma}{\alpha}$} \\ \\
\multicolumn{5}{l}{\makecell[l]{$\sigma_i= \frac{\alpha}{\gamma}\left(-150 \mathbb{R}_{*,i}^4+435 \mathbb{R}_{*,i}^3-252 \mathbb{R}_{*,i}^2\right)+101
   \mathbb{R}_{*,i}^5-610 \mathbb{R}_{*,i}^4+1306 \mathbb{R}_{*,i}^3-1180 \mathbb{R}_{*,i}^2+416 \mathbb{R}_{*,i}-32\neq 0$}}\\ \\
\hline\\
 \end{tabular}
   \end{center}
\end{table*}

Some of the fixed points exist only for specific values of the parameters $\alpha$ and $\gamma$. For example, the existence of $\mathcal{C}$ requires $\alpha>0$ and more complex conditions hold for the points $\mathcal{H}_i$ whose coordinates are determined by the equation
\begin{equation}
3 \alpha  (21-10 \mathbb{R}^*_i) \mathbb{R}^*_i+2 \gamma  (\mathbb{R}^*_i-2) [2\mathbb{R}^*_i (5 \mathbb{R}^*_i-16)+21]=0.
\end{equation}

\begin{figure}   
\centering
\includegraphics[scale=0.5]{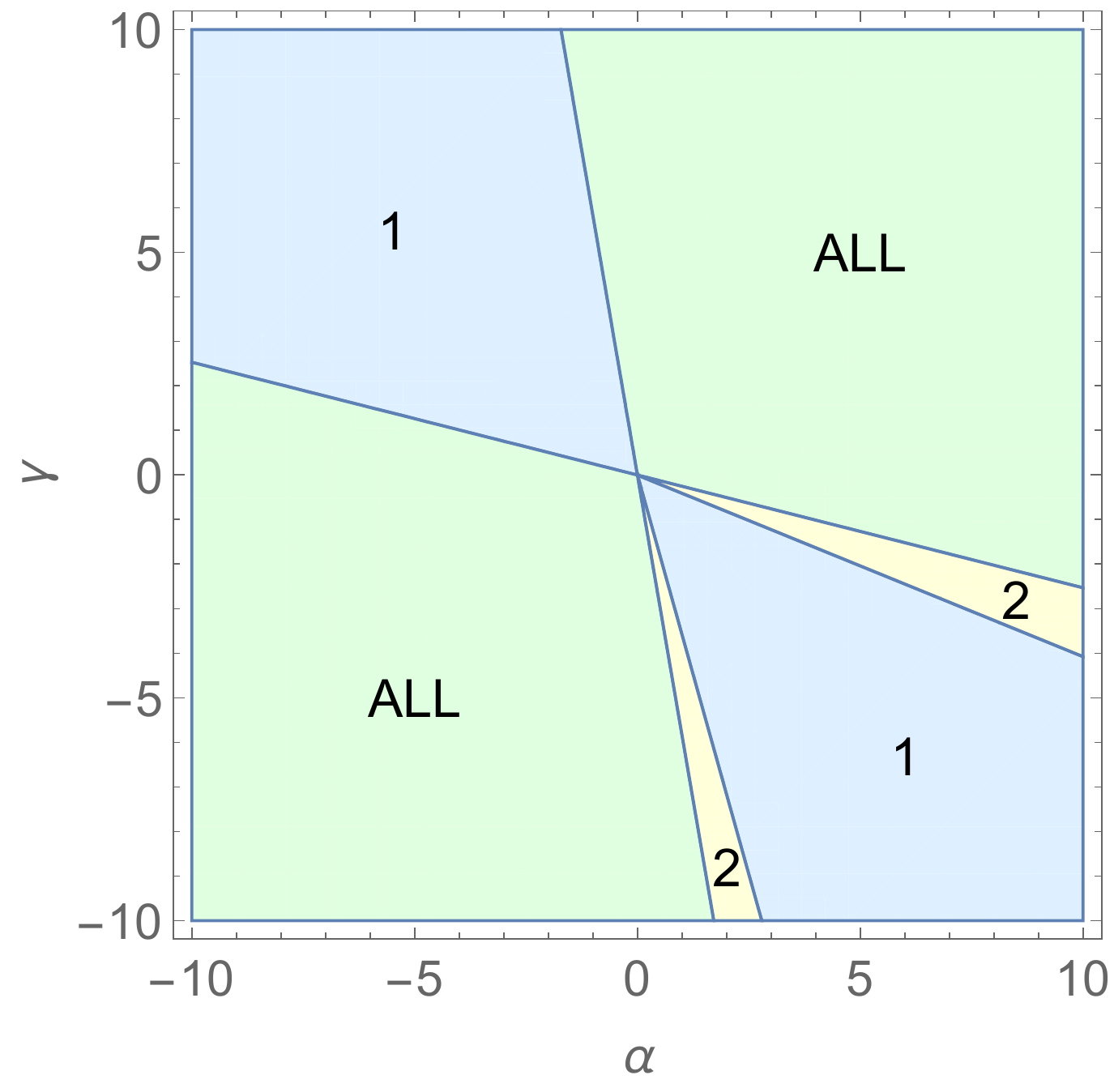}
\vskip -0.3cm
\caption{Region of the parameter space of $\alpha$ and $\gamma$ for which  the fixed points $\mathcal H_i$ in Table\ref{Tab:hogfixed2} exist.}
\label{Fig:hogexistH}
\end{figure}

\begin{figure}
\centering
\includegraphics[scale=0.6]{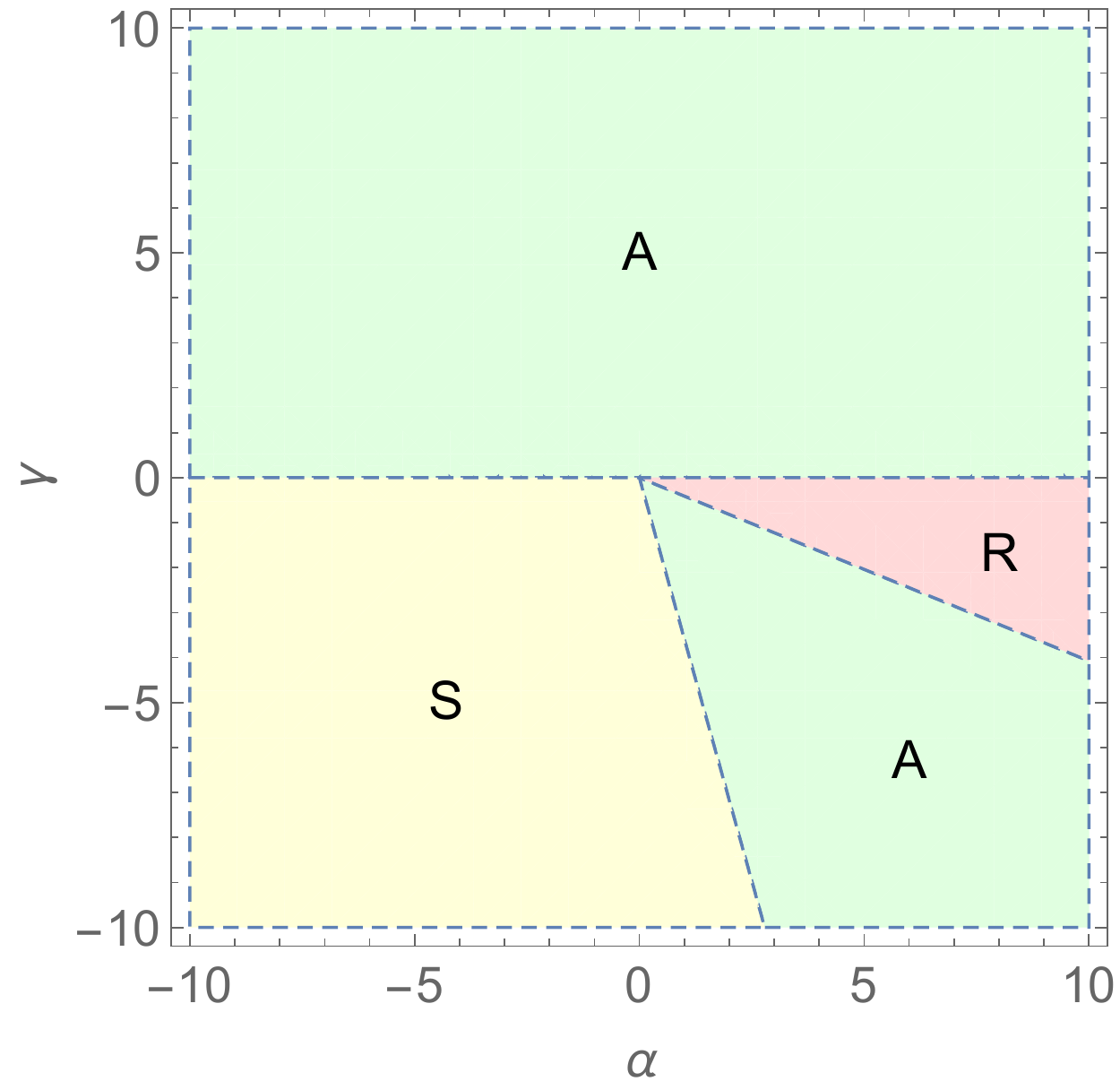}
\vskip -0.3cm
\caption{Stability of the fixed point $\mathcal{H}_1$ in Table\ref{Tab:hogfixed2}. A stands for attractor (green), R stands for repeller (red), and S stands for saddle (yellow).}
\label{Fig:hogstabR1}
\end{figure}

\begin{figure}
\centering
\includegraphics[scale=0.5]{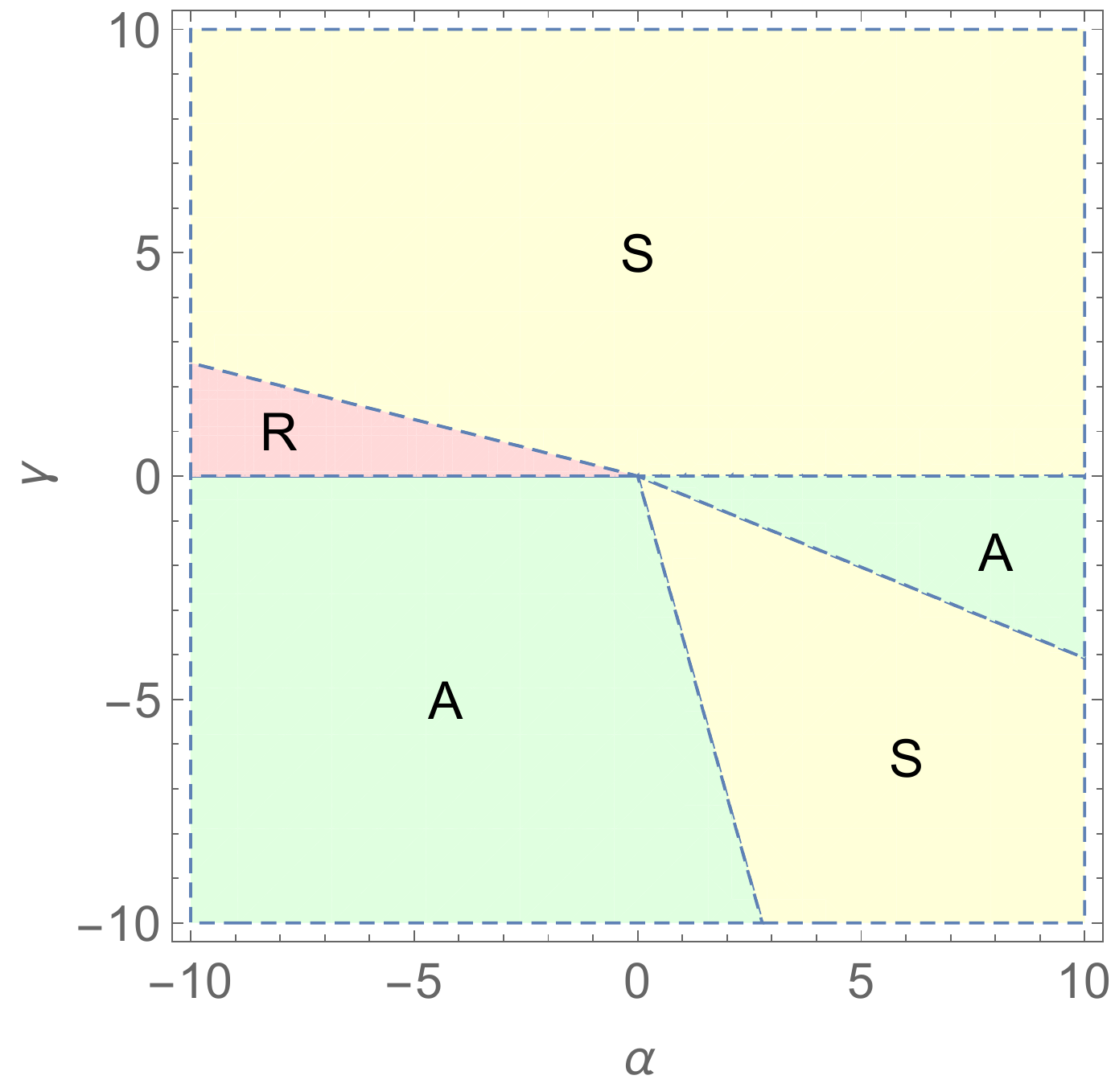}
\vskip -0.3cm
\caption{Stability of the fixed point $\mathcal{H}_2$ in Table\ref{Tab:hogfixed2}. A stands for attractor (green), R stands for repeller (red), and S stands for saddle (yellow).}
\label{Fig:hogstabR2}
\end{figure}

In Fig.~\ref{Fig:hogexistH} we plot the region of existence of these points. With the exception of point $\mathcal{A}$ all the other fixed points are hyperbolic, although their stability depends on the parameters $\alpha$ and $\gamma$. This complex dependence make very complicated to make general statements on the stability of points $\mathcal{H}_i$. We can conclude however that one of these points $\mathcal{H}_3$ is always a saddle. As in the previous case, the stability of point $\mathcal{A}$ can be determined by the analysis of the central manifold. However, from the sign of the other eigenvalues, we can conclude that the point is unstable. In Figs. \ref{Fig:hogstabR1} and \ref{Fig:hogstabR2} we also plot the stability.

\vskip 0.5cm
\vskip 0.5cm

\section{Interaction between fourth and sixth orders}\label{Sec:hogint}

It is useful to compare the results that we have obtained so far with an analysis of fourth order models  made with the same approach (see also Ref. \cite{carloni2} for an equivalent, but slightly different choice of some of the dynamical variables). For simplicity we will consider here a fourth order theory of the form  $f= R+ \alpha R^{q}$. For this choice of $f$ the cosmological equations read
\begin{equation}\label{hogintfriedmann}
 H^{2}+\frac{k}{a^2} =\frac{1}{3(1+\alpha q R^{q-1})}\left[\frac{1}{2}\alpha(q-1)R^q-3\alpha q (q-1) H R^{q-2}\dot{R}+\mu\right]
\end{equation}
\begin{eqnarray}
2\dot{H}+H^{2}+\frac{k}{a^2} =&-&\frac{1}{(1+\alpha q R^{q-1})}\left[\frac{1}{2}\alpha(q-1)R^q-3\alpha q (q-1) H R^{q-2}\dot{R}+\right.\nonumber\\
&+&q(q-1)(q-2)R^{q-3}\dot{R}^2+q(q-1)R^{q-2}R\ddot{R}+\,p\bigg\}\label{hogintraychaudhuri}
\end{eqnarray}
Let us define the following dynamical variables which are a subset of the variable in Eq. \eqref{hogdynvar}
\begin{equation} \label{hogintdynvar}
\mathbb{R}=\frac{R}{6 H^2},\quad \mathbb{K}=\frac{k}{a^2 H^2},\quad \Omega =\frac{\mu }{3H^2},\quad \mathbb{J}=\mathfrak j,\quad \mathbb{Q}={\mathfrak q},\quad \mathbb{A}=\frac{R_0}{H^2}
\end{equation}
With these definitions, the dynamical system can be written as
\begin{eqnarray}
&&\frac{d\mathbb{R}}{dN}=\frac{\mathbb{R}\left[q\left(2 q-3 \right)\K-\left(2 q^2 +3 q+1\right)\mathbb{R} +q \Omega+4 q^2 -5 q\right]}{q(q-1) }-\frac{(K-\Omega +1)}{6^{q-1} \alpha  (q-1) q \A^{q-1}  \mathbb{R}^{q-2}},\nonumber\\
&&\frac{d\Omega}{dN}=\Omega\left(1-3 w+2 \K-2\mathbb{R}\right),\label{hogintdynsys1}\\
&&\frac{d\mathbb{K}}{dN}=2 \K(\K- \mathbb{R}+1),\nonumber\\ 
&&\frac{d\mathbb{A}}{dN}=2\mathbb{A} (2+\K-\mathbb{R})\nonumber .
\end{eqnarray}
with the constraints
\begin{eqnarray}
&&\mathbb{R}=\mathbb{K}+\mathbb{Q}+2\\
&&6\left[(1+\mathbb{K}-\mathbb{R})\left(1+\frac{6^{q-1}\mathbb{R}}{\A^{q-1}}\right)+\mathbb{R}- \Omega\right]+\\
&&+\alpha  6^q \mathbb{A}^{1-q}
   \mathbb{R}^{q-2} \left[q (q-1) \left( \mathbb{J}+\K^2-2 \K-4\right)-2  q (q-1) \K \mathbb{R}+(q^2-q+1) \mathbb{R}^2\right]=0.\nonumber
\end{eqnarray}
The solutions associated to the fixed points can be obtained from the equation
\begin{equation}\label{hoginteqH1}
\mathfrak{s}=\frac{1}{H}\frac{ d^3{H}}{d N^3}
\end{equation}
where $\mathfrak{s}$ is defined in Eq. \eqref{hogdynvar} and its expression in the fixed point can be deduced by Eq.\eqref{hogintraychaudhuri} as we have done for the higher order case. As in the previous sections the solution can be given in general  noting that the characteristic polynomial for this equation has one real root and a pair of complex roots. Hence, we can write an exact solution for $H(N)$:
\begin{equation} \label{hogintsolH1}
H=\exp \left(-p\, N\right)+\exp \left(\frac{1}{2}p N\right)\left[H \cos\left(p\frac{\sqrt{3}}{2} N\right)+\bar{H}\sin\left(p\frac{\sqrt{3}}{2} N\right)\right],\\
\end{equation}
where $p=-\sqrt[3]{\mathfrak{s}^*}$, $H$ and $\bar{H}$ are integration constants. Naturally for $\mathfrak{s}^*=0$ we have the usual equation for the scale factor
\begin{equation}\label{hoginteqscale1}
\dot{a}=a\sum_{i=0}^{2}H_i (\ln a)^i.
\end{equation}
The fixed points for the system in Eq. \eqref{hogintdynsys1} with their stability is presented in Table\ref{Tab:hogintfixed1}.

\begin{table}[h]
\begin{center}
\caption{Fixed points for the system given by Eq.\eqref{hogintdynsys1}.} \label{Tab:hogintfixed1}
\begin{tabular}{llllll} \hline
Point & Coordinates & Existence & Stability & Parameter $\mathfrak s$  \\
&$\{\mathbb{R},\mathbb{J},\mathbb{K},\Omega, \mathbb{A}\}$ &  \\ \hline\\
$\mathcal{A}$ & $\left\{ 0, 1, -1 , 0,0\right\}$ & $\alpha\neq 0$ & Saddle & $-1$ \\ \\
$\mathcal{B}$ & $\left\{ 0,4, 0, 0,0\right\}$ & $\alpha\neq 0$& Reppeler or Saddle & $-8$ \\ \\
$\mathcal{C}$ & $\left\{2, 0, 0, 0, 12\left[\alpha (q-2)\right]^{\frac{1}{q-1}}\right\}$ &\makecell[l]{if $q\in\Re$\\ $\alpha(q-2)>0$} & Attractor if $\frac{32}{25}\lesssim q<2$ & 0\\ \\
$\mathcal{D}$ & $\left\{ 2q(q-1), 1 , 2 (q-1) q-1, 0,0\right\}$  & $q>1$ & Saddle & -1   \\ \\
$\mathcal{E}$ & $\left\{ \frac{(5-4 q) q}{4 q^2-6 q+2}, \left(\frac{q-2}{(q-1)(2q-1)}\right)^3, 0, 0,0\right\}$  &  $q>1$  & Attractor if $q>2$ & $\left(\frac{q-2}{(q-1)(2q-1)}\right)^3$  \\ \\
\hline\\
 \end{tabular}
   \end{center}
\end{table}

Let us now repeat the same analysis for a theory that contains the fourth order term considered above {\it plus} a sixth order term. Consider then the action given by
\begin{equation}
S=\int \sqrt{-g} \left[ R+ \alpha  R_0^{1-q} R^{q}+ \gamma R_0^{-2} R \Box R\right]d^4x+S_m\,,
\end{equation}
which consists of a particular case for which the functions $f_1\left(R\right)$ and $f_2\left(R\right)$ are given by  $f_1=R_0 R+\alpha  R_0^{1-q} R^{q}$ and $f_2= R_0^{-2} R$, respectively. The non zero auxiliary quantities in Eq.\ref{hogdynfun} are
\begin{eqnarray}
&&{\bf X}_1\left(\mathbb{A},\mathbb{R}\right)= 6\mathbb{R}+\alpha 6^q\mathbb{R}^q\mathbb{A}^{1-q},\qquad {\bf X}_2\left(\mathbb{A},\mathbb{R} \right)=\frac{6\gamma \mathbb{R}}{\mathbb{A}^2},\nonumber\\
&&{\bf Y}_1\left(\mathbb{A},\mathbb{R} \right)= 1+6^{q-1}\alpha \mathbb{R}^{q-1}\mathbb{A}^{1-q}, \qquad {\bf Y}_2\left(\mathbb{A},\mathbb{R} \right)=\frac{\gamma}{\mathbb{A}^2},\\
&&{\bf Z}_1\left(\mathbb{A},\mathbb{R} \right) =\alpha q(q-1)6^{q-2} \mathbb{R}^{q-2}\mathbb{A}^{1-q},\qquad {\bf W}_1\left(\mathbb{A},\mathbb{R} \right) =\alpha q(q-1)(q-2)6^{q-3} \mathbb{R}^{q-3}\mathbb{A}^{1-q}\nonumber .
\end{eqnarray}
As before the cosmological equations can be decoupled to give an explicit equation for
$\mathbb{S}_1$ and $\mathbb{S}_2$ and one can construct the dynamical system equations in the form
\begin{eqnarray}
\frac{d\mathbb{R}}{dN}&=&\mathbb{J}+(\mathbb{K}-2) \mathbb{K}-(\mathbb{R}-2)^2 \nonumber \\
\frac{d\mathbb{B}}{dN}&=&\mathbb{B} (3 \mathbb{K}-3 \mathbb{R}+7)+\frac{1}{2} \left(\mathbb{J}+\mathbb{K}^2-2 \mathbb{K}(\mathbb{R}+1)+\mathbb{R}^2-4\right)^2+\frac{\A^2}{12 \gamma } (-\mathbb{K}+\Omega -1)-\nonumber\\
&-&\frac{\alpha }{\gamma } \left\{2^{q-3} 3^{q-2} \A^{3-q} \mathbb{R}^{q-2} \left[q(q-1)  (\mathbb{J}+(\mathbb{K}-2)\mathbb{K}-4)+q \mathbb{R} (\mathbb{K} (3-2 q)+1)+(q-1)^2 \mathbb{R}^2\right]\right\}\nonumber\\
\frac{d\Omega}{dN}&=&\Omega  (1-3w+2 \mathbb{K}-2 \mathbb{R})\label{hogintdynsys2}\\
\frac{d\mathbb{J}}{dN} &=&-\mathbb{B}+\mathbb{J} (5\mathbb{K}-5 \mathbb{R}+3)+(\mathbb{K}-\mathbb{R})\left(\mathbb{K}^2-\mathbb{K} (2 \mathbb{R}+7)+\mathbb{R}(\mathbb{R}+5)\right)-22 \mathbb{K}+20\mathbb{R}-12\nonumber \\
\frac{d\mathbb{K}}{dN}&=&2 \mathbb{K} (\mathbb{K}-\mathbb{R}+1)\nonumber\\
\frac{d\mathbb{A}}{dN}&=&2 \mathbb{A}(\mathbb{K}-\mathbb{R}+2)\nonumber
\end{eqnarray}
In Table \ref{Tab:hogintfixed2} we provide the fixed points and their stability. Although fundamentally different the two phase spaces present some similarities. Points~$\mathcal{A}$, $\mathcal{B}$ and $\mathcal{C}$ have exactly the same coordinates. In Points~$\mathcal{E}$  and Point~$\mathcal{D}$, instead, the relation among the values of some of the coordinates is  the same as the one of Points~$\mathcal{H}$. The difference in the coordinates of these points is probably due to the additional contributions generated in the gravitational field equations by the $R\Box R$ correction. As one could expect, the same additional terms can change the stability of all the fixed points.

\begin{table}[h]
\begin{center}
\caption{Fixed points for the system given by Eq.\eqref{hogintdynsys2} assuming $\{\alpha,\gamma\}\neq0$ and $q\neq 1$.} \label{Tab:hogintfixed2}
\begin{tabular}{llllll} \hline
Point & Coordinates  & Existence &  Stability & Parameter $\mathfrak{s}_2$\\
&$\{\mathbb{R},\mathbb{B},\mathbb{J},\Omega,\mathbb{K},\mathbb{A}\}$ & & \\ \hline\\
$\mathcal{A}$ & $\left\{ 0,0,1,0, -1 , 0\right\}$ & $q\leq 3$ & Saddle & $-1$ \\ \\
$\mathcal{B}$ & $\left\{ 0, 0, 4, 0, 0, 0\right\}$ & always & \makecell[l]{ Reppeler for $w < 1/3$ \\ Saddle for $w > 1/3$} & $-32$\\ \\
$\mathcal{C}$ & $\left\{ 2, 0, 0,0, 0, 12\left[\alpha (q-2)\right]^{\frac{1}{q-1}}\right\}$  &\makecell[l]{if $q\in\Re$\\ $\alpha(q-2)>0$}& Saddle & 0\\ \\
$\mathcal{G}$ & $\left\{ 12+ \frac{2\gamma}{3\alpha}, 0, 1,0, 11+ \frac{2\gamma}{3\alpha}, 0\right\}$ & $q=3$ & Saddle & $\mathfrak{s}_\mathcal{G}$  \\ \\
$\mathcal{H}_1$ & $\left\{\mathbb{R}^*_1 , -6\mathbb{R}^*_1(\mathbb{R}^*_1-1)(\mathbb{R}^*_1-2),(\mathbb{R}^*_1-2)^2,0,0,0\right\}$  & $q=3$, Fig.~\ref{Fig:hogexistH} &   Fig.~\ref{Fig:hogstabR1} & $\sigma_{1}$  \\ \\
$\mathcal{H}_2$ & $\left\{\mathbb{R}^*_2 , -6\mathbb{R}^*_2(\mathbb{R}^*_2-1)(\mathbb{R}^*_2-2),(\mathbb{R}^*_2-2)^2,0,0,0\right\}$ & $q=3$, Fig.~\ref{Fig:hogexistH} &   Fig.~\ref{Fig:hogstabR2} & $\sigma_{2}$  \\ \\
$\mathcal{H}_3$ & $\left\{\mathbb{R}^*_3 , -6\mathbb{R}^*_3(\mathbb{R}^*_3-1)(\mathbb{R}^*_3-2),(\mathbb{R}^*_3-2)^2,0,0,0\right\}$ & $q=3$, Fig.~\ref{Fig:hogexistH} &   Saddle & $\sigma_{3}$  \\ \\
$\mathcal{I}_{\pm}$ & $\left\{a_\mathcal I^\pm ,b_\mathcal I^\pm,c_\mathcal I^\pm,0,0,0\right\}$  & q$<$3 &   Saddle & $\sigma_4$  \\ \\
\hline \\ 
\multicolumn{5}{l}{$\mathfrak{s}_\mathcal{G}=-577-5184\frac{\alpha}{\gamma}-11\frac{\gamma}{\alpha}$} \\ \\
\multicolumn{5}{l}{\makecell[l]{$\sigma_i= \frac{\alpha}{\gamma}\left(-150 \mathbb{R}_{*,i}^4+435 \mathbb{R}_{*,i}^3-252 \mathbb{R}_{*,i}^2\right)+101
   \mathbb{R}_{*,i}^5-610 \mathbb{R}_{*,i}^4+1306 \mathbb{R}_{*,i}^3-1180 \mathbb{R}_{*,i}^2+416 \mathbb{R}_{*,i}-32\neq 0$ }}\\ \\
\multicolumn{5}{l}{ $\sigma_4=5^{-\left(q+1\right)}\left(\pm \frac{4549}{2}-\frac{2689 \sqrt{\frac{23}{2}}}{4}\right)$}\\ \\
\multicolumn{5}{l}{$a_\mathcal I^\pm=\frac{1}{10}\left(16\pm\sqrt{46}\right)\qquad b_\mathcal I^\pm= \frac{1}{50}\left(31\pm4\sqrt{46}\right),\qquad c_\mathcal I^\pm=-\frac{9}{250}\left(74\pm 9\sqrt{46}\right)$} \\ \\
\hline\\
 \end{tabular}
   \end{center}
\end{table}

For our purposes, the most important result of this comparative analysis is the fact that both the phase spaces present the fixed point $\mathcal{C}$. As we have seen, such point is characterised  by the vanishing of the quantity associated to  both $\mathfrak{s}=0$ and $\mathfrak{s}_2=0$, and it can represent a solution with a finite time singularity. Looking at Table~\ref{Tab:hogintfixed1} we see that the fourth order theory  point $\mathcal{C}$ for $32/25<q<2$ is an attractor.  However, in the sixth order theory, it is possible to prove numerically that in the interval $32/25<q<2$ the point $\mathcal{C}$ is always unstable. Therefore we can say that the introduction of the sixth order terms prevents the cosmology to evolve towards $\mathcal{C}$. Effectively, this amounts to "curing" the pathology of the fourth order model as the sixth order terms prevents the occurrence of a finite time singularity. In this sense, we can say that, as the time asymptotic state of sixth order cosmologies is never singular, these models are more "stable" with respect to the appearance of singularities. When we will consider eight order corrections, we will use in  the  results obtained in this section to reach the same conclusion.

\section{Going beyond 6th order}

As mentioned in the previous sections, because of the possibility to transform total divergences in boundary terms the general function $f$ we have considered can be used to investigate the behaviour of gravitational actions in which only the Ricci scalar an its d'Alembertian appear and are of order
$2n+4$, where $n$ is the power that enters into the d'Alembertian of $R$ term. Such possibility implies that the equations we have developed can be employed also to treat these models. The advantage of our approach is that with the variables we will choose we can analyse the dynamics of these models without changing the dimension of the phase space. 

Let us start extending the set of variables used in the previous section, i.e., 
\begin{eqnarray}\label{hogdynvar2}
\mathbb{R}=\frac{R}{6 H^2},\quad \mathbb{B}=\frac{\Box R}{6H^4},\quad  \mathbb{K}=\frac{k}{a^2 H^2},\quad \Omega =\frac{\mu }{3H^2},\quad \mathbb{J}={\mathfrak j},\\
\mathbb{Q}={\mathfrak q},\quad \mathbb{S}={\mathfrak s},\quad \mathbb{S}_1={\mathfrak s}_1 ,\quad \mathbb{S}_2={\mathfrak s}_2, \quad \mathbb{S}_3={\mathfrak s}_3,\quad \mathbb{A}=\frac{R_0}{H^2}.\nonumber
\end{eqnarray}
The Jacobian of this variable definition reads
\begin{equation}
J=-\frac{1}{108 a^2 H^{47}}
\end{equation}
which implies that, as in the sixth-order case, the variables are always regular if $H\neq 0$ and $a \neq 0$. As before, the requirement to have a closed systems of equations requires the introduction of the auxiliary quantities:
\begin{eqnarray}
&& {\bf X}\left(\mathbb{A},\mathbb{R},\mathbb{B}\right)= \frac{f\left(\mathbb{A},\mathbb{R},\mathbb{B}\right)}{H^2},\qquad\qquad {\bf Y}_1\left(\mathbb{A},\mathbb{R},\mathbb{B} \right)= f^{(1,0)}_{R,\Box R}\left(\mathbb{A},\mathbb{R},\mathbb{B}\right),\nonumber\\ 
&&{\bf Y}_2\left(\mathbb{A},\mathbb{R},\mathbb{B} \right)= H^4 f^{(1,1)}_{R,\Box R}\left(\mathbb{A},\mathbb{R},\mathbb{B}\right),\qquad {\bf Y}_3\left(\mathbb{A},\mathbb{R},\mathbb{B} \right)= H^2 f^{(0,1)}_{R,\Box R}\left(\mathbb{A},\mathbb{R},\mathbb{B}\right),\nonumber\\ 
&&{\bf Z}_1\left(\mathbb{A},\mathbb{R},\mathbb{B}\right)= H^2 f^{(2,0)}_{R,\Box R}\left(\mathbb{A},\mathbb{R},\mathbb{B}\right),\qquad {\bf Z}_2\left(\mathbb{A},\mathbb{R},\mathbb{B} \right)=H^6 f^{(2,1)}_{R,\Box R}\left(\mathbb{A},\mathbb{R},\mathbb{B}\right),\nonumber\\
&& {\bf Z}_3\left(\mathbb{A},\mathbb{R},\mathbb{B}\right)= H^{10} f^{(2,2)}_{R,\Box R}\left(\mathbb{A},\mathbb{R},\mathbb{B}\right),\qquad {\bf Z}_4\left(\mathbb{A},\mathbb{R},\mathbb{B} \right)= H^8 f^{(1,2)}_{R,\Box R}\left(\mathbb{A},\mathbb{R},\mathbb{B}\right),\nonumber\\
&&{\bf Z}_5\left(\mathbb{A},\mathbb{R},\mathbb{B} \right)= H^6 f^{(0,2)}_{R,\Box R}\left(\mathbb{A},\mathbb{R},\mathbb{B}\right),\qquad {\bf W}_1\left(\mathbb{A},\mathbb{R},\mathbb{B}\right)=H^4 f^{(3,0)}_{R,\Box R}\left(\mathbb{A},\mathbb{R},\mathbb{B}\right),\label{hogdynfun2}\\ 
&&{\bf W}_2\left(\mathbb{A},\mathbb{R},\mathbb{B} \right)= H^8 f^{(3,1)}_{R,\Box R}\left(\mathbb{A},\mathbb{R},\mathbb{B}\right),\qquad {\bf W}_3\left(\mathbb{A},\mathbb{R},\mathbb{B}\right)= H^{12} f^{(3,2)}_{R,\Box R}\left(\mathbb{A},\mathbb{R},\mathbb{B}\right),\nonumber \\ 
&& {\bf W}_4\left(\mathbb{A},\mathbb{R} ,\mathbb{B}\right)= H^{14} f^{(2,3)}_{R,\Box R}\left(\mathbb{A},\mathbb{R},\mathbb{B}\right),\qquad {\bf W}_5\left(\mathbb{A},\mathbb{R},\mathbb{B}\right)= H^{12} f^{(1,3)}_{R,\Box R}\left(\mathbb{A},\mathbb{R},\mathbb{B}\right),\nonumber\\
&&{\bf W}_6\left(\mathbb{A},\mathbb{R},\mathbb{B} \right)= H^{10} f^{(0,3)}_{R,\Box R}\left(\mathbb{A},\mathbb{R},\mathbb{B}\right),\qquad {\bf T}_1\left(\mathbb{A},\mathbb{R},\mathbb{B}\right)= H^{14} f^{(0,4)}_{R,\Box R}\left(\mathbb{A},\mathbb{R},\mathbb{B}\right),\nonumber\\
&&{\bf T}_2\left(\mathbb{A},\mathbb{R},\mathbb{B} \right)= H^{16} f^{(1,4)}_{R,\Box R}\left(\mathbb{A},\mathbb{R},\mathbb{B}\right),\qquad {\bf T}_3\left(\mathbb{A},\mathbb{R},\mathbb{B}\right)= H^{10} f^{(4,1)}_{R,\Box R}\left(\mathbb{A},\mathbb{R},\mathbb{B}\right),\nonumber\\ 
&&{\bf V}\left(\mathbb{A},\mathbb{R},\mathbb{B}\right)= H^{18} f^{(0,5)}_{R,\Box R}\left(\mathbb{A},\mathbb{R},\mathbb{B}\right),\nonumber
\end{eqnarray}
where, for simplicity, we indicate with $f^{(i,j)}_{R,\Box R}$ the $i$-th  derivative of $f$ with respect to R and the $j$-th derivative of $f$ with respect to $\Box R$. Using the previous definitions, the cosmological dynamics can then be described by the autonomous system 
\begin{eqnarray}
&&\frac{d\mathbb{R}}{dN}=\mathbb{J}-2 \mathbb{K}-2 \mathbb{Q} \mathbb{R}+\mathbb{Q} (\mathbb{Q}+4) ,\nonumber\\ 
&&\frac{d\mathbb{B}}{dN}= -4 \mathbb{B} \mathbb{Q}-4 \mathbb{J}^2+\mathbb{J} (2 \mathbb{K}-\mathbb{Q} (11
\mathbb{Q}+43)-12)-\nonumber\\
&&\qquad\mathbb{Q} (\mathbb{Q} (-2 \mathbb{K}+\mathbb{Q} (\mathbb{Q}+22)+36)+7 \mathbb{S})-4\mathbb{K}-7 \mathbb{S}-\mathbb{S}_1,\nonumber\\ 
&&\frac{d\Omega}{dN}=- \Omega  (2 \mathbb{Q}+3w+3),\nonumber\\ 
&&\frac{d\mathbb{J}}{dN}=\mathbb{S}-\mathbb{J} \mathbb{Q},\nonumber\\
&&\frac{d\mathbb{Q}}{dN}=\mathbb{J}-\mathbb{Q}^2,\label{hogdynsys8o}\\ 
&&\frac{d\mathbb{K}}{dN}=-2 \mathbb{K} (\mathbb{Q}+1),\nonumber\\
&&\frac{d\mathbb{S}}{dN}=\mathbb{S}_1-\mathbb{Q} \mathbb{S},\nonumber\\ 
&&\frac{d\mathbb{S}_1}{dN}=\mathbb{S}_2-\mathbb{Q}  \mathbb{S}_1,\nonumber\\
&&\frac{d\mathbb{S}_2}{dN}=\mathbb{S}_3({\bf X},{\bf Y}_1,..)-\mathbb{Q}  \mathbb{S}_2,\nonumber\\ 
&&\frac{d\mathbb{S}_3}{dN}=\mathbb{S}_4({\bf X},{\bf Y}_1,..)-\mathbb{Q}  \mathbb{S}_3({\bf X},{\bf Y}_1,..), \nonumber\\
&&\frac{d\mathbb{A}}{dN}=-2 A\mathbb{Q},\nonumber
\end{eqnarray}
where  $\mathbb{S}_4={\mathfrak s}_4$. As before, the system above is completed by three constraints: the one coming from the modified Friedmann equation, Eq. \eqref{hogfriedmann}, and the definitions of $R$ and $\Box R$ in Eqs.\eqref{hogdefR} and \eqref{hogdefboxR}, respectively. We choose to use these constraints to eliminate $\mathbb{Q}, \mathbb{S}$, and $\mathbb{S}_3$. The variable $\mathbb{S}_4$ instead, can substituted using the modified Raychaudhuri equation, given by Eq.\eqref{hograychaudhuri}. In Eq. \eqref{hogdynsys8o} these variables are not substituted explicitly in order to give a more compact representation of the system. The substitution of $\mathbb{S}_3$ and $\mathbb{S}_4$ also brings in the system the parameters defined in Eq.\eqref{hogdynfun2}. 

The solutions associated to the fixed points can be found by solving the differential equation
\begin{equation}\label{hoggenequation2}
\frac{1}{H}\frac{ d^7{H}}{d N^7}=\mathfrak{s}^*_4,
\end{equation}
where $\mathfrak{s}^*_4$ is provided by the modified Raychaudhuri equation, Eq. \eqref{hograychaudhuri}. Eq. \eqref{hoggenequation2} can be solved in general to give a result structurally similar to the one of the previous section, i.e., 
\begin{equation} \label{hoggensolH2}
H=\sum_{i=0}^{3}\exp \left(p\,\alpha_i N\right)\left[H_i \cos\left(\beta_i p N\right)+\bar{H}_i \sin\left(\beta_i p N\right)\right], 
\end{equation}
where $p=-\sqrt[7]{\mathfrak{s}^*_4}$, $H_i$ and $\bar{H}_i$ are integration constants and $a_i$ and $b_i$ are the real and imaginary part of the seventh root of the unity. These quantities are expressed by the relation
\begin{equation}
\begin{array}{ll}
\alpha_0=-1, & \beta_0=0,\\
\alpha_i=\frac{r}{2}, &i\neq0,\\
\beta_i=\sqrt{1-\frac{r^2}{4}}, &i\neq0,\,
\end{array}
\end{equation}
where $r$ is the solution of the algebraic equation $r^3+r^2-2r-1=0$. The scale factor is given by the equation
\begin{equation}\label{hoggeneqscale2}
\dot{a}=\sum_{i=0}^{3}a^{1+p\,\alpha_i}\left[H_i \cos\left(\beta_i p \ln a\right)+\bar{H}_i \sin\left(\beta_i p \ln a\right)\right],\\
\end{equation} 
which can be solved numerically. If $\mathfrak{s}^*_4=0$ then Eq.\eqref{hoggeneqscale2} can be written as
\begin{equation}\label{hoggeneqN2}
\dot{N}=\sum_{i=0}^{7}H_i N^i,
\end{equation}
and the existence of a finite time singularity is only possible if the polynomial on the left hand side has  complex roots. In the following we consider two examples of theories of order eight. As in the previous section we will first examine a model in which only contribution of order eight appear and in the second we will explore the interaction of the eight-order terms with the fourth-order ones. In the following examples, the dynamical system of equations given in Eq.\eqref{hogdynsys8o} becomes extremely long, and thus we shall not write the equations explicitly.

\subsection{$R+\gamma\left(\Box R\right)^2$ gravity}\label{Sec:hog81}

We start again by examining a model for which the action consists only of the Einstein-Hilbert term $R$ and one higher order term, which in this case we consider to be the eight-order term $\left(\Box R\right)^2$, so we can analyse how the higher order term affects the results. The action for this theory can be written as 
\begin{equation}\label{hogaction8o1}
S=\int\sqrt{-g} \left[R+ \gamma R_0^{-3}  (\Box R)^2\right] d^4x+S_m.
\end{equation}
For this simple version of the theory, the only non zero auxiliary quantities defined in Eq.\eqref{hogdynfun2} are
\begin{equation}
{\bf X}\left(\mathbb{A},\mathbb{R},\mathbb{B}\right)= 6\left(\mathbb{R}+\frac{6\gamma\mathbb{B}}{\mathbb{A}^2}\right), \qquad{\bf Y}_1\left(\mathbb{A},\mathbb{R},\mathbb{B} \right)= 1,\qquad {\bf Y}_3\left(\mathbb{A},\mathbb{R} ,\mathbb{B}\right)= \frac{12\gamma\mathbb{B}}{\mathbb{A}^3},\qquad {\bf Z}_5\left(\mathbb{A},\mathbb{R},\mathbb{B} \right) =\frac{2\gamma}{\mathbb{A}^3}.
\end{equation}
The cosmological equations can be decoupled to give explicit equations for $\mathbb{S}_3$ and $\mathbb{S}_4$. Due to their size, however, we choose not to write them explicitly, along with the dynamical system in Eq.\eqref{hogdynsys8o}. Instead, we shall just describe that the dynamical system presents the  invariant submanifolds ($\mathbb{A}=0$, $ \Omega=0$, $\mathbb{K}=0$) and therefore no global attractor with coordinates different from $\mathbb{A}=0$, $ \Omega=0$, $\mathbb{K}=0$ can exist. Table \ref{Tab:hogfixed3} summarises the fixed points for this system with the associated solution and their stability.  The system presents a line of fixed points, all unstable, and a global attractor, point $\mathcal{H}_2$, which is associated with a solution with non zero $\mathfrak{s}_4$. Points $\mathcal{A}$ and  $\mathcal{C}$ are non hyperbolic, the latter having two zero eigenvalues, but they can be both considered unstable. A detailed treatment of the stability of $\mathcal{C}$ would require blow up techniques, but for the scope of this thesis it is enough to state that it is a saddle point.

\begin{table*}[h!]
\begin{center}
\caption{Fixed points for the system given by Eq.\eqref{hogdynsys8o} in the particular case given in Eq.\eqref{hogaction8o1}} \label{Tab:hogfixed3}
\begin{tabular}{llllll} \hline
Point & Coordinates &  Existence&  Stability & Parameter $\mathfrak{s}_4$ \\
&$\{\mathbb{R},\mathbb{B},\mathbb{J},\mathbb{S}_1,\mathbb{S}_2,\mathbb{K},\Omega, \mathbb{A}\}$ &  &  \\ \hline\\
$\mathcal{A}$ & $\left\{ 1,0,1,1, -1,0, 0 , 0\right\}$  & $\gamma\neq 0$ & Saddle & $-1$ \\ \\
$\mathcal{B}$ & $\left\{0,0, 4, 16 , -32, 0, 0, 0\right\}$  & $\gamma\neq 0$ & \makecell[c]{Reppeler for $w<1/3$\\Saddle for $w>1/3$} & $-128$\\ \\
$\mathcal{C}$ & $\left\{2,0, 0,0,0, 0, 0, 0\right\}$ & $\gamma\neq 0$ & Saddle & $0$\\ \\
$\mathcal{H}_1$ & \makecell[l]{$\left\{ a_\mathcal{H}^-,  -6a_\mathcal{H}^-(a_\mathcal{H}^--1)(a_\mathcal{H}^--2),(a_\mathcal{H}^--2)^2,\right.$\\$\left.(a_\mathcal{H}^--2)^4,(a_\mathcal{H}^--2)^5,0,0,0\right\}$} & $\gamma\neq 0$ & Saddle &$\approx -7.8\times 10^{-3}$\\ 
&&&&\\\\
$\mathcal{H}_2$ & \makecell[l]{$\left\{ a_\mathcal{H}^+,  -6a_\mathcal{H}^-(a_\mathcal{H}^+-1)(a_\mathcal{H}^+-2),(a_\mathcal{H}^+-2)^2,\right.$\\$\left.(a_\mathcal{H}^+-2)^4,(a_\mathcal{H}^+-2)^5,0,0,0\right\}$} & $\gamma\neq 0$ & Attractor & $\approx 5.6\times 10^{-5}$\\ 
&&&&\\\\
\hline 
Line & Coordinates & Existence&  Stability & Parameter $\mathfrak{s}_4$ \\
&$\{\mathbb{R},\mathbb{B},\mathbb{J},\mathbb{S}_1,\mathbb{S}_2,\mathbb{K},\Omega, \mathbb{A}\}$ &  &  \\ \hline\\
\multirow{2}{*}{$\mathcal{L}$} & \multirow{2}{*}{$\left\{ \mathbb{R}_*, 0, 1, 1, -1,0, \mathbb{R}_*-1, 0\right\}$ } &\multirow{2}{*}{Always}&\multirow{2}{*}{Saddle} & \multirow{2}{*}{$-1$}  \\
  && && \\\hline\\
\multicolumn{5}{c}{\makecell[c]{$a_\mathcal{I}^\pm=\frac{1}{210}\left(373\pm\sqrt{9769}\right) $}}\\\\\hline\\
 \end{tabular}
   \end{center}
\end{table*}

\subsection{$R+ \alpha R^q + \gamma \left(\Box R\right)^2$ gravity}\label{Sec:hog82}

Now, let us consider an example where the action is described by the Einstein-Hilbert term plus two higher order terms, one of 4th order and one of 8th order, in such a way that we can analyze the interaction between these orders. The action for the theory can be written as
\begin{equation}\label{hogaction8o2}
S=\int \sqrt{-g} \left[R+\alpha R_0^{1-q} R^q + \gamma R_0^{-6}  (\Box R)^2\right]d^4x+S_m.
\end{equation}
The structure of the correction shows that we are dealing here with a theory of order ten in the derivative of the metric.  For this theory the only non zero auxiliary quantities defined in Eq.\eqref{hogdynfun2} are
\begin{eqnarray}
&&{\bf X}\left(\mathbb{A},\mathbb{R},\mathbb{B}\right)= 6\left(\mathbb{R}+\frac{6\gamma\mathbb{B}}{\mathbb{A}^2}+\alpha 6^q\mathbb{R}^q\mathbb{A}^{1-q}\right),\qquad {\bf Y}_1\left(\mathbb{A},\mathbb{R},\mathbb{B} \right)= 1+6^{q-1}\alpha \mathbb{R}^{q-1}\mathbb{A}^{1-q},\qquad {\bf Y}_3\left(\mathbb{A},\mathbb{R} ,\mathbb{B}\right)= \frac{12\gamma\mathbb{B}}{\mathbb{A}^3},\nonumber\\
&&{\bf Z}_1\left(\mathbb{A},\mathbb{R},\mathbb{B} \right) =\alpha q(q-1)6^{q-2} \mathbb{R}^{q-2}\mathbb{A}^{1-q},\qquad {\bf Z}_5\left(\mathbb{A},\mathbb{R},\mathbb{B} \right) =\frac{2\gamma}{\mathbb{A}^3},\qquad {\bf W}_1\left(\mathbb{A},\mathbb{R} ,\mathbb{B}\right) =\alpha q(q-1)(q-2)6^{q-3} \mathbb{R}^{q-3}\mathbb{A}^{1-q}.
\end{eqnarray}

As before, the cosmological equations can be decoupled to give an explicit equation for $\mathbb{S}_3$ and  another for $\mathbb{S}_4$. However we will not show them here due to their size. We shall also not write the dynamical system provided in Eq.\eqref{hogdynsys8o} explicitly for the same reason. Instead, let us just comment that the system presents analogies with the ones of the previous examples: the invariant submanifolds  present  in these cases are also $\mathbb{A}=0$, $ \Omega=0$, $\mathbb{K}=0$ and therefore the only possible type global attractor must lay on the intersection of these coordinates. The fixed points with their stability and the parameter $\mathfrak{s}_4$ that characterise the solution is given in Table \ref{Tab:hogfixed4}. In this case none of the fixed points in the finite phase space is an attractor.  

\begin{table*}
\begin{center}
\caption{Fixed points for the system given by Eq.\eqref{hogdynsys8o} in the particular case given in Eq.\eqref{hogaction8o2}}
\label{Tab:hogfixed4}
\begin{tabular}{llllll} \hline
Point & Coordinates & Existence&  Stability & Parameter $\mathfrak{s}_4$ \\ 
&$\{\mathbb{R},\mathbb{B},\mathbb{J},\mathbb{S}_1,\mathbb{S}_2,\mathbb{K},\Omega, \mathbb{A}\}$ &  &  \\ \hline\\
$\mathcal{A}$ & $\left\{ 1,0,1,1, -1,0, 0 , 0\right\}$ & $q\leq3$ & Saddle & $-1$ \\ \\
$\mathcal{B}$ & $\left\{0,0, 4, 16 , -32, 0, 0, 0\right\}$ & $\gamma\neq 0$ & \makecell[c]{Reppeler for $w<1/3$\\Saddle for $w>1/3$} & $-128$\\ \\
$\mathcal{C}_1$ & $\left\{2,0, 0,0,0, 0, 0,12\left[\alpha (q-2)\right]^{\frac{1}{q-1}} \right\}$ &\makecell[l]{if $q\in\Re$\\ $\alpha(q-2)>0$}& Saddle & $0$ \\ \\
$\mathcal{C}_2$ & $\left\{2,0, 0,0,0, 0, 0, 0\right\}$  & $\gamma\neq 0$ & Saddle & $0$\\ \\
$\mathcal{H}_i$ & \makecell[l]{$\left\{\mathbb{R}^*_i , -6\mathbb{R}^*_i(\mathbb{R}^*_i-1)(\mathbb{R}^*_i-2),(\mathbb{R}^*_i-2)^2,\right.$\\$\left.(\mathbb{R}^*_i-2)^4,(\mathbb{R}^*_i-2)^5,0,0,0\right\}$}
& $q=4$ & \makecell[l]{One Attractor for\\$\left|\frac{\alpha}{\gamma}\right|\gtrsim 0.011$\\$\left|\frac{\alpha}{\gamma}\right|\lesssim 0.0035$\\ other points\\ unstable} & $\sigma_{i}$\\ \\
$\mathcal{I}_1$ & \makecell[l]{$\left\{ a_\mathcal{H}^-,  -6a_\mathcal{H}^-(a_\mathcal{H}^--1)(a_\mathcal{H}^--2),(a_\mathcal{H}^--2)^2,\right.$\\$\left.(a_\mathcal{H}^--2)^4,(a_\mathcal{H}^--2)^5,0,0,0\right\}$} & $q\leq3$ & Saddle & $\approx -7.8\times 10^{-3}$\\ 
&&&&\\\\
$\mathcal{I}_2$ & \makecell[l]{$\left\{ a_\mathcal{H}^+,  -6a_\mathcal{H}^+(a_\mathcal{H}^+-1)(a_\mathcal{H}^+-2),(a_\mathcal{H}^+-2)^2,\right.$\\$\left.(a_\mathcal{H}^+-2)^4,(a_\mathcal{H}^+-2)^5,0,0,0\right\}$} & $q\leq3$& Saddle & $\approx 5.6\times 10^{-5}$\\ 
&&&&\\\\
\hline 
Line & Coordinates $\{\mathbb{R},\mathbb{B},\mathbb{J},\mathbb{S}_1,\mathbb{S}_2,\mathbb{K},\Omega, \A\}$ & Solution & Existence&  Stability \\ \hline
\multirow{2}{*}{$\mathcal{L}$} & \multirow{2}{*}{$\left\{ \mathbb{R}_*, 0, 1, 1, -1,0, \mathbb{R}_*-1, 0\right\}$ } &\multirow{2}{*}{$q\leq3$}&\multirow{2}{*}{Saddle} & \multirow{2}{*}{$-1$} \\
  && && \\\hline\\
\multicolumn{5}{c}{\makecell[c]{$a_\mathcal{H}^\pm=\frac{1}{210}\left(373\pm\sqrt{9769}\right) $}}\\\\\hline\hline\\
 \end{tabular}
   \end{center}
\end{table*}

\subsection{$R+ \alpha R^4 + \beta R\Box R+ \gamma \left(\Box R\right)^2$ gravity}\label{Sec:hog83}

To finalize, let us consider an example in which fourth, sicth and eight order corrections are present in the action.For the fourth order term we consider a correction of the type $\alpha R^4$ to reduce the number of the parameters involved in the analysis. The action can thus be written as
\begin{equation}\label{hogaction8o3}
S=\int \sqrt{-g} \left[R+\alpha R_0^{-3} R^4 +\beta R_0^{-2} R \Box R+ \gamma R_0^{-6}  (\Box R)^2\right]d^4x+S_m.
\end{equation}
For this theory the only non zero auxiliary quantities in Eq.~\eqref{hogdynvar2} are
\begin{eqnarray}
&&{\bf X}\left(\mathbb{A},\mathbb{R},\mathbb{B}\right)= 6\left(\mathbb{R}+\alpha\frac{6^3   \mathbb{R}^3}{\mathbb{A}^3}+\beta\frac{6^2 \mathbb{B} \mathbb{R}}{\mathbb{A}^2}+\gamma\frac{6\mathbb{B}}{\mathbb{A}^3}\right),\qquad {\bf Y}_1\left(\mathbb{A},\mathbb{R},\mathbb{B} \right)= 1+4 \alpha\frac{ 6^3 \mathbb{R}^3}{\mathbb{A}^3},\nonumber \\
&&{\bf Y}_2\left(\mathbb{A},\mathbb{R} ,\mathbb{B}\right)= \frac{\beta}{\mathbb{A}^2},\qquad {\bf Y}_3\left(\mathbb{A},\mathbb{R} ,\mathbb{B}\right)= 6\left(\beta\frac{ 6 \mathbb{R}}{\mathbb{A}^2}+\gamma\frac{2\mathbb{B}}{\mathbb{A}^3}\right),\qquad {\bf Z}_1\left(\mathbb{A},\mathbb{R},\mathbb{B} \right) = \frac{6^3 \alpha \mathbb{R}^2}{\mathbb{A}^3},\nonumber \\
&&{\bf Z}_5\left(\mathbb{A},\mathbb{R},\mathbb{B} \right) =\frac{2\gamma}{\mathbb{A}^3},\qquad {\bf W}_1\left(\mathbb{A},\mathbb{R} ,\mathbb{B}\right) =\frac{144 \alpha }{\mathbb{A}^2}.
\end{eqnarray}
As before, the cosmological equations can be decoupled to give an explicit equation for $\mathbb{S}_3$ and  another for $\mathbb{S}_4$. However, we will not show them here due to their size. Also, we shall also avoid writing the resultant dynamical system for the same reason. This system presents the usual invariant submanifolds  $\mathbb{A}=0$, $ \Omega=0$, $\mathbb{K}=0$. The fixed points with their stability and the parameter $\mathfrak{s}_4$ that characterise the solution is given in Table \ref{Tab:hogfixed5}. The dynamics of this case is very similar to the one of the previous case, with the difference that the line of fixed points is not present. The only possible attractor is given by one of the points $\mathcal{H}_i$ whereas all the other points are unstable.

\begin{table*}
\begin{center}
\caption{Fixed points for the system given by Eq.\eqref{hogdynsys8o} in the particular case given in Eq.\eqref{hogaction8o3}}
\label{Tab:hogfixed5}
\begin{tabular}{llllll} \hline
Point & Coordinates & Existence&  Stability & Parameter $\mathfrak{s}_4$  \\ 
& $\{\mathbb{R},\mathbb{B},\mathbb{J},\mathbb{S}_1,\mathbb{S}_2,\mathbb{K},\Omega, \mathbb{A}\}$ & & & \\\hline\\
$\mathcal{A}$ & $\left\{ 1,0,1,1, -1,0, 0 , 0\right\}$ & $\alpha,\beta, \gamma \neq 0$ & Saddle & $-1$ \\ \\
$\mathcal{B}$ & $\left\{0,0, 4, 16 , -32, 0, 0, 0\right\}$ & $\alpha,\beta, \gamma \neq 0$ & \makecell[l]{Reppeler for $w<1/3$\\Saddle for $w>1/3$} & $-128$\\ \\
$\mathcal{C}$ & $\left\{2,0, 0,0,0, 0, 0,12\left[\alpha (q-2)\right]^{\frac{1}{q-1}} \right\}$ &\makecell[l]{if $q\in\Re$\\ $\alpha(q-2)>0$}& Saddle & $0$\\ \\
$\mathcal{I}_1$ & $\left\{24,0, 1,1,-1, 23, 0, 0\right\}$ & $\alpha,\beta, \gamma \neq 0$ & Saddle & $-1$\\ \\
$\mathcal{H}_i$ & \makecell[l]{$\left\{\mathbb{R}^*_i , -6\mathbb{R}^*_i(\mathbb{R}^*_i-1)(\mathbb{R}^*_i-2),(\mathbb{R}^*_i-2)^2,\right.$\\$\left.(\mathbb{R}^*_i-2)^4,(\mathbb{R}^*_i-2)^5,0,0,0\right\} $}& $\alpha,\beta, \gamma \neq 0$ & \makecell[l]{One Attractor for\\$-0.0044\lesssim \frac{\alpha}{\gamma}\lesssim -0.0060$} & $\sigma_{i}$\\
  && && \\\hline\\
\multicolumn{5}{c}{\makecell[l]{$\sigma_{i}=\frac{\alpha}{\gamma}\left(-150 \mathbb{R}_{*,i}^4+435 \mathbb{R}_{*,i}^3-252 \mathbb{R}_{*,i}^2\right)+101
   \mathbb{R}_{*,i}^5-610 \mathbb{R}_{*,i}^4+1306 \mathbb{R}_{*,i}^3-1180 \mathbb{R}_{*,i}^2+416 \mathbb{R}_{*,i}-32\neq 0$}}\\ 
 \\\hline\hline\\
 \end{tabular}
   \end{center}
\end{table*}

\section{Conclusions}

The structure of the phase space has similarities in all of the particular cases studied. For example, all of those cases feature a fixed point which is a past attractor, that we denoted as point $\mathcal B$. This point can be a global feature of the phase space, as it lays in the intersection of all the invariant submanifolds $\mathbb{A}=0$, $ \Omega=0$, $\mathbb{K}=0$. Also, fixed points $\mathcal A$ and $\mathcal C$ exist in all the cases studied and they are always unstable. 

Concerning the attractors of the theory, we find that for the model of Sec. \ref{Sec:hog61} there exists one global attractor, point $\mathcal{I}_2$. This point is charcterised by $\mathbb{B}\neq 0$, i.e., it represents a state in which the higher-order terms $\Box R$ of the theory are dominant. This is an unexpected result, as it is normally assumed that these terms to be less and less important as the curvature becomes smaller and smaller.  The fact that $\mathcal{I}_2$ is an attractor seems to indicate that instead the cosmology of these theories tends to a state with $\mathbb{B}\neq 0$. Such a state is represented by a solution in which the scale factor converges to a constant value asymptotically. Thus, with respect to their four order counterparts this sixth-order model seems to be more stable in the sense that theories of this type do not incur in these singularities. It is interesting that a similar behaviour is found also in the case of Sec. \ref{Sec:hog81} in which the order of the correction of the Hilbert-Einstein term is of order eight.

In Sec.~\ref{Sec:hog62} we have put at test the robustness of the previous result considering a theory which contains a fourth order term on top of the sixth order one. We have that even in this case in the action the time asymptotic phase space is characterised by a $\mathbb{B}\neq 0$ and therefore to a static universe, while the fixed point $\mathcal C$ is unstable. This is an interesting phenomenon as in Ref.~\cite{carloni2} it was shown that points of the type $\mathcal C$ are very often attractors in the phase space for fourth order models. This result suggest that higher order terms might "cure" the pathologies induced by the fourth order ones. 

In Sec.~\ref{Sec:hogint} we have  given an explicit analysis of this possibility. In particular, we have shown that the same fixed point $\mathcal C$ appears in the phase space for the theory  $f= R+ \alpha R^{q}$ and  $f= R+ \alpha R^{q} +\gamma R \Box R$.  For the values of the parameter $q$ for which $\mathcal C$  is an attractor the fourth order model, the same point is unstable (saddle). This indicates that the inclusion of sixth order terms is able to prevent the onset of a singularity that would otherwise plague its fourth order  counterpart. In this sense, the sixth-order theory seems to be "more stable". The final state of the cosmology, however, depends on the value of $q$. In the specific case $q=3$ this endpoint is represented by one of the points $\mathcal H$. However this is not true for all values of $q$, as the points ${\mathcal H}_i$ do not exist for $q\neq 3$. In this case the final state of the cosmology is probably a point in the asymptotic part of the phase space which we have not explored here.  Clearly we have considered here only a particular example and therefore we cannot prove that this behavior is general. However, the fact that any fourth order model analysed in Ref.~\cite{carloni2}  presents one or more points of the type  $\mathcal C$ and that we always expect a change in stability of the corresponding sixth order theory, suggests that we are reporting here a general phenomenon. 
 
The phase space structure is basically the same when one introduces eight order terms. When these terms are added directly to the Hilbert-Einstein Lagrangian the attractor of the new theory is a static cosmology which corresponds to the dominance of the eight order terms. When also fourth order terms are introduced, we observe the same phenomenon observed in the case of sixth order actions: the potentially pathological fixed point that is present in the fourth order gravity phase space becomes unstable for every value of the parameters. We conclude therefore that, like for sixth order terms, also the inclusion eight order therms is able to avoid the onset of singularities. We also considered a model in which fourth, sixth and eight order terms are present in order to estimate their comparative effect. However, in the formalism we have chosen, six and eight-order are indistinguishable. A different set of variables might resolve this degeneracy.

To summarize, we have applied dynamical systems techniques to analyse the structure of the phase space of $f\left(R,\Box R\right)$ gravity. This class of theories can be proven to represent theories of order
$2n+4$, where $n$ is the power that enters into the d'Alembertian of $R$ term. Our choice of dynamical variables allows us to study the cosmological phase space of this entire class of theories by means of a phase space which has at most dimension eight. We have then considered some examples of theories of order six and of order eight designed specifically to highlight the influence that higher-order terms have on the cosmological evolution. We found that there is complex interplay between terms of different order which make the time asymptotic behaviour of these cosmological model non trivial and not easily deducible form their lower order counterpart.

Connecting our results with the ones available in literature, we can state that our analysis confirms the result of the absence of a double inflationary phase in theories of order six in full accord with the results of \cite{amendola1,gottlober1}. Indeed we are able to extend this conclusion also to theories of order eight and there seems to be no reason to believe that this behaviour will change considering higher-orders. This might indicate that no theory of the type $f(R, \Box R)$ is indeed able to generate multiple inflationary phases in spite of their multiple scalar field representation.  
\cleardoublepage

\chapter{Derrick's theorem in modified theories of gravity}
\label{chapter:chapter9}

We extend Derrick's theorem to the case of a generic irrotational curved spacetime adopting a strategy similar to the original proof. We show that a static relativistic star made of real scalar fields is never possible regardless of the geometrical properties of the (static) spacetimes. The generalised theorem offers a tool that can be used to check the stability of localised solutions of various types of scalar field models as well as of compact objects of modified theories of gravity with a non-minimally coupled scalar degree of freedom. Although the method described here is not directly applicable to the GHMPG theory, it still serves as an introductory step in this direction as this theory, as well as $f\left(R\right)$, are theories with extra scalar degrees of freedom.

\section{Introduction}

Derrick's theorem \cite{derrick1} constitutes one of the most important results on localised solutions of the Klein-Gordon equation in Minkowski spacetime. The theorem was developed originally as an attempt to build a model for non point-like elementary particles \cite{wheeler1,enz1} based on the now well known concept of "quasi-particle". Wheeler was the first to suggest the idea of an electromagnetic quasiparticle which he called \textit{Geons}. In spite of the fact that Wheeler's geons do not really exist, other models were proposed (and are still studied) in which Geons are composed of other fields in various setting. There are even (time dependent) formulations of this idea which are based on gravitational waves \cite{brill1,anderson1}.

It is clear that in the exploration of the idea that fundamental particles could be some form of Geons, a crucial problem is to infer the stability of the Geon itself. Derrick's theorem deals specifically with the stability of Geons made of scalar fields. In particular, Derrick found that in flat spacetimes the Klein Gordon equation cannot have static solution with finite energy \cite{manton1}.

In relativistic astrophysics, Derrick's theorem has profound consequences: its proof implies that no stable boson star can be constructed with real scalar fields and therefore that the existence of these objects requires more complex fields. Indeed the term boson stars nowadays is largely used to refer to complex scalar field stars, which are also called \textit{Q-balls} \cite{coleman1}. 

The consequences of Derrick's result span many different field of physics from low energy phenomena to QCD, to non linear phenomena, to pure mathematics (see e.g. the list of papers citing \cite{derrick1}). This is due to the fact that Derricks's results are related to a very general property of a class of differential equations called "Euclidean scalar field equations" to which the static Klein-Gordon equation belongs. In particular Derrick's theorem is a direct consequence of the so called Pohozaev identity \cite{pohozaev1,berestycki1}. This identity is akin to the well known Virial theorem as it relates the kinetic and potential energy of a localised scalar field configuration.

In this chapter, we provide a general proof of the theorem in curved spacetime and also include backreaction. This proof is based on the use of the 1+1+2 covariant approach \cite{clarkson1,betschart1,clarkson2}. This formalism arises as a generalization of the 3+1 formalism, in which one foliates the spacetime with a family of 3-dimensional hypersurfaces and rewrites the equations of motion in terms of a set of variables defined on these hypersurfaces, which has been of extreme usefulness in the development of numerical methods in relativity. In the 1+1+2 formalism, one considers a further decomposition into 2-hypersurfaces. We will start the following by describing the 1+1+2 formalism. Then in the following sections, we first use the 1+1+2 formalism to rederive Derrick's theorem in flat spacetime, which we then extend to curved spacetime and analyse the effect of backreaction. We then apply the Pohozaev identity to relevant models of scalar fields and finalize by discussing the applications of these results to nonminimally couples modified theories of gravity.

\section{the 1+1+2 covariant approach}

In what follows we will make use of the 1+1+2 covariant formalism to construct a proof of Derrick's theorem in curved spacetime and in the context of modified gravity. In the 1+1+2 formalism a generic spacetime is foliated in 2 surfaces, which we will call $\Upsilon$, by the definition of a timelike and a spacelike congruence represented by the vector $u_a$ and $e_a$, respectively. The metric tensor can then be decomposed as 
\begin{equation}
g_{ab}=-u_au_b+e_ae_b+N_{ab},
\end{equation}
where $N_{ab}$ is, at the same time a projector operator and the metric of $\Upsilon$. It will be useful also to define a 3 surface $W$ with metric $h_{ab}=e_ae_b+N_{ab}$.

In line with the above decomposition we can define three differential operators: 
a dot $\left(\ \dot{}\ \right)$ represents the projection of the covariant derivative along $u_a$:
\begin{equation}
\dot{X}^{a..b}{}_{c..d}{} = u^{e} \nabla_{e} {X}^{a..b}{}_{c..d}~,
\end{equation}

a hat $\left(\ \hat{}\ \right) $ denotes the projection of the covariant derivative along $e_a$:
\begin{equation}
\hat{X}_{a..b}{}^{c..d} \equiv  e^{f}D_{f}X_{a..b}{}^{c..d}~,
\end{equation}
$ \delta_a$ represents the covariant derivative projected with $N_{ab}$ e.g. 
\begin{equation}
\delta_lX_{a..b}{}^{c..d} \equiv  N_{a}{}^{f}...N_{b}{}^gN_{h}{}^{c}...
N_{i}{}^{d}N_l{}^jD_jX_{f..g}{}^{h..i}\;.
\end{equation}
At this point the kinematics and dynamics of any spacetime can be described via the definition of some specific quantities constructed with the derivatives of $u_a$, $e_a$ and $N_{ab}$. 

If one consider a spacetime endowed with a local rotational symmetry (LRS), i.e., a spacetime in which a multiply-transitive isometry group acting on the spacetime manifold, the 1+1+2 formalism allows to write the equations in terms only of scalar quantities. In our case only the quantities
\begin{eqnarray}
&&\mathcal{A}=e_a u_b\nabla^b{u}^{a}=e_a \dot{u}^{a}~,  \nonumber\\
&&\phi = N_{ab}\nabla^b e^{a}= \delta_a e^a~,\nonumber\\
&&\mathcal A_b=N_{ab}\dot{u}^{a},\label{dt211vars}\\
&& a_b= \hat{e}_b,\nonumber\\
&& \zeta_{ab}=\left( N^{c}{}_{(a}N_{b)}{}^{d} - \frac{1}{2}N_{ab} N^{cd}\right) \nabla_{c}e_{d},\nonumber
\end{eqnarray}
will be necessary.

Notice that our treatment will not involve vorticity as vortical spacetimes are inherently stationary and we are interested here only in static spacetimes. In the following, for sake of simplicity, we will call such general irrotational spacetimes "curved". 

It is important to clarify the limits of the approach that we will follow to  extend Derrick's idea. We first assume that our curved spacetime is such that at any point the quantities $u_a$, $e_a$ and $N_{ab}$ can be defined as $C(1)$ tensor fields. In other words, the spacetime must be regular enough to be consistent with those fields. 

\section{Covariant Derrick's theorem in flat spacetime}

The equation of motion for a real scalar field $\varphi$ minimally coupled to gravity is a Klein-Gordon equation of the form
\begin{equation}\label{dtkleingordon}
\Box\varphi-V_{\varphi}=0,
\end{equation}
where $\Box=\nabla^a\nabla_a$ is the D'Alembert operator, $\nabla_a$ is the covariant derivative with respect to the metric $g_{ab}$, $V=V\left(\varphi\right)$ is the scalar field potential, and $V_{\varphi}$ denotes a derivative with respect to the scalar field $\varphi$. Let us start by considering the simplest case of spherically symmetric spacetimes before proceeding to more general static spacetimes.

\subsection{Spherically symmetric spacetimes}

In Derrick's original approach \cite{derrick1}, to explore stability one expresses the variation of the action deforming the spatial coordinates with a constant parameter $\lambda$ in the Klein Gordon action. We will use here the properties of the covariant approach to perform an equivalent operation. For simplicity, let us consider first the spherically symmetric case in a Minkowski spacetime. In the covariant language, a deformation like the one used by Derrick in \cite{derrick1} can be represented by the quasi-conformal transformation
\begin{eqnarray}
&& u_a\Rightarrow \bar{u}_a=u_a,\nonumber\\
&& e_a\Rightarrow \bar{e}_a=\frac{1}{\lambda} e_a,\label{dtqctrans}\\
&& N_{ab} \Rightarrow \bar{N}_{ab}=\frac{1}{\lambda^2} N_{ab},\nonumber
\end{eqnarray}
where $\lambda$ is assumed to be a generic positive function. 
Under Eq. \eqref{dtqctrans} the D'Alembertian $\Box\varphi$ (see Sec. \ref{Sec:dalembert}) transforms as
\begin{equation}
\Box\varphi\Rightarrow \lambda^2\Box\varphi-\lambda \lambda^{,a}\varphi_{,a},
\end{equation}
which in static flat spacetime can be written as
\begin{equation}\label{dtkgtransformflat}
\Box\varphi\Rightarrow \lambda^2\varphi_{,qq}-\lambda \lambda_{,q}\varphi_{,q},
\end{equation}
where $q$ is a parameter associated to the congruence $e_a$, i.e., a derivative with respect to $q$ corresponds to a derivative along the curves of this congruence. Using the relation above, Eq. \eqref{dtkleingordon} becomes
\begin{equation}\label{dtkgflat}
\lambda^2\varphi_{,qq}-\lambda \lambda_{,q}\varphi_{,q}-V_\varphi=0.
\end{equation}
Equations of this type do not satisfy in general the Helmholtz conditions \cite{starlet1} and therefore they cannot be directly obtained as the Euler-Lagrange equations of any Lagrangian. However, Darboux showed \cite{darboux1,baldiotti1} that in one dimension there is an equivalent second-order equation for which a variational principle can be found, namely,
\begin{equation}\label{dtkghelmoltz}
e^{\Phi}\left(\lambda^2\varphi_{,qq}-\lambda \lambda_{,q}\varphi_{,q}-V_\varphi\right)=
0,
\end{equation}
where $e^{\Phi}$ is known as the integrator multiplier. The form of the integrator multiplier in the case of an equation with the structure of Eq. \eqref{dtkghelmoltz} can be found via the relation
\begin{equation}\label{dtauxrelflat}
\frac{d}{d\varphi_{,q}}\mathcal{Q}- \frac{d}{d q}\left[\frac{d}{d\varphi_{,qq}}\mathcal{Q}\right]=0,
\end{equation}
where $\mathcal{Q}$ represents Eq.\eqref{dtkghelmoltz} and we assumed that $\Phi$ does not depend on the derivatives of $\varphi$. In our case it turns out that $\Phi=-3\ln \lambda+ \Phi_0$ where $\Phi_0$ is a constant. We will choose here $\Phi_0=0$ so that $\Phi=0$ for $\lambda=1$ and we recover the original action. With this choice the action for Eq. \eqref{dtkgflat} is given by  
\begin{equation}\label{dtactionflat}
S(\lambda)=-\frac{1}{2}\int \frac{1}{\lambda^3}\left[\lambda^2\varphi_{,q}^2+2V\left(\varphi\right)\right]dq.
\end{equation}
If the solution of Eq.\eqref{dtauxrelflat} is localised, this integral will be well defined and finite. Derrick's deformation is given by $\lambda=\lambda_0$ for some constant $\lambda_0$, which implies that Eq.\eqref{dtactionflat} can be written as  
\begin{equation}\label{dtactionflatdef}
S(\lambda)=-\frac{1}{2}\left(\frac{I_1}{ \lambda}+\frac{I_2}{\lambda^3}\right),
\end{equation}
where $I_1$ and $I_2$ are integrals defined as
\begin{eqnarray}
&&I_1=\int \varphi_{,q}^2\, d q ,\\ 
&&I_2=2\int V\left[\varphi(q)\right]\, d q.\nonumber
\end{eqnarray}
We now check that Eq. \eqref{dtkleingordon} corresponds to an extremum of the action requiring that  
\begin{equation}
\frac{\partial S(\lambda)}{\partial \lambda}=0 \rightarrow \frac{I_1}{ \lambda^2}+3\frac{I_2}{\lambda^4}=0.
\end{equation}
Setting $\lambda=1$, we obtain that 
\begin{equation}\label{dtpohozaevflat}
I_2=-\frac{I_1}{3},
\end{equation}
i.e. the Pohozaev identity. This relation tells us that the Klein Gordon equation can be an extremum of action \eqref{dtactionflatdef} only if the integral of the potential is negative. This implies that, for example, a mass potential, which is defined positive, would never lead to an equilibrium. We can determine the character of the extremum by considering the second order derivative of $S(\lambda)$:
\begin{equation}
\frac{\partial^2 S(\lambda)}{\partial \lambda^2}=-\frac{I_1}{ \lambda^3}-6\frac{I_2}{\lambda^5}.
\end{equation}
Substituting Eq. \eqref{dtpohozaevflat} and setting $\lambda=1$ we obtain
\begin{equation}
\frac{\partial^2 S(\lambda)}{\partial \lambda^2}=I_1>0.
\end{equation}
Hence Eq. \eqref{dtkleingordon} is a minimum for the action provided that the integral of the potential $V$ is negative.   

Now, in the static case the energy function of $\varphi$ can be related to the action via the relation (This relation can be verified calculating directly the $(0,0)$ component of the stress energy density for the scalar field, which corresponds to the Hamiltonian or, more precisely, to the Lagrangian energy, and integrating over the entire spacetime)
\begin{equation}
E= -2S,
\end{equation}

 which implies
\begin{equation}
\frac{\partial^2 E(\lambda)}{\partial \lambda^2}\bigg|_{\lambda=1}=-2I_1<0.
\end{equation}
Therefore a minimum of the action corresponds to a maximum of the energy, and a localised  solution $\varphi(q)$ of the Eq.\eqref{dtkleingordon} must be unstable.

\subsection{General irrotational spacetimes}

We can generalise this reasoning to the non spherically symmetric case in which also $\delta$ derivatives appear. From the expression of $\Box\varphi$ (see Sec. \ref{Sec:dalembert}), using the parameters $w_2$ and $w_3$ to map the 2-surface $\Upsilon$ (for example, if one chooses $u^a$ to be directed along the time coordinate $t$ and $e^a$ to be directed along a radial coordinate $r$, then $w_i$ will correspond to the angular coordinates), the d'Alembertian can be written as
\begin{equation}
 \Box\varphi\Rightarrow \lambda^2\varphi_{,qq}-\lambda \lambda_{,q}\varphi_{,q}+\sum_{i=2}^{3}\left(\lambda^2\varphi_{,w_i w_i}-\lambda \lambda_{,w_i}\varphi_{,w_i}\right).
\end{equation}
Hence, the Klein-Gordon equation becomes
\begin{equation}\label{dtkgflatnss}
\lambda^2\varphi_{,qq}-\lambda \lambda_{,q}\varphi_{,q}+\sum_{i=2}^{3}\left(\lambda^2\varphi_{,w_i w_i}-\lambda \lambda_{,w_i}\varphi_{,w_i}\right)-V_\varphi=0.
\end{equation}
Multiplying the previous equation by $e^{\Phi}$, we obtain the equivalent equation
\begin{equation}\label{dtkghelmoltznss}
e^{\Phi}\left[\lambda^2\varphi_{,qq}-\lambda \lambda_{,q}\varphi_{,q}+\sum_{i=2}^{3}\left(\lambda^2\varphi_{,w_i w_i}-\lambda \lambda_{,w_i}\varphi_{,w_i}\right)-V_\varphi\right]=0.
\end{equation}
where the integrator multiplier can be calculated using a condition similar to Eq.\eqref{dtauxrelflat}
\begin{equation}\label{dtauxrelflatnss}
\sum_{i=1}^{3}\left\{\frac{d}{d\varphi_{,p_i}}\mathcal{Q}- \frac{d}{d p_i}\left[\frac{d}{d\varphi_{,p_i  p_i}}\mathcal{Q}\right]\right\}=0\;,
\end{equation}
where $p_i=(0,q,w_2,w_3)$, $\mathcal{Q}$ is Eq.\eqref{dtkghelmoltznss} and we assumed again that $\Phi$ does not depend on the derivatives of $\varphi$. Here we appear to force the original approach by Darboux, which works only for one dimensional actions. However, the integrator multiplier can be associated to the volume form for the scalar field action (see Sec.\ref{Sec:211} for details) and therefore it can be determined with the Darboux procedure also in multidimensional actions (at least in our specific case). Indeed it will become clear that the form of the integrator multiplier is actually irrelevant for our purpose, because its transformation properties can be deduced in general. This relation amounts to the partial differential equation
\begin{equation}
\sum_{i=1}^{3}\left[\Phi_{,p_i}+3\frac{\lambda_{,p_i}}{\lambda}\right]=0.
\end{equation}
Using the method of the characteristics we can find the solutions 
\begin{equation}
\Phi=- 3\ln\lambda + C(w_2-q,w_3-q) \;.
\end{equation}
Since we want  to return to the standard action $e^\Phi=1$ for $\lambda=1$ we can set $C=0$.
Thus the action can be written as
\begin{equation}
S=-\frac{1}{2}\int \frac{1}{\lambda^3}\left[\lambda^2\varphi_{,q}^2+\lambda^2\sum_{i=2}^{3}\varphi_{,w_i}^2+2V\left(\varphi\right)\right]d \Omega,
\end{equation}
where the $d\Omega=\prod_{a=1}^{3} d p_i$. Repeating the procedure above, we obtain again the action in the form of Eq. \eqref{dtactionflatdef}, and the Pohozaev identity in Eq.\eqref{dtpohozaevflat}, where the integrals $I_1$ and $I_2$ in this case are defined by
\begin{eqnarray}
&&I_1=\int \left[\varphi_{,q}^2+\sum_{i=2}^{3}\varphi_{,w_i}^2\right]\, d\Omega, \nonumber\\ 
&&I_2=2\int V\left[\varphi(q)\right]\,  d\Omega,
\end{eqnarray}
which implies that
\begin{eqnarray}
&&\frac{\partial E(\lambda)}{\partial \lambda}\bigg|_{\lambda=1}=0\rightarrow I_1=- 3I_2,\nonumber \\
&&\frac{\partial^2 E(\lambda)}{\partial \lambda^2}\bigg|_{\lambda=1}=-2I_1<0,
\end{eqnarray}
and shows that the solution $\varphi$ is unstable. 
 
This result is called in literature the Derrick's theorem and it is the main reason why localised solutions of real scalar fields are generally  considered unphysical.  In the next sections we will give a generalisation of this result in the case of irrotational LRS spacetimes and explore its validity in more general spacetimes and in the context of modified gravity.

\section{Covariant Derrick theorem in curved spacetimes}

Let us now prove Derricks theorem in curved spacetimes. As before, for simplicity we will start with the spherically symmetric case and then we will consider more complex cases.

\subsection{Spherically symmetric spacetimes}

Decomposing Eq.\eqref{dtkleingordon} in the 1+1+2 variables and considering spherically symmetric LRSII spacetimes, the transformed Klein-Gordon equation reads
\begin{equation}\label{dtkgcurved}
\lambda^2\varphi_{,qq}-\lambda \lambda_{,q}\varphi_{,q}+\left[\mathcal A(\lambda)+\phi(\lambda)\right]\lambda\varphi_{,q}-V_\varphi=0.
\end{equation}
This equation can be generated by the action
\begin{equation}\label{dtactioncurved}
S(\lambda)=-\frac{1}{2}\int e^{\Phi(\lambda)}\left[\lambda^2\varphi_{,q}^2+2V\left(\varphi\right)\right]dq,
\end{equation}
\begin{equation}\label{dtphicurved}
\Phi(\lambda)=\int\left\{\frac{1}{\lambda}\left[\mathcal A(\lambda)+\phi(\lambda)\right]-3\frac{\lambda_{,q}}{\lambda}\right\}dq .
\end{equation}
The integral in Eq. \eqref{dtphicurved} can be simplified by noting that under the transformation in Eq. \eqref{dtkgtransformflat} we have
\begin{equation}\label{dtqcAphi}
\mathcal{A}(\lambda)=\lambda e_a\dot{u}^a=\lambda \mathcal{A},\qquad
\phi(\lambda) =\lambda \delta_a e^a =\lambda\phi,
\end{equation}
which implies that
\begin{equation}\label{dttransphi}
\Phi(\lambda)-\Phi_0=\int\left[\mathcal A+\phi\right]dq- 3\ln \lambda,,
\end{equation}
and thus
\begin{equation}\label{dtexptransform}
e^{\Phi(\lambda)}\Rightarrow\frac{e^{\Phi}}{\lambda^3},
\end{equation} 
where we have chosen $\Phi_0=0$ so that $\lambda=1$ implies $e^{\Phi(\lambda)}=e^{\Phi}$. 
In this way Eq. \eqref{dtkgcurved} can be derived from the action
\begin{equation}
S=-\frac{1}{2}\int \frac{e^{\Phi}}{\lambda^3}\left[\lambda^2\varphi_{,q}^2+2V\left(\varphi\right)\right]dq.
\end{equation}
The above expression is consistent with the interpretation of $e^{\Phi}$ as the volume form for the action in Eq. \eqref{dtactioncurved} (see Sec.\ref{Sec:211} for details). In this perspective the choice that we made for $\Phi_0$ corresponds to a choice of the asymptotic properties of the metric. This fact can be understood bearing in mind that, by definition, $\mathcal A$  and $\phi$ are identically zero when the spacetime is Minkowskian. As we consider localised solutions for $\varphi$, it is only natural to choose an "asymptotically flat" $\Phi$ by choosing $\Phi_0=0$. Setting $\lambda=\lambda_0$, we can write again the action in the form of Eq.\eqref{dtactionflatdef}, and the Pohozaev identity in Eq.\eqref{dtpohozaevflat}, where the integrals $I_1$ and $I_2$ are given by 
\begin{eqnarray}
&&I_1=\int e^{\Phi}\varphi_{,q}^2\, d q, \\ 
&&I_2=2\int e^{\Phi}V\left[\varphi(q)\right]\, d q.
\end{eqnarray}
The Pohozaev identity is exactly of the same form and thus leads to the same conditions. This implies that Derrick's theorem is valid also in the curved spherically symmetric case. It is important to stress that the Darboux procedure we have used so far to deduce the action is valid only if the integrator multiplier is different form zero. One can prove that this condition implies that the spacetime we are considering does not contain a perfect or Killing horizon. Such constraint excludes the case of spacetimes describing for example black holes and trapped surfaces.

\subsection{General irrotational spacetimes}

Let us now turn to more complex spacetimes. If vorticity is zero, 
upon the transformations provided in Eq. \eqref{dtkgtransformflat} the Klein-Gordon equation  reads
\begin{equation}
\lambda^2\varphi_{,qq}-\lambda \lambda_{,q}\varphi_{,q}+\left[\mathcal A(\lambda)+\phi(\lambda)\right]\lambda\varphi_{,q}+\sum_{b=2}^{3}\left[\lambda^2\varphi_{,w_i w_i}+\lambda \lambda_{,w_i}\varphi_{,w_i}+\mathcal A_b(\lambda)\lambda^2\varphi_{,w_i}+a_b(\lambda)\lambda^2\varphi_{,w_i}\right]-V_\varphi=0.
\end{equation}
It is clear that in non spherical irrotational LRS spacetimes where $\mathcal A_b$ and $a_b$ are identically zero (which still belong to the LRSII class), Derrick's theorem holds. We have the same action and Pohozaev identity as shown in Eqs. \eqref{dtactionflatdef} and \eqref{dtpohozaevflat} respectively, where the integrals $I_1$ and $I_2$ are given by
\begin{eqnarray}
&&I_1=\int e^{\Phi}\left[\varphi_{,q}^2+\sum_{i=2}^{3}\varphi_{,w_i}^2\right]\, d\Omega,\nonumber \\ 
&&I_2=2\int e^{\Phi}V\left[\varphi(q)\right] d\Omega,\label{dtintegralcurv}
\end{eqnarray}
 and $\Phi$ is given by Eq. \eqref{dtphicurved} for $\lambda=\lambda_0$. If we consider more general spacetimes ($\mathcal A_b\neq0$ and $a_b\neq0$), we have to explore the transformation of the acceleration vectors under the transformations in Eq. \eqref{dtqctrans}. We thus have
 \begin{eqnarray}
&& \mathcal A_b(\lambda)=N_{b}{}^c \dot{u}_c=\mathcal A_b,\label{dtqcvectors}\\
 && a_b(\lambda)=N_b{}^c \hat{e}_c=N_b{}^c \lambda \widehat{\left(\frac{e_c}{\lambda}\right)}=a_b.\nonumber
 \end{eqnarray}

Defining the four vector
\begin{equation}
V_a =(\mathcal A+\phi)e_a+\left(\mathcal A_c+a_c\right)N^c_b,
\end{equation}

the condition in Eq. \eqref{dtauxrelflatnss} for this case takes the form of the partial differential equation involving the components of $V_a$ as
\begin{equation}\label{dtdarbouxcurv}
\sum_{i=1}^{3}\left[\Phi_{,p_i}+3\frac{\lambda_{,p_i}}{\lambda}+ V_i\right]=0.
\end{equation}

We can use the method of characteristics again to solve the above equation and, as in all partial differential equations, the existence and properties of the solutions will depend critically from the boundary conditions. As we have seen in the spherically symmetric case, the boundary conditions are strictly related to the  asymptotic properties of the specific metric which one is considering. Since we have assumed that the scalar field is localised, it is natural to assume asymptotic flatness. However, as far as it generates the correct field equations, the exact form of $\Phi$ is irrelevant for our purposes. We only need to determine the transformation of the quantity $\Phi$ under \eqref{dtqctrans}. From Eq. \eqref{dtdarbouxcurv} it is evident that $\Phi$ will transform such that 
\begin{equation}
\Phi(\lambda)=\Phi -3\ln\lambda.
\end{equation}
Using the above result, we recover the action in the form of Eq.\eqref{dtactionflatdef} and the Pohozaev identity of Eq. \eqref{dtpohozaevflat}, whre the integrals $I_1$ and $I_2$ are of the same form as Eq.\eqref{dtintegralcurv}, the difference being only in the form of $\Phi$. This is the same result obtained in the flat case.

\section{Introducing backreaction}

In the previous section we have made the tacit assumption that the mass of the confined scalar field solution would not perturb the assigned metric of the spacetime. In other words we have neglected backreaction. Let us now generalize the strategy above to the case in which the localised scalar field solution also contributes to the spacetime metric. In this case, one should add the Hilbert Einstein term to the action for the scalar field, that is
\begin{equation}\label{dtbraction}
S=\frac{1}{2}\int d x^{4}\sqrt{-g}\left[R-\nabla_a\varphi\nabla^a\varphi - 2V(\varphi)\right]
\end{equation}
and derive its transformation under Eqs. \eqref{dtqctrans}.

It should be pointed out that at present there is no general consensus on the definition of the energy of the gravitational field. One should ask, then, if it makes sense to extend Derrick's results also to the backreaction case. A positive answer can be provided thinking that we are considering a very special case.  First of all, in order to keep finite the action/energy integral, we have to assume an asymptotically flat background. In addition, since our choice of the vector field $u_a$ corresponds to a timelike Killing field for the spacetimes we consider, the class of observers we consider is static. 

From the results of \cite{katz1,katz2} we have that in stationary spacetimes the energy of the gravitation field can be written as the scalar
\begin{equation}\label{dtdefegrav}
E_G=\int \sqrt{-g} \left(t^m{}_n u^n+ \sigma^{[mn]}{}_p\partial_n u^p\right)u_m d\Omega, 
\end{equation}
where $t^m{}_n $ is the Einstein pseudotensor and $\sigma^{[mn]}{}_p$ is Freud's complex \cite{freud1} given by
\begin{equation}
\sigma^{[mn]}{}_p= \frac{1}{g}g_{pr}\left(g\, g^{r [m}\,g^{n] s}\right)_{,s}.
\end{equation}
The Eq. \eqref{dtdefegrav} can be written, in our assumptions, as
\begin{equation}
E_G= -\frac{1}{2}
\int \sqrt{-g} \mathcal{L}^{(3)}_{\bar{\Gamma}\bar{\Gamma}} d\Omega,
\end{equation}
\begin{equation}
\mathcal{L}^{(3)}_{\bar{\Gamma}\bar{\Gamma}}=h^{ab}\left(\bar{\Gamma}_{a d}{}^{c}\bar{\Gamma}_{c b}{}^{d}-\bar{\Gamma}_{ab}{}^{c}\bar{\Gamma}_{c d}{}^{d}\right),
\end{equation}
where $\bar{\Gamma}$ are the Christoffel symbols of the three surface $W$.  Now, starting from the Hilbert-Einstein action in the static case we can write
\begin{equation}
S_G= \frac{1}{2}\int \sqrt{-g}\,R\, dt d\Omega=\frac{T_0}{2}\int \sqrt{-g}\,R \,d\Omega,
\end{equation}
where $T_0$ is a constant, which we can set to one without loss of generality. Using the contracted Gauss-Codazzi equation, we have
\begin{equation}\label{dtactionenergy}
S_G= \frac{1}{2}\int \sqrt{-g}\left[R^{(3)}+ K^2 - K_{a}{}^{b}K^{a}{}_{b}+\nabla_a\left(\dot{u}^a+ u^a K\right)\right] d\Omega.
\end{equation}
where $R^{(3)}$ is the Ricci scalar for  submanifold $W$,
\begin{equation}
K_{ab} = h_a{}^c h_b{}^d \nabla_{c}u_{d},
\end{equation}
is the second fundamental form of $W$ and $K=K^{a}{}_{a}$. Using the Gauss theorem we can integrate out the last factor in \eqref{dtactionenergy}. In addition, in static irrotational spacetimes the terms associated to the extrinsic curvature are identically zero. Now, $R^{(3)}$ can be decomposed in as similar way as $R$ in \eqref{dtactionenergy}, employing the extrinsic curvature of $\Upsilon$. Integrating out the second (projected) derivatives and rewriting the expression in terms of the Christoffel symbols, we arrive to 
\begin{equation}
S_G=  S_{0}+\frac{1}{2}
\int \sqrt{-g}\,\mathcal{L}^{(3)}_{\bar{\Gamma}\bar{\Gamma}}\,d\Omega= -\frac{E_0}{2}- E_G.
\end{equation}
where $S_0$ is a constant and we have defined $E_0=-2S_0$. Thus in our case (and only in this case), modulus an irrelevant constant, the energy of the gravitational field  can linked to the Hilbert Einstein action:
\begin{equation}
E_G= -\frac{E_0}{2}-\frac{1}{2}\int e^{\Phi}R \;d\Omega.
\end{equation}
Here we have connected the volume form to $e^\Phi$ as described in Sec.\ref{Sec:211} in the Appendix.

The next task is to evaluate how $E_G$ transforms under Eqs. \eqref{dtqctrans}. Using the Gauss Codazzi equation also on $R^{(3)}$ gives, in static irrotational spacetimes, 
\begin{equation}
R =R^{(2)}-2 \hat{\phi}-\frac{3}{2}\phi^2- 2 \zeta ^2 -2 a_b a^b+ 2\delta_b a^b.
\end{equation}
Now, since $R^{(2)}=2K_G$, where $K_G$ is the gaussian curvature, the Brioschi formula implies that, under Eqs. \eqref{dtqctrans}, $R^{(2)}$ transforms as
\begin{equation}\label{dtR2trans}
 R^{(2)}\Rightarrow R^{(2)}(\lambda)=\lambda^2 R^{(2)}+...
\end{equation}
where the dots (...) represent terms which contain derivatives of $\lambda$. In addition, from definitions in Eqs. \eqref{dt211vars} one finds
\begin{eqnarray}
&&\zeta_{ab}\Rightarrow\zeta_{ab}(\lambda)=\frac{\zeta_{ab}}{\lambda }\nonumber\\
&&\zeta\Rightarrow\zeta(\lambda)=\lambda^2 \zeta.
\end{eqnarray}

Using also the transformation in Eqs. \eqref{dtqcAphi} and \eqref{dtqcvectors} we arrive at
\begin{equation}
 R (\lambda)=\lambda^2 R +...
\end{equation}
where, again, the dots represent additional terms which contain derivatives of $\lambda$. As we will eventually set up $\lambda=1$, these terms are irrelevant and can be neglected.

The total action derived from Eq. \eqref{dtbraction} thus transforms as 
\begin{equation}\label{dtbractiondef}
S(\lambda)=\frac{S_G}{\lambda}-\frac{1}{2}\int \frac{e^{\Phi}}{\lambda^3}\Bigg[ \lambda^2\varphi_{,q}^2
+\lambda^2\sum_{i=2}^{3}\varphi_{,w_i}^2+2V\left(\varphi\right)\Bigg] d\Omega\,.
\end{equation}
In the case $\lambda=\lambda_0$, we now define the integrals $I_1$ to $I_3$ as
\begin{eqnarray}
&&I_1=\int e^{\Phi}\left[\varphi_{,q}^2+\sum_{i=2}^{3}\varphi_{,w_i}^2\right]\,  d\Omega , \nonumber\\ 
&&I_2=2\int e^{\Phi}V\left(\varphi\right)\,  d\Omega,\\
&&I_3=-E_0-2 E_G,\nonumber 
\end{eqnarray}
one can write the action as
\begin{equation}
E(\lambda)=-2 S(\lambda)=
\left(\frac{I_1}{ \lambda}-\frac{I_3}{ \lambda}+\frac{I_2}{\lambda^3}\right),
\end{equation}
which no longer provides the Pohozaev identity. Since, as we have seen, relation $E=-2S$ between the energy and the action is still valid, we can examine the stability of the backreacting solution with the same strategy as the previous section. We have in this case
\begin{eqnarray}
&&\frac{\partial E(\lambda)}{\partial \lambda}\bigg|_{\lambda=1}=0\rightarrow I_3=I_1+ 3I_2,\nonumber\\
&&\frac{\partial^2 E(\lambda)}{\partial \lambda^2}\bigg|_{\lambda=1}=6I_2\,.\label{dtbrderivatives}
\end{eqnarray}
Now, note the trace of the field equations provides another relation that should be taken in consideration. This relation is
\begin{equation}\label{dtefetrace}
R=\nabla_a\varphi\nabla^a\varphi+4V,
\end{equation}
which upon integration gives $I_3=I_1+2I_2$. Combining the above results with the first of Eqs. \eqref{dtbrderivatives} gives $I_2=0$. This yields 
\begin{equation}\label{dtbrresult}
\frac{\partial^2 E(\lambda)}{\partial \lambda^2}\bigg|_{\lambda=1}=6I_2=0.
\end{equation}
Since the above quantity has opposite signs if we consider $I_2\rightarrow I_2\pm\epsilon$, where $\epsilon$ is a small constant, the solution corresponds to an inflection point. Hence the presence of gravity has weakened the instability but cannot eliminate it completely. 

The weakest point of the reasoning given above is, undoubtedly, the definition of the gravitational energy of the system. One might object that even with our specific assumptions the definition of energy we have used might miss some crucial aspect of the physics of these systems. We can argue here that this is not the case going around the problem of the definition of $E_G$ by eliminating the Hilbert Einstein term from the action using the field equations i.e. considering the on-shell action. 
  
For example, using the relation in Eq. \eqref{dtefetrace} we have   
\begin{equation}
E^{tot}(\lambda)=-2\int \frac{e^{\Phi}}{\lambda^3}V\left(\varphi\right) d\Omega,
\end{equation}
which immediately implies Eq. \eqref{dtbrresult}. This result shows that our previous argument is correct and at the same time suggests a easy shortcut to prove Derrick's theorem with backreaction. In the following we will make ample use of this shortcut, especially in dealing with more complex settings. 

We can use the on shell action to probe further in the validity of Derrick's theorem by considering, for example, the case in which the scalar field backreacts with a spacetime with non zero cosmological constant $\Lambda$. In this case, Eq. \eqref{dtbractiondef} would become
\begin{equation}
E(\lambda)=
\left(\frac{I_1}{ \lambda}-\frac{I_3}{ \lambda}+\frac{I_2}{\lambda^3}+\frac{2 I_4}{\lambda^3}\right),
\end{equation}
\begin{equation}
I_4=\Lambda \int e^{\Phi}  d\Omega.
\end{equation}
We obtain, on shell,
\begin{eqnarray}
&&\frac{\partial E(\lambda)}{\partial \lambda}\bigg|_{\lambda=1}=0\rightarrow I_3=I_1- I_2,\\
&&\frac{\partial^2 E(\lambda)}{\partial \lambda^2}\bigg|_{\lambda=1}=3(I_1+2I_2)\,.\nonumber
\end{eqnarray}
Hence in this case stability is possible if
\begin{equation}
I_2>-\frac{1}{2}I_1.
\end{equation}
Therefore the presence of a cosmological constant can lead to stable solutions. However these solutions make sense physically only at scales in which $\Lambda$ is relevant, and therefore exclude microscopic or astrophysical objects. Yet, the picture that emerges is that Derrick's instability cannot be avoided by minimal modifications of the model. In the following we will explore further the validity of Derrick's theorem looking at the effect of scalar field coupling, non canonical scalar field and modified gravity.

\section{Applications to scalar-field models}

Let us now analyze how different scalar field models with different couplings are affected  by the previous results. We start by analyzing the coupling between minimally coupled scalar fields with each other, we then study models of non-canonical minimally coupled scalar fields such as phantom, quintom, and k-essence fields, and we finalize by providing applications to scalar-tensor theories of gravity. Unfortunately, the same procedure is not appliable to geometrical theories of gravity such as $f\left(R\right)$ and hybrid metric-Palatini theories, since the quasi-conformal transformation ceases to be trivial and also there is no general definition of gravitational energy in these cases. One could, in principle, make a transformation to the scalar-tensor representation and apply the quasi-conformal transformation there, but the equivalence between the two representations is not guaranteed after performing such transformation. This comes from the fact that the scalar fields we are dealing with here are invariant under the quasi-conformal transformation. However, if the scalar fields are defined as derivatives of a function $f\left(R\right)$, since $R$ is not invariant under these transformations one can not guarantee that the scalar field will still be invariant.  

\subsection{Scalar field couplings}

Derrick's instability is very robust. No additional standard coupling of the scalar field with matter or other fields can prevent its appearance. A  coupling with another scalar field of the type $f(\varphi)g(\psi)$  would just make more complicated the definition of the integral $I_2$. In fact, starting from the corresponding Klein-Gordon equations 
\begin{equation}
\Box\varphi-V_{,\varphi}-f_{,\varphi}g(\psi)=0,
\end{equation}
and proceeding as in the previous section we obtain an action of the same form of Eq.\eqref{dtactionflatdef}, with the integrals $I_1$ and $I_2$ in this case being
\begin{eqnarray}
&&I_1=\int e^{\Phi}\left[\varphi_{,q}^2+\sum_{i=2}^{3}\varphi_{,w_i}^2\right]\,  d\Omega\,,\nonumber \\ 
&&I_2=2\int e^{\Phi}\left\{V\left[\varphi(q)\right]+ f\left[\varphi(q)\right] g\left[\psi(q)\right]\right\}\,  d\Omega\,.
\end{eqnarray}
It follows that we can prove Derrick's theorem  also in this case. This conclusion is independent from the sign of the terms appearing in the above integral. The presence of the coupling, however, changes the physical significance of the Pohozaev equilibrium condition. The same happens when we introduce backreaction.

The strategy of the proof we have presented shows that, whatever the coupling, the key point in the determination of the stability of localised scalar field configurations depends on the $\lambda$-dependence of the transformation of the integrator multiplier. If the transformation of $\exp(\Phi)$ is such that the action can be written as a combination of $\lambda$ terms and $\lambda$-independent integrals, like e.g. in Eq \eqref{dtexptransform}, there will be a chance to prove (in)stability. In other cases, Derrick's approach does not lead to a definite answer. 

A simple example is the case of  derivative coupling of the type $ a \hat{\varphi} g(\psi)$. For this coupling the transformation of the integrator multiplier is given by 
\begin{equation}
 \Phi_{g}=\Phi+ a \int\varphi_{,q} \frac{g(\psi)}{\lambda^2} dq\,,
\end{equation}
and $\lambda$ is not factorisable. This fact makes it impossible to find a form of the action similar to the usual Eq.\eqref{dtactionflatdef}. Instead, considering a coupling of the type $ a \hat{\varphi} \hat{\psi}^2$ will yield
\begin{equation}
e^{\Phi_{gd}(\lambda)}\Rightarrow\frac{e^{\Phi_{gd}}}{\lambda},\qquad \Phi_{gd}=\Phi+ a \int\varphi_{,q} (\psi_{,q})^2 dq\,,
\end{equation}
which leads to an action similar to Eq.\eqref{dtactionflatdef},  and thus implies instability.

\subsection{Non-canonical scalar fields}

In the context of cosmology and in particular when dealing with the problem of dark energy, various non canonical scalar fields have been introduced. Using the strategy above, we can extend Derrick's theorem to these cases. In the following, we will consider the case of: Phantom fields \cite{caldwell1,caldwell2}, Quintom fields \cite{cai1} and k-essence \cite{chiba1}. In these models, as in the ones of next Section, for the sake of brevity, we will make Derrick's deformation directly in the action, rather than prove that the transformed action comes from the modified Klein-Gordon equation. This connection is, however, always valid. We will also consider only the case of LRSII spacetimes as the generalisation to more complicated geometries can be derived easily from the considerations above.

\subsubsection{Phantom fields}

Phantom fields are scalar fields whose action contains a kinetic term with opposite sign with respect to the canonical one:
\begin{equation}\label{dtpfaction}
S=\frac{1}{2}\int d x^{4}\sqrt{-g}\left[R+\nabla_a\varphi\nabla^a\varphi - 2V(\varphi)\right].
\end{equation}
Excluding backreaction, we have an energy function associated to the action in Eq.\eqref{dtpfaction}
\begin{equation}\label{dtpfenergy}
 E(\lambda)=-\left(\frac{I_1}{ \lambda}-\frac{I_2}{\lambda^3}\right),
\end{equation}
where the integrals $I_1$ and $I_2$ are in this case
\begin{eqnarray}
&&I_1=\int e^{\Phi}\left[\varphi_{,q}^2+\sum_{i=2}^{3}\varphi_{,w_i}^2\right]\,  d\Omega\,, \nonumber\\ 
&&I_2=2\int e^{\Phi}V\left[\varphi(q)\right]\,  d\Omega\,.
\end{eqnarray}
Eq. \eqref{dtpfenergy} yields
\begin{eqnarray}
&&\frac{\partial E(\lambda)}{\partial \lambda}\bigg|_{\lambda=1}=0\rightarrow I_1= 3I_2\nonumber,\\
&&\frac{\partial^2 E(\lambda)}{\partial \lambda^2}\bigg|_{\lambda=1}=2I_1>0,
\end{eqnarray}
which implies that a localised solution of phantom fields is actually stable.  This result reveals that a key element of Derrick's instability is the sign of the scalar field kinetic terms.

The inclusion of backreaction, however, introduces instability. On shell, the energy can be written as
\begin{equation}
 E(\lambda)=\frac{I_2}{\lambda^3},
\end{equation}
 as in the standard case,
 \begin{eqnarray}
&&\frac{\partial E(\lambda)}{\partial \lambda}\bigg|_{\lambda=1}=0\rightarrow -3I_2= 0,\nonumber\\
&&\frac{\partial^2 E(\lambda)}{\partial \lambda^2}\bigg|_{\lambda=1}=12I_2=0\,,
 \end{eqnarray}
which corresponds again to an inflection point and, thus, to an instability. 

\subsubsection{Quintom fields}

In the case of Quintom fields, we have two interacting fields: one canonical and the other non canonical. The action reads thus
\begin{equation}\label{dtqfaction}
S=\frac{1}{2}\int d x^{4}\sqrt{-g}\left[R-\nabla_a\varphi\nabla^a\varphi+\nabla_a\psi\nabla^a\psi- 2V(\varphi, \psi)
\right].
\end{equation}
Excluding backreaction,  the energy function associated to the action in Eq. \eqref{dtqfaction} becomes
\begin{equation}
 E(\lambda)=\left(\frac{I_1}{ \lambda}-\frac{I_4}{ \lambda}+\frac{I_2}{\lambda^3}\right),
\end{equation}
where the integrals $I_1$ to $I_3$ are defined as
\begin{eqnarray}
&&I_1=\int e^{\Phi}\left[\varphi_{,q}^2+\sum_{i=2}^{3}\varphi_{,w_i}^2\right]\,  d\Omega \,,\nonumber\\ 
&&I_2=2\int e^{\Phi}V\left[\varphi(q)\right]\,  d\Omega\,,\\
&&I_4=\int e^{\Phi}\left[\psi_{,q}^2+\sum_{i=2}^{3}\psi_{,w_i}^2\right]\,  d\Omega.\nonumber
\end{eqnarray}
This leads to the relations
\begin{eqnarray}
&&\frac{\partial E(\lambda)}{\partial \lambda}\bigg|_{\lambda=1}=0\rightarrow I_4= 3I_2+I_1,\nonumber\\
&&\frac{\partial^2 E(\lambda)}{\partial \lambda^2}\bigg|_{\lambda=1}=6I_2>0\,,
\end{eqnarray}
which, as in the case of the phantom field, can be stable if $I_2$ is positive. Again, using the trace of the gravitational field equations to include backreaction, we can write the action above on shell, which leads to instability. This was an expected result, which confirms the general conclusions we have drawn before: a multi-field system becomes unstable if one of its components presents instability.

\subsubsection{k-essence fields}

In the case of k-essence fields, the action is generalised as 
\begin{equation}\label{dtkeaction}
S=\frac{1}{2}\int d x^{4}\sqrt{-g}\left[R+
P(\varphi,X)\right],
\end{equation}
where $X=\nabla_a\varphi\nabla^a\varphi$. Using the fact that for $\lambda=1$, ${ P}_{,\lambda}=2X{P}_{,X}$, under the quasi-conformal transformations in Eq. \eqref{dtqctrans} and without backreaction one has
\begin{eqnarray}
&&\frac{\partial E(\lambda)}{\partial \lambda}\bigg|_{\lambda=1}=0 \rightarrow 2 X \partial_{X}P-3P=0\,,\nonumber\\
&&\frac{\partial^2 E(\lambda)}{\partial \lambda^2}\bigg|_{\lambda=1}=0.\label{dtkederivatives}
\end{eqnarray}
From the first of Eq. \eqref{dtkederivatives} we obtained that $P=P_0(\varphi)X^{3/2}$ for some constant $P_0$, which thus implies the second of Eq. \eqref{dtkederivatives}. We again recover an inflection point and thus we always have instabilities.  

Considering backreaction, we have
\begin{eqnarray}
&&\frac{\partial E(\lambda)}{\partial \lambda}\bigg|_{\lambda=1}=0 \rightarrow  X \partial_{X}P-P=0,\nonumber \\
&&\frac{\partial^2 E(\lambda)}{\partial \lambda^2}=0,
\end{eqnarray}
which again is an inflection point and implies instabilities.

On top of it intrinsic value, these results show that conditions of Derrick's theorem can be used to constrain modifications of general relativity in which undetermined functions are present. In the next section we will look at some examples of such constraints.   

\subsection{Scalar-tensor gravity}

Using the results from the previous sections  we can proceed to the generalisation of Derrick's theorem to non minimal couplings. Let us consider, for example, the case of scalar tensor theories. This class of theories of gravity are characterised an action of the form
\begin{equation}\label{dtstaction}
S=\frac{1}{2}\int d x^{4}\sqrt{-g}\left[F(\varphi)R-\nabla_a\varphi\nabla^a\varphi- 2V(\varphi)\right],
\end{equation}
whose variation gives the field equations for the metric and the Klein-Gordon equation
\begin{equation}
FG_{ab}=\nabla_{a}\varphi\nabla_{b}\varphi-g_{ab}\left[\frac{1}{2}\nabla_a\varphi\nabla^a\varphi + V(\varphi)\right]+ \nabla_{a}\nabla_{b}F -g_{ab}\Box F,
\end{equation}
\begin{equation}\label{dtstkleingordon}
\Box\varphi+\frac{1}{2} R F_{,\varphi}-V_{,\varphi}=0,
\end{equation}
respectively, were $F$ represents the non minimal coupling of the geometry (the Ricci scalar) with the field $\varphi$. Notice that, since the Ricci scalar naturally enters in Eq. \eqref{dtstkleingordon}, there is no need to add backreaction by hand. We will then treat the full case writing the action on shell. Using the trace of the gravitational field equations and the Klein-Gordon equation, Eq. \eqref{dtstaction} can be written on-shell as
\begin{equation}\label{dtstactionshell}
S=\frac{1}{2}\int d x^{4}\sqrt{-g}\left[K(\varphi)\nabla_a\varphi\nabla^a\varphi+ W(\varphi)\right],
\end{equation}
\begin{eqnarray}
&&K(\varphi)=-\frac{3(F_{,\varphi}^2-2 F F_{,\varphi\varphi})}{2F+3F_{,\varphi}^2},\nonumber\\
&&W(\varphi)=\frac{VF_{,\varphi}^2- FF_{,\varphi}V_{\varphi}-2 F V}{2F+3F_{,\varphi}^2}.
\end{eqnarray}

At this point, defining the integrals $I_1$ and $I_2$ as
\begin{eqnarray}
&&I_1=\int e^{\Phi}K(\varphi)\left[\varphi_{,q}^2+\sum_{i=2}^{3}\varphi_{,w_i}^2\right]\,  d\Omega ,\nonumber \\ 
&&I_2=2\int e^{\Phi}W\left(\varphi\right)\, d \Omega, 
\end{eqnarray}
we can write the energy function associated to this case as
\begin{equation}
 E(\lambda)=\left(\frac{I_1}{ \lambda}+\frac{I_2}{\lambda^3}\right),
\end{equation}
which leads to the relations
\begin{eqnarray}
&&\frac{\partial E(\lambda)}{\partial \lambda}\bigg|_{\lambda=1}=0\rightarrow I_1+3I_2= 0,\nonumber\\
&&\frac{\partial^2 E(\lambda)}{\partial \lambda^2}\bigg|_{\lambda=1}=2I_1\,,
\end{eqnarray}
therefore stability with a non minimal coupling is possible if $I_1>0$ and  $I_2<0$. These results allow us to provide limits on the functions $F$ and $V$. In particular, we must have
\begin{eqnarray}
&&-\frac{1}{6}\varphi^2<F\leq \frac{\beta ^2 }{4 \alpha }\varphi ^2 +\beta  \varphi+\alpha,\\
&&V_0\varphi^4<V\leq V_0 \left(\varphi-\frac{2\alpha}{\beta}\right)^{2-\frac{4\alpha}{3\beta^2}},
\end{eqnarray}
where  $V_0<0$, $\alpha>0$ and $\beta$ can be chosen freely.

\subsubsection{Particular case of the coupling}

Notice that Eq. \eqref{dtstactionshell} can only be used if we exclude the coupling 
\begin{equation}
F_C=-\frac{1}{6}\varphi^2.
\end{equation}
In order to obtain a result also in this case we can construct an action on shell using the Klein Gordon equation only. We thus obtain
\begin{equation}
S=-\frac{1}{2}\int d x^{4}\sqrt{-g}\left[\varphi\Box\varphi+\nabla_a\varphi\nabla^a\varphi+2 W_C(\varphi)\right]\;
\end{equation}
\begin{equation}
W_C(\varphi)=V- \varphi V_{,\varphi}.
\end{equation}
The above is in principle a non canonical action. However, it can be converted in a canonical one integrating by parts the higher order term. Indeed we have
\begin{equation}
\varphi\Box\varphi= \nabla_a\left(\varphi\nabla^a\varphi\right)-\nabla_a\varphi\nabla^a\varphi,
\end{equation}
and the action on shell can be written as
\begin{equation}
S=-\frac{1}{2}\int d x^{4}\sqrt{-g}\Big[2\nabla_a\varphi\nabla^a\varphi+ W_C(\varphi)\Big]\;.
\end{equation} 
In this way we can employ the usual procedure to explore this case. The energy function associated to this particular case becomes
\begin{equation}
 E(\lambda)=\left(2\frac{I_1}{ \lambda}+\frac{I_2}{\lambda^3}\right),
\end{equation}
where the integrals $I_1$ and $I_2$ are defined to be
\begin{eqnarray}
&&I_1=\int e^{\Phi}\left[\varphi_{,q}^2+\sum_{i=2}^{3}\varphi_{,w_i}^2\right]\,  d\Omega , \nonumber \\ 
&&I_2=2\int e^{\Phi}W_C\left(\varphi\right)\, d \Omega, 
\end{eqnarray}
which leads to the relations
\begin{eqnarray}
&&\frac{\partial E(\lambda)}{\partial \lambda}\bigg|_{\lambda=1}=0\rightarrow 2I_1+3I_2= 0,\nonumber\\
&&\frac{\partial^2 E(\lambda)}{\partial \lambda^2}\bigg|_{\lambda=1}=-4I_1<0\,,
\end{eqnarray}
and implies instability. Therefore, only very specific combinations of the coupling and the potential can lead to stable scalar field configurations.

\subsubsection{Horndeski theory}

It is widely believed today that the most general model with a single additional scalar degree of freedom  and second order field equations is given by the so called Horndeski theory \cite{horndeski1}. This theory has been proved to be equivalent to the  curved spacetime generalisation of a scalar field theory with Galilean shift symmetry in flat spacetime  \cite{deffayet1,kobayashi1}. We will now apply the equilibrium  and stability conditions we have derived above to this class of theories.

It is useful for our purposes to write their action in the form
\begin{equation}
S=\sum_{i=1}^{4}\int d^{4}x\sqrt{-g}{\cal L}_i\,,
\end{equation}
\begin{eqnarray}
{\cal L}_1=&&G_1\,,\nonumber\\
{\cal L}_2=&&-G_{2}\Box\varphi\,,\\
{\cal L}_3=&&G_{3}R+G_{3,X}\left[(\Box\varphi)^2-\varphi_{; mn}\varphi^{; mn}\right]\,,\nonumber\\
{\cal L}_4=&&G_{4}\,\mathds{G}_{m n}\varphi^{; mn}-\frac{{G}_{4,X}}{6}\left[(\Box\varphi)^3 +2\varphi_{; mn}\varphi^{;m}{}_{;a} \varphi^{; an}-3\varphi_{; mn}\varphi^{; mn}\Box\varphi\right]\,.\nonumber
\end{eqnarray}
Here $\mathds{G}_{m n}$ is the Einstein tensor, $G_{i}$ are functions of $X$ and $\varphi$, and the subscript $;$ denotes covariant derivatives. Under Eqs. \eqref{dtqctrans} we have, together with Eq. \eqref{dtR2trans}, the trnsformations
\begin{eqnarray}
&&\Box\varphi\Rightarrow \lambda^2 \Box\varphi + ...\nonumber\\
&&\varphi_{; mn}\Rightarrow\varphi_{; mn}+ ...\\
&&\mathds{G}_{m n}\varphi^{; mn}\Rightarrow \lambda^6 \mathds{G}_{m n}\varphi^{; mn} + ...\nonumber
\end{eqnarray}
where, as before, the dots represent additional terms which contain derivatives of $\lambda$ and therefore are irrelevant to the stability study.

Before proceeding we should point out that this theory has a non canonical Lagrangian and therefore the energy function is not the standard one. However, the Ostrogradsky approach \cite{ostrogradsky1} can be used to show that also in this case one can define a function that has the characteristics of the energy of the system (i.e. it is conserved and generates a time evolution).

The transformed action, in the static case and assuming for brevity $\lambda=\lambda_0$, can be written as
\begin{equation}
S(\lambda)=\int d^{4}x\frac{\sqrt{-g}}{\lambda^3}\left[{\cal L}_1+\lambda^2{\cal L}_2+\lambda^4{\cal L}_3+\lambda^6{\cal L}_4\right],
\end{equation}
where ${\cal L}_i={\cal L}_i(\lambda)$. The Pohozaev identity for $\lambda=1$ reads then
\begin{equation}\label{dthtpohozaev}
2X{\cal L}_{1,X}-3{\cal L}_1+2X{\cal L}_{2,X}-{\cal L}_2+2X{\cal L}_{3,X}+{\cal L}_3+2X{\cal L}_{4,X}+3{\cal L}_4=0\,.
\end{equation}
The Eq. \eqref{dthtpohozaev} contains derivatives of the scalar field and therefore, without further assumptions, it cannot be used to obtain general constraints of the functions $G_i$ as we have done in the previous section. When such assumptions are provided, one can find a number different combinations of these functions which can lead to stability. The trivial ones are the ones corresponding to  general relativity ($G_{1}=-\frac{1}{2}X-V$,$G_{2}=0$, $G_{3,X}=0$ and $G_{4}=0$) and scalar tensor gravity ($G_{1}=-\frac{1}{2}C_1(\varphi)X-V$,$G_{2}=0$, $G_{3}=C_3(\varphi)$ and $G_{4}=0$). A more general analytical condition can be obtained, for example, assuming only $G_{4}=0$. In this case stability is possible for
\begin{eqnarray}
&&G_2=C_{2,1}(\varphi)X^{1/2}+C_{2,2}(\varphi)X,\nonumber\\
&&G_3=C_{3,1}(\varphi)X^{3/2}+C_{3,2}(\varphi),
\end{eqnarray}
and $G_1$ within the functions
\begin{eqnarray}
&&G^*_{1}=C_{1,1}(\varphi)X^{5/2}+C_{1,2}(\varphi)X^{3/2}+C_{1,3}(\varphi)X\nonumber\\
&&G^*_2=C_{1,5}X^{3/2}+C_{1,6}(\varphi)X+C_{1,4}(\varphi)\ln X+C_{1,7}(\varphi).
\end{eqnarray}

Note that this is only a particular solution  for the $G_i$. Obtaining a general solution would require the resolution of a non linear third order differential equation for $G_3$ which cannot be achieved in general. 

As said, there is a number of other combinations of forms of the function $G_i$ which can lead to stability. The fact that such a big number of different conditions is possible is to be ascribed to the high level of generality of this class of theories. Members of Horndeski group of theories can have a wildly different physical behaviour and this reflects also in the stability of localised solution of their scalar degree of freedom.

\section{Conclusions}

In this chapter we have presented a complete proof of Derrick's theorem in curved spacetime. 
The proof follows the same strategy of the original paper by Derrick, but has the advantage to be completely covariant and to not require any assumption on the potential for the scalar field nor  on the underlying geometry (other than the absence of vorticity).  This is made possible by combining the 1+1+2 covariant formalism and a technique firstly developed by Darboux which allows to write down Lagrangians for dissipative systems.

For scalar fields in a fixed background, i.e., non backreacting, we have been able to prove that  no stable localised solution of the Klein-Gordon equation is possible.  In addition, we found  that the coupling of the scalar field with other types of matter can only change the conditions necessary to achieve equilibrium. 

The results we have obtained can be understood recognising that a scalar field can be represented macroscopically as an effective fluid with negative pressure (tension). Such fluids will tend naturally to collapse in flat spacetime. In the non flat case, if the energy of the scalar field is not enough to appreciably influence the curvature of spacetime (i.e. without backreaction) a localised solution will be again unstable. When couplings are considered, the  interaction influences the tension of the scalar field, changing its magnitude.

In the case of real scalar field stars, as the metric of these compact objects is determined by its matter distribution, backreaction cannot be neglected. Our generalisation of Derrick's theorem, therefore, implies that in curved spacetimes stable relativistic stars made of real scalar fields cannot exist, even when considering couplings. Indeed the introduction of gravity ``mitigates'' the instability in the sense that the maximum of the energy of the scalar field solution is turned in an inflection point.

In terms of the fluid interpretation this is also clear: if the scalar field is able to influence the spacetime in which it is embedded, its tension will induce a repulsion, much in the same way as dark energy. As a result, the configuration is less unstable and, in principle, with enough tension stability could be possible. Our result suggests, however, that no sufficient level of tension can be achieved to support a localised solution. 

It should be remarked, however, that full stability is not a necessary condition for physical validity. In order to be able to consider scalar field stars as possible astrophysical objects one can require the weaker conditions of ``long term'' stability (e.g. longer/comparable to the age of the Universe). It is not possible with sole the tools of Derrick's proof to evaluate this aspect of the stability of scalar field stars.

Differently from full stability, the equilibrium condition still constitutes a strong physical constraint. Its  application in the context of non canonical scalar field and some classes of modified gravity reveals that the extended Pohozaev identity can be used to select potentials and classes of theories which present an equilibrium. This is particularly relevant in the case of modified gravity, as for these theories all compact stars are also real scalar field stars in which the scalar field coincides with the gravitational scalar degrees of freedom. We found some very stringent criteria for scalar tensor gravity and Horndeski-type theories.  Also in this case, coupling with other fields as well as standard matter might be able to modify these constraints.

We conclude pointing out that our analysis was performed in a purely classical setting: we have neglected completely the quantum nature of the scalar field and the corresponding modification to its action. These corrections, which are necessary to build more realistic models of scalar field stars, could lead to modifications of the equilibrium conditions we have derived and even to stability.  
\cleardoublepage


\chapter{Conclusions}
\label{chapter:conclusions}

In this thesis we have studied a wide variety of applications of the recently proposed generalized hybrid metric-Palatini gravity theory in the context of cosmology and astrophysics as well as a few extansions. We shall now summarize the most important conclusions on the topics explored throughout this work.

In Chapter 3, we have used the so-called reconstruction methods to obtain cosmological solutions representing FLRW universes in the scalar-tensor representation of the GHMPG. The most important contrast with GR and $f\left(R\right)$ theories is that a given behavior of the scale-factor is not anymore associated with a specific distribution of matter. In fact, we show that for every vacuum solution there is always a perfect-fluid counterpart. The relevance of this kind of solutions stands on the fact that a negligible matter distribution can still provide expansion laws coherent with observations without requiring the total baryonic matter to be a relevant percentage of the total energy density, nor the explicit appearence of a cosmological constant in the action.

Using the dynamimcal system formalism, in chapter 4 we explored the cosmological phase space of the GHMPG. Due to the structure of the phase space, i.e., the existence of invariant submanifolds associated with zero spacial curvature ($K=0$), vacuum ($\Omega=0$) and a vanishing Palatini scalar $\mathcal R$ ($Z=0$), and the fact that no attractors lie in the intersection of all three submanifolds, there can not be global attractors in the cosmological phase space. Eitherway, and due to the fourth-order nature of the theory, we have shown that for specific choices of the action there are stable cosmological solutions associated with scale factors that diverge in a finite time or tend assimptotically to a constant value. Both these solutions are also able to model phenomena like the big bang, inflation, and late-time cosmic acceleration. However, we did not study how specific orbits approach these solutions, and thus both behaviors might occur only for timescales $t\to\infty$.

In the pursuit of analytical solutions of the GHMPG describing compact objects, we derived the junction conditions of the theory in chapter 5 for both the geometrical and the scalar-tensor representations. The results are coherent, thus emphasizing the expected equivalence between the two formalisms. In this chapter, we also provide two examples of application of these conditions. The most important differences between GR and the GHMPG stand on the fact that for the matching to be smooth both the Ricci scalar $R$ and the Palatini scalar $\mathcal R$ must also be continuous across the matching hypersurface, which was not necessary in GR, and also that the matching with a thin shell must also preserve the continuity of the trace of the extrinsic curvature, $K$. In general, this implies that the matching must be done at an hypersurface with a specific radius $r$. Curiously, these extra conditions correspond to well-known values for the radial coordinate in the two specific examples considered, namely the Buchdhal radius $r=9M/4$ and the light-ring radius $r=3M$. However, due to the extra junction conditions this theory can not reproduce other well-known results such as the Oppenheimer-Snyder collapse.

We have also studied the existence of analytical traversable wormhole solutions in chapter 6. Using the junction conditions previously derived, we obtained an analytical solution for a traversable wormhole where the throat is supported by a non-exotic perfect fluid and a thin-shell, and an exterior described by a Schwarzschild-AdS solution. This wormhole satisfies the NEC not only near the throat but also for the entire spacetime, being thus the first solution of its kind. For this solution to exist, the matching has to be done at a very specific radius and the free parameters of the theory must be fine-tuned. This might imply that these kind of analytical solutions are scarse. However, more complicated and realistic solutions might be obtained via numerical integrations of the field equations and the free parameters of these solutions might not be so strongly constrained.

In chapter 7 we have studied the stability of the massive scalar degree of freedom of the GHMPG in a Kerr-BH background under the assumption of the usual Lorentz gauge. We have shown that the equation that describes the perturbation in the Ricci scalar $\delta R$ can be written in the form of a fourth order massive scalar-field equation which can be factorized in the form of two modified Klein-Gordon equations. These equations have been extensively studied and are known to be separable in a Kerr-BH background and might give rise to superradiant instabilities if the superradiant modes are confined by a potential well. We show that there are specific regular forms of the action for which the superradiant condition and the confinement of the superradiant modes never occur simultaneously, and thus the Kerr-BH is stable against perturbations in the massive scalar degree of freedom of the GHMPG.

As an intermediate step to study sixth and eight-order hybrid metric-Palatini theories, we used the dynamical system formalism in chapter 8 to study the cosmological phase space of quantum-gravity motivated theories where the Einstein-Hilbert term in the action interacts with higher order terms of the form $\Box^n R$. Although we did not consider the Palatini scalar $\mathcal R$ in this chapter, we can still enumerate important conclusions about these models, namely the fact that, unlike expected, the higher-order terms are not negligible with respect to the lower-order ones, and their presence affects not only the structure of the phase space itself but also the stability of the fixed points it presents.

Finally, in chapter 9 we have provided a new formalism to approach the Derrick's theorem. We have rederived this theorem in flat spacetime, generalized the theorem to curved spacetimes and also included the effects of backreaction. We have shown that no stable localized solutions of real scalar fields minimally coupled to gravity can exist except for the particular case of Phantom fields. Also, in general, scalar-tensor theories of gravity also present this instability except for very particular cases. In the context of Horndeski theories, stable models can be found for specific forms of the functions $G_i$. Unfortunately, due to the geometrical nature of $f\left(R\right)$ and hybrid metric-Palatini gravity theories, the same approach used in this work can not be applied because the quasi-conformal transformation in the functions $f$ ceases to be trivial.

We have shown that the GHMPG is able to succesfully describe a wide variety of phenomena with relevant implications in cosmology and astrophysics. However, there are still some milestones to achieve. In this work, we have not considered the weak-field slow-motion regime of the theory to study the solar system dynamics, which is of crucial importance to the validity of the theory. Also, our results of chapter 5 seem to indicate that no regular perfect-fluid star solutions satisfying the NEC can exist in this theory, which requires further study. A deeper analysis of the Derrick's theorem in geometrical modified theories of gravity must be done in the pursuit of prooving the results of chapter 10 in this kind of theories and, in particular, in the GHMPG. 
\cleardoublepage

%
\appendix



\chapter{Variational calculations}
\label{appendixC}

In this appendix, we compute explicitly the variations of the quantities needed to compute the equations of motion of the generalized hybrid metric-Palatini gravity, more precisely the metric determinant $g$, the volume element $\sqrt{-g}$, metric inverse $g^{ab}$, the Levi-Civita connection $\Gamma^c_{ab}$, and the Ricci tensor $R_{ab}$ in both the isolated form and multiplied by a scalar quantity $\phi$.

\section{Variations of the metric quantities}

Let us start by computing the variation of the metric quantities $g$, $\sqrt{-g}$ and $g^{ab}$. Before we start, we have to define the determinant of a matric. Let $A^{ab}$ be the cofactor matrix of the metric $g_{ab}$. We define the determinant $g$ of the matrix $g_{ab}$ as follows:
\begin{equation}
g=g_{ab}A^{ab},
\end{equation}
where the index $a$ is fixed i.e. we are performing this calculation along a given line of the matric $g_{ab}$, and we sum over the index $b$. The variation of the determinant $g$ as a function of the cofactor matrix $A^{ab}$ thus becomes
\begin{equation}\label{Cdetcofactor}
\delta g=\frac{\partial g}{\partial g_{ab}} \delta g_{ab} = A^{ab} \delta g_{ab}.
\end{equation}
The metric inverse $g^{ab}$ can also be written in terms of the cofactor matrix in the form
\begin{equation}\label{Cinvcofactor}
g^{ab}=\frac{1}{g}(A^{ab})^T=\frac{1}{g}A^{ba}.
\end{equation}
Multiplying both sides of Eq.\eqref{Cinvcofactor}, using the fact that the metric $g_{ab}$ is a symmetric matrix and thus also the cofactor matrix $A^{ab}$ is also symmetric, and using the result to cancel the term depending on $a^{ab}$ in Eq.\eqref{Cdetcofactor} we obtain
\begin{equation}\label{Cvardetg}
\delta g=g g^{ab}\delta g_{ab}.
\end{equation}

It is actually more useful to us in the computation of the equations of motion the variation of the volume element $\sqrt{-g}$, which can be obtained from Eq. \eqref{Cvardetg} as
\begin{equation}\label{Cvarsqrtg}
\delta \sqrt{-g}=-\frac{1}{2\sqrt{-g}}\delta g = -\frac{1}{2\sqrt{-g}}g g^{ab} \delta g_{ab}=\frac{1}{2}\sqrt{-g}g^{ab}\delta g_{ab}.
\end{equation}

Furthermore, we can compute the variation of the metric inverse, as follows:
\begin{equation}
\delta g^{ab}=\delta(g^{ac}g^{bd}g_{cd})=g^{bd}g_{cd}\delta g^{ac}+g^{ac}g_{cd}\delta g^{bd}+g^{ac}g^{bd}\delta g_{cd}=\delta^{b}_{c} \delta g^{ac}+\delta^{a}_{d}\delta g^{bd}+g^{ac}g^{bd}\delta g_{cd},
\end{equation}
where we used the definition of the Kronecker delta, $g^{ac}g_{cb}=\delta^{a}_{b}$. Finally, using the fact that the metric is symmetric and canceling the terms $\delta g^{ab}$ we obtain
\begin{equation}\label{Cvarinvgab}
\delta g^{ab}=-g^{ac}g^{bd}\delta g_{cd}.
\end{equation}

\section{Variation of the Levi-Civita connection}

Let us now turn to the variation of the metric connection i.e. the Levi-Civita connection $\Gamma^c_{ab}$. This connection can be written in terms of the metric tensor $g_{ab}$ as
\begin{equation}\label{Cdeflevicivita}
\Gamma^a_{bc}=\frac{1}{2}g^{ad}\left(\partial_bg_{cd}+\partial_cg_{db}-\partial_dg_{bc}\right).
\end{equation}
Taking the derivative of Eq.\eqref{Cdeflevicivita} yields
\begin{equation}\label{Cauxvarlevicivita}
\delta\Gamma^a_{bc}=\frac{1}{2}\delta g^{ad}\left(\partial_bg_{cd}+\partial_cg_{db}-\partial_dg_{bc}\right)+\frac{1}{2}g^{ad}\left(\partial_b\delta g_{cd}+\partial_c\delta g_{db}-\partial_d\delta g_{bc}\right).
\end{equation}
The term $\delta g^{ab}$ can be simplified using the result of Eq.\eqref{Cvarinvgab}. Also, using the definition of $\Gamma^c_{ab}$ in Eq.\eqref{Cdeflevicivita}, we can rewrite Eq.\eqref{Cauxvarlevicivita} in the form
\begin{eqnarray}
\delta\Gamma^a_{bc}&=&-g^{ae}\Gamma^f_{bc}\delta g_{ef}+\frac{1}{2}g^{ad}\left(\partial_b\delta g_{cd}+\partial_c\delta g_{db}-\partial_d\delta g_{bc}\right)=\nonumber \\
&=&\frac{1}{2}g^{ad}\left(\partial_b\delta g_{cd}+\partial_c\delta g_{db}-\partial_d\delta g_{bc}-2\Gamma^e_{bc}\delta g_{de}\right),\label{Cauxvarlevicivita2}
\end{eqnarray}
where the last equality is obtained from index manipulation. Consider now the following results for the covariant derivative of the variation of the metric $g_{ab}$, which can be obtained from each other via index manipulation:
\begin{equation}\label{Cauxcovgab1}
\nabla_b\delta g_{cd}=\partial_b\delta g_{cd}-\Gamma^e_{bc}\delta g_{ed}-\Gamma^e_{bd}\delta g_{ce},
\end{equation}
\begin{equation}
\nabla_c\delta g_{db}=\partial_c\delta g_{db}-\Gamma^e_{cd}\delta g_{eb}-\Gamma^e_{cb}\delta g_{de},
\end{equation}
\begin{equation}\label{Cauxcovgab3}
\nabla_d\delta g_{bc}=\partial_d\delta g_{bc}-\Gamma^e_{db}\delta g_{ec}-\Gamma^e_{dc}\delta g_{be}.
\end{equation}
Inserting the definitions given in Eqs.\eqref{Cauxcovgab1} to \eqref{Cauxcovgab3} into the partial derivative terms of Eq.\eqref{Cauxvarlevicivita2} we obtain
\begin{eqnarray}
\delta\Gamma^a_{bc}&=&\frac{1}{2}g^{ad}\left(\nabla_b\delta g_{cd}+\nabla_c\delta g_{db}+\nabla_d\delta g_{bc}+\Gamma^e_{bc}\delta g_{ed}+\Gamma^e_{bd}\delta g_{ce}+\right.\nonumber\\
&+&\left.\Gamma^e_{cd}\delta g_{eb}+\Gamma^e_{cb}\delta g_{de}-\Gamma^e_{db}\delta g_{ec}-\Gamma^e_{dc}\delta g_{be}-2\Gamma^e_{bc}\delta g_{de}\right).
\end{eqnarray}
Finally, manipulating the indeces and using the fact that both the metric and the connection are symmetric in the covariant indeces, the connection terms cancel out and we obtain the final result
\begin{equation}\label{Cvarlevicivita}
\delta\Gamma^a_{bc}=\frac{1}{2}g^{ad}\left(\nabla_b\delta g_{cd}+\nabla_c\delta g_{db}+\nabla_d\delta g_{bc}\right).
\end{equation}

\section{Variation of the isolated Ricci tensor}

To compute the variation of the Ricci tensor, we shall write it explicitly in terms of the Levi-Civita connection $\Gamma_{ab}^c$ in the form
\begin{equation}\label{Cdefricci}
R_{ab}=R^{c}_{acb}=\partial_c \Gamma^{c}_{ab}-\partial_b \Gamma^{c}_{ac}+\Gamma^{e}_{ab}\Gamma^{c}_{ec}-\Gamma^{e}_{ac}\Gamma^{c}_{eb}.
\end{equation}
A variation of the previous definition yields
\begin{equation}
\delta R_{ab}=\partial_c \delta \Gamma^{c}_{ab}-\partial_b \delta \Gamma^{c}_{ac} +\delta \Gamma^{e}_{ab}\Gamma^{c}_{ec}+\Gamma^{e}_{ab}\delta \Gamma^{c}_{ec}-\delta \Gamma^{e}_{ac}\Gamma^{c}_{eb}-\Gamma^{e}_{ac}\delta \Gamma^{c}_{eb}.
\end{equation}
This result can be simplified if we notice that these eight terms correspond to two covariant derivatives of the Levi-Civita connection $\Gamma^c_{ab}$, given by
\begin{equation}\label{Cauxcovlevicivita1}
\nabla_c \delta \Gamma^{c}_{ab}=\partial_c \delta \Gamma^{c}_{ab} + \Gamma^{c}_{ec}\delta \Gamma^{e}_{ab}-\Gamma^{e}_{ac}\delta \Gamma^{c}_{eb}-\Gamma^{e}_{bc} \delta \Gamma^{c}_{ea},
\end{equation}
\begin{equation}\label{Cauxcovlevicivita2}
\nabla_b \delta \Gamma^{c}_{ac}=\partial_b \delta \Gamma^{c}_{ac} + \Gamma^{c}_{eb}\delta \Gamma^{e}_{ac}-\Gamma^{e}_{ab}\delta \Gamma^{c}_{ec}-\Gamma^{e}_{cb} \delta \Gamma^{c}_{ea},
\end{equation}
and so, using the fact that the Levi-Civita connection is symmetric in the covariant indeces, we can sum and subtract the last term of Eqs.\eqref{Cauxcovlevicivita1} and \eqref{Cauxcovlevicivita2}, leading us to the result
\begin{equation}\label{Cvarricci}
\delta R_{ab}=\nabla_c \delta \Gamma^{c}_{ab}-\nabla_b \delta \Gamma^{c}_{ac}.
\end{equation}
Eq.\eqref{Cvarricci} gives us the form of the variation of the Ricci tensor in terms of covariant derivatives of the variations of the Levi-Civita connection. Let us now see how these terms contribute to the equations of motion from the action principle. We can write the variation of the term with the Ricci tensor in the action as
\begin{equation}\label{Cauxactionrab}
\delta S=\int_\Omega \sqrt{-g}g^{ab}\delta R_{ab}d^4x=\int_\Omega \sqrt{-g}g^{ab} \left( \nabla_c \delta \Gamma^{c}_{ab}-\nabla_b \delta \Gamma^{c}_{ac}\right)d^4x.
\end{equation}
Now, applying the Leibniz rule to $\nabla_c (g^{ab}\delta\Gamma^{c}_{ab})$, noting that the covariant derivative of the metric $g_{ab}$ vanishes by definition, and manipulating the indeces so we are able to write both resulting terms inside the same covariant derivative, the integral in Eq.\eqref{Cauxactionrab} takes the form
\begin{equation}
\delta S=\int_\Omega \sqrt{-g}\nabla_c \left(g^{ab}\delta\Gamma^{c}_{ab}-g^{ac}\delta\Gamma^{b}_{ab}\right)d^4x.
\end{equation}
Let us now define a new vector $A^c\equiv g^{ab}\delta\Gamma^{c}_{ab}-g^{ac}\delta\Gamma^{b}_{ab}$ so that we can write the integral as
\begin{equation}
\delta S=\int_\Omega \sqrt{-g}\nabla_c A^c d^4x,
\end{equation}
and using the result given in Eq.\eqref{Ddefbox} for the D'Alembert operator of a vector field, the action integral takes the final form
\begin{equation}
\delta S=\int_\Omega \sqrt{-g}\frac{1}{\sqrt{-g}}\partial_c \left( \sqrt{-g} g^{cb}A_b\right) d^4x=\int_\Omega \partial_c \left( \sqrt{-g} A^c\right) d^4x.
\end{equation}
Finally, by the Stokes theorem, the result of the previous integration is
\begin{equation}\label{Cvarrab}
\delta S=\int_{\partial\Omega}\sqrt{-g} A^c d^3x=0,
\end{equation}
which vanishes identically since, by definition, the variation vanishes at the boundary $\partial\Omega$ which in this case corresponds to the infinity. Therefore, this term does not contribute to the equation of motion for the metric tensor.

\section{Variation of the Ricci tensor multiplied by a scalar}

In this case we want to study the variation $\phi\delta R_{ab}$, where $\phi$ is a scalar field. Due to the presense of the scalar field, this term can no longer be manipulated into a boundary term and, therefore, does not vanish like the variation of the isolated Ricci tensor. Using Eq.\eqref{Cvarricci}, the action term becomes
\begin{equation}\label{Cauxactionrabphi}
\delta S=\int_\Omega \sqrt{-g}\phi g^{ab}\delta R_{ab}d^4x=\int_\Omega \sqrt{-g}\phi g^{ab} \left( \nabla_c \delta \Gamma^{c}_{ab}-\nabla_b \delta \Gamma^{c}_{ac}\right)d^4x.
\end{equation}
Inserting the results for the variation of the Levi-Civita connection given in Eq.\eqref{Cvarlevicivita} into the previous result, we obtain
\begin{equation}
\delta R_{ab}=\nabla_c \left[\frac{1}{2}g^{cd}\left(\nabla_a\delta g_{db}+\nabla_b\delta g_{da}-\nabla_d\delta g_{ab}\right)\right]-\nabla_b \left[\frac{1}{2}g^{cd}\left(\nabla_a\delta g_{dc}+\nabla_c\delta g_{da}-\nabla_d\delta g_{ac}\right)\right].
\end{equation}
Noting that the covariant derivative of the metric $g_{ab}$ vanishes, we can take out the metric inside the covariant derivative in the first four terms and raise the index of the derivative. The fifth and sixth terms cancel out and we obtain
\begin{equation}
\delta R_{ab}=\frac{1}{2}\left(\nabla^d\nabla_a\delta g_{db}+\nabla^d\nabla_b\delta g_{da}-\nabla^d\nabla_d\delta g_{ab}-g^{cd}\nabla_b\nabla_a\delta g_{dc}\right).
\end{equation}
Replacing this result into the action term given in Eq.\eqref{Cauxactionrabphi} and contracting with the metric inverse it yields
\begin{equation}
\delta S=\int_\Omega \sqrt{-g}\phi\frac{1}{2}\left(\nabla^d\nabla^b\delta g_{db}+\nabla^d\nabla^a\delta g_{da}-g^{ab}\Box\delta g_{ab}-g^{cd}\Box\delta g_{dc}\right)d^4x.
\end{equation}
Manipulating the indeces, the previous result can be further simplified to
\begin{equation}
\delta S=\int_\Omega \sqrt{-g}\phi\left(\nabla^a\nabla^b\delta g_{ab}-g^{ab}\Box\delta g_{ab}\right)d^4x.
\end{equation}
Finally, using the definition of the D'Alembert operator given in Eq.\eqref{Ddefbox} and integrating by parts twice each of the terms inside the integral, we can isolate the variation of the metric and obtain the final result as
\begin{equation}\label{Cauxactionrabphi2}
\delta S=\int_\Omega \sqrt{-g} \left(\nabla^a\nabla^b\phi-g^{ab}\Box\phi\right) \delta g_{ab}d^4x.
\end{equation}
Comparing Eq.\eqref{Cauxactionrabphi2} with Eq.\eqref{Cauxactionrabphi}, we verify that we can write the variation of the Ricci tensor multiplied by a scalar field $\phi$ as
\begin{equation}\label{Cvarrabphi}
g^{ab}\phi\delta R_{ab}=\left(\nabla^a\nabla^b\phi -g^{ab}\Box\phi \right)\delta g_{ab}.
\end{equation} 
\cleardoublepage

\chapter{Useful results in Riemannian geometry}

Throughout this thesis, there were a few results used without demonstration, more specifically how Eq.\eqref{frpeomcon} implies the existance of a metric $h_{ab}=f'\left(\mathcal R\right)g_{ab}$ (and the same for Eq.\eqref{ghmpgeomcon} in the generalized hybrid metric-Palatini gravity), the explicit expression used for the D'alembertion operator used for example to derive the equations of motion in chapter \ref{appendixA}, and the relation between Ricci tensors of conformally related metrics used to compute the relation between $R_{ab}$ and $\mathcal R_{ab}$. In this appendix, we provide brief demonstrations of these results.

\section{Condition for $\Gamma$ to be the Levi-Civita connection}\label{sec:Dlevicivita}

In this section we show that the covariant derivative $\nabla_c\left(\sqrt{-g}g^{ab}\right)$ vanishes if and only if the connection $\Gamma^c_{ab}$ is the Levi-Civita connection for the metric tensor $g_{ab}$. Let us start by prooving the inverse implication. Since the metric tensor $g_{ab}$ is a rank-2 covariant tensor, by taking the determinant of the coordinate transformation we can show that the determinant $g$ transforms like a scalar density of weight 2:
\begin{equation}
{g}'_{ab}\left({x}'\right)=\frac{\partial x^c}{\partial {x}'^a}\frac{\partial x^d}{\partial {x}'^b}g_{cd}\left(x\right) \Leftrightarrow {g}'=J^2g \Leftrightarrow \sqrt{-{g}'}=J\sqrt{-g},
\end{equation}
where $J$ is the jacobian of the coordinate change matrix. Noting that the covariant derivative of the metric tensor $g_{ab}$ vanishes, from $\nabla_c g_{ab}=0$ we can derive the relation
\begin{equation}
\partial_cg_{ab}=\Gamma^d_{ac}g_{db}+\Gamma^d_{bc}g_{ad}.
\end{equation}
Inserting this result into Eq.\eqref{Cvardetg} we can write 
\begin{equation}
\partial_cg=gg^{ab}\left(\Gamma^d_{ac}g_{db}+\Gamma^d_{bc}g_{ad}\right)=2g\Gamma^a_{ac},
\end{equation}
where we have used the definition of the Kronecker delta $\delta^a_b=g^{ac}g_{cb}$. It is now clear that, applying the expression for the covariant derivative of a scalar density of weight 2, we get
\begin{equation}
\nabla_cg=\partial_cg-2g\Gamma^a_{ac}=0.
\end{equation}
Using the same approach in Eq.\eqref{Cvarsqrtg} we find 
\begin{equation}
\partial\sqrt{-g}=\sqrt{-g}\Gamma^a_{ac} \implies \nabla_c\sqrt{-g}=0.
\end{equation}
Combining this result with the fact that the covariant derivative of the metric tensor $g_{ab}$ vanishes, we arrive at the expected result
\begin{equation}
\nabla_c\left(\sqrt{-g}g_{ab}\right)=g_{ab}\nabla_c\sqrt{-g}+\sqrt{-g}\nabla_cg_{ab}=0.
\end{equation}
Let us now prove the equivalence by assuming that this relation hold and showing that $\Gamma^c_{ab}$ must be the Levi-Civita connection for the metric $g_{ab}$. The covariant derivative of $\sqrt{-g}$ can be written as, noticing that this is a scalar density of weight 1,
\begin{equation}\label{Dnablasqrtg}
\nabla_c\sqrt{-g}=\partial_c\sqrt{-g}-\Gamma^d_{dc}\sqrt{-g}=\frac{1}{2}\sqrt{-g}g^{ab}\partial_c g_{ab}-\Gamma^d_{dc}\sqrt{-g},
\end{equation}
where in the second equality we have used the result given in Eq.\eqref{Cvarsqrtg}. Notice that
\begin{equation}
0=\partial_c\delta^a_a=\partial_c\left(g^{ab}g_{ab}\right)=g_{ab}\partial_cg^{ab}+g^{ab}\partial_cg_{ab},
\end{equation}
and we can use this result to change the sign of the first term in Eq.\eqref{Dnablasqrtg} to obtain a term $\partial_cg^{ab}$. We are also going to use the fact that the following covariant derivative vanishes:
\begin{equation}
\nabla_c\left(\sqrt{-g}g^{ab}\right)=\partial_c\sqrt{-g}g^{ab}+\sqrt{-g}\partial_cg^{ab}-\sqrt{-g}g^{ab}\Gamma^d_{dc}+\sqrt{-g}\Gamma^a_{cd}g^{db}+\sqrt{-g}\Gamma^b_{cd}g^{ad}=0.
\end{equation}
Replacing this result in the first term of the covariant derivative of $\sqrt{-g}$ in Eq.\eqref{Dnablasqrtg} we obtain
\begin{equation}
\nabla_c\sqrt{-g}=\frac{1}{2}g_{ab}\left(\partial_c\sqrt{-g}g^{ab}-\sqrt{-g}g^{ab}\Gamma^d_{dc}+\sqrt{-g}\Gamma^a_{cd}g^{db}+\sqrt{-g}\Gamma^b_{cd}g^{ad}\right),
\end{equation}
and applying the definition of the Kronecker's delta we are lead to the result
\begin{equation}
\nabla_c\sqrt{-g}=2\partial_c\sqrt{-g}-2\Gamma^d_{dc}\sqrt{-g}=2\sqrt{-g}\Leftrightarrow\nabla_c\sqrt{-g}=0.
\end{equation}
Combining this result with our first assumption $\nabla_c\left(\sqrt{-g}g_{ab}\right)=0$, it yields directly
\begin{equation}
0=\nabla_c\left(\sqrt{-g}g^{ab}\right)=\sqrt{-g}\nabla_c g^{ab}+g^{ab}\nabla_c\sqrt{-g}=\sqrt{-g}\nabla_c g^{ab}
\end{equation}
which finally implies that
\begin{equation}
\nabla_c g^{ab}=0.
\end{equation}
This result means that if our assumption holds, then $\Gamma^c_{ab}$ must be the Levi-Civita connection for the metric tensor $g_{ab}$. Combining this result with the inverse desmonstration, it is shown that there is an equivalence between $\Gamma^c_{ab}$ being the metric connection and the fact that the covariant derivative vanishes.

\section{Explicit expression for the D'Alembert operator}\label{Sec:dalembert}

Let us start by recalling the result given in Eq. \eqref{Cvarsqrtg}, which can be written, when we consider a derivative of the metric determinant in order to the coordinates $x^a$, as
\begin{equation}
\partial_c g = g g^{ab} \partial_c g_{ab}.
\end{equation}
The fact that the covariant derivative of the metric tensor $g_{ab}$ vanishes gives us a relation between the partial derivative of the metric tensor and the Christoffel symbols $\Gamma^c_{ab}$ given by
\begin{equation}
\partial_c g_{ab}=\Gamma^{d}_{ac}g_{db}+\Gamma^{d}_{bc}g_{ad},
\end{equation}
which can be inserted in the derivative of the metric determinant to obtain
\begin{equation}
\partial_c g = g g^{ab} \left(\Gamma^{d}_{ac}g_{db}+\Gamma^{d}_{bc}g_{ad}\right).
\end{equation}
Manipulating the indeces and using the definition of the Kronecker delta $\delta^a_b$, we can derive the following result
\begin{equation}
\partial_c g = g \left(\Gamma^{d}_{ac}\delta^a_d+\Gamma^{d}_{bc}\delta^b_d\right)=g \left(\Gamma^{d}_{dc}+\Gamma^{d}_{dc}\right)=2g\Gamma^{a}_{ac}.
\end{equation}
This results enables us of writing the partial derivative of $\sqrt{-g}$ as
\begin{equation}
\partial_a \sqrt{-g} = -\frac{1}{2\sqrt{-g}}\partial_a g =-\frac{1}{2\sqrt{-g}}2g\Gamma^{b}_{ba}=\sqrt{-g}\Gamma^{b}_{ba}.
\end{equation}
Using this relation, one can write the covariant derivative of a vector field $A^a$ in the form
\begin{equation}
\nabla_a A^a=\Gamma^{b}_{ba}A^a+\partial_a A^a=\frac{1}{\sqrt{-g}}A^a\partial_a\sqrt{-g}+\partial_a A^a=\frac{1}{\sqrt{-g}}\partial_a\left(\sqrt{-g}g^{ab}A_b\right),
\end{equation}
where we have used the Leibniz rule in order to write the final equality. In particular, if we take the vector $A^a$ to be $\nabla_a\phi=\partial_a\phi$, we find a general expression for the D'Alembert operator of $\phi$ given by
\begin{equation}\label{Ddefbox}
\Box\phi=\nabla^a\nabla_a\phi=\frac{1}{\sqrt{-g}}\partial_a \left( \sqrt{-g} g^{ab}\partial_b \phi \right).
\end{equation}

\section{Relation between Ricci tensors of conformally related metrics}\label{Sec:Brelproof}

Let us consider two metric tensors conformally related by $t_{ab}=\phi g_{ab}$ and compute how the Ricci tensors $R_{ab}$ written in terms of these two metrics relate to each other. Define $\Gamma^c_{ab}$ and $\hat\Gamma^c_{ab}$ as the Levi-Civita connections , and $R_{ab}$ and $\mathcal R_{ab}$ as the Ricci tensors of the metrics $g_{ab}$ and $t_{ab}$ repectively. The connection for the metric $t_{ab}$ can be written in terms of the metric $g_{ab}$ as
\begin{equation}
\hat\Gamma^a_{bc}=\frac{1}{2}t^{ab}\left(\partial_c t_{db}+\partial_b t_{dc}-\partial_d t_{bc}\right)=\frac{1}{2\phi}g^{ad}\left[\partial_c\left(\phi g_{db}\right)+\partial_b\left(\phi g_{dc}\right)-\partial_d\left(\phi g_{bc}\right)\right].
\end{equation}
Expanding the derivatives, re-grouping the terms of $\Gamma^a_{bc}$, and using the definition of the Kronecker delta $g^{ac}g_{cb}=\delta^a_b$ we obtain
\begin{equation}
\hat\Gamma^a_{bc}=\Gamma^a_{bc}+\frac{1}{2\phi}\left(\delta^a_b\partial_c\phi+\delta^a_c\partial_b\phi-g^{ad}g_{bc}\partial_d\phi\right).
\end{equation}
Using the previous expression for $\hat\Gamma^c_{ab}$, we can rearrange the terms of $\Gamma^a_{bc}$ in the definition of the Ricci tensor $\mathcal R_{ab}$ in order to find $R_{ab}$ and so we write
\begin{eqnarray}
\mathcal{R}_{ab}=R_{ab}+
\partial_c\left[\frac{1}{2\phi}\left(\delta^c_a\partial_b\phi+\delta^c_b\partial_a\phi-g^{cd}g_{ab}\partial_d\phi\right)\right]-
\partial_b\left[\frac{1}{2\phi}\left(\delta^c_a\partial_c\phi+\delta^c_c\partial_a\phi-g^{cd}g_{ac}\partial_d\phi\right)\right]+ \nonumber \\
+\frac{1}{2\phi}\Gamma^c_{ce}\left(\delta^e_a\partial_b\phi+\delta^e_b\partial_a\phi-g^{ed}g_{ab}\partial_d\phi\right)+
\frac{1}{2\phi}\Gamma^e_{ab}\left(\delta^c_c\partial_e\phi+\delta^c_e\partial_c\phi-g^{cd}g_{ce}\partial_d\phi\right)-\nonumber \\
-\frac{1}{2\phi}\Gamma^c_{ae}\left(\delta^e_c\partial_b\phi+\delta^e_b\partial_c\phi-g^{ed}g_{cb}\partial_d\phi\right)-
\frac{1}{2\phi}\Gamma^e_{cb}\left(\delta^c_a\partial_e\phi+\delta^c_e\partial_a\phi-g^{cd}g_{ae}\partial_d\phi\right)+\nonumber \\
+\frac{1}{4\phi^2}\left(\delta^c_c\partial_e\phi+\delta^c_e\partial_c\phi-g^{cd}g_{ce}\partial_d\phi\right)
\left(\delta^e_a\partial_b\phi+\delta^e_b\partial_a\phi-g^{ed}g_{ab}\partial_d\phi\right)-\nonumber \\
-\frac{1}{4\phi^2}\left(\delta^c_a\partial_e\phi+\delta^c_e\partial_a\phi-g^{cd}g_{ae}\partial_d\phi\right)
\left(\delta^e_c\partial_b\phi+\delta^e_b\partial_c\phi-g^{ed}g_{cb}\partial_d\phi\right).
\end{eqnarray}
Expanding the derivatives, using $\delta^c_c=4$, and noting that $\partial_a\partial_b\phi=\partial_b\partial_a\phi$ we obtain
\begin{eqnarray}
&&\mathcal{R}_{ab}=R_{ab}+\frac{3}{2\phi^2}\partial_a\phi\partial_b\phi-\frac{1}{\phi}\left(\partial_a\partial_b\phi-\Gamma^c_{ab}\partial_c\phi\right)-\nonumber \\
&&-\frac{1}{2\phi}\left[\partial_c\left(g^{cd}g_{ab}\partial_d\phi\right)+\Gamma^c_{ce}g^{ed}g_{ab}\partial_d\phi-\Gamma^c_{ae}g^{ed}g_{cb}\partial_d\phi-\Gamma^e_{cb}g^{cd}g_{ae}\partial_d\phi\right].
\end{eqnarray}
The third and fourth terms correspond to the definition of the covariant derivative of a covariant tensor. Also, noticing that the covariant derivative of a scalar field reduces to its partial derivative, we can write those terms as $\nabla_a\nabla_b\phi$. The fifth and sixth terms can be expanded as follows:
\begin{equation}
\partial_c\left(g^{cd}g_{ab}\partial_d\phi\right)+\Gamma^c_{ce}g^{ed}g_{ab}\partial_d\phi=g^{cd}\partial_d\phi\partial_c g_{ab}+g_{ab}\Box\phi,
\end{equation}
where $\Box\phi=\nabla_a\left(g^{ab}\partial_b\phi\right)$ is the d'Alembert operator. Finally, the first term on the right hand side of the previous equation can be combined with the seventh and eight terms of $\mathcal{R}_{ab}$ in order to obtain the covariant derivative of the metric tensor $g_{Ab}$
\begin{equation}
\left(\partial_c g_{ab}-\Gamma^e_{ca}g_{eb}-\Gamma^e_{cb}g_{ae}\right)g^{cd}\partial_d\phi=\nabla_c g_{ab}g^{cd}\partial_d\phi=0.
\end{equation}
We then obtain the equation for $\mathcal{R}_{ab}$ as a function of $R_{ab}$ and $\phi$, that is the equation that relates the Ricci tensors of two conformally related metric tensors, as
\begin{equation}
\mathcal{R}_{ab}=R_{ab}-\frac{1}{\phi}\left(\nabla_a\nabla_b\phi+\frac{1}{2}g_{ab}\Box\phi\right)+\frac{3}{2\phi^2}\partial_a\phi\partial_b\phi.
\end{equation}

\section{Volume form in the 1+1+2 formalism}\label{Sec:211}

In this section we derive the relation between the volume form $\sqrt{-g}$ and the integrator multiplied defined in chapter \ref{chapter:chapter9} as $\exp\left(\Phi\right)$. Let us start with the covariant divergence of the vector $u_a$. It is well known that
\begin{equation}
\nabla_a u^a= \frac{1}{\sqrt{-g}}\partial_a\left(\sqrt{-g} u^a\right).
\end{equation}
On the other hand, by definition, $\nabla_a u^a=\Theta$, i.e.,  the left hand side of the previous equation corresponds to the expansion in the 1+1+2 formalism. In this way one can write
\begin{equation}
u^a\partial_a \ln |g|= 2(\Theta-\partial_a u^a).
\end{equation}
The same procedure can be applied to the quantity $\nabla_a e^a$ to obtain
\begin{equation}
 e^a\partial_a \ln |g|= 2(\A+\phi-\partial_a e^a),
\end{equation}
Instead, for $N_{ca}\nabla_{b}N^{ab}$, since
\begin{equation}
\nabla_b N^{ab}= \frac{1}{\sqrt{-g}}\partial_b\left(\sqrt{-g} N^{ab}\right)+\Gamma_{bc}^a N^{bc},
\end{equation}
we have
\begin{equation}
N_c{}^a \partial_a \ln |g|=  2(\A_c+a_c)+N_c{}^b \partial_b \ln{N},
\end{equation}
where we call $N$ the determinant of the non zero minor of $N_{ab}$. In this way we have
\begin{equation}\label{Dderivlng}
\partial_a \ln |g| = 2\Big\{(\partial_c u^c-\Theta)u_a+(\A+\phi-\partial_c e^c)e_a +\left(\A_c+a_c\right)N^c_a\Big\}+N_a{}^b \partial_b\ln{N}.
\end{equation} 
Let us now define
\begin{eqnarray}
&&V_a = -\Theta u_a+(\A+\phi)e_a+\left(\A_c+a_c\right)N^c_a,\\ 
&&W_a = -\left(\partial_b u^b\right) u_a+\left(\partial_b e^b\right) e_a+N_a{}^b \partial_b \ln{\sqrt{N}},
\end{eqnarray}
so that 
\begin{equation}\label{Dderivlng2}
\partial_a \ln{|g|} =2(V_a- W_a).
\end{equation}
The partial differential equation given in Eq. \eqref{Dderivlng} is equivalent to the one we have encountered in the main text to determine $\Phi$ (see Eq. \eqref{dttransphi} for $\lambda=1$). 

In the case of a static and spherically symmetric spacetime and choosing $e_a$ normalised to one we have $\Theta=0$, $\A_c=0$, $a_c=0$ and $\partial_a u^a= 0$. Inserting this result in Eq. \eqref{Dderivlng2} and integrating out the total divergences in $W_a$ we obtain
\begin{equation}
\sqrt{|g|}= g_0\exp\left[\int \left(\A+\phi\right)d q\right]\,,
\end{equation}
i.e., modulus an irrelevant constant, we recover Eq. \eqref{dtphicurved}. This suggests that we can write in a $(-,+,+,+)$ signature
\begin{equation}
e^\Phi =\sqrt{-g}.
\end{equation}
In more general spacetimes, Eq. \eqref{Dderivlng2} is a system of partial differential equations which one must solve in order to find the expression of the volume form. As the case of Eq. \eqref{dtdarbouxcurv}, we can use the method of characteristics to obtain some solutions, but we need an accurate description of the boundary conditions to determine a solution. However, as discussed for $\Phi$, the exact form of the metric tensor is irrelevant for our discussion, the only important thing is the transformation of $\sqrt{-g}$. 

Let us then look at the transformation of $g$. It is easy to see that under a conformal transformation $g_{ab}(\lambda)=\lambda g_{ab}$ one has
\begin{eqnarray}
&&V_a(\lambda) = V_a,\nonumber\\ 
&&W_a(\lambda) = W_a + \partial_a \ln\lambda,\label{DVWtrans}
\end{eqnarray}
and this implies 
\begin{eqnarray}
&&\Phi (\lambda)=\Phi- 4 \ln \lambda\,,\nonumber \\
&&g (\lambda)=\frac{g}{\lambda^4},
\end{eqnarray}
which is consistent with the known conformal transformation of a tensor density.

Under the transformations in Eq. \eqref{dtqctrans} one sees that, again, $V_a$ and $W_a$ transform as in Eqs. \eqref{DVWtrans}. However, these results now imply
\begin{eqnarray}
&&\Phi (\lambda)=\Phi- 3 \ln \lambda,\nonumber \\
&&g (\lambda)=\frac{g}{\lambda^3},
\end{eqnarray}
i.e. the same transformation for $\Phi$ obtained in \eqref{dtexptransform}. 
\cleardoublepage

\chapter{Equations of motion of the generalized hybrid metric-Palatini gravity}
\label{appendixA}

In this appendix we shall apply the variational method to both the geometrical and scalar-tensor representations of the generalized hybrid metric-Palatini gravity and obtain the equations of motion step by step, namely the field equations and the modified Klein-Gordon equations.

\section{Geometrical representation}

Let us start by studying the geometrical representation of the generalized hybrid metric-Palatini gravity. Consider the following form of the action of the theory:
\begin{equation}\label{Aghmpgaction}
S=\frac{1}{2\kappa^2}\int_\Omega\sqrt{-g}f\left(R,\cal{R}\right)d^4x+S_m,
\end{equation}
where $\kappa^2\equiv 8\pi G$, $S_m$ is the matter action, $R$ is the metric Ricci scalar written in terms of the connection $\Gamma_{aB}^c$ which is the Levi-Civita connection of the metric $g_{ab}$, $\mathcal{R}\equiv\mathcal{R}^{ab}g_{ab}$ is the Palatini curvature which is defined in terms of an independent connection $\hat\Gamma^c_{ab}$ as
\begin{equation}\label{Adefpalatini}
\mathcal{R}_{ab}=\partial_c\hat\Gamma^c_{ab}-\partial_b\hat\Gamma^c_{ac}+\hat\Gamma^c_{cd}\hat\Gamma^d_{ab}-\hat\Gamma^c_{ad}\hat\Gamma^d_{cb}.
\end{equation}

There are two independent variables in this action, namely the metric $g_{ab}$ and the independent connection $\hat\Gamma_{ab}^c$, so we will have to do the variational method twice, once for each variable, and obtain two equations of motion. Let us first vary the action with respect to the independent connection. Since the only variable that deppends on the independent connection is $\mathcal{R}$, the variation becomes
\begin{equation}
\delta\mathcal{L}_{\hat\Gamma}=\frac{1}{2\kappa^2}\sqrt{-g}\frac{\partial f}{\partial\cal{R}}g^{ab}\delta\mathcal{R}_{ab}.
\end{equation}
Defining an auxiliary 2-tensor $A^{ab}$ with the form $A^{ab}\equiv\frac{1}{2\kappa^2}\sqrt{-g}\frac{\partial f}{\partial\cal{R}} g^{ab}$ and using the definition of the Palatini curvature $\mathcal R$ given in Eq.\eqref{Adefpalatini} we obtain
\begin{equation}
\delta\mathcal{L}_{\hat\Gamma}=A^{ab}\left(\partial_c\delta\hat\Gamma^c_{ab}-\partial_b\delta\hat\Gamma^c_{ac}+\delta\hat\Gamma^c_{cd}\hat\Gamma^d_{ab}+\hat\Gamma^c_{cd}\delta\hat\Gamma^d_{ab}-\delta\hat\Gamma^c_{ad}\hat\Gamma^d_{cb}-\hat\Gamma^c_{ad}\delta\hat\Gamma^d_{cb}\right).
\end{equation}
The first two terms in this expression can be integrated by parts. Notice that the primitive evaluated at the boundary of the integral vanishes because, by definition, the variation of the action at the boundary points is zero. Now, using the definitions of the covariant derivative of $A^{ab}$ bearing in mind that this is a tensor density of weight 1 due to the factor $\sqrt{-g}$,
\begin{equation}
\hat\nabla_cA^{ab}=\partial_cA^{ab}-A^{ab}\hat\Gamma^d_{dc}+A^{db}\hat\Gamma^a_{cd}+A^{ad}\hat\Gamma^b_{cd},
\end{equation}
some of the terms of the variation can be grouped together and others cancel out, leading to
\begin{eqnarray}
\delta\mathcal{L}_{\hat\Gamma}&=&\hat\nabla_bA^{ab}\delta\hat\Gamma^c_{ac}-\hat\nabla_cA^{ab}\delta\hat\Gamma^c_{ab}+A^{db}\hat\Gamma^a_{cd}\delta\hat\Gamma^c_{ab}+A^{ad}\hat\Gamma^b_{cd}\delta\hat\Gamma^c_{ab}-\nonumber\\
&-&A^{de}\hat\Gamma^a_{ed}\delta\hat\Gamma^c_{ac}-A^{ab}\hat\Gamma^d_{cb}\delta^c_{ad}-A^{ab}\hat\Gamma^c_{ad}\delta\hat\Gamma^d_{cb}+A^{ab}\hat\Gamma^d_{ab}\delta\hat\Gamma^c_{cd}.
\end{eqnarray}
Now, my index manipulation we can rewrite every term as a function of the variation $\delta\hat\Gamma^c_{ab}$, simplifying the result further to
\begin{equation}
\left(\hat\nabla_dA^{ad}\delta^b_c-\hat\nabla_cA^{ab}+A^{ed}\hat\Gamma^b_{ed}\delta^a_c-A^{de}\hat\Gamma^a_{ed}\delta^b_c\right)\delta\hat\Gamma^c_{ab}=0.
\end{equation}
Now, let us define the tensor $X^{ab}_{c}=A^{ed}\hat\Gamma^b_{ed}\delta^a_c-A^{de}\hat\Gamma^a_{ed}\delta^b_c$. This tensor is an anti-symmetric tensor in the indeces $a$ and $b$. Therefore, since $\hat\Gamma^c_{ab}$ is symmetric in the same pair of indeces, the last two terms in the previous equation correspond to the multiplication of a symmetric tensor with an anti-symmetric tensor, which vanishes identically. On the other hand, by the same reason, only the symmetric part of the remaining terms shall be considered, and we are left with
\begin{equation}
\hat\nabla_cA^{ab}=\frac{1}{2}\left(\hat\nabla_dA^{ad}\delta^b_c+\hat\nabla_dA^{bd}\delta^a_c\right).
\end{equation}
Finally, replacing back the definition of the tensor $A^{ab}$ and performing the contraction with the Kronecker delta $\delta_a^b$ yields finally the equation of motion for the independent connection $\hat\Gamma_{ab}^c$
\begin{equation}
\hat\nabla_c\left(\sqrt{-g}\frac{\partial f}{\partial \cal{R}}g^{ab}\right)=0,
\end{equation}
which implies, by the result of Sec.\ref{sec:Dlevicivita}, that there is a metric tensor $h_{ab}$ conformally related to the metric $g_{ab}$, given by $h_{ab}=\frac{\partial f}{\partial \cal{R}}g_{ab}$ for which the independent connection $\hat\Gamma_{ab}^c$ corresponds to the Levi-Civita connection, i.e.
\begin{equation}
\hat\Gamma^a_{bc}=\frac{1}{2}h^{ad}\left(\partial_b g_{dc}+\partial_c g_{bd}-\partial_d g_{bc}\right).
\end{equation}

Let us now turn to the variation with respect to the metric $g_{ab}$ from which we will obtain the modified field equations. Since both $R$ and $\cal{R}$ depend on the metric, the variation with respect to $g_{ab}$ takes the form
\begin{equation}
\delta\mathcal{L}_g=\frac{1}{2\kappa^2}\left[f\left(R,\cal{R}\right)\delta\sqrt{-g}+\sqrt{-g}\left(\frac{\partial f}{\partial R}\delta R+\frac{\partial f}{\partial \cal{R}}\delta\cal{R}\right)\right].
\end{equation}
The first term on the right hand side corresponds to a variation of $\sqrt{-g}$ which is computed in Eq.\eqref{Cvarsqrtg}. On the other hand, the remaining terms can be expanded in the form
\begin{equation}
\frac{1}{2\kappa^2}\sqrt{-g}\left(\frac{\partial f}{\partial R}\delta R+\frac{\partial f}{\partial \cal{R}}\delta\cal{R}\right)=\frac{1}{2\kappa^2}\sqrt{-g}\left[\frac{\partial f}{\partial R}\left(\delta g^{ab} R_{ab}+g^{ab}\delta R_{ab}\right)+\frac{\partial f}{\partial \cal{R}}\left(\delta g^{ab} \mathcal{R}_{ab}+g^{ab}\delta \mathcal{R}_{ab}\right)\right].
\end{equation}

Since $\mathcal{R}_{ab}$ only depends on the independent conection, the factor $\delta\mathcal{R}_{ab}=0$ and the last term vanishes. The variations of the metric inverse $g^{ab}$ can be written in terms of variations of the metric $g_{ab}$, see Eq.\eqref{Cvarinvgab}. The variation $\delta R_{ab}$, which in GR corresponds to a boundary term that vanishes due to the Stokes theorem, in this case does not vanish because of the scalar factor $\frac{\partial f}{\partial R}$. The variation $\delta R$ multiplied by a scalar factor is computed in Eq.\eqref{Cvarrabphi}, and upon using this result with $\phi=\frac{\partial f}{\partial R}$, we obtain
\begin{eqnarray}
\delta\mathcal{L}_g&=&\frac{1}{2\kappa^2}\left(\frac{1}{2}f\sqrt{-g}g^{ab}\delta g_{ab}-\sqrt{-g}\frac{\partial f}{\partial R}g^{ac}g^{bd}R_{ab}\delta g_{cd}+\right.\nonumber\\
&&\left.+\sqrt{-g}\left(\nabla^a\nabla^b-g^{ab}\Box\right)\frac{\partial f}{\partial R}\delta g_{ab}-\sqrt{-g}\frac{\partial f}{\partial \mathcal{R}}g^{ac}g^{bd}\mathcal{R}_{ab}\delta g_{cd}\right).
\end{eqnarray}
Taking the factor $\sqrt{-g}$ outside of the parenthesis and performing the contractions with the metric tensors yields
\begin{equation}
\delta \mathcal{L}_g=\frac{1}{2\kappa^2}\sqrt{-g}\left[\frac{1}{2}fg^{ab}-\frac{\partial f}{\partial R}R^{ab}-\frac{\partial f}{\partial \mathcal{R}}\mathcal{R}^{ab}+\left(\nabla^a\nabla^b-g^{ab}\Box\right)\frac{\partial f}{\partial R}\right]\delta g_{ab}.
\end{equation}
From the previous equation and the usual definition of the stress-energy tensor $T_{ab}$ provided in Eq.\eqref{deftab} we finally obtain the modified field equations as
\begin{equation}
\frac{\partial f}{\partial R}R_{ab}+\frac{\partial f}{\partial \mathcal{R}}\mathcal{R}_{ab}-\frac{1}{2}g_{ab}f\left(R,\cal{R}\right)-\left(\nabla_a\nabla_b-g_{ab}\Box\right)\frac{\partial f}{\partial R}=\kappa^2 T_{ab}.
\end{equation}

\section{Scalar-tensor representation}

Let us now turn to the scalar-tensor representation of the theory and obtain the equations of motion for the metric $g_{ab}$ and the scalar fields $\varphi$ and $\psi$ from the action
\begin{equation}\label{Aghmpgstaction}
S=\frac{1}{2\kappa^2}\int \sqrt{-g}\left[\left(\varphi-\psi\right) R-\frac{3}{2\psi}\partial^a\psi\partial_a\psi-V\left(\varphi,\psi\right)\right]d^4x.
\end{equation} 
In this case the action is a function of these three variables and the variational method must be applied three times. Consider first the variation with respect to the metric $g_{ab}$ which is
\begin{equation}
\delta\mathcal{L}_g=\frac{1}{2\kappa^2}\left\{\delta\sqrt{-g}\left[\left(\varphi-\psi\right) R-\frac{3}{2\psi}\partial^c\psi\partial_c\psi-V\right]+\sqrt{-g}\left[\left(\varphi-\psi\right)\delta R-\frac{3}{2\psi}\partial_a\psi\partial_b\psi\delta g^{ab}\right]\right\}.
\end{equation}
Again, all the terms correspond to variations of the metric determinant in the form $\sqrt{-g}$, the metric inverse $g^{ab}$, and the Ricci scalar $R$ multiplied by a scalar factor which in this case takes the form $\left(\varphi-\psi\right)$. All these variations have been computed explicitly in Appendix \ref{appendixC}. Writing $\delta R=R_{ab}\delta g^{ab}+g^{ab}\delta R_{ab}$ and using the variations computed in Appendix \ref{appendixC} we obtain directly.
\begin{eqnarray}
\delta\mathcal{L}_g&=&\frac{1}{2\kappa^2}\sqrt{-g}\left\{\frac{1}{2}g^{ab}\left[\left(\varphi-\psi\right) R-\frac{3}{2\psi}\partial^c\psi\partial_c\psi-V\right]+\frac{3}{2\psi}\partial^a\psi\partial^b\psi+\right.\nonumber\\
&&\left.+\nabla^a\nabla^b\left(\varphi-\psi\right)-g^{ab}\Box\left(\varphi-\psi\right)-\left(\varphi-\psi\right) R^{ab}\right\}\delta g_{ab},
\end{eqnarray}
which, along with the definition of the stress-energy tensor $t_{ab}$ procided in Eq.\eqref{deftab}  yields the equation of motion
\begin{equation}
-\frac{1}{2}\left[\left(\varphi-\psi\right) R-\frac{3}{2\psi}\partial^c\psi\partial_c\psi-V\right]g_{ab}-\frac{3}{2\psi}\partial_a\psi\partial_b\psi-\left(\nabla_a\nabla_b-g_{ab}\Box-R_{ab}\right)\left(\varphi-\psi\right)=\kappa^2T_{ab}.
\end{equation}
The previous equation can be re-written in a more convenient form by making use of the definition of the Einstein's tensor $G_{ab}=R_{aB}-\frac{1}{2}Rg_{ab}$. Rearranging the resultant terms we obtain the modified field equations as
\begin{equation}\label{Aghmpgstfield}
\left(\varphi-\psi\right) G_{ab}=\kappa^2T_{ab}-\left(\Box\left(\varphi-\psi\right)+\frac{1}{2}V+\frac{3}{4\psi}\partial^c\psi\partial_c\psi\right)g_{ab}+\frac{3}{2\psi}\partial_a\psi\partial_b\psi+\nabla_a\nabla_b\left(\varphi-\psi\right).
\end{equation}

Consider now the variation of Eq.\eqref{Aghmpgstaction} with respect to the scalar fields, starting with the scalar field $\psi$. The variation equation becomes
\begin{equation}
\delta\mathcal{L}_\psi=\frac{1}{2\kappa^2}\sqrt{-g}\left(\frac{3}{2\psi^2}\partial^c\psi\partial_c\psi\delta\psi-\frac{3}{\psi}\partial^c\psi\partial_c\delta\psi-V_\psi\delta\psi-R\delta\psi\right).
\end{equation}
All terms except for the second in the right hand side of the previous equation are already in the needed form to obtain the equation of motion. Integrating the second term by parts we obtain
\begin{equation}
\int_\Omega\sqrt{-g}\frac{3}{\psi}\partial^a\psi\partial_a\delta\psi d^4x=\int_\Omega\left[\partial_a\left(\sqrt{-g}\frac{3}{\psi}\delta\psi\partial^a\psi\right)-\partial_a\left(\sqrt{-g}\frac{3}{\psi}\partial^a\psi\right)\delta\psi\right]d^4x.
\end{equation}
The first term in the integral corresponds to a boundary term and thus vanishes by definition. The second term can be expanded by the application of the derivative of a product as
\begin{equation}
\partial_a\left(\sqrt{-g}\frac{3}{\psi}\partial^a\psi\right)=-\frac{3}{\psi^2}\sqrt{-g}\partial_a\psi\partial^a\psi+\frac{3}{\psi}\partial_a\left(\sqrt{-g}\partial^a\psi\right).
\end{equation}
Multiplying and dividing the second term in the previous equation by $\sqrt{-g}$ allows us to use the result of Sec.\ref{Sec:dalembert} to write the term as a D'Alembert operator. Inserting this result into the variation equation yields
\begin{equation}
\delta\mathcal{L}_\psi=\frac{1}{2\kappa^2}\sqrt{-g}\left(-R-\frac{3}{2\psi^2}\partial_a\psi\partial^a\psi+\frac{3}{\psi}\Box\psi-V_\psi\right)\delta\psi,
\end{equation}
which leads to the modified Klein-Gordon equation for the scalar field $\psi$ in the form
\begin{equation}\label{Akgpsiaux}
-R-\frac{3}{2\psi^2}\partial_a\psi\partial^a\psi+\frac{3}{\psi}\Box\psi-V_\psi=0.
\end{equation}
This is not the most useful form of the equation of motion for $\psi$ since it depends on the Ricci scalar $R$ later in this section, we shall rewrite this equation in a more appropriate form. The only equation left to obtain is the modified Klein-Gordon equation for the field $\varphi$. Varying the action with respect to $\varphi$ gives directly the result
\begin{equation}
\delta\mathcal{L}_\varphi=\frac{1}{2\kappa^2}\sqrt{-g}\left(R-V_\varphi\right),
\end{equation}
from which we immediatly conclude that
\begin{equation}\label{Akgphiaux}
R-V_\varphi=0.
\end{equation}

Now, we are intrested in taking the Ricci scalar $R$ out of Eqs.\eqref{Akgpsiaux} and \eqref{Akgphiaux} to obtain more suitable forms of these equations. To do so, we trace the field equations given in Eq.\eqref{Aghmpgstfield} to obtain
\begin{equation}
\left(\varphi-\psi\right)R=-\kappa^2T+3\Box\varphi-3\Box\psi+2V+\frac{3}{2\psi}\partial^a\psi\partial_a\psi.
\end{equation}
This equation can be solved with respect to $R$ and the result can be inserted into Eqs.\eqref{Akgpsiaux} and \eqref{Akgphiaux} which yield, after rearrangements
\begin{equation}
\Box\psi-\frac{\psi}{\varphi}\Box\varphi-\frac{1}{2\psi}\partial^a\psi\partial_a\psi-\frac{\psi}{3\varphi}\left[2V+\left(\varphi-\psi\right)V_\psi\right]=-\frac{\kappa^2\psi}{3\varphi}T,
\end{equation}
\begin{equation}
\Box\varphi-\Box\psi+\frac{1}{2\psi}\partial^a\psi\partial_a\psi+\frac{1}{3}\left[2V-\left(\varphi-\psi\right)V_\varphi\right]=\frac{\kappa^2}{3}T,
\end{equation}
respectively. These equations provide the dynamics of combinations of $\varphi$ and $\psi$ coupled to the matter fields via the trace of the stress-energy tensor $T$. These two equation can be manipulated in order to obtain equations that depend solely on $\Box\psi$ and $\Box\varphi$ of the form
\begin{equation}
\Box\psi-\frac{1}{2\psi}\partial^a\psi\partial_a\psi-\frac{\psi}{3}\left(V_\varphi+V_\psi\right)=0,
\end{equation}
\begin{equation}
\Box\varphi+\frac{1}{3}\left(2V-\psi V_\psi-\varphi V_\varphi\right)=\frac{\kappa^2T}{3}.
\end{equation}
\cleardoublepage

\chapter{Useful results in General Relativity}

In this appendiz we provide a few useful results in GR that are useful throughout this thesis, namely important quantities such as the covariant derivative $\nabla_c$, the D'Alembert operator $\Box$, and the Einstein tensor $g_{ab}$ for both the $FLRW$ spacetime and a general spherically spacetime that can be used to described compact objects such as stars, black-holes and wormholes. We also provide a brief analysis about the definition of the stress-energy tensor $T_{ab}$ and its conservation equation in the FLRW spacetime. To finalize, we show that the two Friedmann equations and the conservation equation for the stress-energy tensor $T_{ab}$ are not independent and, in fact, one can obtain the second Friedmann equation by differenciating the first Friedmann equation with respect to time and using the conservation equation for $T_{ab}$ to cancel the terms depending on $\dot\rho$.

\section{Results in the FLRW spacetime}\label{sec:FLRW}

The Friedmann-Lemaître-Robertson-Walker (FLRW) spacetime is an homogeneous and isotropic solution of the Einstein's field equations and it is the most useful metric to study the cosmological dynamics of a given gravitational model. The FLRW spacetime is described by the metric tensor, in spherical coordinates, given by:
\begin{equation}\label{EmetricFLRW}
ds^2=-dt^2+a^2\left(t\right)\left[\frac{dr^2}{1-kr^2}+r^2d\theta^2+r^2\sin^2\theta d\phi^2\right],
\end{equation}
where $a\left(t\right)$ is the scale factor of the universe, also known as the radius of the universe, and $k$ is the curvature. For $k=0$ the universe has a flat geometry, for $k=1$ an open (hyperbolic) geometry, and for $k=-1$ a closed (spheric) geometry. With this metric tensor, the Einstein's tensor $G_{ab}$ is diagonal and it is given by
\begin{equation}
G_{ab}=\left[3\left(\frac{k+\dot a^2}{a^2}\right),-\frac{k+\dot a^2+2a\ddot a}{1-kr^2},-r^2\left(k+\dot a^2+2a\ddot a\right),-r^2\sin^2\theta\left(k+\dot a^2+2a\ddot a\right)\right].
\end{equation}
The Ricci scalar is given by
\begin{equation}
R=-6\left(\frac{\ddot a}{a}+\frac{\dot a^2}{a^2}+\frac{k}{a^2}\right).
\end{equation}
Some useful results that are needed throughout this work on the FLRW metric are related to the differential operators of the scalar fields. When we consider that the scalar fields $\psi$ are only functions of time and not functions of the spacial coordinates, the d'Alembert operator and the kinetic terms take the forms
\begin{equation}
\Box\psi=-\left(\frac{3\dot a}{a}\dot\psi+\ddot\psi\right),
\end{equation}
\begin{equation}
\partial^c\psi\partial_c\psi=-\dot\psi^2.
\end{equation}
Also, the double covariant derivative is also a diagonal matrix given by
\begin{equation}
\nabla_a\nabla_b\psi=\left(\ddot\psi,-\frac{a\dot a\dot\psi}{1-kr^2},-r^2a\dot a\dot\psi,-r^2\sin^2\theta a\dot a\dot\psi\right).
\end{equation}
It is clear from this results that the field equations that arise from the spacial coordinates $r$, $\phi$ and $\psi$ are linearly dependent, and so there are only two independent field equations which are $G_{tt}$ and $G_{rr}$.

\section{Results in a spherically symmetric spacetime}

Wormholes are exotic solutions of the Einstein's field equations which connect two different assymptotically flat regios of spacetime, and Black-Holes described by the Schwarzschild solution are solutions to the Einstein's field equations that describe extremely compact objects featuring horizons and a singularity. These two objects, along with compact stars in geeral and spherical shells can be described generally by a spherically symmetric metric tensor of the form
\begin{equation}
ds^2=-e^{\zeta\left(r\right)}dt^2+\left(1-\frac{b\left(r\right)}{r}\right)^{-1}dr^2+r^2d\theta^2+r^2\sin^2\theta d\phi^2,
\end{equation}
where $\zeta\left(r\right)$ is the redshift function and $b\left(r\right)$ is the shape function, which respects a boundary condition of $b\left(r_0\right)=r_0$, in the case of a wormhole, where $r_0$ is the radius of the throat. With this metric tensor, the Einstein's tensor takes the diagonal form
\begin{equation}
G_{tt}=e^{\zeta}\frac{b'}{r^2}, 
\end{equation}
\begin{equation}
G_{rr}=\frac{1}{r^2}\left[1-\left(1-\frac{b}{r}\right)^{-1}\right]+\frac{\zeta'}{r},
\end{equation}
\begin{equation}
G_{\theta\theta}=r^2\left[\left(1-\frac{b}{r}\right)\left(\frac{\zeta''}{2}+\frac{\zeta'^2}{4}+\frac{\zeta'}{2r}\right)+\left(\frac{b-rb'}{2r^3}\right)\left(1+\frac{r\zeta'}{2}\right)\right],
\end{equation}
\begin{equation}
G_{\phi\phi}=r^2\sin^2\theta\left[\left(1-\frac{b}{r}\right)\left(\frac{\zeta''}{2}+\frac{\zeta'^2}{4}+\frac{\zeta'}{2r}\right)+\left(\frac{b-rb'}{2r^3}\right)\left(1+\frac{r\zeta'}{2}\right)\right].
\end{equation}
We shall also compute the differential operators for the scalar fields. To do so, we assume in this case that the scalar fields are only functions of the radial coordinate and independent of time, to preserve the staticity of the spacetime itself. With these assumptions, we obtain
\begin{equation}
\Box\psi=\left(1-\frac{b}{r}\right)\left(\psi''+\frac{\zeta'\psi'}{2}+\frac{3\psi'}{2r}\right)+\frac{\psi'}{2r}\left(1-b'\right),
\end{equation}
\begin{equation}
\partial^c\psi\partial_c\psi=\left(1-\frac{b}{r}\right)\psi'^2.
\end{equation}
Also, the double covariant derivative is also a diagonal matrix and it is given by
\begin{equation}
\nabla_a\nabla_b\psi=\left[-e^\zeta\left(1-\frac{b}{r}\right)\frac{\zeta'\psi'}{2},\psi''+\frac{\psi'}{2r^2}\left(b-rb'\right)\left(1-\frac{b}{r}\right)^{-1},r^2\frac{\psi'}{r}\left(1-\frac{b}{r}\right),r^2\sin^2\theta\frac{\psi'}{r}\left(1-\frac{b}{r}\right)\right].
\end{equation}
As we can see from the previous results, there are only three independent field equations since the equations $G_{\theta\theta}$ and $G_{\phi\phi}$ are linearly dependent. The same happens for the components of the double covariant derivative.

\section{Stress-energy tensor}

The stress-energy tensor $T_{ab}$ is a tensor that describes the density and flux of energy and momentum in a given spacetime. There are many definitions of the stress-energy tensor, but we are going to use the Hilbert's definition, given by
\begin{equation}
T_{ab}=-\sqrt{2}{\sqrt{-g}}\frac{\delta\left(\sqrt{-g}\mathcal{L}_m\right)}{\delta g^{ab}}=-2\frac{\delta\mathcal{L}_m}{\delta g^{ab}}+g_{ab}\mathcal{L}_m,
\end{equation}
where $\mathcal{L}_M$ is the non-gravitational matter Lagrangian. This way, as we apply the variational method to the matter action $S_m$, it yields
\begin{equation}
\delta S_m=\int_\Omega\frac{\delta\left(\sqrt{-g}\mathcal{L}_m\right)}{\delta g^{ab}}\delta g_{ab}=-\int_\Omega\frac{1}{2}\sqrt{-g}T_{ab}\delta g_{ab}.
\end{equation}
In general relativity, it is usual to describe the matter as a perfect fluid. The stress-energy tensor for a perfect fluid in equilibrium is of the form
\begin{equation}
T^{ab}=\left(\rho+p\right)u^au^b+pg^{ab},
\end{equation}
where $\rho$ is the energy density, $p$ is the pressure, and $u^a$ are the 4-velocity vectors. It is often useful to write this tensor in the mixed index form because it becomes a diagonal matrix given by $T^a_b=\left(-\rho,p,p,p\right)$, from which we obtain the trace of the tensor as $T=3p-\rho$. In case we are intrested in dealing with dust matter instead of a perfect fluid, we just have to take the limit where $p$ vanishes. On the other hand, if we are interested in working with anisotropic fluids, the radial pressure $p_r$ and the trasverse pressure $p_t$ become different and we use $T^a_b=\left(-\rho,p_r,p_t,p_t\right)$ and $T=p_r+2p_t-\rho$ instead.

The stress-energy tensor can be used to obtain the equations of continuity and the Navier-Stokes equation. In general relativity, one can obtain the conservation equations simply by computing the divergence of $T^{ab}$, which vanishes identically. Therefore, using the FLRW metric given in Eq.\eqref{EmetricFLRW}, we obtain: 
\begin{equation}
\nabla_aT^{ab}=0 \Leftrightarrow \frac{d}{dt}\rho+\frac{3\dot a}{a}\left(\rho+p\right)=0 \Leftrightarrow \dot\rho=-3H\rho\left(1+\frac{p}{\rho}\right).
\end{equation}
The number of free parameters in this equation can be reduced by imposing an equation of state of the form $p=w\rho$. In this case, the previous equation simplifies to
\begin{equation}\label{Econstab}
\dot\rho=-3H\rho\left(1+w\right).
\end{equation}
Integrating the previous equation with respect to time using the definition of the Hubble parameter $H=\dot a/a$ it yields directly the relation
\begin{equation}
\rho=\rho_0a^{-3\left(1+w\right)} \implies p=p_0a^{-3\left(1+w\right)},
\end{equation}
where $p_0$ and $\rho_0$ are constants of integration related by $p_0=w\rho_0$. These steps are needed if we want to study cosmological dynamics of a universe populated by matter.

\section{Equivalence of Friedmann equations}
\label{Etabequiv}

In general relativity, as explained in Sec.\ref{sec:FLRW}, there are two independent field equations when one considers the FLRW spacetime, and these are called the Friedmann equations, the second of which is also known as the Raychaudhuri equation. The two Friedmann equations in GR are given by
\begin{equation}
3\frac{\dot a^2+k}{a^2}-\Lambda=8\pi\rho,
\end{equation}
\begin{equation}
\frac{2a\ddot a+\dot a^2+k}{a^2}-\Lambda=-8\pi p,
\end{equation}
respectively, where $\rho$ and $p$ are related by the conservation of the stress-energy tensor, given by Eq.\eqref{Econstab}, with $H=\dot a/a$. Differentiating the first Friedmann equation with respect to time yields directly
\begin{equation}
2\left(\frac{\dot a\ddot a}{a^2}-\frac{\dot a^3}{a^3}-\frac{k\dot a}{a^3}\right)=8\pi\dot\rho.
\end{equation}
Using the conservation of the stress-energy tensor in Eq.\eqref{Econstab} to cancel the term $\dot\rho$ and dividing through by $H=\frac{\dot a}{a}$ we obtain
\begin{equation}
2\left(\frac{\ddot a}{a}-\frac{\dot a^2}{a^2}-\frac{k}{a^2}\right)=-8\pi\left(\rho+p\right).
\end{equation}
Using again the first Friedmann equation to cancel the term $8\pi\rho$ we obtain
\begin{equation}
\frac{2a\ddot a+\dot a^2+k}{a^2}-\Lambda=-8\pi p,
\end{equation}
which corresponds to the second Friedmann equation. This implies that from the three equations, namely the Friedmann equation, the Raychaudhuri equation and the conservation equation for $T_{ab}$, one only needs two of these equations to fully describe the dynamics of the universe. For simplicity, one usually considers the first Friedmann equation and the conservation equation for $T_{ab}$, thus dropping the redundant Raychaudhuri equation. 

\cleardoublepage

\chapter{Particle physics potential}

The scalar fields that appear in the scalar-tensor representation of the generalized hybrid metric-Palatini gravity do not need to have any physical interpretation in principle, for they are only auxiliary fields one introduces to simplify the representation of the theory and not physical fields i.e. particles. There are also no particular forms of the potential that must be used to mantain the physical realism of the scalar fields. However, there is a particular form of the potential that considerably simplifies the equations of motion in the scalar-tensor representation, see chapters \ref{chapter:chapter3} and \ref{chapter:chapter6}, which is $V=V_0\left(\varphi-\psi\right)^2$. This form of the potential not only has the particularity of simplifying the modified Klein-Gordon equations given by Eqs.\eqref{ghmpgstkgphi} and \eqref{ghmpgstkgpsi}, but can also be rewritten in a physically relevant form for particle physics.

Let us analyse where this potential comes from and how to obtain a physical meaning from it. Consider the scalar field equations given by Eqs.\eqref{ghmpgstkg1} and \eqref{ghmpgstkg2}. The simplest constraint one can impose into these equations is that the potential terms vanish, i.e.
\begin{equation}
2V+\left(\varphi-\psi\right)V_\psi=0;
\end{equation}
\begin{equation}
2V-\left(\varphi-\psi\right)V_\varphi=0.
\end{equation}
These two equations form a system of partial differential equations in $V\left(\varphi,\psi\right)$ that can be solved to obtain the simplified expression for the potential. The two equations are separable in the variables $V$, $\psi$ and $\varphi$ and can be written as
\begin{equation}
\frac{1}{2V}dV=-\frac{1}{\varphi-\psi}d\psi,
\end{equation}
\begin{equation}
\frac{1}{2V}dV=\frac{1}{\varphi-\psi}d\varphi.
\end{equation}
Assuming the independence of the scalar field, one can integrate these equations directly and obtain the results
\begin{equation}
\frac{1}{2}\log V=\log\left(\varphi-\psi\right)+C_1\left(\varphi\right)\Leftrightarrow V=C_1'\left(\varphi\right)\left(\varphi-\psi\right)^2,
\end{equation}
\begin{equation}
\frac{1}{2}\log V=\log\left(\varphi-\psi\right)+C_2\left(\psi\right)\Leftrightarrow V=C_2'\left(\psi\right)\left(\varphi-\psi\right)^2,
\end{equation}
where the functions $C_i$ are functions of just one of the fields and the functions $C_i'=\exp\left(C_i\right)$ are their exponentials. Comparing the two results it is immediate to see that these functions must be constant, because the remaining factor of the result is the same in both solutions. Denoting this constant by $V_0$, we can write the potential in the general form 
\begin{equation}
V\left(\varphi,\psi\right)=V_0\left(\varphi-\psi\right)^2=V_0\left(\varphi^2+\psi^2-2\varphi\psi\right).
\end{equation}
As we can see, this potential features mass terms in $\varphi$ and $\psi$ given by the quadratic terms in these fields, but there is one extra term which features a product of the form $\varphi\psi$, which indicate that these fields are not the physical i.e. they are not eigenstates of mass. To obtain the physical fields, we shall compute the eigenvalues and eigenstates of mass:
\begin{equation}
V\left(\varphi,\psi\right)=
\begin{bmatrix}
\varphi & \psi
\end{bmatrix}
V_0
\begin{bmatrix}
1 & -1 \\[0.5em]
-1 & 1
\end{bmatrix}
\begin{bmatrix}
\varphi \\[0.5em]
\psi
\end{bmatrix}
\equiv \Psi^T \textbf{V} \Psi.
\end{equation}
The eigenvalues $\lambda_i$ of the matrix $\textbf{V}$ are $\lambda_1=2V_0$ and $\lambda_2=0$. This means that one of the physical fields is going to be massless. Now, to compute the eingenstates $\phi_i$, we apply directly the definition $\textbf{V}\phi_i=\lambda_i\phi_i$ which yields
\begin{equation}
\phi_1=A\left(\varphi-\psi\right), \ \ \ \ \ \phi_2=A\left(\varphi+\psi\right),
\end{equation}
where A is the constant of renormalization. To compute this constant of renormalization, we compare the two potentials obtained in this combination of fields and the previous one:
\begin{equation}
V=2V_0\phi_1^2=2V_0A^2\left(\varphi-\psi\right)^2=V_0\left(\varphi-\psi\right)^2\implies A=\frac{1}{\sqrt{2}}.
\end{equation}
Also, we can compute the mass of the fields by writing the potential with the mass terms as
\begin{equation}
V=\frac{1}{2}m_{\phi_1}^2\phi_1^2=2V_0\phi_1^2,
\end{equation}
which implies that $m_{\phi_1}=2\sqrt{V_0}$ and $m_{\phi_2}=0$, as expected from the eigen values $\lambda_i$. The two physical fields are thus given by
\begin{equation}\label{Bfield1}
\phi_1=\frac{1}{\sqrt{2}}\left(\varphi-\psi\right), \ \ \ m_{\phi_1}=2\sqrt{V_0};
\end{equation}
\begin{equation}\label{Bfield2}
\phi_2=\frac{1}{\sqrt{2}}\left(\varphi+\psi\right), \ \ \ m_{\phi_2}=0.
\end{equation}
We can also invert these definitions to obtain the non-physical fields $\varphi$ and $\psi$ as functions of the physical fields $\phi_1$ and $\phi_2$. To do so, we sum and subtract Eqs.\eqref{Bfield1} and \eqref{Bfield2} and solve with respect to $\varphi$ and $\psi$. The result is as follows:
\begin{equation}
\varphi=\frac{1}{\sqrt{2}}\left(\phi_1+\phi_2\right),\ \ \ \ \ \psi=\frac{1}{\sqrt{2}}\left(\phi_2-\phi_1\right).
\end{equation}
This computation was not meant to find new fields on which to base the calculations upon. The use of this demonstration is to show that the potential that arises from the constraints imposed a priori does have a physically significant meaning, and so it is justified to continue working with the result with possible applications to particle physics. 
\cleardoublepage

\cleardoublepage


\phantomsection
\addcontentsline{toc}{chapter}{\bibname}

\cleardoublepage

\end{document}